\documentclass[12pt,a4paper] {article}
\usepackage{epsfig}
\usepackage{rotating}
\usepackage{cite}
\begin{document}
\newcommand{\omclb}      {\mbox{${\mathrm {1-CL_b}}$}}
\newcommand{\clb}        {\mbox{${\mathrm {CL_b}}$}}
\newcommand{\cls}        {\mbox{${\mathrm {CL_s}}$}}
\newcommand{\clsb}       {\mbox{${\mathrm {CL_{s+b}}}$}}
\newcommand{\Z}{\mbox{$\mathrm Z$}}
\newcommand{\Zo}{\mbox{$\mathrm Z$}}
\newcommand{\W}{\mbox{$\mathrm W$}}
\newcommand{\Wp}{\mbox{$\mathrm{W}^+$}}
\newcommand{\Wm}{\mbox{$\mathrm{W}^-$}}
\newcommand{\Wpm}{\mbox{$\mathrm{W}^\pm$}}
\newcommand{\Zg}{\mbox{$\mathrm{Z}\gamma$}}
\newcommand{\ZZ}{\mbox{$\mathrm{Z}\mathrm{Z}$}}
\newcommand{\WW}{\mbox{$\mathrm{W}\mathrm{W}$}}
\newcommand{\Zs}{\mbox{$\mathrm{Z}^{*}$}}
\newcommand{\h}{\mbox{$\mathrm{h}$}}
\newcommand{\Ho}{\mbox{$\mathrm{H}$}}
\newcommand{\calH}{\mbox{${\cal H}$}}
\newcommand{\calHa}{\mbox{${\cal H}_1$}}
\newcommand{\calHb}{\mbox{${\cal H}_2$}}
\newcommand{\calHc}{\mbox{${\cal H}_3$}}
\newcommand{\caG}{\mbox{${\cal H}_3$}}
\newcommand{\Ha}{\mbox{$\mathrm{H}_1$}}
\newcommand{\Hb}{\mbox{$\mathrm{H}_2$}}
\newcommand{\Hc}{\mbox{$\mathrm{H}_3$}}
\newcommand{\Hi}{\mbox{$\mathrm{H}_i$}}
\newcommand{\Hj}{\mbox{$\mathrm{H}_j$}}
\newcommand{\ho}{\mbox{$\mathrm{h}$}}
\newcommand{\Hp}{\mbox{$\mathrm{H}^{+}$}}
\newcommand{\Hm}{\mbox{$\mathrm{H}^{-}$}}
\newcommand{\Hsm}{\mbox{$\mathrm{H}^{0}_{SM}$}}
\newcommand{\A}{\mbox{$\mathrm{A}$}}
\newcommand{\Hpm}{\mbox{$\mathrm{H}^{\pm}$}}
\newcommand{\X}{\mbox{${\tilde{\chi}^0}$}}
\newcommand{\ko}{\mbox{${\tilde{\chi}^0}$}}
\newcommand{\ee}{\mbox{$\mathrm{e}^{+}\mathrm{e}^{-}$}}
\newcommand{\bee}{\mbox{$\boldmath {\mathrm{e}^{+}\mathrm{e}^{-}} $}}
\newcommand{\mm}{\mbox{$\mu^{+}\mu^{-}$}}
\newcommand{\nn}{\mbox{$\nu \bar{\nu}$}}
\newcommand{\qq}{\mbox{$\mathrm{q} \bar{\mathrm{q}}$}}
\newcommand{\pb}{\mbox{$\mathrm{pb}^{-1}$}}
\newcommand{\ra}{\mbox{$\rightarrow$}}
\newcommand{\br}{\mbox{$\boldmath {\rightarrow}$}}
\newcommand{\tautau}{\mbox{$\tau^+\tau^-$}}
\newcommand{\ga}{\mbox{$\gamma$}}
\newcommand{\gamgam}{\mbox{$\gamma\gamma$}}
\newcommand{\tp}{\mbox{$\tau^+$}}
\newcommand{\tm}{\mbox{$\tau^-$}}
\newcommand{\tpm}{\mbox{$\tau^{\pm}$}}
\newcommand{\uu}{\mbox{$\mathrm{u} \bar{\mathrm{u}}$}}
\newcommand{\dd}{\mbox{$\mathrm{d} \bar{\mathrm{d}}$}}
\newcommand{\bb}{\mbox{$\mathrm{b} \bar{\mathrm{b}}$}}
\newcommand{\cc}{\mbox{$\mathrm{c} \bar{\mathrm{c}}$}}
\newcommand{\mumu}{\mbox{$\mu^+\mu^-$}}
\newcommand{\csbar}{\mbox{$\mathrm{c} \bar{\mathrm{s}}$}}
\newcommand{\cbars}{\mbox{$\bar{\mathrm{c}}\mathrm{s}$}}
\newcommand{\nunu}{\mbox{$\nu \bar{\nu}$}}
\newcommand{\nubar}{\mbox{$\bar{\nu}$}}
\newcommand{\mQ}{\mbox{$m_{\mathrm{Q}}$}}
\newcommand{\mZ}{\mbox{$M_{\mathrm{Z}}$}}
\newcommand{\mH}{\mbox{$m_{\mathrm{H}}$}}
\newcommand{\mcalH}{\mbox{$m_{\cal H}$}}
\newcommand{\mcalHa}{\mbox{$m_{{\cal H}_1}$}}
\newcommand{\mcalHb}{\mbox{$m_{{\cal H}_2}$}}
\newcommand{\mcalHc}{\mbox{$m_{{\cal H}_3}$}}
\newcommand{\mHrec}{\mbox{$m_{\mathrm{H}}^{\mathrm{rec}}$}}
\newcommand{\mHp}{\mbox{$m_{\mathrm{H}^+}$}}
\newcommand{\mh}{\mbox{$m_{\mathrm{h}}$}}
\newcommand{\mA}{\mbox{$m_{\mathrm{A}}$}}
\newcommand{\mHpm}{\mbox{$m_{\mathrm{H}^{\pm}}$}}
\newcommand{\mHa}{\mbox{$m_{\mathrm{H}_1}$}}
\newcommand{\mHb}{\mbox{$m_{\mathrm{H}_2}$}}
\newcommand{\mHsm}{\mbox{$m_{\mathrm{H}^0_{SM}}$}}
\newcommand{\mW}{\mbox{$m_{\mathrm{W}^{\pm}}$}}
\newcommand{\mt}{\mbox{$m_{\mathrm{t}}$}}
\newcommand{\mb}{\mbox{$m_{\mathrm{b}}$}}
\newcommand{\lpm}{\mbox{$\ell ^+ \ell^-$}}
\newcommand{\G}{\mbox{$\mathrm{GeV}$}}
\newcommand{\Gc}{\mbox{${\rm GeV}/c$}}
\newcommand{\Gcs}{\mbox{${\rm GeV}/c^2$}}
\newcommand{\gevcs}{\mbox{${\rm GeV}/c^2$}}
\newcommand{\tevcs}{\mbox{${\rm TeV}/c^2$}}
\newcommand{\Mcs}{\mbox{${\rm MeV}/c^2$}}
\newcommand{\sba}{\mbox{$\sin ^2 (\beta -\alpha)$}}
\newcommand{\cba}{\mbox{$\cos ^2 (\beta -\alpha)$}}
\newcommand{\tanb}{\mbox{$\tan \beta$}}
\newcommand{\sqrts}{\mbox{$\sqrt {s}$}}
\newcommand{\sqrtsp}{\mbox{$\sqrt {s'}$}}
\newcommand{\msusy}{\mbox{$M_{\rm SUSY}$}}
\newcommand{\mg}{\mbox{$m_{\tilde{\rm g}}$}}
\newcommand{\hp} {\mbox{$ {\mathrm H}^+ \,$}}
\newcommand{\hm} {\mbox{$ {\mathrm H}^- $}}
\newcommand{\hpm}{\mbox{${\mathrm H}^{\pm}$}}
\newcommand{\MZ} {\mbox{$ m_{\mathrm Z} \, $}}
\newcommand{\MW} {\mbox{$ m_{\mathrm W} \, $}}
\newcommand{\MA} {\mbox{$ m_{\mathrm A} $}}
\newcommand{\MH} {\mbox{$ m_{\mathrm H} \, $}}
\newcommand{\mhp }{\mbox{$ m_{{\mathrm H}^+} \, $}}
\newcommand{\ffbar}{\mbox{${\mathrm f}\bar{\mathrm f}$}}
\newcommand{\qqbar}{\mbox{${\mathrm q}\bar{\mathrm q}$}}
\newcommand{\bbbar}{\mbox{${\mathrm b}\bar{\mathrm b}$}}
\newcommand{\ccbar}{\mbox{${\mathrm c}\bar{\mathrm c}$}}
\newcommand{\nunubar}{$\nu \bar{\nu}\;$}
\newcommand{\ton}{$\tau \nu_{\tau} \;$ }
\newcommand{\toto}{\mbox{$\tau^+ \tau^-$}}
\newcommand{\ttqq}{$\tau^+\tau^- {\mathrm q \bar{q}}$ }
\newcommand{\nnqq}{$\nu\bar{\nu}{\mathrm q \bar{q}}$ }
\newcommand{\tautauqq}{\mbox{$\tau^+ \tau^- $\qqbar }}
\newcommand{\tautaubb}{\mbox{$\tau^+ \tau^- $\bbbar }}
\newcommand{\hAtt}{${\mathrm{hA}}\rightarrow$\toto \qqbar}
\newcommand{\hAbb}{hA$\rightarrow $\bbbar \bbbar}
\newcommand{\Abb}{\calH$_1$$\rightarrow $ \bbbar}
\newcommand{\Acc}{\calH$_1$$\rightarrow $ \ccbar}
\newcommand{\Aany}{\calH$_1$$\rightarrow$ any}
\newcommand{\Agg}{\calH$_1$$\rightarrow \gamma \gamma $}
\newcommand{\App}{\calH$_1$$\rightarrow $ 2 prongs}
\newcommand{\Ahad}{\calH$_1$$\rightarrow $ hadrons}
\newcommand{\Att}{\calH$_1$$\rightarrow $ \toto}
\newcommand{\Zqq} {${\mathrm{Z}} \rightarrow$ \qqbar}
\newcommand{\Zany} {Z $\rightarrow$ any}
\newcommand{\Zanyt} {Z $\rightarrow$ any but \toto}
\newcommand{\Zll} {${\mathrm{Z}} \rightarrow$ \llb}
\newcommand{\Ztt} {${\mathrm{Z}} \rightarrow$ \toto}
\newcommand{\Zem} {${\mathrm{Z}} \rightarrow$ \ee, \mm}
\newcommand{\Znn} {${\mathrm{Z}} \rightarrow$ \nunubar}
\newcommand{\hAA} {h $\rightarrow {\mathrm{AA}}$}
\newcommand{\cascg} {\calH$_1$\calH$_1$$\rightarrow \gamma \gamma $}
\newcommand{\cascv} {\calH$_1$\calH$_1$$\rightarrow {\mathrm V}^0 {\mathrm V}^0$}
\newcommand{\cascp} {\calH$_1$\calH$_1$$\rightarrow$ 4 prongs}
\newcommand{\casca} {\calH$_1$\calH$_1$$\rightarrow $ any}
\newcommand{\casch} {\calH$_1$\calH$_1$$\rightarrow $ hadrons}
\newcommand{\casct} {\calH$_1$\calH$_1$$\rightarrow $ \toto \toto}
\newcommand{\cascb} {\calH$_1$\calH$_1$$\rightarrow $ \bbbar \bbbar}
\newcommand{\cascc} {\calH$_1$\calH$_1$$\rightarrow$ \ccbar \ccbar}
\newcommand{\hbb}{\calH $\rightarrow $ \bbbar}
\newcommand{\hqq}{\calH $\rightarrow $ \qqbar}
\newcommand{\htt}{\calH $\rightarrow $ \toto}
\newcommand{\hvz}{\calH $\rightarrow {\mathrm V}^0$}
\newcommand{\hpp}{\calH $\rightarrow $ 2 prongs}
\newcommand{\hmono}{\calH $\rightarrow $ jet}
\newcommand{\hjj}{\calH $\rightarrow $ jet jet}
\newcommand{\rgr}{$\rightarrow$}
\newcommand{\MeV} {\mbox{${\mathrm{MeV}} $}}
\newcommand{\MeVc} {\mbox{${\mathrm{MeV}}/c $}}
\newcommand{\MeVcc} {\mbox{${\mathrm{MeV}}/c^2 $}}
\newcommand{\GeV} {\mbox{${\mathrm{GeV}} $}}
\newcommand{\GeVc} {\mbox{${\mathrm{GeV}}/c $}}
\newcommand{\GeVcc} {\mbox{${\mathrm{GeV}}/c^2 $}}
\newcommand{\TeVcc} {\mbox{${\mathrm{TeV}}/c^2 $}}
\newcommand{\dgree} {\mbox{$^{\circ}$}}
\newcommand{\mydeg} {$^{\circ}$}
\newcommand{\pbinv} {\mbox{pb$^{-1}$}}
\newcommand{\hptn }{\mbox{$ \hp \rightarrow \tau^+ \nu_{\tau} \,$}}
\newcommand{\hpcs }{\mbox{$ \hp \rightarrow c \bar{s} \,$}}
\newcommand{\hpcb }{\mbox{$ \hp \rightarrow c \bar{b} \,$}}
\newcommand{\cscs }{\mbox{$c \bar{s} \bar{c} s \,$}}
\newcommand{\tntn }{\mbox{$\tau^+ \nu_{\tau} \tau^- {\bar{\nu}}_{\tau} \,$}}
\newcommand{\cstn }{\mbox{$c s \tau \nu_{\tau} \,$}}
\newcommand{\hpmtntn }{\mbox{$ \HH \rightarrow \tntn \,$}}
\newcommand{\hpmcstn }{\mbox{$ \HH \rightarrow \cstn \,$}}
\newcommand{\hpmcscs }{\mbox{$ \HH \rightarrow \cscs \,$}}
\newcommand{\mhrec }{\mbox{$ m_{{\mathrm H}^+}^{rec} \,$}}
\newcommand{\lumi}{\mbox{$\cal L$}}
\newcommand{\like}{\mbox{$\cal L$}}
\newcommand{\likear}{\mbox{$\cal Q$}}
\newcommand{\clsinf}{\mbox{$\langle \rm CL_{\rm s} \rangle$}}
\newcommand{\CLb}{\mbox{$\rm CL_{\rm b}$}}
\newcommand{\CLs}{\mbox{$\rm CL_{\rm s}$}}
\newcommand{\CLsb}{\mbox{$\rm CL_{\rm s + b}$}}
\newcommand{\rs}{\mbox{$\sqrt{s}$}}
\newcommand{\AZ} {\mbox{$ {\mathrm A} {\mathrm Z} \, $}}
\newcommand{\hH} {\mbox{$ {\mathrm h} {\mathrm H} \, $}}
\newcommand{\tbeta} {\mbox{$\tan \beta$}}
\newcommand{\sinab} {\mbox{$\sin (\alpha-\beta)$}}
\newcommand{\cosab} {\mbox{$\cos (\alpha-\beta)$}}
\newcommand{\alp} {\mbox{$\alpha$}}

\def\EPJ#1#2#3{{ Eur. Phys. J.} {\bf{C#1}} (#2) #3}
\def\CPC#1#2#3{{ Comp. Phys. Comm.} {\bf{#1}} (#2) #3}
\def\NPB#1#2#3{{ Nucl.~Phys.} {\bf{B#1}} (#2) #3}
\def\PLB#1#2#3{{ Phys.~Lett.} {\bf{B#1}} (#2) #3}
\def\PRD#1#2#3{{ Phys.~Rev.} {\bf{D#1}} (#2) #3}
\def\PRL#1#2#3{{ Phys.~Rev.~Lett.} {\bf{#1}} (19#2) #3}
\def\ZPC#1#2#3{{ Z.~Phys.} {\bf C#1} (#2) #3}
\def\PTP#1#2#3{{ Prog.~Theor.~Phys.} {\bf#1} (19#2) #3}
\def\MPL#1#2#3{{ Mod.~Phys.~Lett.} {\bf#1} (19#2) #3}
\def\PR#1#2#3{{ Phys.~Rep.} {\bf#1} (#2) #3}
\def\RMP#1#2#3{{ Rev.~Mod.~Phys.} {\bf#1} (#2) #3}
\def\HPA#1#2#3{{ Helv.~Phys.~Acta} {\bf#1} (#2) #3}
\def\NIMA#1#2#3{{ Nucl.~Instr.~and~Meth.} {\bf A#1} (#2) #3}
\def\etal{\mbox{{\it et al.}}}
\begin{titlepage}
\begin{center}
\vspace{-0.5cm}
{\Large EUROPEAN ORGANIZATION FOR NUCLEAR RESEARCH}
\end{center}
\bigskip

\begin{flushright}
CERN-PH-EP/2006-001\\
$\phantom{...}$\\
17 January 2006 \\
\end{flushright}


\bigskip
\begin{center}{\Large \bf Search for Neutral MSSM Higgs Bosons at LEP}
\end{center}
\begin{center}
      {\Large  ALEPH, DELPHI, L3 and OPAL Collaborations\\
      The LEP Working Group for Higgs Boson Searches\footnote{See Appendix C for the list of authors} }
\end{center}
\begin{center}{\Large  Abstract}\end{center}
The four LEP collaborations, ALEPH, DELPHI, L3 and OPAL, have searched for the neutral Higgs bosons 
which are  predicted by the Minimal Supersymmetric Standard Model (MSSM).
The data of the four collaborations are statistically combined and examined 
for their consistency with the background 
hypothesis and with a possible Higgs boson signal. 
The combined LEP data show no significant excess of events which would indicate the production 
of Higgs bosons.
The search results are used to set upper bounds on the cross-sections
of various Higgs-like event topologies.
The results are interpreted within the MSSM in a number of ``benchmark" models, 
including CP-conserving and CP-violating scenarios. These interpretations 
lead in all cases to large exclusions in the MSSM parameter space. 
Absolute limits are set on the parameter \tanb\ and, in some scenarios, on the masses 
of neutral Higgs bosons.\\
$\phantom{.....}$\\

\begin{center}
{\it To be submitted to Eur. Phys. Journal C}
\end{center}

$\phantom{.....}$\\
%
\end{titlepage}
%
\section{Introduction}
One of the outstanding questions in particle physics is that of 
electroweak symmetry breaking and the origin of mass. The leading candidate for an answer is 
the Higgs mechanism~\cite{higgs} whereby
fundamental scalar Higgs fields acquire nonzero vacuum expectation
values and spontaneously break the electroweak symmetry. 
Gauge bosons and fermions obtain their 
masses by interacting with the resulting
vacuum Higgs fields. Associated with this description is the 
existence of massive scalar particles,
the Higgs bosons. 

The Standard Model~\cite{sm} requires one complex Higgs field doublet 
and predicts a single neutral Higgs boson of unknown mass.
After extensive searches at LEP, a lower bound of 114.4~\Gcs\ has been established for 
the mass of the Standard Model Higgs boson, at the 95\% confidence level (CL)~\cite{lep-sm}.  

Supersymmetric (SUSY)~\cite{susy} extensions of the Standard Model are of interest since they
provide a consistent framework for the unification of the gauge interactions 
at a high energy scale and for the stability of the electroweak scale. 
Moreover, their predictions are compatible with existing high-precision data~\cite{deboer}.
The Minimal Supersymmetric Standard Model (MSSM)~(reviewed, {\it e.g.}, in~\cite{mssm}) 
is the SUSY extension with minimal new particle content. It requires two Higgs 
field doublets 
and predicts the existence of three neutral and two charged Higgs bosons.
The lightest of the neutral Higgs bosons is predicted to have a mass 
less than about 140~\Gcs\ including radiative
corrections~\cite{radcor}. This prediction provided 
a strong motivation for the searches at LEP energies.

Most of the experimental investigations carried out in the past at LEP and elsewhere
were interpreted in MSSM scenarios where
CP conservation in the Higgs sector was assumed. 
In such scenarios the neutral Higgs 
bosons are CP eigenstates.
However, CP violation in the Higgs sector cannot be {\it a priori} excluded~\cite{cpviol-1}. 
Scenarios with CP violation are theoretically appealing since they provide 
one of the ingredients needed to explain the observed cosmic matter-antimatter 
asymmetry. The observed size of CP violation in B and K meson systems is not sufficient to drive this
asymmetry. In the MSSM, however, substantial CP violation can be induced by complex  
phases in the soft SUSY-breaking sector, through radiative 
corrections, especially from third-generation scalar quarks~\cite{cpviol-2}. 
In such scenarios the three neutral Higgs mass 
eigenstates are mixtures of CP-even and CP-odd fields,
with production and decay properties different from those in the CP-conserving scenarios. 
Hence, the experimental exclusions published so far 
for the CP-conserving MSSM scenarios may be weakened by CP-violating effects. 
There is currently one publication on searches interpreted in 
CP-violating scenarios~\cite{mssm-o}.

In this paper we describe the results of a statistical combination based on the searches of
the four LEP collaborations~\cite{a2000b,mssm-d,mssm-l,mssm-o}, which was carried out by 
the LEP Working Group for Higgs Boson Searches.
These searches include 
all LEP2 data up to the highest energy, 209~GeV; in the case of Refs.~\cite{mssm-d,mssm-o} they
also include the LEP1 data collected at energies in the vicinity of 91~GeV (the Z boson resonance).
The combined LEP data show no significant signal for Higgs boson production.
The search results are used to set upper bounds on topological cross-sections for 
a number of Higgs-like final states. Furthermore, they are interpreted in a set of representative MSSM 
``benchmark" models, with and without CP-violating effects in the Higgs sector. 
%
\section{The MSSM framework}
The LEP searches and their statistical combination presented in this paper
are interpreted in a
constrained MSSM model. At tree level, two parameters are sufficient (besides the 
known parameters of the Standard Model fermion and gauge sectors) to
fully describe the Higgs sector. A convenient choice is one Higgs boson mass
(\mA\ is chosen in CP-conserving scenarios and \mHpm\ in CP-violating scenarios), and the ratio
$\tanb=v_2/v_1$  of the vacuum expectation values of the two Higgs fields
($v_2$ and $v_1$ refer to the fields which couple to the up- and down-type 
fermions). Additional parameters, \msusy, $M_2$, $\mu$, $A$ and \mg, 
enter at the level of radiative corrections. 
\msusy\ is a soft SUSY-breaking mass parameter and represents a common mass
for all scalar fermions (sfermions) at the electroweak scale.  
Similarly, $M_2$ represents a common SU(2) gaugino mass at the electroweak scale.
The ``Higgs mass parameter" $\mu$ is the strength of the supersymmetric Higgs mixing;
$A = A_{\rm t} = A_{\rm b}$ is a common trilinear Higgs-squark coupling at the electroweak scale and \mg\ 
the gluino mass. 
Three of these parameters define the stop and sbottom mixing
parameters $X_{\rm t}=A-\mu\cot\beta$ and $X_{\rm b}=A-\mu\tan\beta$. 
In CP-violating scenarios, the complex phases related to
$A$ and \mg, $\arg (A)$ and $\arg (\mg)$, are supplementary parameters.  
In addition to all these MSSM parameters, the top quark mass also has a strong impact on the predictions
through radiative corrections. In this paper, four fixed values are used in the calculations:   
$m_{\rm t}$ = 169.3, 174.3, 179.3 and 183.0~\Gcs. For the purposes of illustration, 
$m_{\rm t}$ = 174.3~\Gcs\ is used in producing the figures (unless explicitly
specified otherwise), which is a previous world-average value~\cite{groom} and which is within the current experimental 
range of 172.7$\pm$2.9~\Gcs~\cite{topmass}. The influence of the top quark mass on the exclusion 
limits is discussed in Sections 5 and 6 along with the other results.

The combined LEP data are compared to the predictions of a number of MSSM ``benchmark" models
~\cite{benchmarks}.
Within each of these models,
the two tree-level parameters, \tanb\ and \mA\ (in the CP-conserving scenarios) or
\mHpm\ (in the CP-violating scenarios) are scanned while the other parameters are set 
to fixed values. Each scan point thus represents a specific MSSM model. The ranges of the scanned
parameters and the values of the fixed parameters are listed in
Table~\ref{tab:benchmarks} for the main scenarios studied. The first five models represent the main benchmarks for CP-conserving 
scenarios while the
last model, labelled {\it CPX}, is a benchmark model for CP-violating scenarios. Some variants of these benchmark scenarios,
which are also investigated, are presented in the text below. 
%
\begin{table}[ht] 
{\small
\begin{center}
\begin{tabular}{|l||c|c|c|c|c||c|}
\hline 
\multicolumn{7}{|c|}{Benchmark parameters}\\
\hline 
          &  (1) & (2) & (3) & (4) & (5) & (6)    \\ 
         & {\it \mh-max} & {\it no-mixing}   & {\it large-$\mu$}  & {\it gluophobic} & {\it small-$\alpha_{eff}$} &  {\it CPX} \\
\hline
\hline
\multicolumn{7}{|c|}{Parameters varied in the scan} \\
\hline
$\tanb$                       & 0.4--40 & 0.4--40 & 0.7--50   & 0.4--40   & 0.4--40     & 0.6--40 \\
$m_{\mathrm{A}}$ (\Gcs)        & 0.1--1000 & 0.1--1000 & 0.1--400  & 0.1--1000   & 0.1--1000     & -- \\
$m_{\mathrm{H}^{\pm}}$ (\Gcs)  & --      &  --     &  --     &  --       &  --         & 4--1000 \\
\hline
\multicolumn{7}{|c|}{Fixed parameters} \\
\hline
$M_{\mathrm{SUSY}}$ (GeV)    & 1000    & 1000    & 400   & 350   & 800                & 500   \\
$M_2$ (GeV)                  & 200     & 200     & 400   & 300   & 500                & 200   \\
$\mu$ (GeV)                  & $-200$    & $-200$    & 1000  & 300   & 2000               & 2000  \\
$m_{\tilde\mathrm{g}}$ (\Gcs) & 800     & 800     & 200   & 500   & 500                & 1000  \\
\hline
$X_{\rm t}$  (GeV)   & $2\,M_{\mathrm{SUSY}}$ & 0 &  $-300$ & $-750$  & $-1100$            & $A-\mu\cot\beta$ \\
\hline
$A$ (GeV)    & $X_{\rm t}$+$\mu\cot\beta$ & $X_{\rm t}$+$\mu\cot\beta$ &  $X_{\rm t}$+$\mu\cot\beta$     
&  $X_{\rm t}$+$\mu\cot\beta$   & $X_{\rm t}$+$\mu\cot\beta$ & 1000 \\
$\arg (A)$=$\arg (m_{\tilde\mathrm{g}})$  & - & -  & -  & - & -                          & $90^{\circ}$ \\  
\hline 
\end{tabular}
\end{center}
}
\caption{\sl Parameters of the main benchmark scenarios investigated in this paper. The values of \tanb\ and the
mass parameters \mA\ (in the CP-conserving scenarios) or \mHpm\ (in the CP-violating scenarios) are scanned 
within the indicated ranges. For the definitions of $A$ and $X_{\rm t}$, the Feynman-diagrammatic
on-shell renormalisation scheme is used 
in the CP-conserving scenarios and the 
$\overline{\mathrm{MS}}$ renormalisation scheme in the CP-violating scenarios.}
\label{tab:benchmarks}
\end{table} 

The scan range of \tanb\ is limited by the following considerations.
For values of \tanb\ below the indicated lower bounds, the calculations of the observables in the 
Higgs sector (masses, cross-sections and decay branching ratios) 
become uncertain; for values above the upper bounds,
the decay width of the Higgs bosons may become larger than the experimental mass resolution
(typically a few \Gcs) and the modelling of the kinematic distributions of the signal becomes inaccurate\footnote{
The DELPHI Collaboration included the variation of the Higgs boson decay width with \tanb\ in their simulation
for \tanb\ between 30 and 50. With increasing \tanb, DELPHI observed an increase of the mass resolutions and hence
a loss in the signal detection efficiencies; but this was compensated by the increase of the cross-sections, such that 
DELPHI found no significant drop in the overall sensitivity.}.
The scan range of \mA\ is limited in most cases to less than 1000~\Gcs; at higher values the Higgs phenomenology is 
insensitive to the choice of \mA.

For a given scan point, the observables in the Higgs sector 
are calculated
using two theoretical approaches, both including one- and two-loop corrections. 
The {\tt FeynHiggs2.0} code~\cite{Heinemeyer:CFeyn} is based
on a Feynman-diagrammatic approach and uses the {\it on-shell} renormalization scheme.
The {\tt SUBHPOLE} calculation and its CP-violating
variant {\tt CPH}~\cite{Carena:2000ks} are based on a renormalization-group improved effective potential
calculation~\cite{MSSMMHBOUND5} and use the $\overline{\mathrm{MS}}$
scheme\footnote{New developments in this approach are implemented in the code
{\tt CPsuperH}~\cite{cpsuperh}.}. 

In the CP-conserving case, the {\tt FeynHiggs} calculation is retained for the 
presentation of the results since it yields slightly more conservative
results (the theoretically allowed parameter space is wider) than {\tt SUBHPOLE} does.
Also, {\tt FeynHiggs} is preferred on theoretical grounds since its radiative  
corrections are more detailed than those of {\tt SUBHPOLE}. 

In the CP-violating case, neither of the two calculations is preferred 
on theoretical grounds. While {\tt FeynHiggs} contains more advanced one-loop corrections,
the {\tt CPH} code has a more precise phase dependence at the two-loop level.
We opted therefore for a solution where, in each scan point, the 
{\tt CPH} and {\tt FeynHiggs} calculations are compared and the
calculation yielding the weaker exclusion (more conservative) is retained. However, we also 
discuss in Section 6 the 
effect of using separately either one or the other of the two calculations.
Rather large discrepancies between the two codes are found in
calculating the partial width for the Higgs boson cascade decay
$\Gamma(\calHb\to\calHa\calHa)$ (\calHa\ and \calHb\ are the lightest and the second-lightest 
neutral MSSM Higgs bosons). 
Aiming at conservative exclusion limits, therefore, the {\tt CPH} formula for this decay 
was also used within the {\tt FeynHiggs} code.

All codes are implemented in a modified version of
the {\tt HZHA} program package~\cite{Janot:1996hzha}, which takes into account
initial-state radiation and the interference between identical final states from Higgsstrahlung and 
boson fusion processes.

%
\subsection{CP-conserving scenarios}
Assuming CP conservation, the spectrum of MSSM Higgs 
bosons consists of two CP-even neutral scalars, \h\ and \Ho\ (\h\ is defined to be the 
lighter of the two), one CP-odd neutral scalar, \A, and one pair of charged Higgs bosons, \Hpm.
The following ordering of masses is valid at tree level:  
\mh$<$(\mZ, \mA)$<$\mH\ and \mW$<$\mHpm. This ordering
may be substantially modified by radiative corrections~\cite{radcor} where
the largest contribution arises from the incomplete
cancellation between top and scalar top (stop) loops. The corrections affect mainly the
neutral Higgs boson masses and decay branching ratios. 

In \ee\ collisions at LEP energies, the main production processes of \h, \Ho\ and \A\ are the\\
Higgsstrahlung processes \ee\ra~\h\Z\ and \Ho\Z\ 
and the pair production processes \ee\ra~\h\A\ and \Ho\A\ (in most of the MSSM parameter space 
only the hZ and hA processes are possible by kinematics). 
The fusion processes \mbox{\ee\ra~(\WW\ra~\h)$\nu_{\rm e}\bar{\nu_{\rm e}}$} and
\mbox{\ee\ra~(\Z\Z\ra~\h)\ee}  
play a marginal role at LEP energies but they are also taken into account in the derivation of the results.

The cross-sections for Higgsstrahlung and pair production can be expressed in terms of the 
Standard Model Higgs boson production cross-section $\sigma_{\rm{HZ}}^{\rm SM}$.
The following expressions hold for the processes involving the lightest scalar boson \h: 
\begin{equation}
\sigma_{{\rm h}{\rm Z}} = \sin^2(\beta - \alpha)~\sigma_{\rm{HZ}}^{\rm SM}
\end{equation}
\begin{equation}
\sigma_{{\rm h}{\rm A}} = \cos^2(\beta - \alpha){\bar \lambda}~\sigma_{\rm{HZ}}^{\rm SM}.
\end{equation}
Here $\alpha$  is the mixing angle which diagonalises the CP-even Higgs mass matrix 
(at lowest order it can be expressed in terms of \mA, $M_{\rm Z}$ and \tanb) and $\bar\lambda$ is
a kinematic factor:
\begin{equation}
{\bar\lambda}=\lambda_{{\rm A}{\rm h}}^{3/2}/[\lambda_{{\rm Z}{\rm h}}^{1/2}
(12M_{\rm Z}^2/s + \lambda_{{\rm Z}{\rm h}})] 
\end{equation}
with  
\begin{equation}
\lambda_{ij} = [1-(m_i+m_j)^2/s][1-(m_i-m_j)^2/s], 
\end{equation}
where $s$ is the square of the centre-of-mass energy.
The cross-sections for the processes involving the heavy scalar boson \Ho\ are obtained by interchanging 
the MSSM suppression factors \sba\ and \cba\ 
in Eqs.~1 and 2 and replacing the index \h\ by \Ho\ in Eqs.~1,~2 and 3.
The Higgsstrahlung and pair production cross-sections are complementary, as seen from Eqs.~1 and 2. 
At LEP energies, the process \ee\ra~\h\Z\ is typically more abundant at small \tanb\ and \ee\ra~\h\A\ 
at large \tanb, but the latter process can be suppressed also by the kinematic factor ${\bar\lambda}$.

The following decay features are relevant to the neutral MSSM Higgs bosons.
The \h\ boson decays mainly to fermion pairs, with only a small fraction of WW$^*$ and ZZ$^*$ decays, 
since its mass is below the threshold of the on-shell processes \h\ra~\W\W\ and \h\ra~\Z\Z. 
However, for particular choices of the parameters,
the fermionic final states may be strongly suppressed.   
The \A\ boson also decays predominantly to fermion pairs, independently of its mass, 
since its coupling to vector bosons
is zero at leading order. For \tanb$>$1,
decays of \h\ and \A\ to \bb\ and $\tau^+\tau^-$ pairs are preferred while the decays to \cc\ and 
gluon pairs are suppressed. Decays to \cc\ may become 
important for \tanb$<$1.  
The decay \h\ra~\A\A\ may be dominant if allowed by kinematics~\cite{htoaa}.
Higgs boson decays into SUSY particles, such as sfermions, charginos or invisible neutralinos, are 
suppressed due to the high values of the SUSY-breaking scale $M_{\rm SUSY}$ which have been chosen.

In the following we describe the CP-conserving benchmark scenarios~\cite{benchmarks} which are 
examined in this paper. The corresponding parameters are listed in Table~\ref{tab:benchmarks}.
\subsubsection{The {\it \mh-max} scenario}
In the {\it \mh-max} scenario the stop mixing
parameter is set to a large value, 
$X_{\rm t} = 2 \msusy$. 
This model
is designed to maximise the theoretical upper bound on \mh\ for a given \tanb\ and fixed $m_{\rm t}$ 
and \msusy\ (uncertainties due to unknown higher-order corrections are ignored).   
This model thus provides the largest parameter space in the \mh\ direction and 
conservative exclusion limits for \tanb.

We also examine a variant of this scenario where the sign of $\mu$ is changed to positive, since this is favoured by presently 
available results on $(g-2)_\mu$~\cite{g-2,moroi}. This variant is labelled {\it \mh-max}~(a) below.
Furthermore, we examine the case where, besides changing the sign of $\mu$ to positive, 
the sign of the mixing parameter $X_{\rm t}$
is changed to negative. This choice of parameters gives better agreement with measurements of the branching ratios and of the
CP- and isospin-asymmetries for the process b\ra~s$\gamma$~\cite{btosg,benchmarks}. This variant is labelled 
{\it \mh-max}~(b) below.
\subsubsection{The {\it no-mixing} scenario}
In the {\it no-mixing} scenario the stop mixing
parameter $X_{\mathrm{t}}$ is set to zero, giving rise to a relatively restricted MSSM 
parameter space.
We also examine a variant of this scenario where the sign of $\mu$ is changed to positive, 
for a better agreement with recent measurements of $(g-2)_\mu$~\cite{g-2,moroi}, 
and \msusy\ is raised to 2~TeV in order to enlarge the parameter space of the
standard {\it no-mixing} scenario~\cite{benchmarks}.
This variant is labelled {\it no-mixing}~(a) below. In this case, \tanb\ is scanned only from 0.7 upward
due to numerical instabilities at lower values in the diagonalisation of the mass matrix.
\subsubsection{Special scenarios}
Some scenarios were designed to illustrate choices of the MSSM parameters for which the detection of Higgs bosons 
at LEP, at the Tevatron and at the LHC is expected to be difficult {\it a priori} due to the suppression of some 
main discovery channels~\cite{benchmarks}.
\begin{itemize}
\item
The {\it large-$\mu$} scenario is constructed in such a way that, while the h boson is accessible by kinematics 
at LEP for all scan points, 
the 
decay \h\ra~\bb, on which most of the searches at LEP and at the Tevatron are based, is typically
strongly suppressed. For many of the scan points the decay h\ra~\tautau\ is also suppressed, such that
the dominant decay modes are h\ra~\cc, gg and WW$^*$. The detection of Higgs bosons thus relies
mainly on flavour- and decay-mode-independent searches. Moreover, for some of the scan points, 
the \ee\ra~\h\Z\ process is suppressed altogether by a small value of \sba.
In such cases, however, the heavy neutral scalar H is within reach
($m_{\rm H} < 111$~\Gcs) and the cross-section for \ee\ra~HZ, proportional to \cba, is large;
the search may thus proceed {\it via} the heavy Higgs boson H.
\item
The {\it gluophobic} scenario is constructed in such a way that the Higgs boson coupling to gluons is suppressed due 
to a cancellation between the top and the stop loops at the hgg vertex. Since at the LHC the 
searches will rely heavily on producing the Higgs boson 
in gluon-gluon fusion, and since the mass determination will rely in part on the decays into gluon pairs, 
such a scenario may present experimental difficulties. 
\item
In the {\it small-$\alpha_{eff}$} scenario the couplings governing
the decays h\ra~\bb\ and h\ra~\tautau\ are suppressed with respect to their Standard Model values by a factor
$-\sin\alpha_{\rm eff}/\cos\beta$ 
($\alpha_{\rm eff}$ is the effective mixing angle
of the neutral CP-even Higgs sector including radiative corrections).
The suppression occurs mainly for large \tanb\ and moderate $m_{\rm A}$. 
\end{itemize}

%
\subsection{CP-violating scenarios}
In CP-violating MSSM scenarios the three neutral Higgs mass eigenstates \calH$_i$~($i=1,2,3$) do not have 
well defined CP quantum numbers. Each of them can thus be produced by Higgsstrahlung
(\ee\ra~\calH$_i$\Z) via the CP-even field component and in pairs (\ee\ra~\calH$_i$\calH$_j$~~($i\neq j$)). 
The relative rates depend on the choice of the parameters describing the CP-even/odd mixing.

Experimentally, the CP-violating scenarios are 
more challenging than the CP-conserving scenarios.
For a wide range of model parameters, the coupling of the lightest Higgs boson
\calHa\ to the \Z\ boson may be suppressed. Furthermore, 
the second- and third-lightest \calHb\ and \calHc\ bosons may both have masses close to or beyond 
the kinematic reach of LEP. 
Also, in CP-violating scenarios, the decays to the main ``discovery channels", \calHa\ra~\bb, \calHb\ra~\bb\ and
\calHb\ra~\calHa\calHa\ra~\bb\bb~\footnote{Regarding the decay properties, the CP-violating scenarios 
maintain a certain similarity to the CP-conserving scenarios although the branching ratios are, in general, different.
The lightest mass eigenstate \calHa\ 
predominantly decays to \bb\ if allowed by kinematics, with a small fraction decaying to $\tau^+\tau^-$ and \cc. 
The second-lightest Higgs boson \calHb\ may decay to \calHa\calHa\ when  
allowed by kinematics; otherwise it decays preferentially to \bb.}, may have lower branching ratios.
One therefore anticipates less search sensitivity in the CP-violating scenarios than in the
CP-conserving scenarios.
An example illustrating this situation is given in Table~\ref{tab:cpv-examples}.
\begin{table}[ht]
\begin{center}
{\small
\begin{tabular}{|l|cc|} \hline
&&\\
Parameters                                           & {\tt FeynHiggs} & {\tt CPH}  \\      
&&\\
\hline
H$^+$ (\Gcs)                                          & 129.0  & 129.0    \\
\tanb                                                 & 5.0    &  5.0     \\
\hline
\mcalHa~(\Gcs)                                        & 38.1   & 33.4     \\
\mcalHb~(\Gcs)                                        & 105.4  & 102.4    \\
$\sigma$(\calHa\Z\ra~\bb\Z)~(pb)                      & 0.0051 & 0.0019   \\
$\sigma$(\calHb\Z\ra~\bb\Z)~(pb)                      & 0.0156 & 0.0197   \\
$\sigma$(\calHb\Z\ra~\calHa\calHa\Z\ra~\bb\bb\Z)~(pb) & 0.0866 & 0.0978   \\
$\sigma$(\calHa\calHb\ra~\bb\bb)~(pb)                 & 0.0066 & 0.0094   \\
\hline
\end{tabular}
}
\vspace{0.5cm}
\caption{\label{tab:cpv-examples} \sl
A typical parameter set which is difficult to address by the present searches.
The results of the two calculations, {\tt FeynHiggs} and {\tt CPH}, are given for a
centre-of-mass energy of 206~GeV.
The main input parameters are listed in the first two lines; all other input parameters correspond to the {\it CPX}
benchmark scenario and are listed in the last column of Table~\ref{tab:benchmarks}. The output masses \mcalHa, \mcalHb\ and the 
relevant topological cross-sections 
are listed below the second horizontal line.}
\end{center}
\end{table}

The cross-sections for Higgsstrahlung and pair production
are given by~\cite{cpviol-2}
\begin{equation}
\sigma_{{\tiny\calH}_i{\rm Z}} = g^2_{{\tiny\calH}_i{\rm{ZZ}}}~\sigma_{\rm{HZ}}^{\rm SM}
\end{equation}
\begin{equation}
\sigma_{{\tiny\calH}_i{\tiny\calH}_j} = g^2_{{\tiny\calH}_i{\tiny\calH}_j{\rm Z}}~{\bar \lambda}
          ~\sigma_{\rm{HZ}}^{\rm SM}
\end{equation}
(in the expression for ${\bar \lambda}$, Eq.~3, the indices 
\h\ and \A\ have to be replaced by \calH$_i$ and \calH$_j$). 
The couplings
\begin{equation}
g_{{\tiny\calH}_i{\rm{ZZ}}} = \cos\beta{\cal O}_{1i} + \sin\beta {\cal O}_{2i}
\end{equation}
\begin{equation}
g_{{\tiny\calH}_i{\tiny\calH}_j{\rm Z}} = {\cal O}_{3i}(\cos\beta{\cal O}_{2j} - 
\sin\beta{\cal O}_{1j}) -  {\cal O}_{3j}(\cos\beta{\cal O}_{2i} - 
\sin\beta{\cal O}_{1i})
\end{equation}
obey the complementarity relation
\begin{equation}
\sum_{i=1}^{3} g^{2}_{{\tiny\calH}_i{\rm ZZ}} =  1
\end{equation}
\begin{equation}
g_{{\tiny\calH}_{\it k}\rm{ZZ}} =  \varepsilon_{ijk} 
g_{{\tiny\calH}_i{\tiny\calH}_j{\rm Z}}
\end{equation}
where $\varepsilon_{ijk}$ is the usual Levi-Civita symbol.

In CP-violating scenarios, the orthogonal matrix ${\cal O}_{ij}$ ($i,j = 1,2,3$) relating the weak CP
eigenstates to the mass eigenstates has non-vanishing off-diagonal elements. These elements, giving rise to CP-even/odd 
mixing, are proportional to
\begin{equation}
  \frac{m^4_t~\mathrm{Im}(\mu A)}{v^2~M^2_{\mathrm{SUSY}}} 
\end{equation}
with $v=\sqrt{v_1^2 + v_2^2}$.
Substantial deviations from the CP-conserving scenarios are thus expected for small \msusy\ and
large Im($\mu A$), which are obtained if the CP-violating phase $\arg (A)$ takes values close to $90^\circ$. 
Furthermore, the effects from CP violation strongly depend on the precise value of 
the top quark mass~\cite{topmass}.

The parameters of the benchmark model {\it CPX}
  have been chosen~\cite{Carena:2000ks} to maximise the phenomenological differences with respect to 
  the CP-conserving scenarios. Constraints from measurements of the electron and neutron electric dipole 
  moments~\cite{Commins:gv} were also taken into account.
  The basic set of parameters is listed in the last column of Table~\ref{tab:benchmarks}.
  Note that the scan of $m_{\mathrm{H}^{\pm}}$ started at 4~\Gcs\ but values less than about 100~\Gcs\ give unphysical
  results and are thus considered as theoretically inaccessible. 

The parameters which follow have been varied one-by-one while all the other parameters
were kept at their standard {\it CPX} value.
\begin{itemize}
\item
Top quark mass: $m_{\rm{t}} = 169.3, 174.3, 179.3$ and $183.0$~\Gcs, embracing the current experimental value,
$m_{\rm{t}} = 172.7\pm~2.9$~\Gcs~\cite{topmass}. 
\item
The CP-violating phases: $\arg (A) = \arg(\mg) = 0^\circ,~30^\circ,~60^\circ,~90^\circ$~({\it CPX} value), 
$135^\circ$ and $180^\circ$ (the values $0^\circ$ and $180^\circ$ correspond to CP-conserving limits).  
\item
The Higgs mass parameter: $\mu$ = 0.5, 1.0, 2.0~({\it CPX} value) and 4.0~TeV.
\item
The SUSY-breaking scale: \msusy~=~0.5~TeV~({\it CPX} value) and 1.0~TeV. 
The proposal of the {\it CPX} scenario~\cite{Carena:2000ks} predicts a weak dependence on \msusy\ if the relations 
$|A| = |m_{\tilde {\rm g}}| = \mu/2 = 2\msusy$ are preserved. This behaviour is examined by studying  
a model where \msusy\ is increased from 0.5~TeV to 1~TeV and the values of $A,~m_{\tilde {\rm g}}$ and $\mu$ 
are scaled to 2000~GeV, 2000~GeV and 4000~GeV, respectively.
\end{itemize}
%
\section{Experimental searches}
The searches carried out by the four LEP collaborations are based on \ee\ collision data which span a 
large range of centre-of-mass energies, from 91~GeV to 209~GeV. 
The searches include the Higgsstrahlung and pair production processes,
ensuring by their complementarity a high sensitivity over the accessible MSSM parameter space. It is important to note that 
the kinematic properties of the signal processes are to a large extent independent of the CP composition 
of the Higgs bosons. This implies that the same topological searches can be applied 
to study the CP-conserving and CP-violating scenarios. For Higgsstrahlung this is natural since only the CP-even 
components of the Higgs fields couple to the \Z\ boson.
In pair production involving CP-even and CP-odd  
field components, the similarity of the kinematic properties ({\it e.g.}, angular
distributions) arises from the scalar nature of the Higgs bosons. 
Small differences may occur from spin-spin correlations between final-state particles but these were found to 
have no noticeable effect on the signal detection efficiencies.
We therefore adopt in the following a common notation for the CP-conserving and CP-violating processes in which 
${\cal H}_i~(i=1,2,3)$ designate 
three generic neutral Higgs bosons of increasing mass, with undefined CP properties; in the CP-conserving limit 
($\arg (A) = \arg(\mg) = 0^\circ$), these become the CP eigenstates \h, \A, H (the correspondence depends on the mass hierarchy).

In each of the four LEP experiments, the data analysis is done in several steps.
A preselection is applied to reduce some of the largest backgrounds, in particular, from two-photon processes.
The remaining background, mainly from production of fermion pairs and WW or ZZ 
(possibly accompanied by photon or gluon radiation), is further
reduced by more selective cuts or by applying multivariate techniques such as likelihood analyses 
and neural networks. The identification of b-quarks in the decay of the Higgs bosons plays an important role
in the discrimination between signal and background, as does the kinematic reconstruction 
of the Higgs boson masses. The detailed implementation of these analyses, as well as the data samples used
by the four collaborations, are described in the individual publications.
A full catalog of the searches provided by the four LEP collaborations for this combination, with corresponding
references to the detailed descriptions, is given in Appendix A.
\subsection{Search topologies}
Searches have been carried out for the two main signal processes, the Higgsstrahlung process \ee\ra~\calHa\Z\
(which also apply in some cases to \ee\ra~\calHb\Z) and the pair production process \ee\ra~\calHb\calHa. 

\noindent (a)~~Considering first the Higsstrahlung process \ee\ra~\calHa\Z,
the principal signal topologies are those used in the 
search for the Standard Model Higgs boson at LEP~\cite{lep-sm}, namely:
\begin{itemize}
\item the four-jet topology,
(\calHa\ra~\bb)(\Z\ra~\qq), in which the invariant mass of two jets is close to the \Z\ boson mass \mZ\ while 
the other two jets contain b-flavour; 
\item  the missing energy topology, (\calHa\ra~\bb,~\tautau)(\Z\ra~\nn), in which the event consists of two b-jets
or identified tau decays
and substantial missing momentum and missing mass, compatible with \mZ; 
\item  the leptonic final states, (\calHa\ra~\bb)(\Z\ra~\ee, \mm), in which the invariant mass of the two leptons 
is close to \mZ;
\item  the final states with tau-leptons, 
(\calHa\ra~\tautau)(\Z\ra~\qq) and (\calHa\ra~\bb,~\tautau)(\Z\ra~\tautau), 
in which either the \tautau\ or the \qq\ pair has an invariant mass close to \mZ.
\end{itemize}
Most of these signatures are relevant for Higgs boson masses above the \bb\ threshold and rely on the identification 
of b-quarks in the final state. Searches for lighter Higgs bosons, listed in
Appendix A, use signatures which are described in the specific publications.
In some regions of the MSSM parameter space, the \calHa\ra~\bb\ decay may be suppressed while 
decays into other quark flavours
or gluon pairs are favoured. The above searches are therefore complemented or replaced\footnote{The replacement is
necessary whenever the overlap in terms of selected events is important, in order to avoid double-counting.} 
by flavour-independent searches for (\calHa\ra~\qq)\Z\ in which there is no requirement on the quark-flavour 
of the jets.
Finally, the searches for Higgsstrahlung also include the Higgs cascade decay
\ee\ra~\calHb\Z\ra~(\calH$_1$ \calH$_1$)\Z, giving rise to a new class of event topologies. These processes may 
play an important role in those regions of the parameter space where they are allowed by kinematics.

\noindent (b)~~In the case of the pair production process, \ee\ra~\calHb\calHa, 
the principal signal topologies at LEP are: 
\begin{itemize}
\item the four-b final state (\calHb\ra~\bb)(\calHa\ra~\bb);
\item the mixed final states (\calHb\ra~\tautau)(\calHa\ra~\bb) and (\calHb\ra~\bb)(\calHa\ra~\tautau);
\item the four-tau final state (\calHb\ra~\tautau)(\calHa\ra~\tautau).
\end{itemize}
The Higgs cascade decay,
\ee\ra~\calHb\calHa\ra~(\calHa\calHa)\calHa, gives rise to event topologies ranging from six b-jets 
to six tau-leptons. Most of
these searches are relevant for Higgs boson masses above the \tautau\
threshold. Similarly to the Higgsstrahlung case, the above searches for pair production
are complemented or replaced, whenever more efficient, by 
flavour-independent searches.
\subsection{Additional experimental constraints}
If the combination of the above searches is not sufficiently sensitive for excluding a given model point, 
additional constraints are applied; these are listed below.
\begin{itemize}
\item
Constraint from the measured decay width of the $\Zo$ boson, $\Gamma_{\rm Z}$, and its possible
   deviation, $\Delta\Gamma_{\rm Z}$, from the Standard Model prediction. The model point is regarded 
   as excluded if the
   following relation between the relevant cross-sections is found to be true:
   \begin{equation}
     \sum_i \sigma_{{\tiny\calH}_i{\rm Z}}(m_{\rm Z})+ \sum_{i,j} \sigma_{{\tiny\calH}_i{\tiny\calH}_j}(m_{\rm Z}) >
     \frac{\Delta\Gamma_{\rm Z}}{\Gamma_{\rm Z}}\cdot\sigma_{\rm Z}^{\rm tot}(m_{\rm Z}),
   \end{equation}
  where $\Delta\Gamma_{\rm Z} = 2.0$~MeV~\cite{zpole} stands for the 95\% CL upper bound on
  the possible additional decay width of the Z boson, beyond the Standard Model prediction, and $\sigma_{\rm Z}^{\rm tot}$ 
  is the \Z\ pole cross-section.
%
\item
Constraint from a decay mode independent search for \ee\ra~\calHa\Z~\cite{decindep-o}. The model point is 
    regarded as excluded if the condition
   \begin{equation}
     \sigma_{{\tiny\calH}_i{\rm Z}}> k(m_{{\tiny\calH}_i})\cdot\sigma^{\rm SM}_{\rm{HZ}}
   \end{equation}
   is fulfilled, where $k(m_{{\tiny\calH}_i})$ is a mass-dependent factor which scales the Standard Model Higgs 
   production cross-section to the value that is excluded at the 95\% CL. 
\item  
Constraint from a search for light Higgs bosons produced by the Yukawa process\footnote{Note that,
in the case of DELPHI, the Yukawa channels are not used as external constraints
but are combined with the other search channels.}.  
    The model point is regarded as excluded if the predicted Yukawa enhancement factor $\xi(m_{\tiny\calHa})$, 
    defined in~\cite{bib:yukawa}, 
    is excluded by this search. 
    To be conservative, the weaker of the two enhancement factors, for CP-even and CP-odd
    couplings, is used.

\end{itemize}
These additional constraints are particularly useful at small 
\mcalHa\ and \mcalHb, below the \bb\ threshold. 
\subsection{Statistical combination of search channels}
The statistical method by which the
topological searches are combined is described 
in Refs.~\cite{lep-sm,junk}.

After selection,
the combined data configuration (distribution of all selected events in several discriminating variables)
is compared in a frequentist approach to a large number of simulated configurations generated separately
for two hypotheses: the background ($b$) hypothesis 
and the signal-plus-background ($s+b$) hypothesis.
The ratio 
\begin{equation}
   Q={\cal L}_{s+b}/{\cal L}_b
\end{equation}   
of the corresponding likelihoods is used as the test statistic.
The predicted, normalised,
distributions of $Q$ (probability density functions) are integrated to obtain the 
$p$-values~\cite{pdg-stat} $1-CL_b = 1-{\cal P}_b(Q\leq Q_{\rm observed})$ and $CL_{s+b} = {\cal P}_{s+b}(Q\leq Q_{\rm observed})$;
these measure the compatibility of the observed data configuration with the two hypotheses.
Here ${\cal P}_b$ and ${\cal P}_{s+b}$ are the probabilities for a single experiment to obtain a value of $Q$ smaller 
than or equal to the observed value, given the background or the signal-plus-background hypothesis. 
More details can be found in Ref.~\cite{lep-sm}.

Systematic errors are incorporated in the calculation of the likelihoods by randomly varying the signal and 
background estimates in each channel\footnote{
The word ``channel" designates any subset of the data in which a search has been carried out. These subsets may correspond to specific final-state
topologies, to data sets collected at different centre-of-mass energies or to the subsets of data collected by different experiments.}
according to Gaussian error distributions and widths corresponding to the systematic errors.
For a given source of uncertainty, correlations are addressed by applying these random variations simultaneously
to all those channels for which the source of uncertainty is relevant. 
Errors which are correlated among the experiments arise mainly from using the same Monte Carlo generators 
and cross-section calculations for the signal and background processes.
The uncorrelated errors arise mainly from the limited statistics of the simulated background event samples.

In a purely frequentist approach, the exclusion limit is computed from the confidence $CL_{s+b}$ for the 
signal-plus-background hypothesis: a signal is regarded as excluded at the 95\% CL, for example, if 
an observation is made such that $CL_{s+b}$ is lower than 0.05. However, this procedure may lead to the 
undesired situation in which a large downward fluctuation of the background would exclude a signal hypothesis 
for which the experiment has no sensitivity since the expected signal rate is too small. 
This problem is avoided by using the ratio 
\begin{equation}
   CL_s = CL_{s+b}/CL_b
\end{equation}   
instead of $CL_{s+b}$. We adopt this quantity for setting exclusion limits and consider a given model 
to be excluded at the 95\% CL if the
corresponding value of $CL_s$ is less than 0.05. 
Since $CL_b$ is a positive number less than one, $CL_s$ is always larger than $CL_{s+b}$ and the 
limits obtained in this way are therefore conservative. 
\subsection{Comparisons of the data with the expected background}
The distribution of the {\it p}-value $1-CL_b$ over the parameter space covered by the searches provides a convenient way of
studying the agreement between the data and the expected background and of discussing the statistical significance of any local
excess in the data. While a purely background-like behaviour\footnote{Single, background-like,
experiments have values of $1-CL_b$  uniformly distributed between zero and one.} would yield $p$-values close to 0.5, 
much smaller values are expected in the case of a signal-like excess.
For example, a local excess of three or five standard deviations would give rise to a $p$-value $1-CL_b$ of 
$2.7\times10^{-3}$ or $5.7\times10^{-7}$, respectively. 

One has to be careful, however, when interpreting these numbers as probabilities for local excesses occurring 
over the extended domains covered by the searches. For example, the probability for a fluctuation of three standard deviations to occur 
{\it anywhere} in the parameter space is much larger than the number $2.7\times10^{-3}$ just quoted. A multiplication
factor has to be applied to the probability $1-CL_b$
which reflects the number of independent ``bins" of the parameter space;
this factor can be estimated from the total size of the parameter space and the experimental resolutions.
For example, the searches for the Higgsstrahlung process
\ee\ra~\calHa\Z, covering the range 0$<$\mcalHa$<$120~\Gcs\ with a mass resolution $\Delta\mcalHa$~of 
about 3~\Gcs, would yield about twenty
fairly independent mass-bins of width $2 \Delta\mcalHa$; hence, a multiplication factor of about twenty. 
Much bigger multiplication factors
are expected in the searches for the pair production process \ee\ra~\calHb\calHa\ 
with two independent search parameters (masses).

These simple considerations do not take into account, for example, possible correlations from resolution
tails extending over several adjacent bins or correlations between different searches sharing candidate events. 
A more elaborate evaluation of the multiplication factor has therefore been performed.
A large number of background experiments was simulated, covering the whole parameter space, 
using realistic resolution functions and taking correlations into account. From these random experiments, the
probability to obtain $1-CL_b$ smaller than a given value, {\it anywhere} in the parameter space of a given scenario, 
has been determined (the {\it \mh-max} scenario was taken for this study).
A scale factor 
of at least 60
was obtained in this manner. According to this estimate, the probability of observing
a background fluctuation of three standard deviations {\it anywhere} in the parameter space of a given scenario
(e.g., {\it \mh-max}) can be 
16\% or more.
Also, to observe two fluctuations with two standard
deviations turns out to be more likely than to observe only one.

\begin{figure}[p]
\centerline{
\epsfig{file=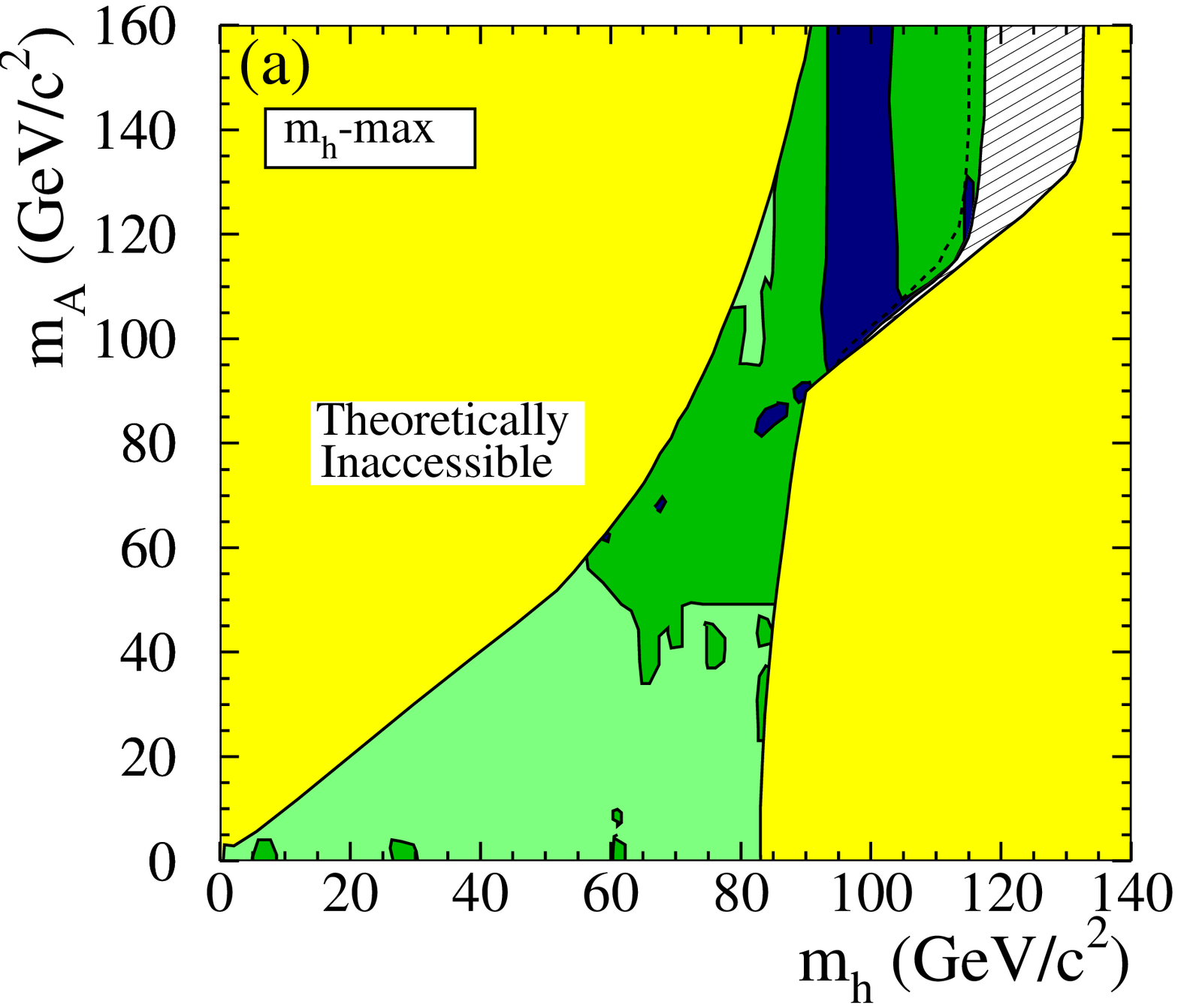,width=0.49\textwidth}
\epsfig{file=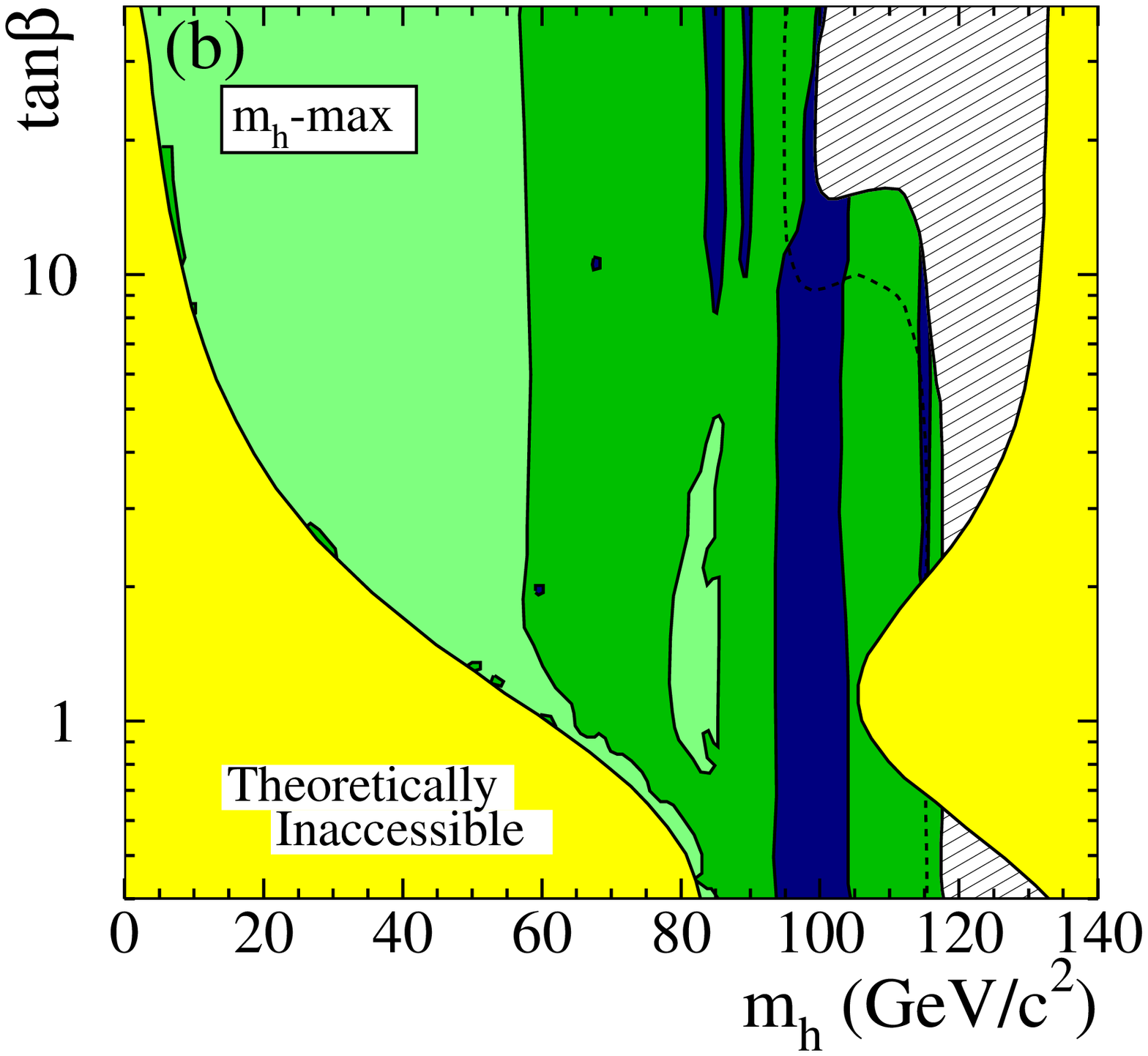,width=0.49\textwidth}
}
\centerline{
\epsfig{file=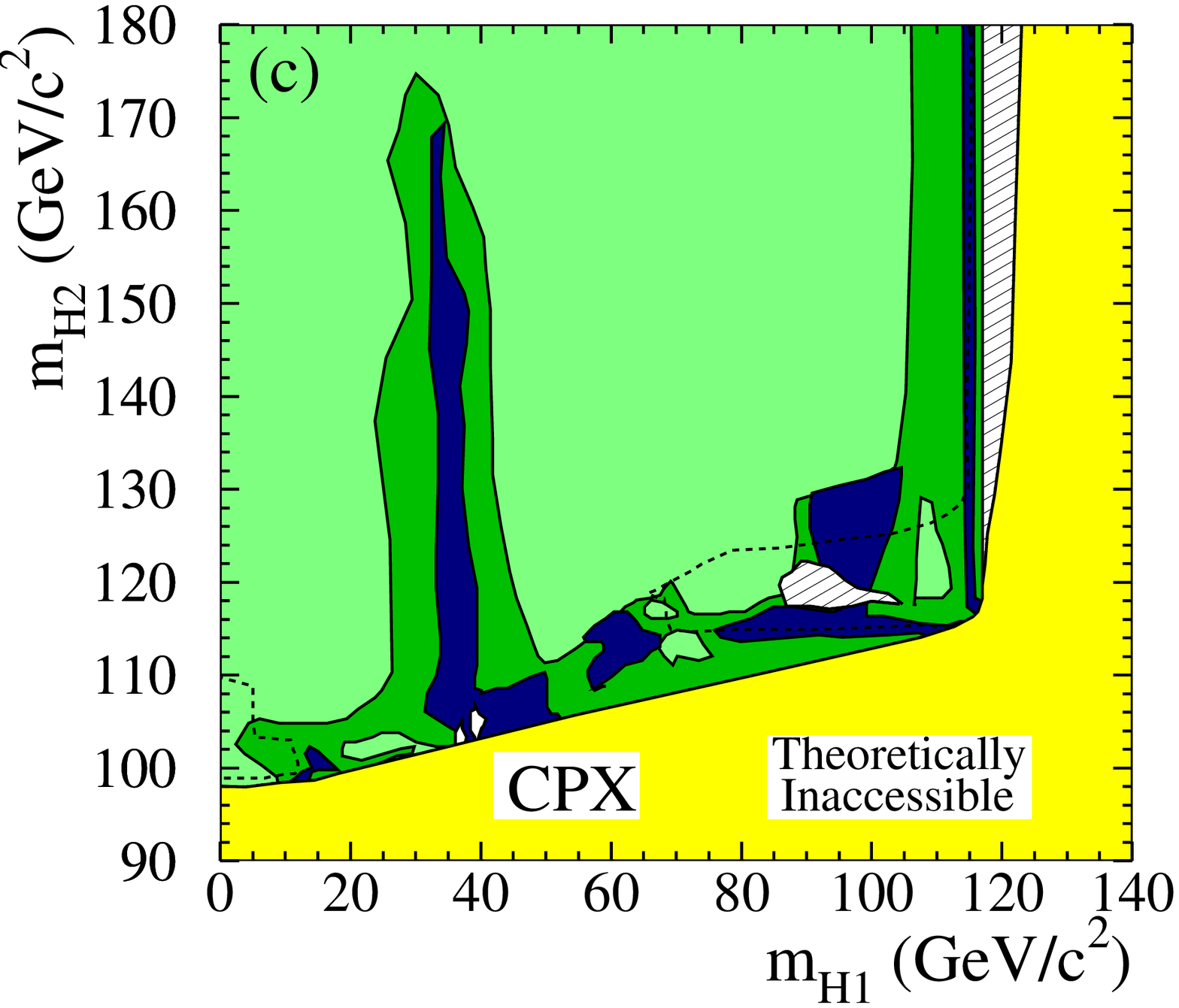,width=0.49\textwidth}
\epsfig{file=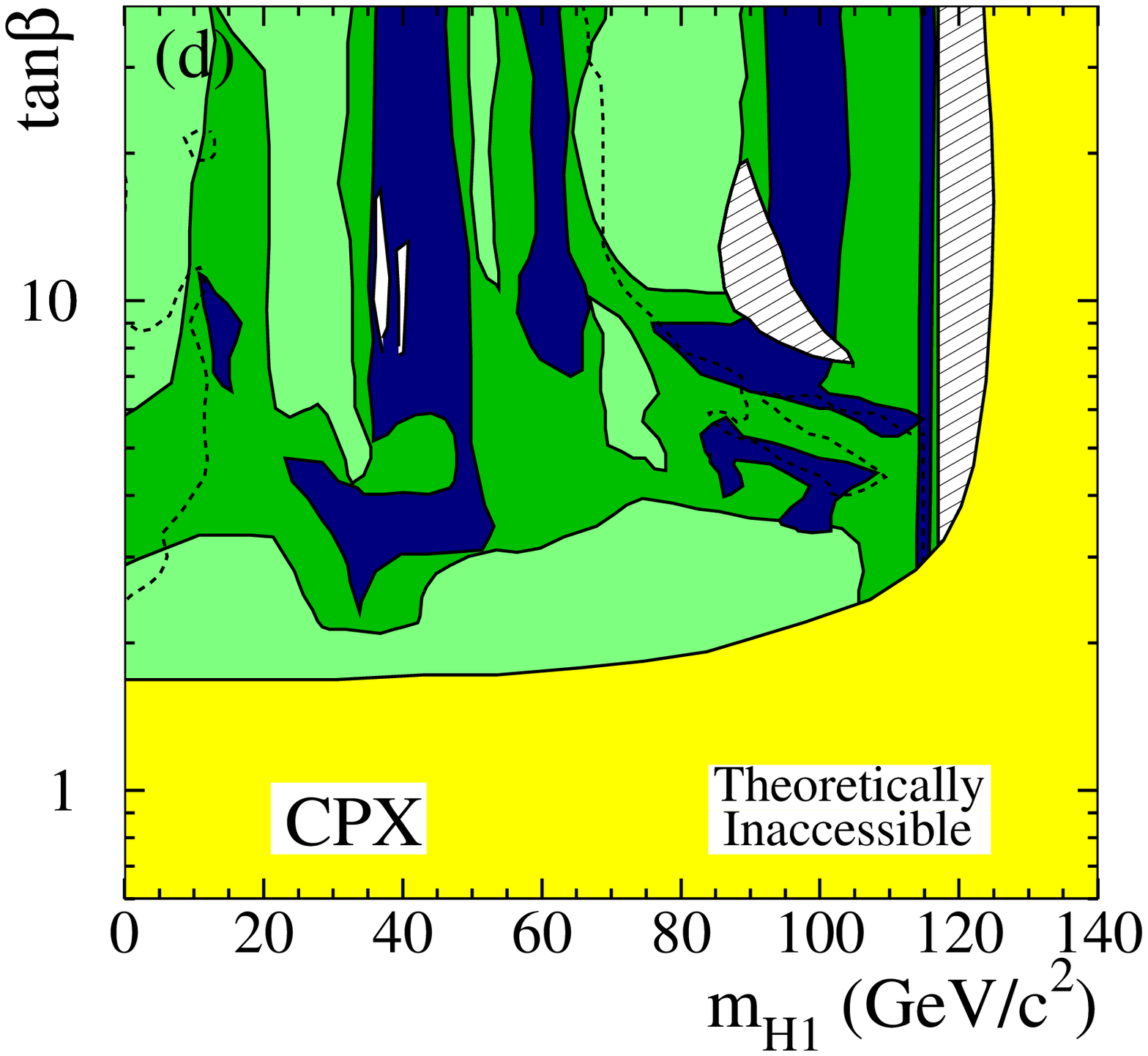,width=0.49\textwidth}
}
\caption[]{\sl
  Contours of the observed $p$-values, $1-CL_b$, indicating the statistical significances 
  of local excesses in the data.
  Plots (a) and (b) refer to the CP-conserving 
  MSSM benchmark scenario {\it \mh-max} and plots (c) and (d) to the CP-violating scenario {\it CPX}.
  For each scenario, the parameter space  is shown in two projections.  
  Regions which are not part of the parameter space (labelled ``Theoretically Inaccessible") are shown in light-grey 
  or yellow.
  In the medium-grey or light-green regions 
  the data show an excess of less than one standard deviation
  above the expected background. Similarly, 
  in the dark-grey or dark-green regions the excess is between one and two standard 
  deviations while in the darkest-grey or blue regions it is between two and three standard deviations.
  In plots (c) and (d), two small regions with excesses larger than three standard deviations are shown in white.
  The dashed lines show the expected exclusion limit at 95\% CL. The hatched areas represent regions where the 
  median expected value of $CL_s$ 
  in the background hypothesis is larger than 0.4; apparent excesses in these regions would not be
  significant.
\label{fig:hotspot}}
\end{figure}

Figure~\ref{fig:hotspot} shows the distribution of the $p$-value $1-CL_b$, determined from the present combined searches, 
for the CP-conserving benchmark scenario {\it \mh-max} and the CP-violating scenario {\it CPX}.
Over the largest part of the parameter space, the local excesses are smaller than two standard 
deviations. 
In the {\it \mh-max} scenario, the lowest value, $1-CL_b = 1.3\times10^{-2}$, lies within the vertical band 
at \mh\ around 100~\Gcs\ and 
corresponds to 2.5 standard deviations. 
This excess, and a less significant excess at about 115~\Gcs, come from the Higgsstrahlung search; both are 
discussed in Ref.~\cite{lep-sm} in the context of the search for the Standard Model Higgs boson. 
In the {\it CPX} scenario, one observes two small regions at
\mcalHa~$\approx$~35-40~\Gcs, \mcalHb~$\approx$~105~\Gcs\ and \tanb~$\approx$~10,  
where the significance exceeds three standard deviations; they arise from the search for the pair production process.

The exact position and size of these fluctuations may vary from one scenario to the other.
In Tables~\ref{tab:cpc-hottest} and~\ref{tab:cpv-hottest} we list the parameters of the most significant excesses 
for all CP-conserving and CP-violating benchmark scenarios considered in this paper. The largest fluctuation of all 
has a significance of 3.5 standard deviations; its probability is estimated as 3.6\% at least, when the scale factor 
of 60 or more is applied. 

From these studies one can conclude that there is a reasonable agreement between the data and the simulated background,
with no compelling evidence for a Higgs boson signal, and that the excesses observed are compatible with  
random fluctuations of the background.

\begin{table}[ht]
\begin{center}
\begin{tabular}{|l|ccccc|cc|} \hline
&&&&&&&\\
Benchmark   & \mh\  & \mH\ & \mA\  & \mHpm\  & \tanb\ & $1-CL_b$   & $\sigma$  \\
            &  &    &     &   &        &            &                (st.dev.) \\
&&&&&&&\\
\hline
{\it \mh-max}              & 99 & 253 & 169 & 184 & 0.7 & 1.3$\times10^{-2}$ & 2.5 \\ 
{\it \mh-max (a)}          & 99 & 277 & 156 & 171 & 0.6 & 1.4$\times10^{-2}$ & 2.5 \\ 
{\it \mh-max (b)}          & 99 & 345 & 310 & 319 & 0.9 & 1.6$\times10^{-2}$ & 2.4 \\ 
{\it no-mixing}            & 99 & 165 & 152 & 171 & 3.7 & 1.4$\times10^{-2}$ & 2.5 \\ 
{\it no-mixing (a)}        & 99 & 134 & 114 & 138 & 5.4 & 1.1$\times10^{-2}$ & 2.5 \\ 
{\it large-$\mu$}          & 59 & 108 & 67  & 104 & 3.1 & 1.0$\times10^{-2}$ & 2.6 \\ 
{\it gluophobic}           & 56 & 124 & 69  & 105 & 4.1 & 5.5$\times10^{-3}$ & 2.8 \\ 
{\it small-$\alpha_{eff}$} & 60 & 121 & 75  & 109 & 5.5 & 2.4$\times10^{-3}$ & 3.0 \\ 
\hline
\end{tabular}
\vspace{0.5cm}
\caption{\label{tab:cpc-hottest} \sl
The most significant excesses with respect to the predicted background, 
for each of the CP-conserving benchmark scenarios. 
Columns 2 to 6 show the mass parameters (in \Gcs) and 
\tanb\ at which the excess occurs. Column 7 gives the corresponding 
p-values $1-CL_b$. In the last column, the significances of the excesses, 
in standard deviations, are listed.}
\end{center}
\end{table}
\begin{table}[ht]
\begin{center}
{\small
\begin{tabular}{|l|ccccc|ccc|} \hline
&&&&&&&&\\
  & \mcalHa\  & \mcalHb\  & \mcalHc\ & \mHpm\  & \tanb\ & $1-CL_b$  & $1-CL_b$       & $\sigma$  \\
  &     &     &   &   &        & ({\tt CPH})     &  ({\tt FeynH.})          & (st.dev.) \\
&&&&&&&&\\
\hline
{\it CPX} scenario      & 35-40   & 105 &  120    & 120 &   10   & $1\times10^{-3}$  & $2\times10^{-3}$ &  3.1 \\
\hline
\mt~= 169~\Gcs\       & 40   & 100 &     125       & 120 & 10-15 & $8\times10^{-4}$  & $9\times10^{-4}$ &  3.3 \\  
\mt~= 179~\Gcs\       & 95   & 125  &   145       & 155 &   3   & $4\times10^{-3}$  & $4\times10^{-3}$ &  2.9 \\  
\mt~= 183~\Gcs\      & 95   & 130    &    150     & 155 &   3   & $4\times10^{-3}$  & $4\times10^{-3}$ &  2.9 \\
$\arg (A)$=$\arg (m_{\tilde {\rm g}})$=$0^\circ$   & 40   & 95     & 125   & 115 &  12   & $8\times10^{-4}$  & $1\times10^{-3}$ &  3.1 \\  
$\arg (A)$=$\arg (m_{\tilde {\rm g}})$=$ 30^\circ$  & 45   & 100   &  125  & 110 & 10-20 & $1\times10^{-3}$  & $1\times10^{-3}$ &  3.1 \\
$\arg (A)$=$\arg (m_{\tilde {\rm g}})$=$ 60^\circ$  & 45   & 95   &  130   & 115 &  5-20 & $5\times10^{-4}$  & $6\times10^{-4}$ &  3.5 \\
$\arg (A)$=$\arg (m_{\tilde {\rm g}})$=$ 135^\circ$ & 40   & 105   &  120  & 110 & $>$20 & $2\times10^{-3}$  & $3\times10^{-3}$ &  3.0 \\
$\arg (A)$=$\arg (m_{\tilde {\rm g}})$=$ 180^\circ$ & 95   & 130   & 170   & 170 &   6   & $4\times10^{-3}$  & $4\times10^{-3}$ &  2.9 \\
$\mu = 500$~GeV       & 95   & 100    &  125      & 130 &   1   & $4\times10^{-3}$  & $4\times10^{-3}$ &  2.9 \\
$\mu = 1000$~GeV      & 95   & 110     &  125     & 135 &   2   & $5\times10^{-3}$  & $5\times10^{-3}$ &  2.8 \\
$\mu = 4000$~GeV      & 95   & 180   &  330       & 300 &   4   & $5\times10^{-3}$  & $5\times10^{-3}$ &  2.8 \\
\msusy=1~TeV          & 95   & 105   &    145     & 130 &   2   & $4\times10^{-3}$  & $4\times10^{-3}$ &  2.9 \\
\msusy=1~TeV, scaled  & 40   & 105    &   120     & 130 &  10   & $2\times10^{-3}$  & $2\times10^{-3}$ &  3.1 \\
\hline
\end{tabular}
\vspace{0.5cm}
\caption{\label{tab:cpv-hottest} \sl
The most significant excesses with respect to the predicted background in the CP-violating
benchmark scenario {\it CPX} and its variants. 
The first column indicates either the {\it CPX} scenario or the parameter value which differs from the standard 
{\it CPX} set listed in the last column of Table~\ref{tab:benchmarks}. Columns 2 to 6 show the mass parameters (in \Gcs) and 
\tanb\ at which the excesses occur (the more conservative of the {\tt CPH} and {\tt FeynHiggs} calculations is used).
Columns 7 and 8 give the corresponding $p$-values, $1-CL_b$, using in turn the {\tt CPH} and
{\tt FeynHiggs} codes (note the overall agreement of the two calculations in this respect). 
In the last column, the significances of the excesses, in standard deviations, are listed.}
}
\end{center}
\end{table}
%
\section{Limits on topological cross-sections}
In this section we present upper bounds on the cross-sections for the most
important final-state topologies expected from the Higgsstrahlung process \ee\ra~\calHa\Z\ and 
the pair production process~\ee\ra~\calHb\calHa. These can be used to test a wide range 
of specific models. 

We define the scaling factor 
\begin{equation}
S_{95} = \sigma_{\rm max}/\sigma_{\rm ref},
\end{equation}
where $\sigma_{\rm max}$  is the largest cross-section compatible with the data, at the 95\% CL, and
$\sigma_{\rm ref}$ is a reference cross-section. For the topologies motivated by Higgsstrahlung,
$\sigma_{\rm ref}$ is taken to be the Standard Model Higgs production cross-section; 
for final states motivated 
by the pair production process, $\sigma_{\rm ref}$ is taken to be the MSSM Higgs production cross-section 
of Eq.~2 with the MSSM suppression factor set to 1. 
Numerical values for the cross-section limits are listed in Appendix B.

Figure~\ref{sm-xi2} shows the upper bound $S_{95}$ for final states motivated by the Higgsstrahlung process 
\ee\ra~\calHa\Z~(the figure is reproduced from Ref.~\cite{lep-sm}).
In part (a), the Higgs boson is assumed to decay into fermions and bosons with branching ratios as given by 
the Standard Model. Contributions
from the fusion processes WW\ra~\calHa\ and ZZ\ra~\calHa, according to the Standard Model,
corrected for initial-state radiation, are assumed 
to scale with energy like the Higgsstrahlung process.
In part (b) it is assumed that the Higgs boson decays exclusively to \bb\ and in part (c) exclusively to \tautau.
Besides representing bounds on topological cross-sections,
this figure also illustrates the overall agreement between the data and the expected background from Standard
Model processes. The largest deviations observed barely exceed two standard deviations. 

\begin{figure}[htb]
\begin{center}
\epsfig{figure=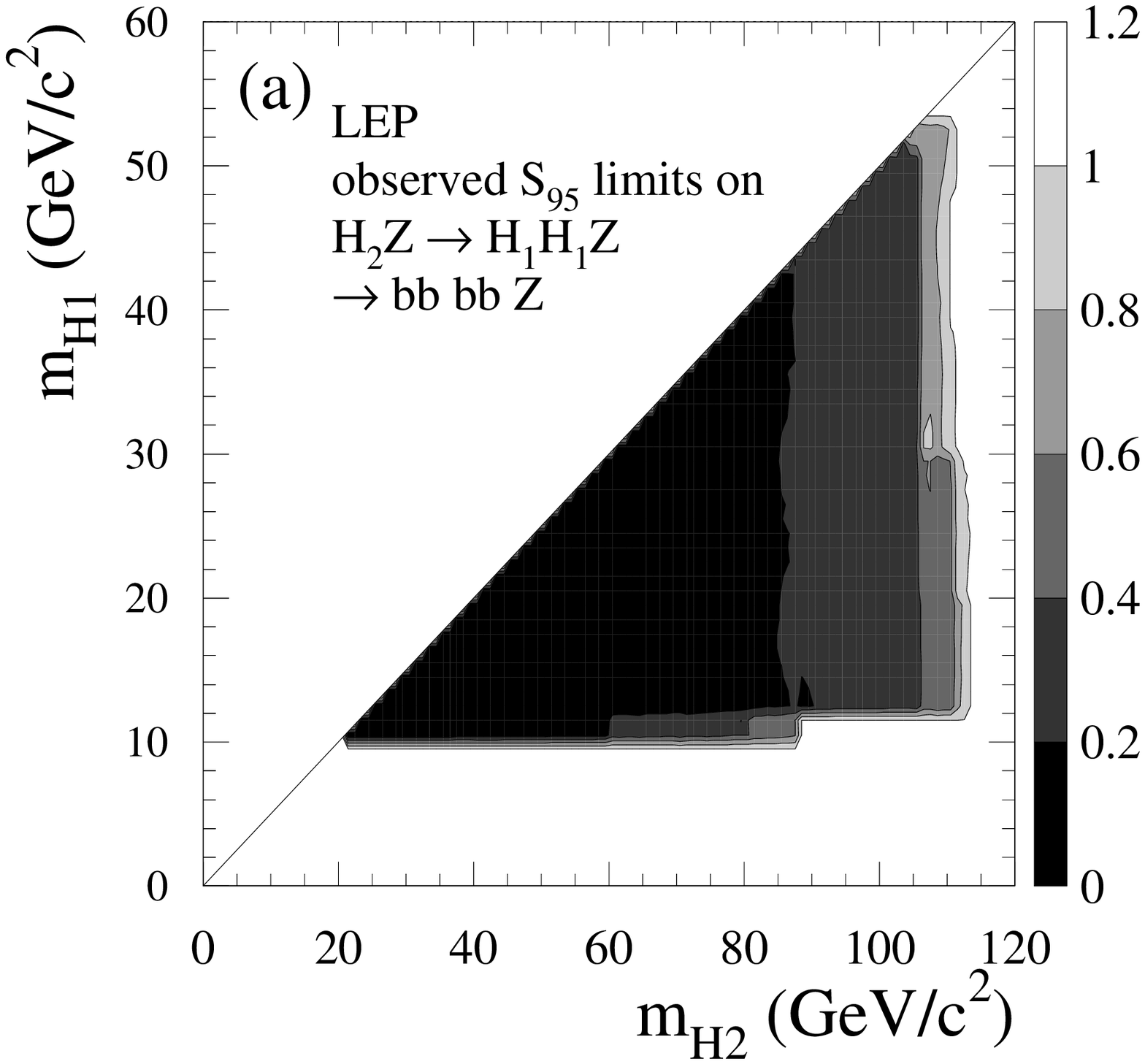,width=0.49\textwidth} 
\epsfig{figure=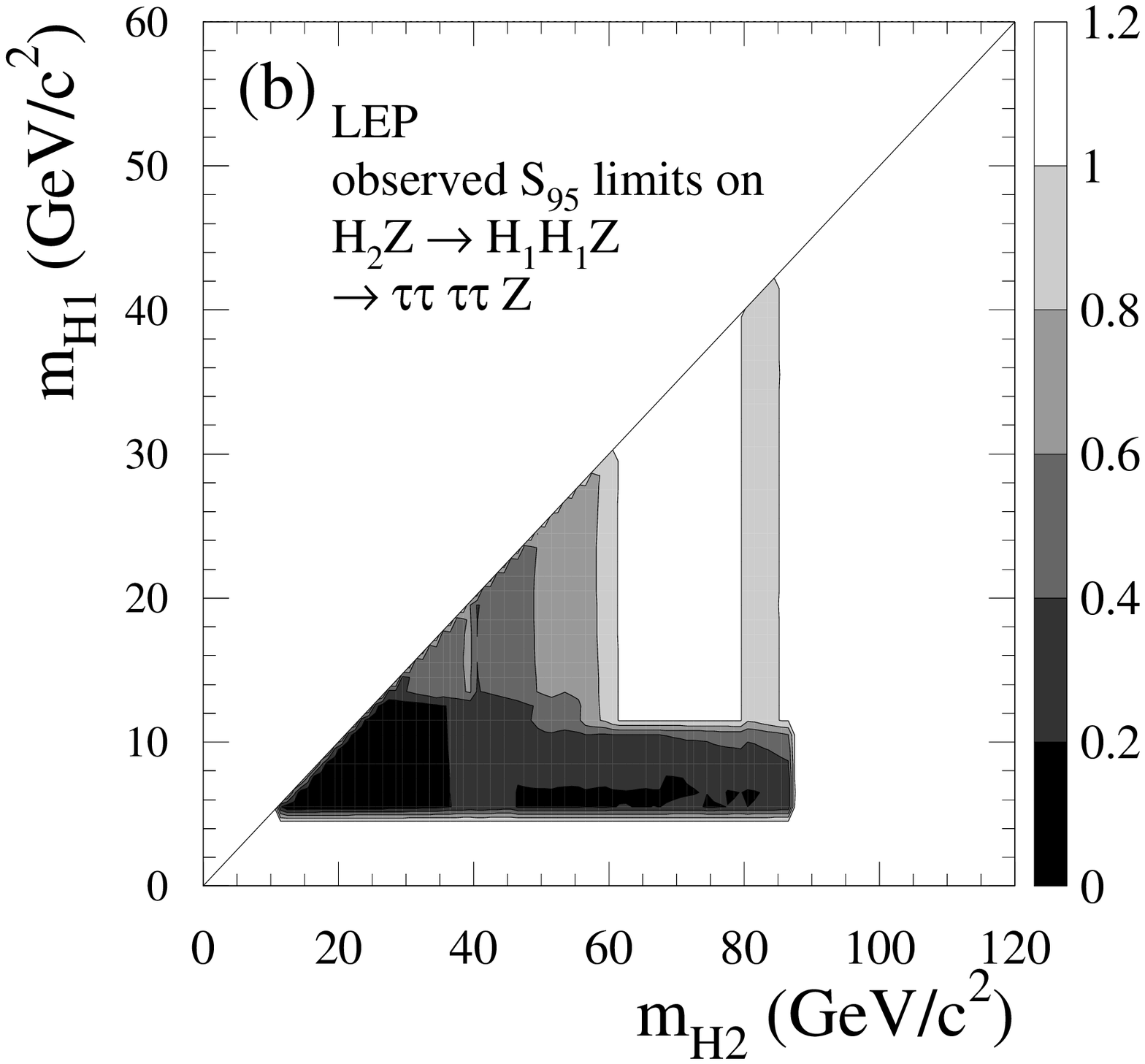,width=0.49\textwidth} \\
\epsfig{figure=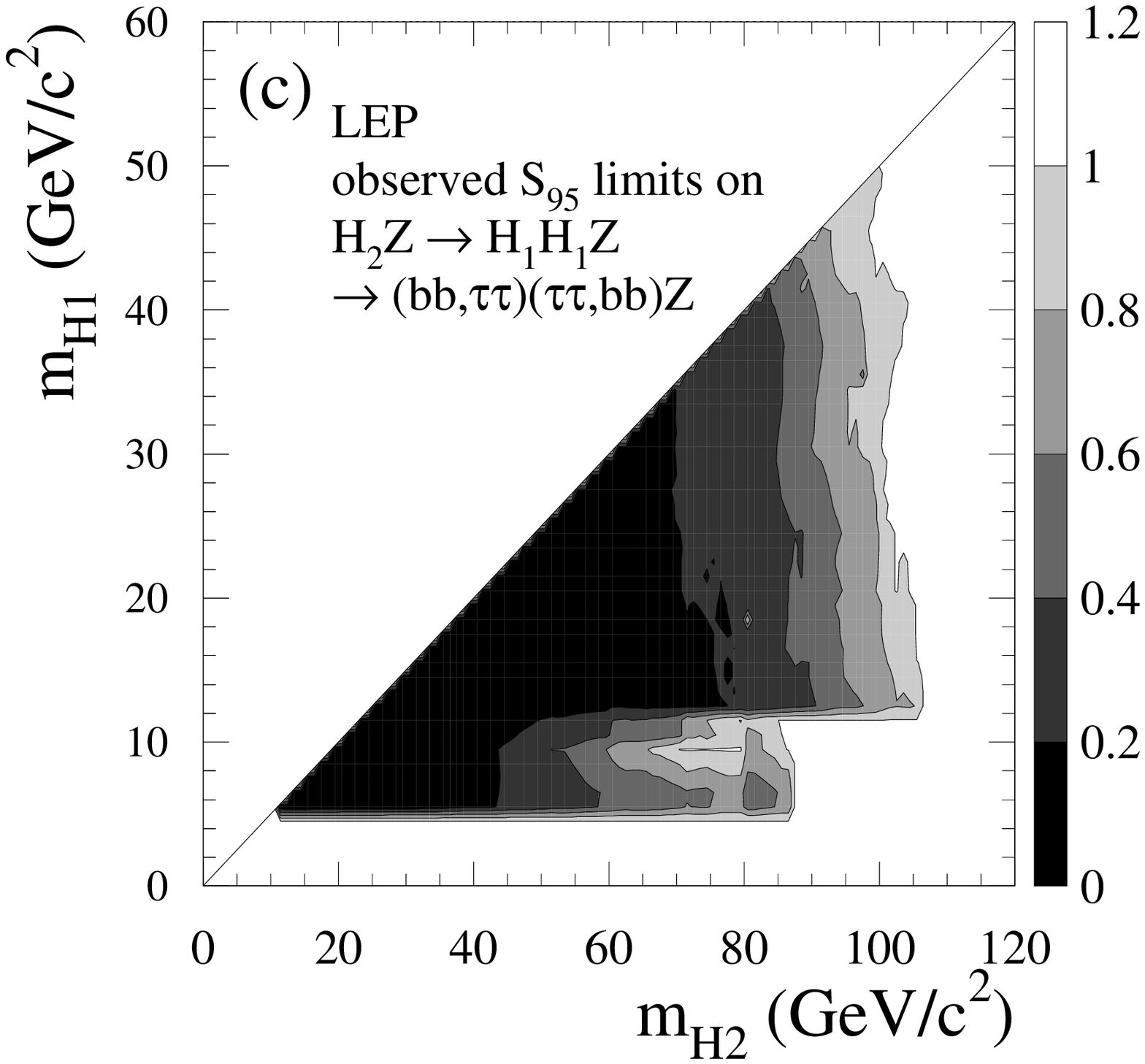,width=0.49\textwidth} 
\end{center}
\caption[]{\sl Contours of the 95\% CL upper bound, $S_{95}$ (see text), 
for various topological cross-sections motivated by the Higgsstrahlung cascade process \ee\ra~(\calHb\ra~\calHa\calHa)\Z, 
projected onto the (\mcalHb, \mcalHa) plane. The scales for the shadings are given on the right-hand side of each plot. 
In plot (a) the \calHa\ boson is assumed to decay exclusively to \bb\ and 
in plot (b) exclusively to \tautau; in plot (c)
it is assumed to decay with equal probabilities to \bb\ and to \tautau.
\label{s2-h2h1}}
\end{figure}

Figure~\ref{s2-h2h1} shows contours of $S_{95}$ for the cascade process
\mbox{\ee\ra~\calHb\Z\ra~(\calHa\calHa)\Z}, projected onto the (\mcalHb, \mcalHa) plane, assuming that the \calHb\ boson 
decays exclusively to \calHa\calHa. 
In part (a) it is assumed that the \calHa\ boson decays exclusively to \bb\ and in part (b) exclusively to \tautau.
In part (c), as an example, an equal mixture of \calHa\ra~\bb\ and \calHa\ra~\tautau\ is assumed,
which implies 25\% \bb\bb\Z, 25\% \tautau\tautau\Z\ and 50\% \bb\tautau\Z\ final states.
The sensitivity of the \bb\bb\Z\ channel starts at the \bb\ threshold and extends almost to the kinematic limit.
In the 
\tautau\tautau\Z\ channel the sensitivity is altogether weaker (the discontinuities reveal the limited and inhomogeneous mass 
coverage of the four experiments in this channel).

\begin{figure}[htb]
\begin{center}
\epsfig{figure=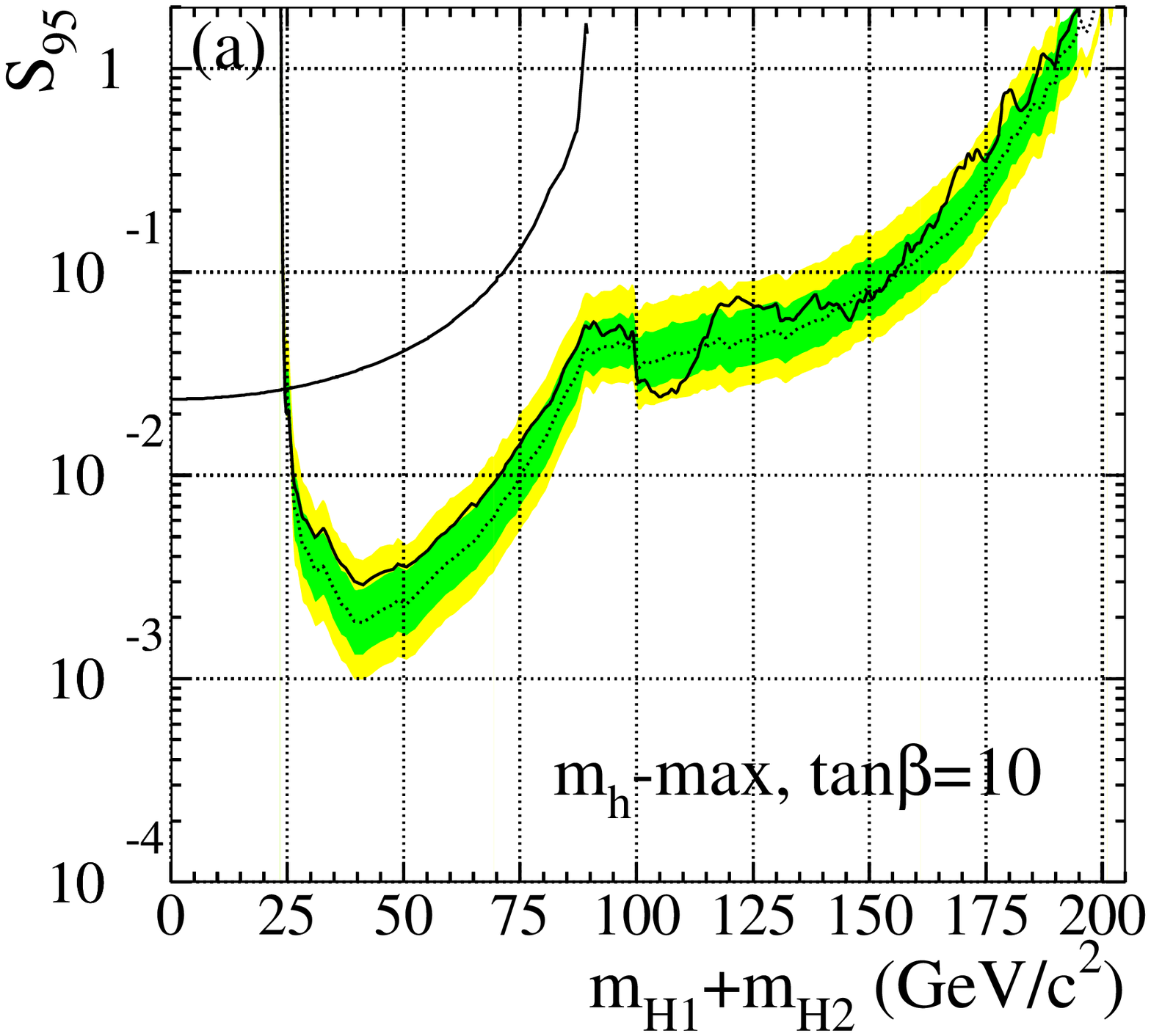,width=0.49\textwidth} 
\epsfig{figure=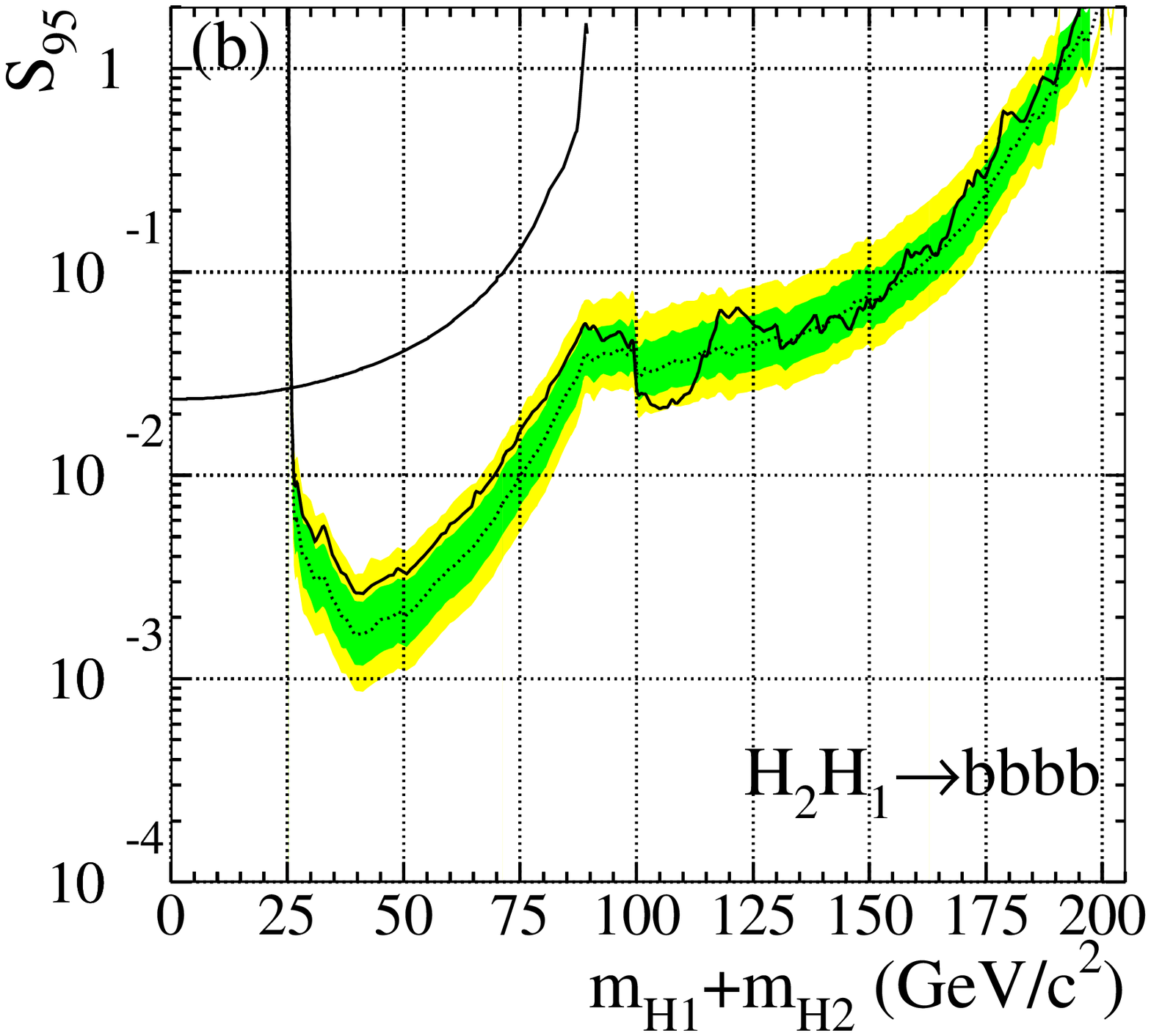,width=0.49\textwidth}\\     
\epsfig{figure=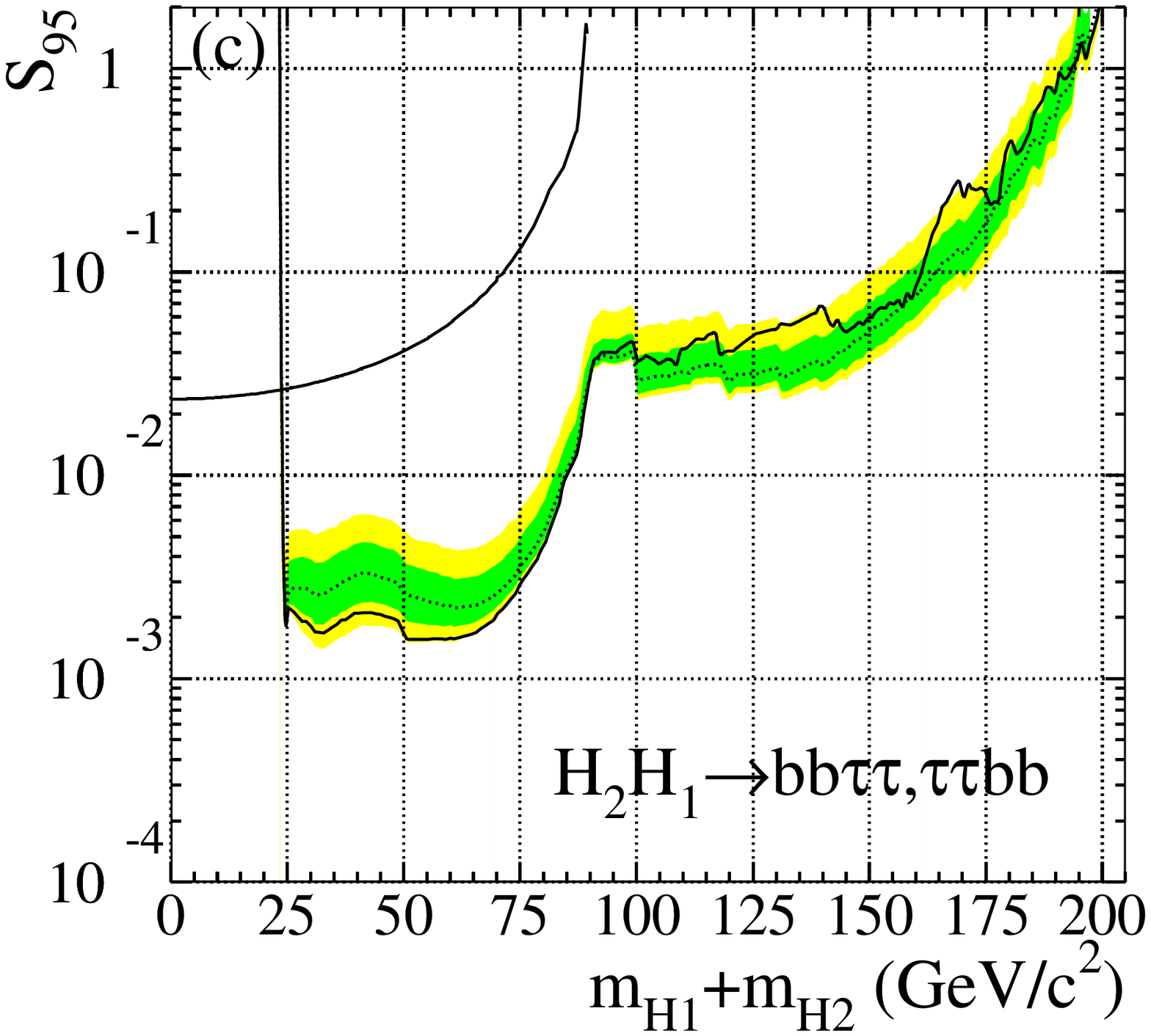,width=0.49\textwidth} 
\epsfig{figure=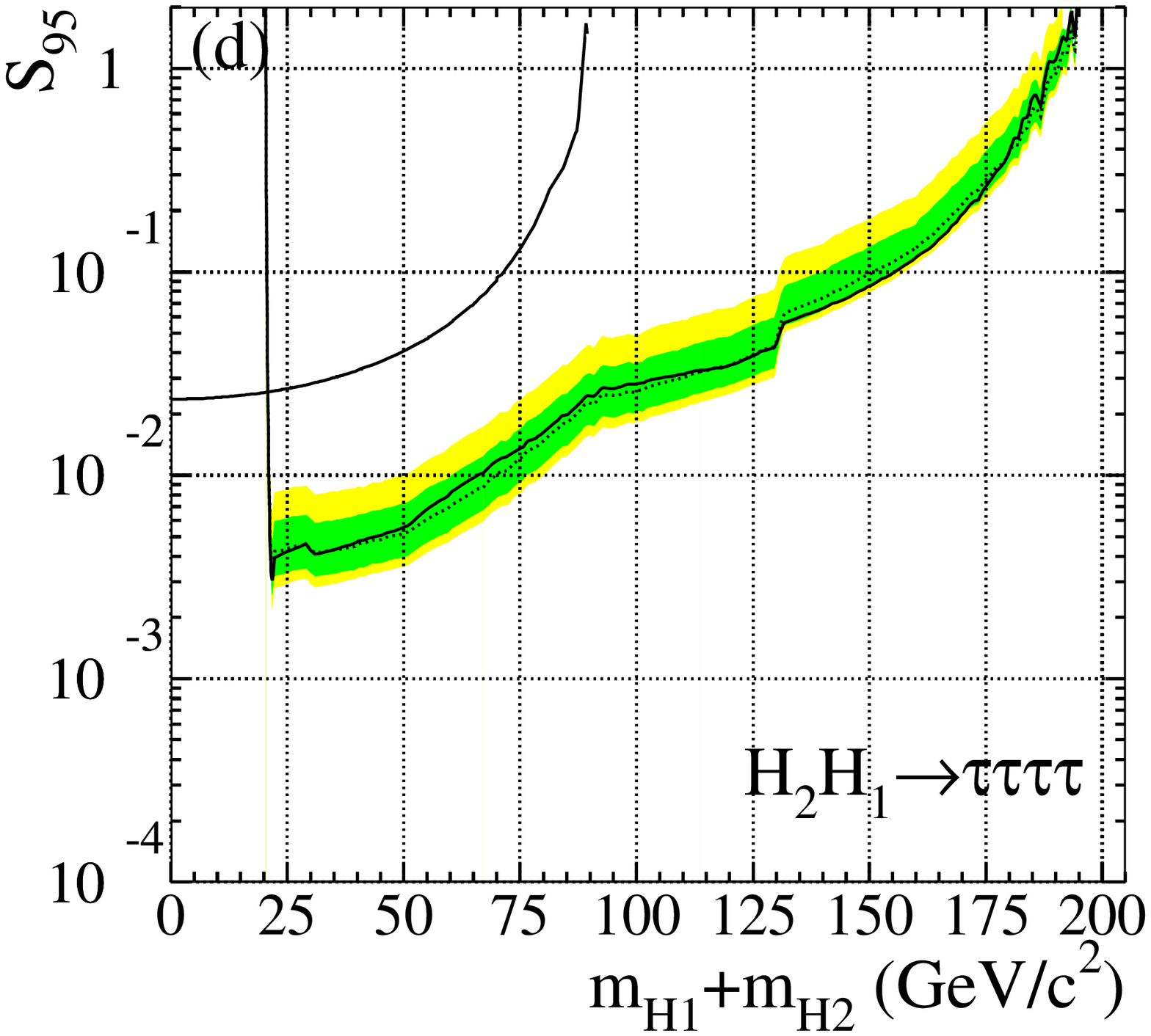,width=0.49\textwidth} 
\end{center}
\caption[]{\sl The 95\% CL upper bounds, $S_{95}$ (see text), 
for various topological cross-sections motivated by the pair production process \ee\ra~\calHb\calHa.
The bounds are obtained 
for the particular case where $\mcalHb$ and $\mcalHa$ are approximately equal. Such is the case, for example, 
in the CP-conserving MSSM scenario {\it \mh-max} for \tanb\ greater than 10 and small $\mcalHb$ ($\equiv$~\mA). 
The abscissa is the sum of the two Higgs boson masses.
The full lines represent the observed limits.
The dark (green) and light (yellow) shaded bands around the median
expectations (dashed lines) correspond to the 68\% and 95\% probability bands. The curves which complete
the exclusions at low masses are obtained using the constraint from the measured decay width of the Z boson, see Section
3.2.  
Plot (a): the Higgs boson decay branching ratios correspond to the {\it \mh-max} benchmark scenario with \tanb=10, namely
94\% \calHa\ra~\bb, 6\% \calHa\ra~\tautau, 92\% \calHb\ra~\bb\
and 8\% \calHb\ra~\tautau; 
plot (b): both Higgs bosons are assumed to decay exclusively to \bb; 
plot (c): one of the Higgs bosons is assumed to decay exclusively to \bb\ and the other exclsively to
\tautau; plot (d): both Higgs bosons are assumed to decay exclusively to \tautau.
\label{eta-h2h1}}
\end{figure}

Figure~\ref{eta-h2h1} shows $S_{95}$ for final states motivated by the pair-production process \ee\ra~\calHb\calHa, 
for the
particular case where the masses \mcalHb\ and \mcalHa\ are approximately equal. 
Such is the case, for example, in the CP-conserving MSSM scenario {\it \mh-max} for \tanb\ larger than 
about 10 and small \mcalHb~($\equiv$~\mA).
In part (a), the \calHb\ and \calHa\ decay
branching ratios correspond to the {\it \mh-max} benchmark scenario with \tanb~= 10 
(see the caption for the exact values); 
in part (b), both \calHb\ and \calHa\
are assumed 
to decay exclusively to \bb; in part (c), one Higgs boson is
assumed to decay exclusively to \bb\ while the other exclusively to \tautau; in part (d), \calHb\ and \calHa\
are both assumed to decay exclusively to \tautau.
At low masses, the exclusion limits are completed using the constraint from the measured decay width of the Z boson
(see Section 3.2). 
This figure also illustrates the overall agreement between the data and the expected background from
Standard Model processes since the largest deviations are within two standard deviations.

\begin{figure}[htb]
\begin{center}
\epsfig{figure=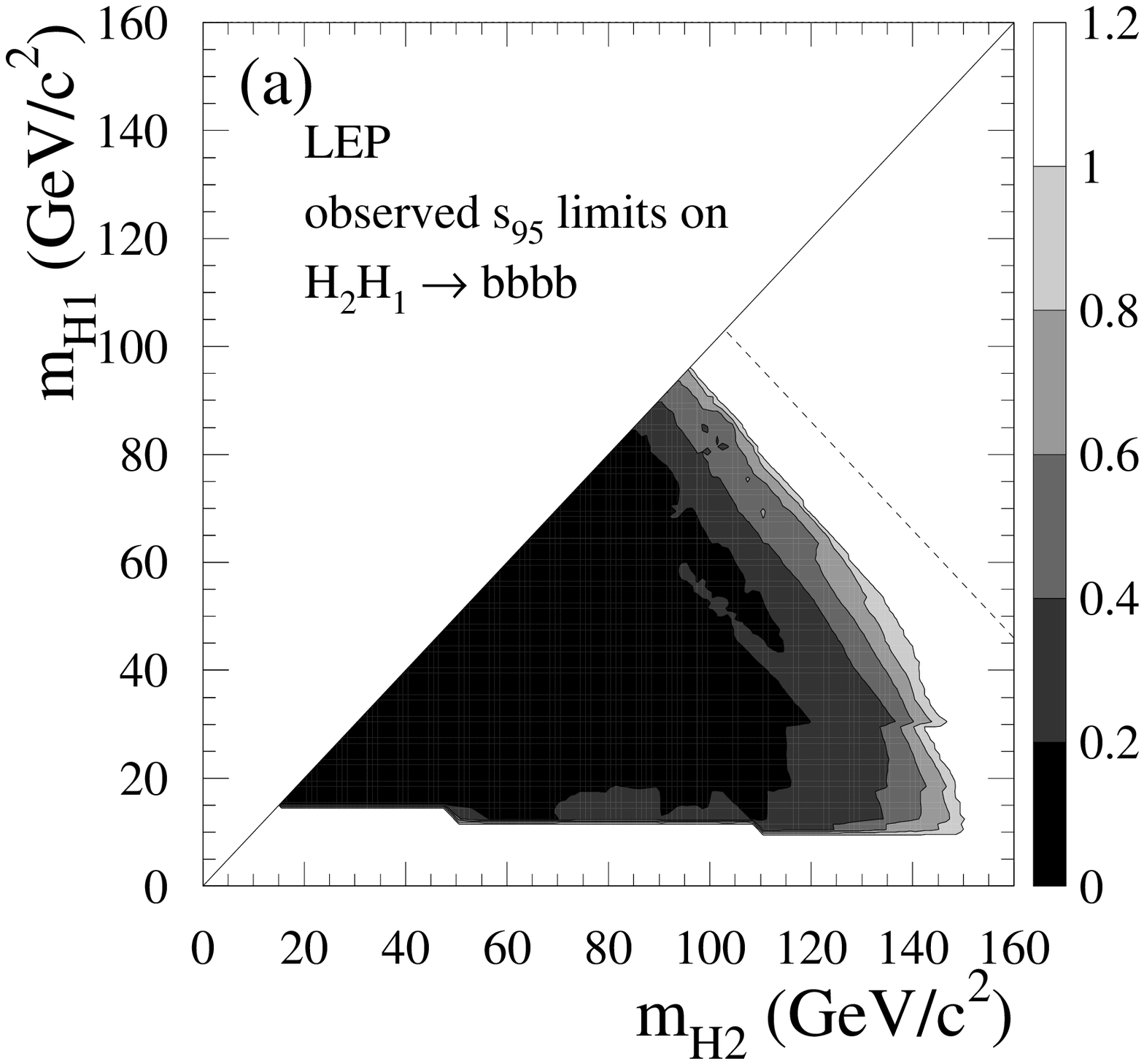,width=0.49\textwidth} 
\epsfig{figure=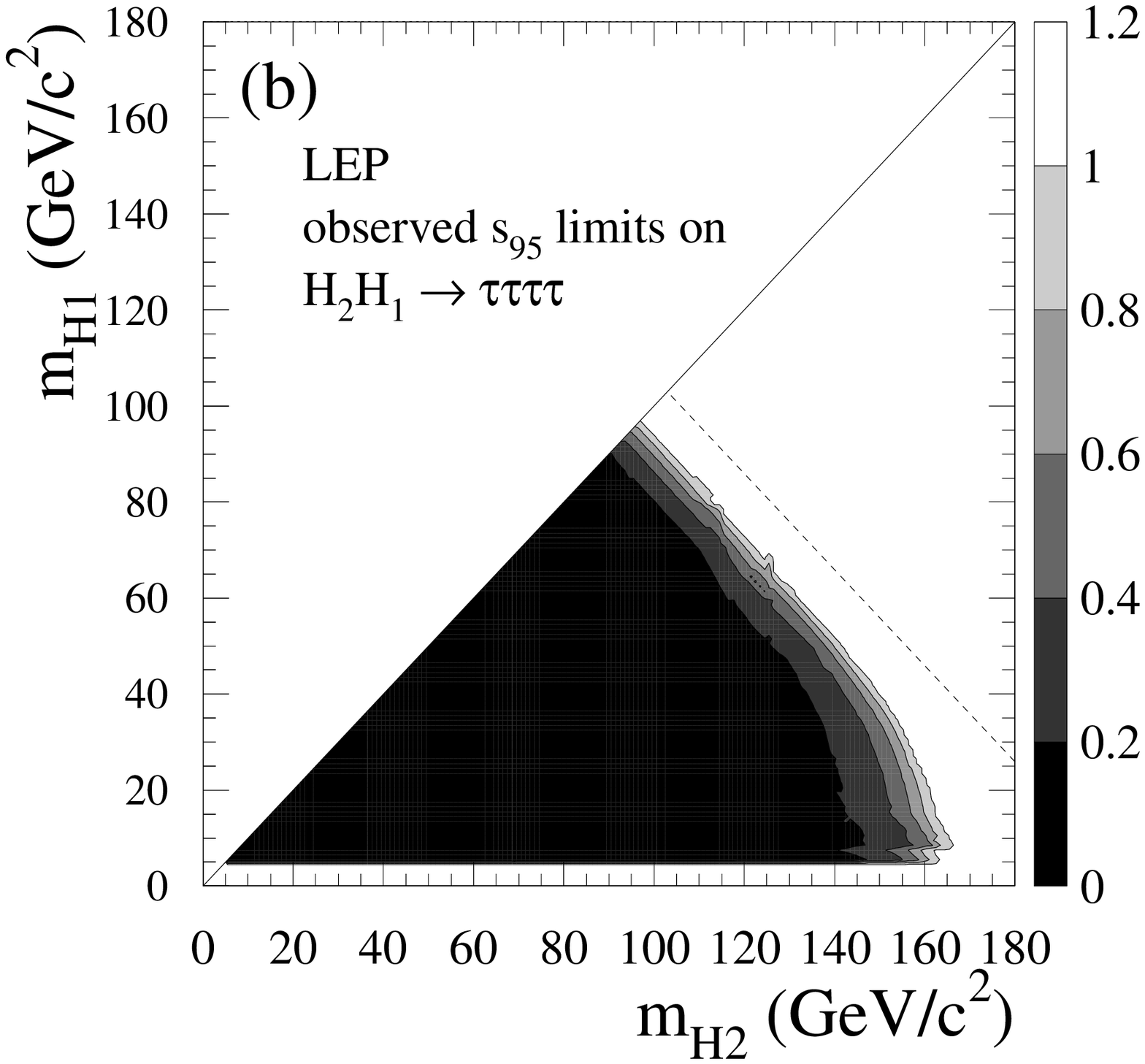,width=0.49\textwidth}\\
\epsfig{figure=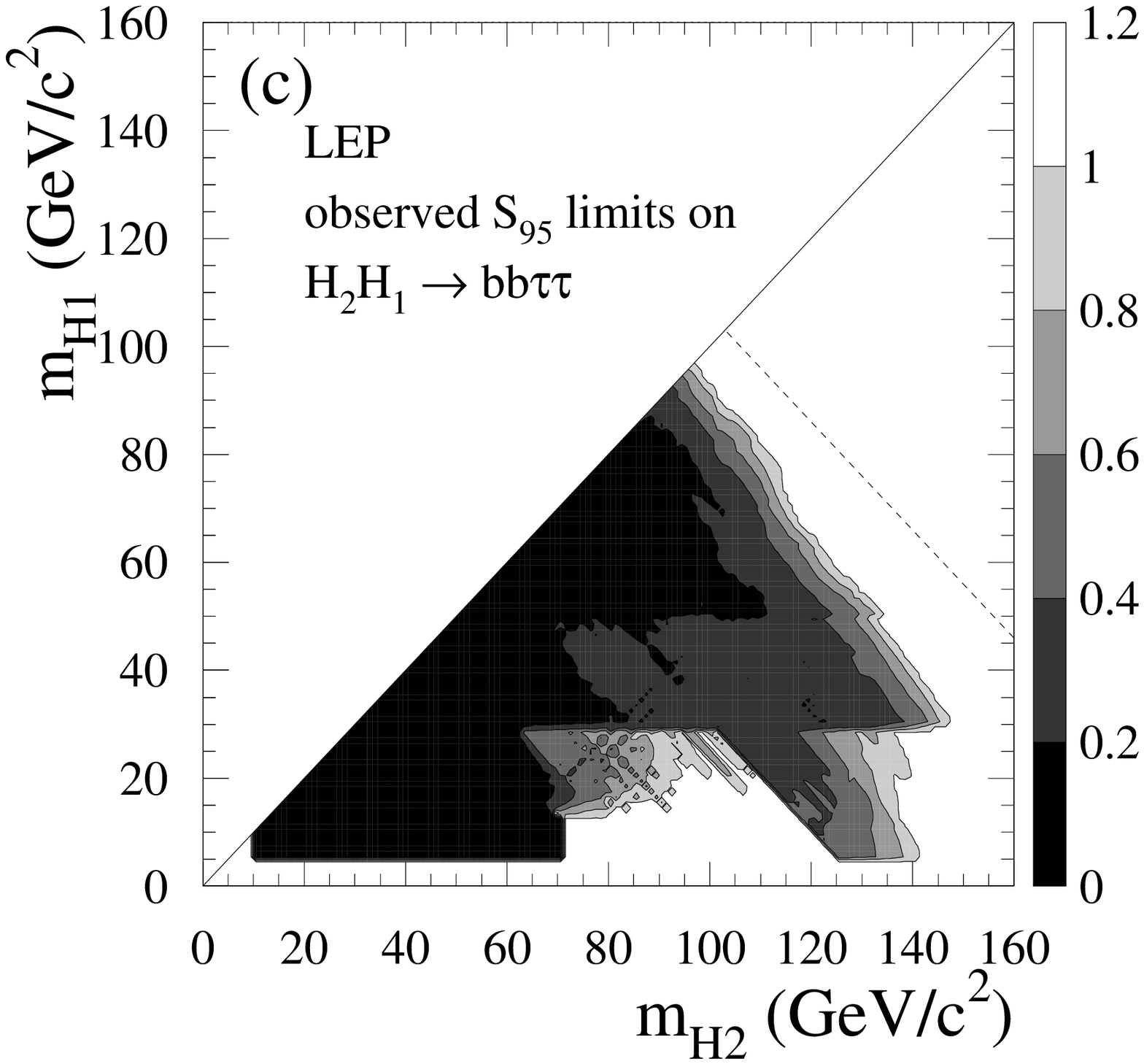,width=0.49\textwidth}
\epsfig{figure=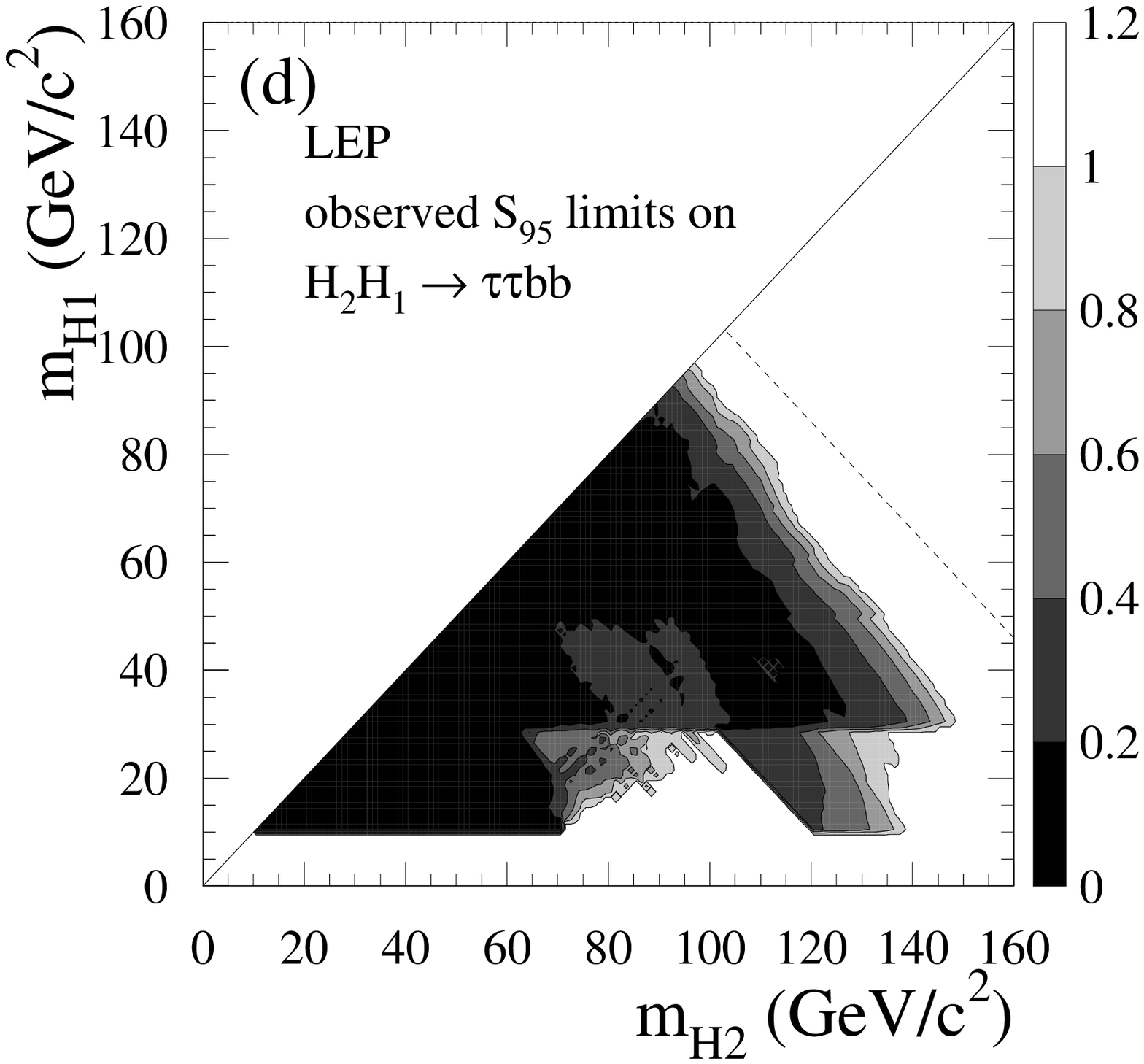,width=0.49\textwidth}
\end{center}
\caption[]{\sl Contours of the 95\% CL upper bound, $S_{95}$ (see text), 
for various topological cross-sections motivated by the pair production 
process \ee\ra~\calHb\calHa, projected onto the (\mcalHb, \mcalHa) plane.  
The scales in terms of the shadings are given on the right-hand side of each plot.
In plot (a) both Higgs bosons are assumed to decay exclusively to \bb\ and in plot (b)
exclusively to \tautau. In plot (c)
the \calHb\ boson is assumed to decay exclusively to \bb\ and the \calHa\ boson exclusively to \tautau\ and in plot (d)
the \calHa\ boson is assumed to decay exclusively to \bb\ and the \calHb\ boson exclusively to \tautau.
The dashed lines represent the approximate kinematic limits of the processes.
\label{c2-h2h1}}
\end{figure}

Figure~\ref{c2-h2h1} shows contours of $S_{95}$  for final states motivated by the
process \ee\ra~\calHb\calHa, projected onto the (\mcalHb, \mcalHa) plane. In part (a), both Higgs bosons are assumed
to decay exclusively to \bb\ and in part (b) exclusively to \tautau. In parts (c) / (d), the \calHb\ / \calHa\ boson 
is assumed to decay exclusively to \bb\ while the other boson is assumed to decay exclusively to \tautau.

\begin{figure}[htb]
\begin{center}
\epsfig{figure=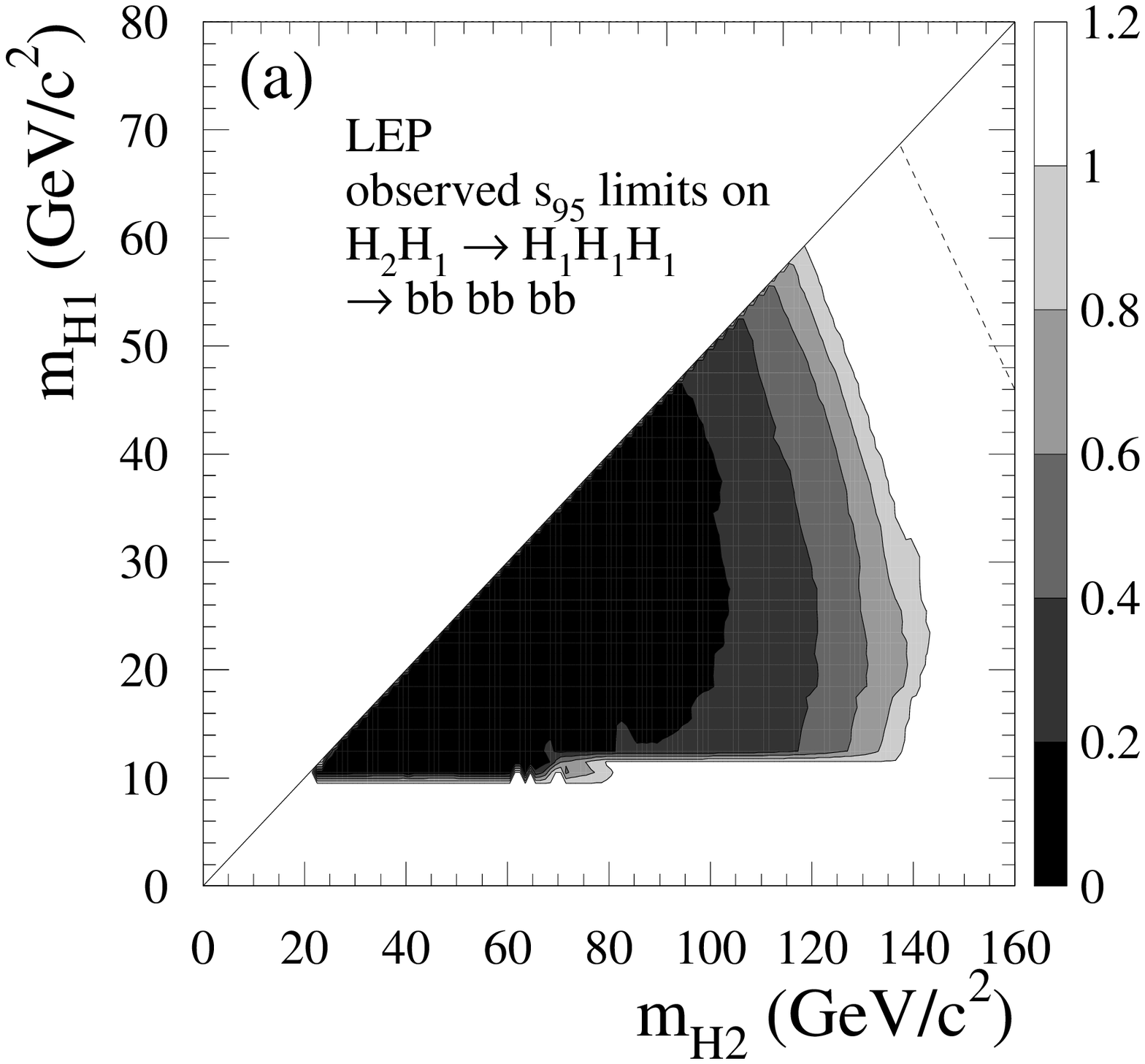,width=0.49\textwidth} 
\epsfig{figure=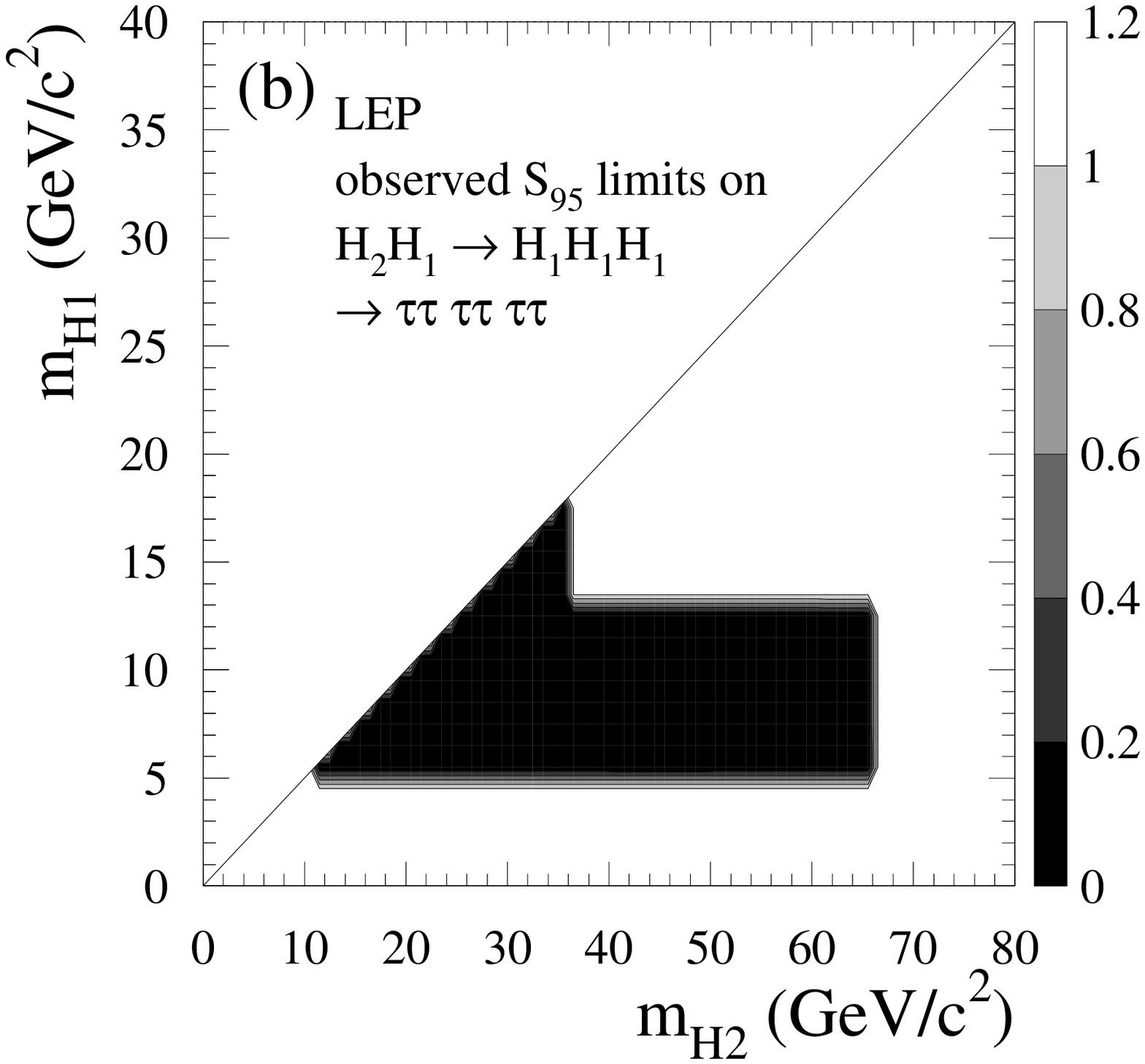,width=0.49\textwidth} \\
\epsfig{figure=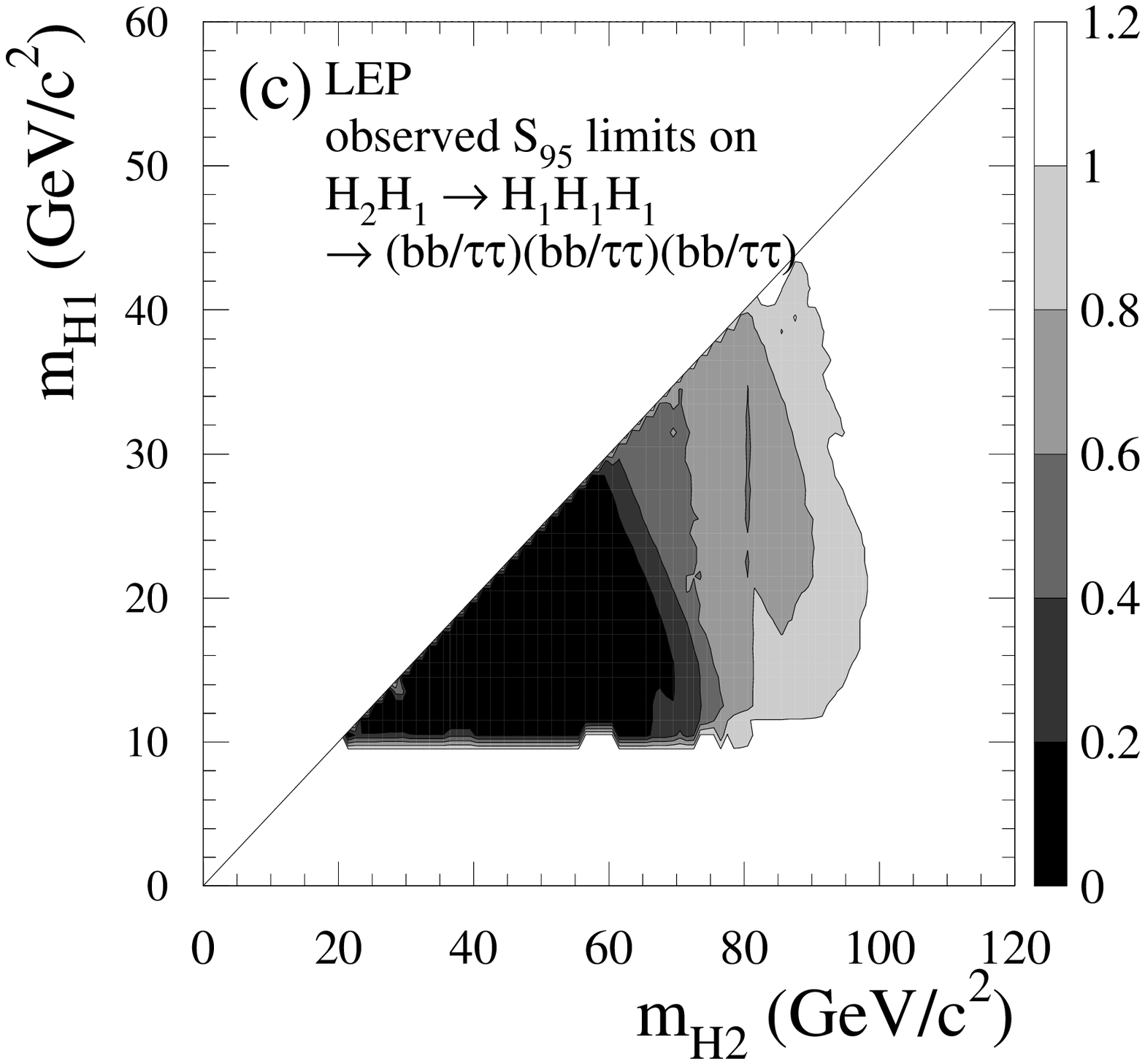,width=0.49\textwidth} 
\end{center}
\caption[]{\sl Contours of the 95\% CL upper bound, $S_{95}$ (see text), 
for various topological cross-sections motivated by the 
pair production cascade process \ee\ra~(\calHb\ra~\calHa\calHa)\calHa, projected onto the (\mcalHb, \mcalHa) plane.
The scales in terms of the shadings are given on the right-hand side of each plot.
In plot (a) the \calHa\ boson is assumed to decay exclusively to \bb\ and in plot (b)
exclusively to \tautau. In plot (c) the \calHa\ boson is assumed to decay with equal probability to \bb\ and to \tautau.
The dashed line in part (a) represents the approximate kinematic limit of the process.
\label{c2-h1h1h1}}
\end{figure}

Figure~\ref{c2-h1h1h1} shows contours of $S_{95}$  
for the cascade process \ee\ra~\calHb\calHa\ra~(\calHa\calHa)\calHa, projected onto the (\mcalHb, \mcalHa) plane,
assuming that the \calHb\ boson decays exclusively to \calHa\calHa. 
In part (a), the \calHa\ boson is assumed to decay exclusively to \bb\ and in part (b) exclusively to \tautau.
In part (c), as an example, an equal mixture of \calHa\ra~\bb\ and \calHa\ra~\tautau\ is assumed,
which implies 12.5\% \bb\bb\bb, 
37.5\% \bb\bb\tautau, 37.5\% bb\tautau\tautau\ and 12.5\% \tautau\tautau\tautau\
final states.

A word of caution is in place concerning the correlations which exist between some of the above cross-section limits
which arise from overlapping candidates in the corresponding selections.
Such correlations are present, for example, between b-tagged and flavour-independent searches 
of a given experiment or between searches addressing direct decays (e.g., \calHa\Z\ra~\bb\bb) 
and cascade decays (e.g., (\calHb\ra~\calHa\calHa)\Z\ra~\bb\bb\bb); they may be a source of problems
if several of the cross-section limits are used in conjunction to test a given model. Note, however, 
that these correlations are properly taken into account in the model interpretations which follow. 
\section{Results interpreted in CP-conserving MSSM scenarios}
In this section, the search results are interpreted in the CP-conserving benchmark scenarios  
presented in Section~2.1.
The exclusion limits, which are shown in the figures below at the 95\% CL and the 99.7\% CL, are obtained from the 
values of $CL_s$ (see Eq.~15), for an assumed top quark mass of \mt~=~174.3~\Gcs. The exclusion limits are presented in 
four projections of the MSSM parameter space.  
The limits expected on the 
basis of Monte Carlo simulations with no signal, at the 95\% CL, are also indicated.
The exact mass bounds and exclusions for \tanb\ are listed in Table~\ref{tab:cpc-results}, for four values of \mt.

\begin{figure}[p]
\centerline{
\epsfig{file=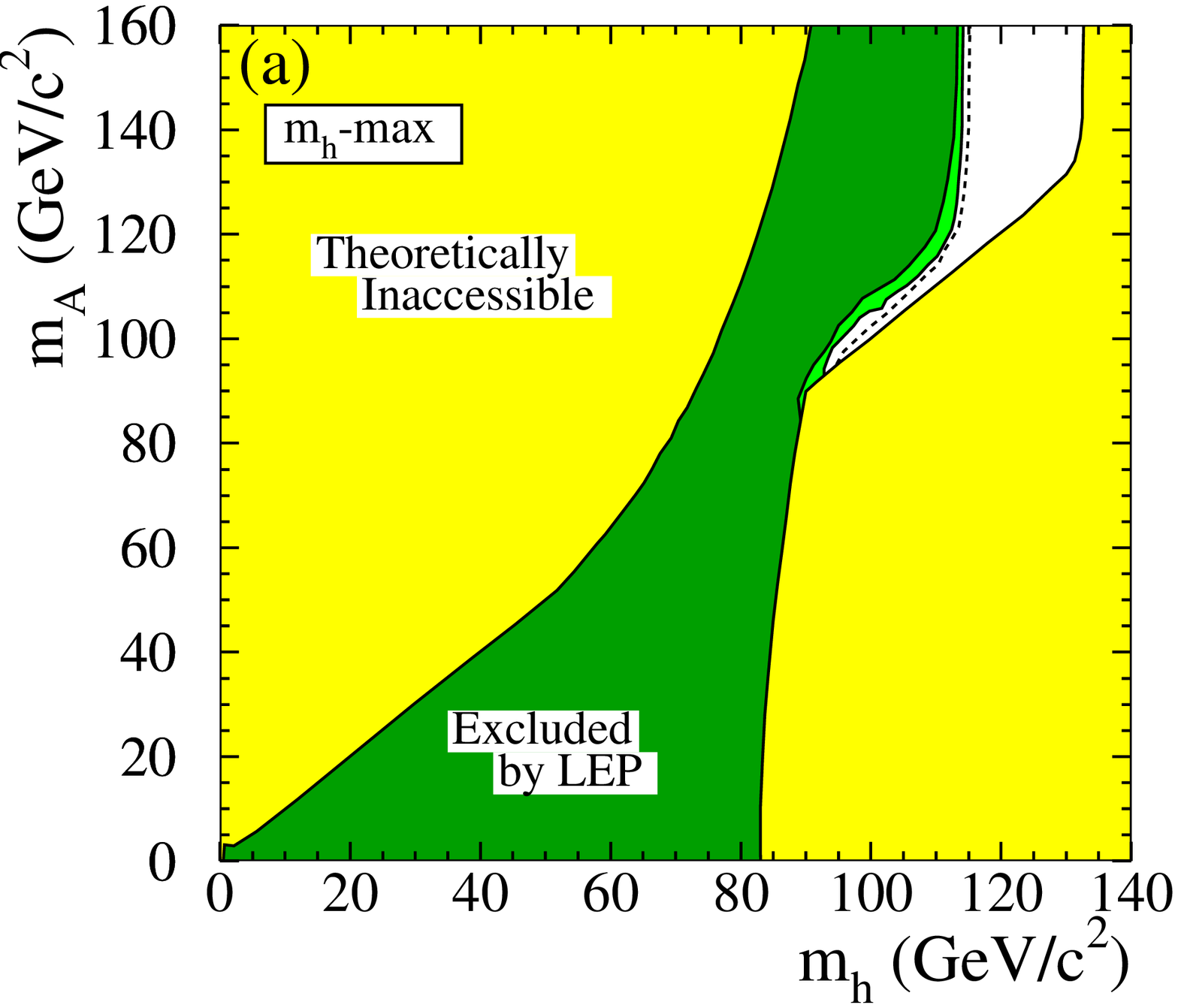,width=0.49\textwidth}
\epsfig{file=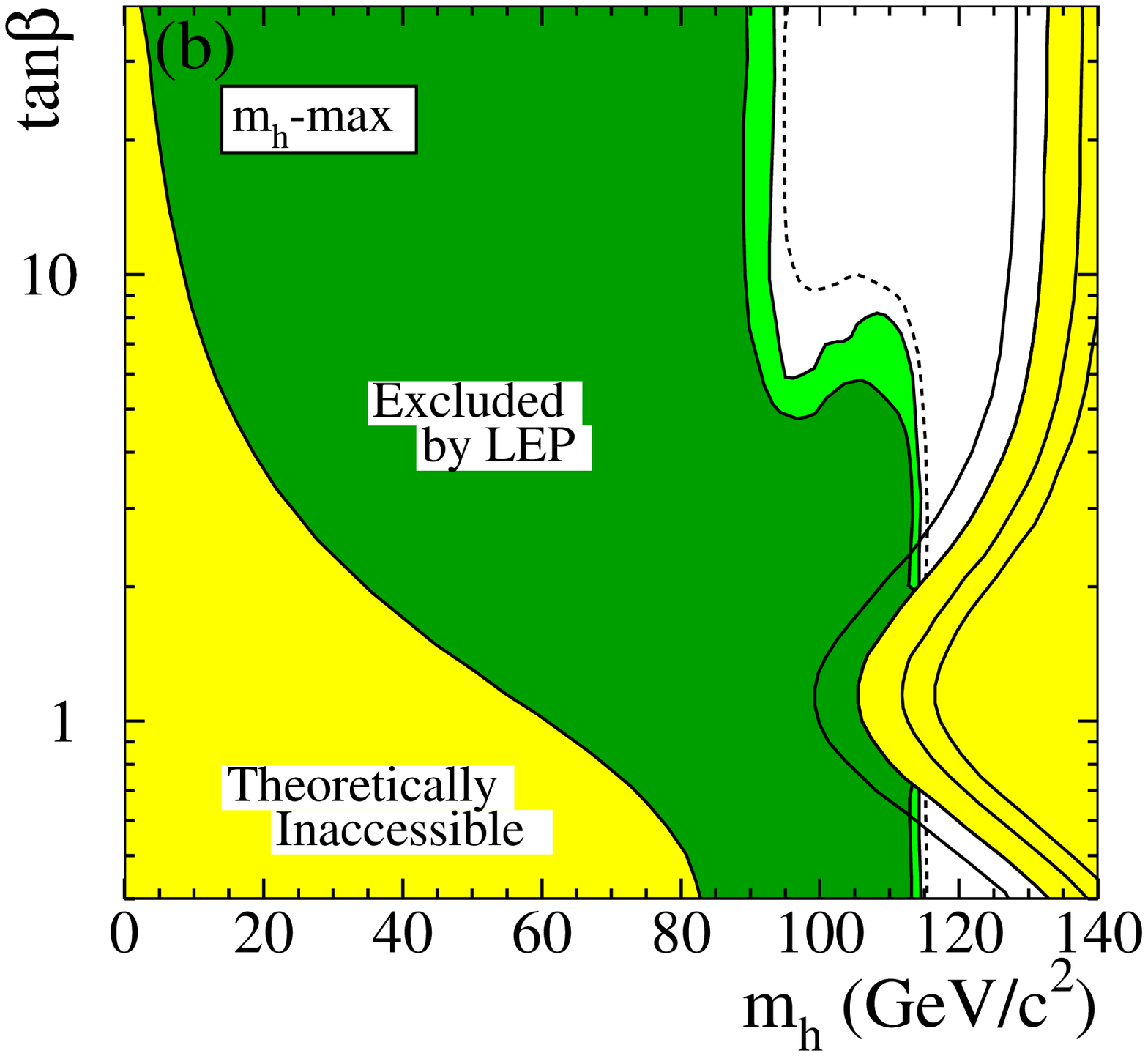,width=0.49\textwidth}
}
\centerline{
\epsfig{file=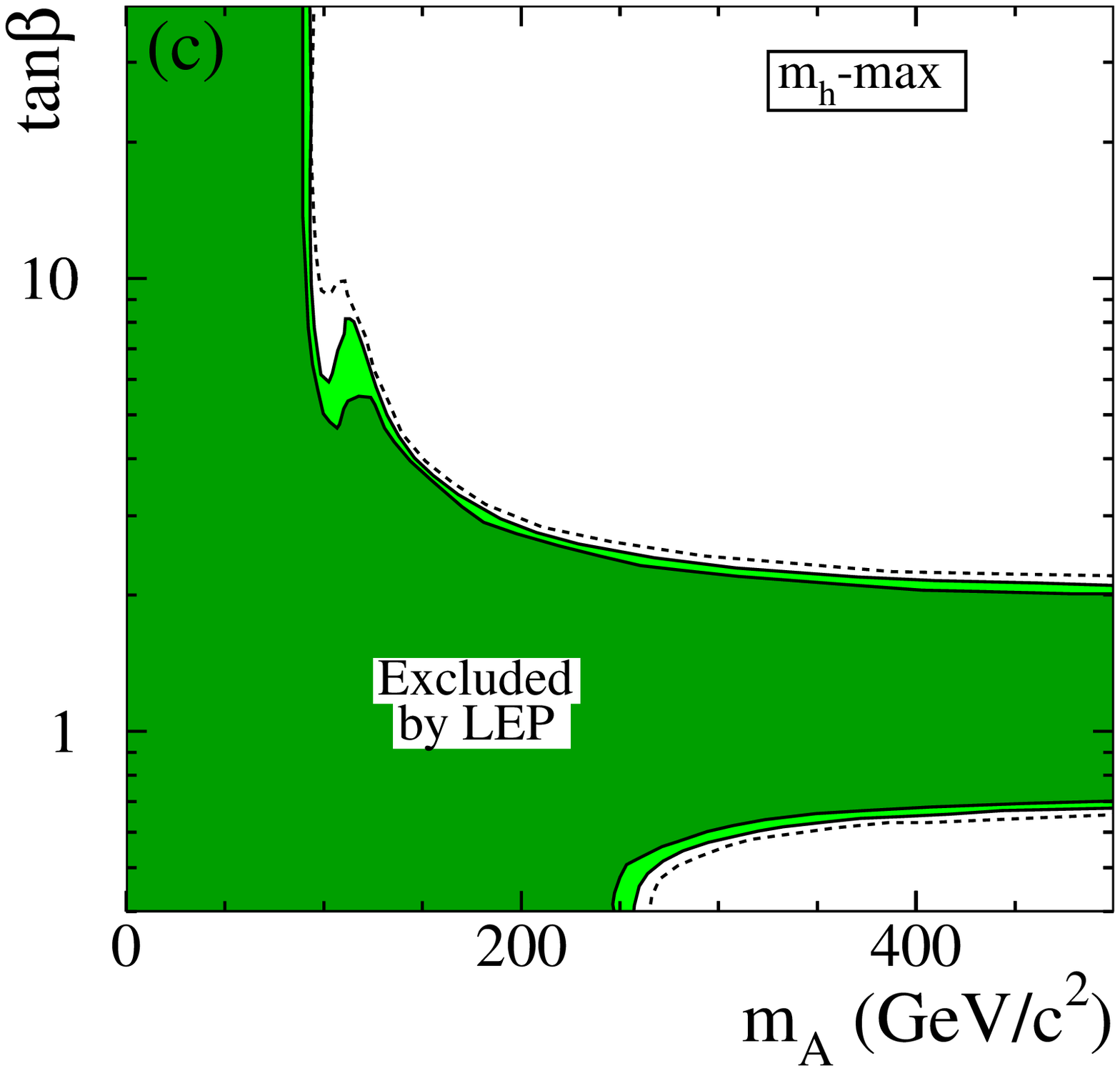,width=0.49\textwidth}
\epsfig{file=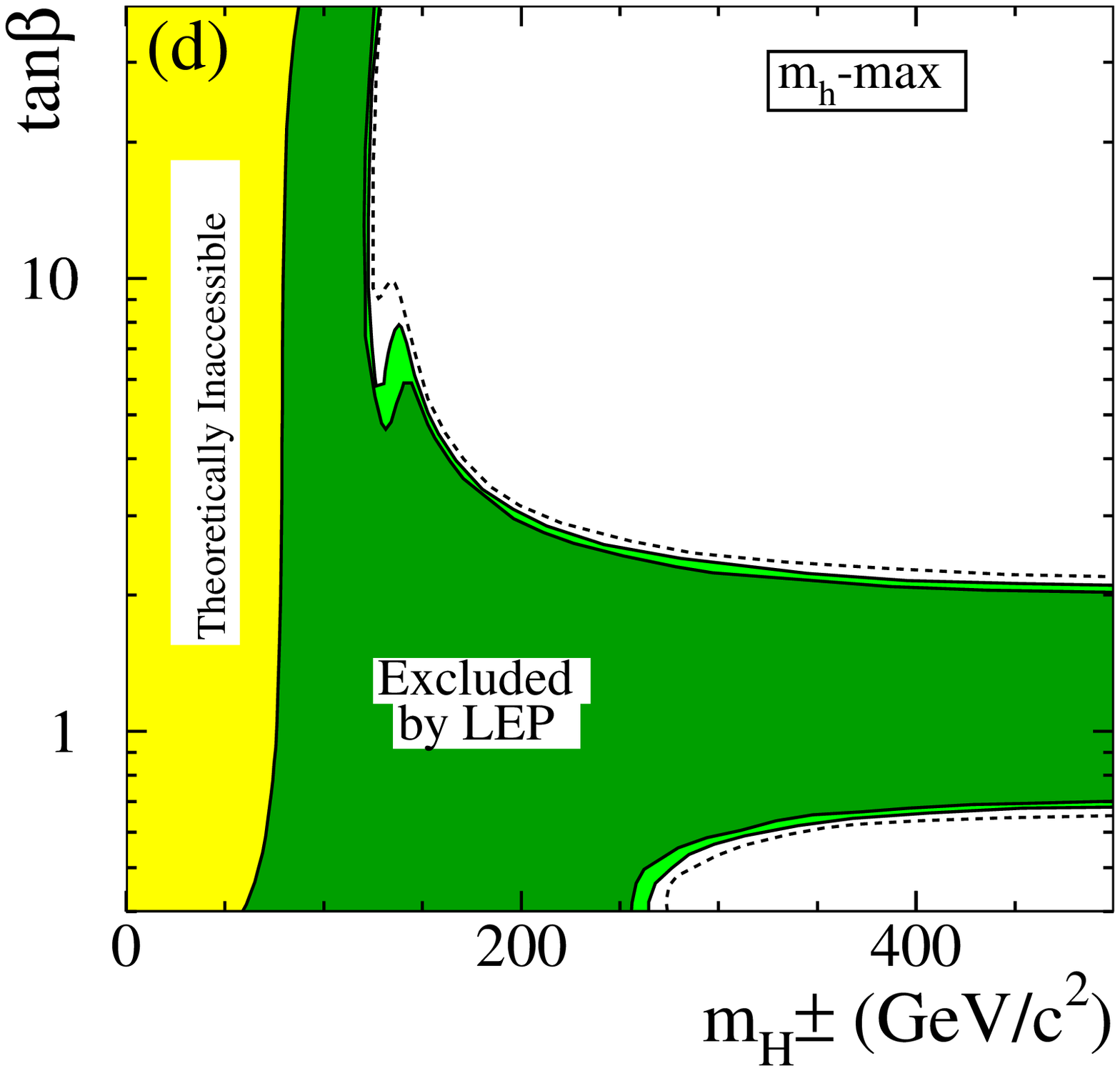,width=0.49\textwidth}
}
\caption[]{\label{fig:mhmax}
\sl  Exclusions, at 95\% CL (medium-grey or light-green) and the 99.7\% CL 
(dark-grey or dark-green), in the case of the CP-conserving {\it \mh-max} benchmark scenario,
  for $m_{\rm t}=174.3$~\Gcs.
  The figure shows the theoretically inaccessible domains (light-grey or yellow)
  and the regions excluded by this search, in four projections of the MSSM parameters:
         (a): (\mh,~\mA); (b): (\mh,~\tanb); (c): (\mA,~\tanb); (d): (\mHpm,~\tanb).
         The dashed lines indicate the boundaries of the
         regions which are expected to be excluded, at 95\% CL,  on the basis of Monte Carlo simulations 
	 with no signal. In the (\mh,~\tanb) projection (plot (b)), the
	 upper boundary of the parameter space is indicated for four values of the top quark mass; from 
	 left to right: $m_{\rm t}$ = 169.3, 174.3, 179.3 and 183.0~\Gcs.
}
\end{figure}

The exclusions for the {\it \mh-max} benchmark scenario are shown in Figure~\ref{fig:mhmax}.
In the region with \tanb\ less than about five, the exclusion is provided mainly by the Higgsstrahlung process,
giving a lower bound of about 114~\Gcs\ for \mh. At high \tanb, the pair production process is most useful, 
providing limits in the vicinity of 93~\Gcs\ for both \mh\ and \mA.
For \mh\ in the vicinity of 100~\Gcs, one observes a deviation between the expected and the experimental exclusions. 
This deviation, which is also present in other CP-conserving scenarios, is due to 
the excess in the Higgsstrahlung channel which was discussed in Ref.~\cite{lep-sm} and gives rise to the vertical bands
in Figures~\ref{fig:hotspot} (a) and (b). Note that the mass bounds obtained are largely insensitive to the top quark mass. 

\begin{figure}[p]
\centerline{
\epsfig{file=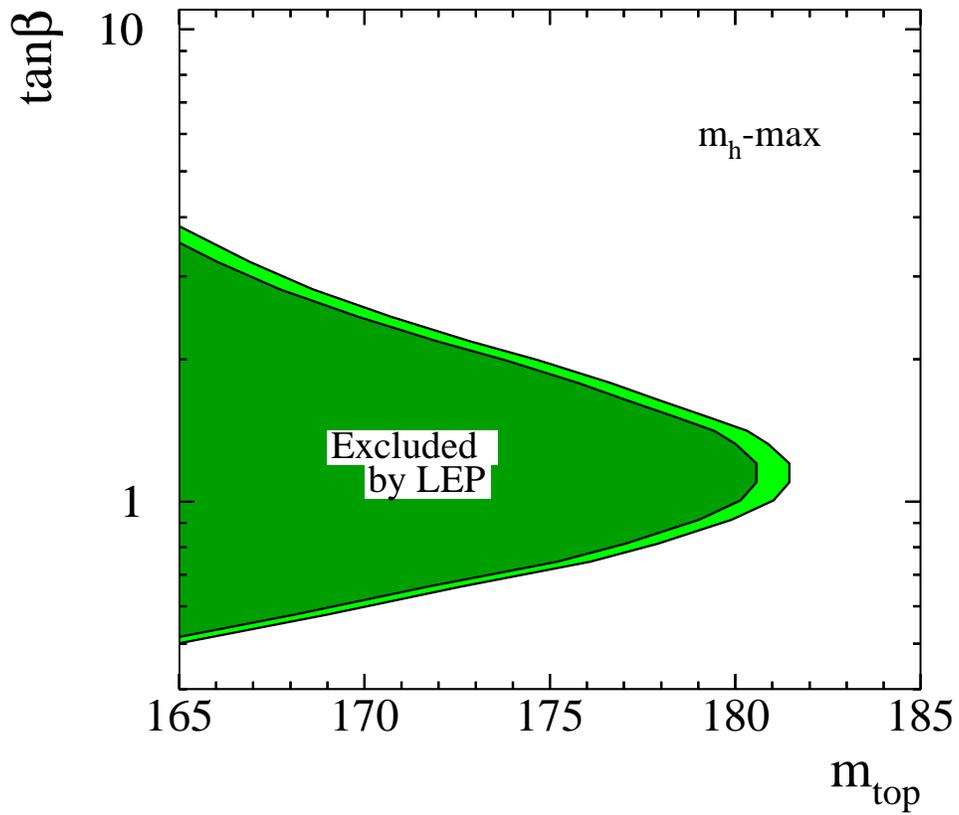,width=0.8\textwidth}
}
\caption[]{\label{mhmax-tanbmtop}
\sl  Domains of \tanb\ which are excluded at the 95\% CL (light-grey or light-green) and the 99.7\% CL (dark-grey or dark-green), 
 in the case of the CP-conserving {\it \mh-max} benchmark scenario, as a function of the assumed top quark mass.
}
\end{figure}

The data also exclude certain domains of \tanb. This is best illustrated in the (\mh, \tanb) projection (plot (b)) 
where the upper boundary of the parameter space along \mh\ is indicated for four values of \mt; the intersections of these 
boundaries with the experimental exclusions define the regions of \tanb\ which are excluded. 
The exclusion in \tanb, as a function of the assumed top quark mass, is
summarised in Figure~\ref{mhmax-tanbmtop};
for \mt\ larger than about 181.5~\Gcs, no 95\% CL limit on \tanb\ can be set in this scenario.

One should be aware that the upper boundary of the parameter space along \mh\ also depends moderately on the choice 
of \msusy. 
For example, changing \msusy\ from 1~TeV to 2~TeV would broaden the parameter space 
by about 2~\Gcs\ along \mh, with corresponding effects on the exclusions in \tanb. This observation holds 
for all CP-conserving scenarios which follow.

\begin{figure}[p]
\centerline{
\epsfig{file=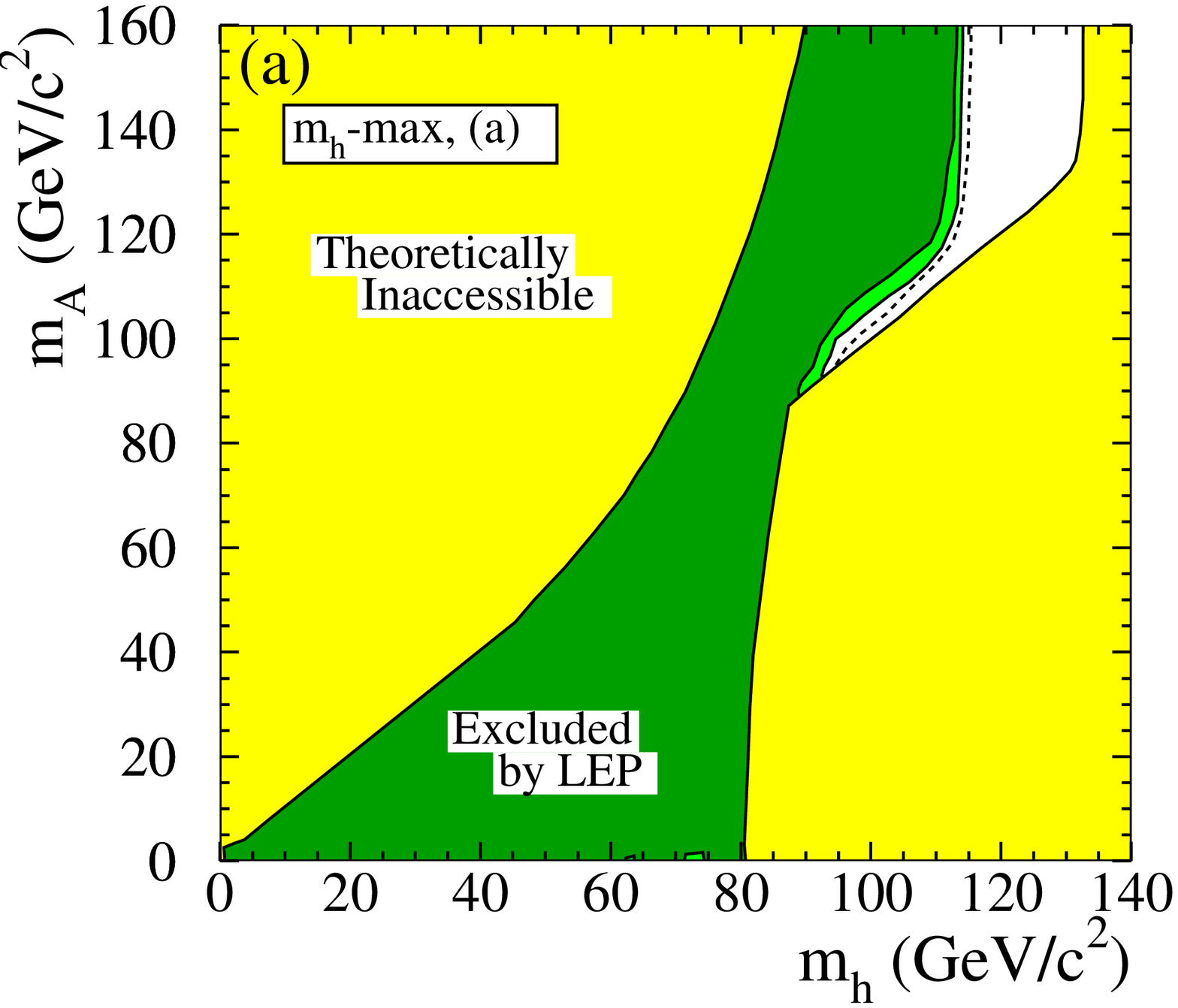,width=0.49\textwidth}
\epsfig{file=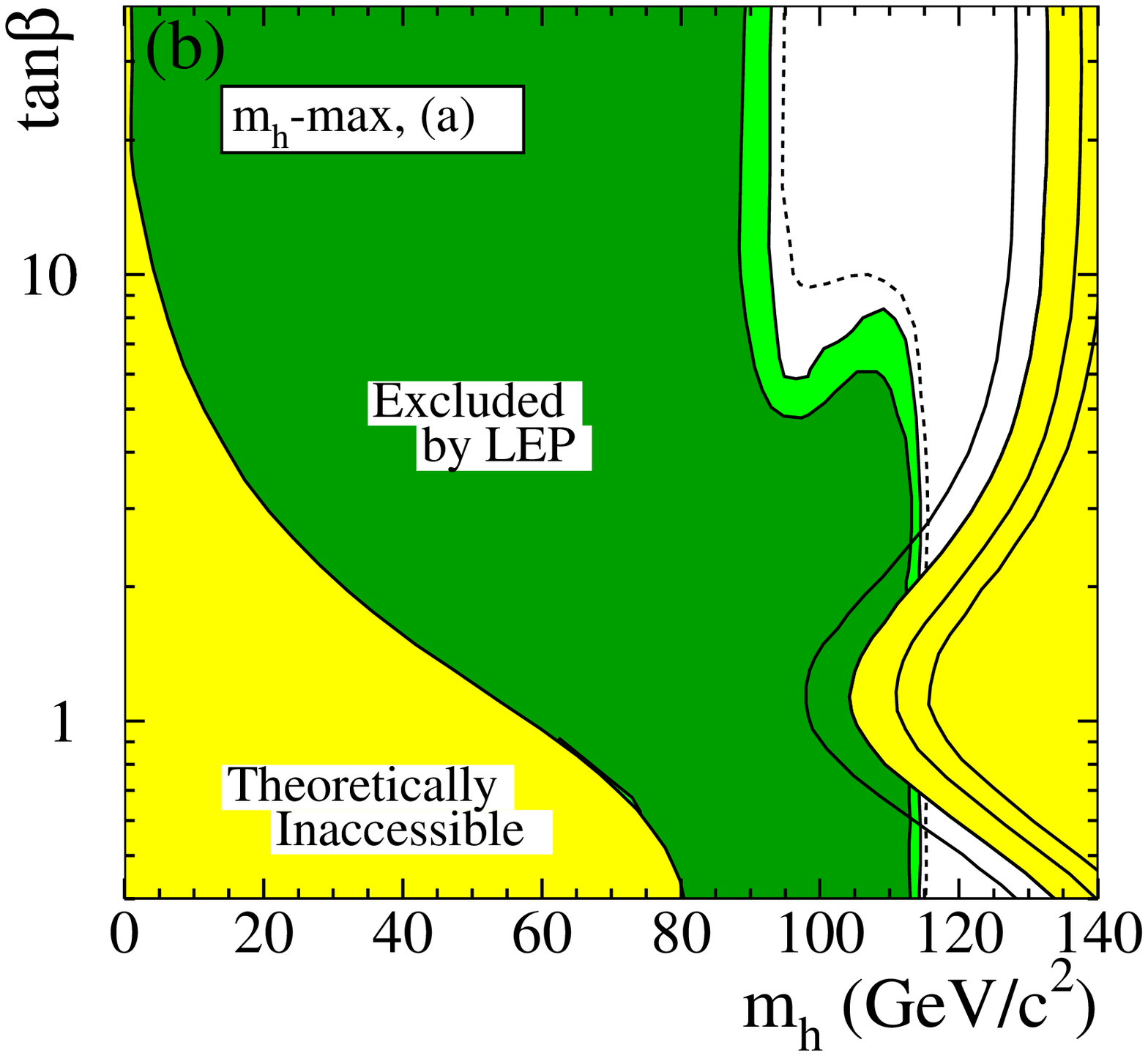,width=0.49\textwidth}
}
\centerline{
\epsfig{file=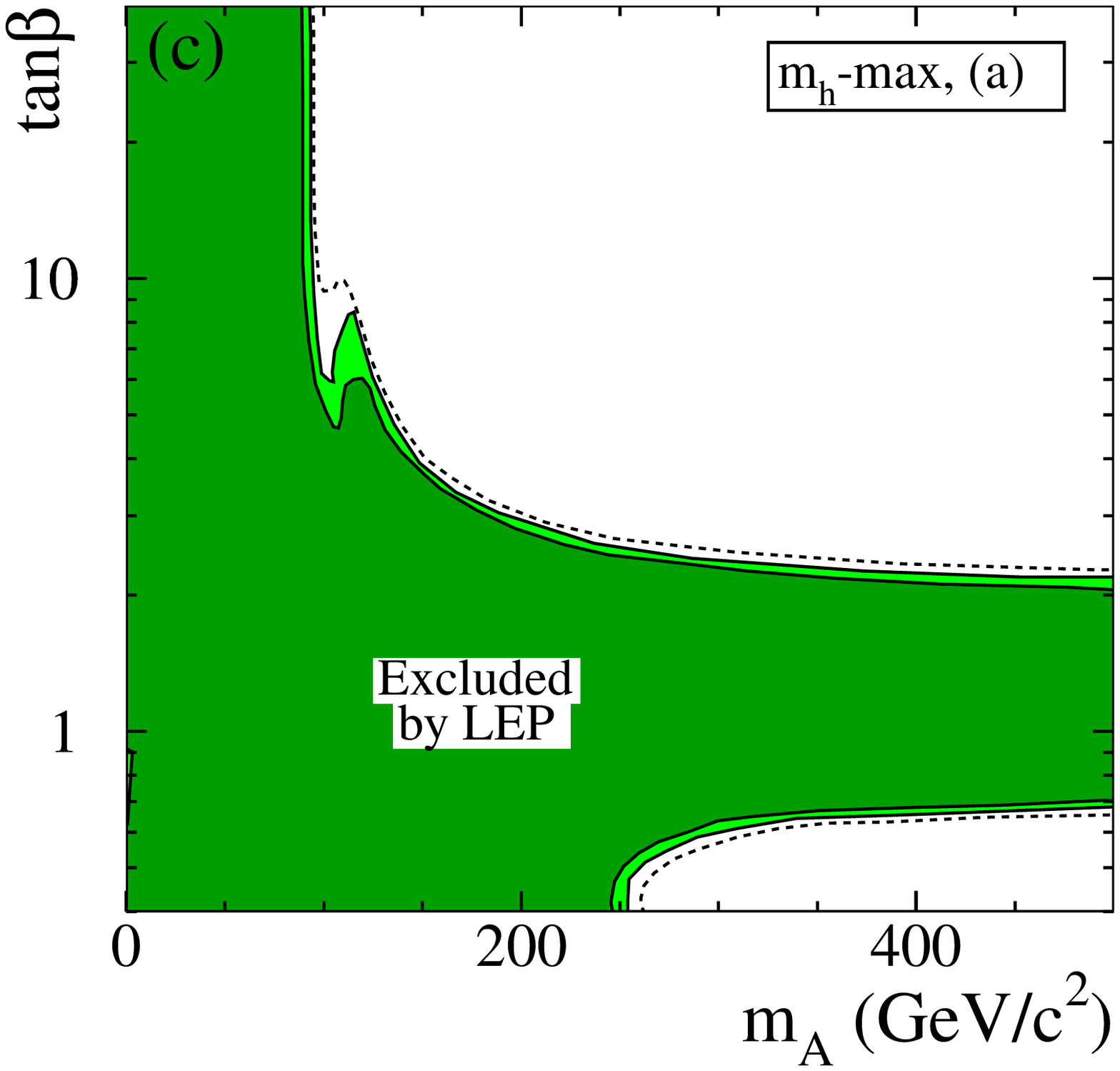,width=0.49\textwidth}
\epsfig{file=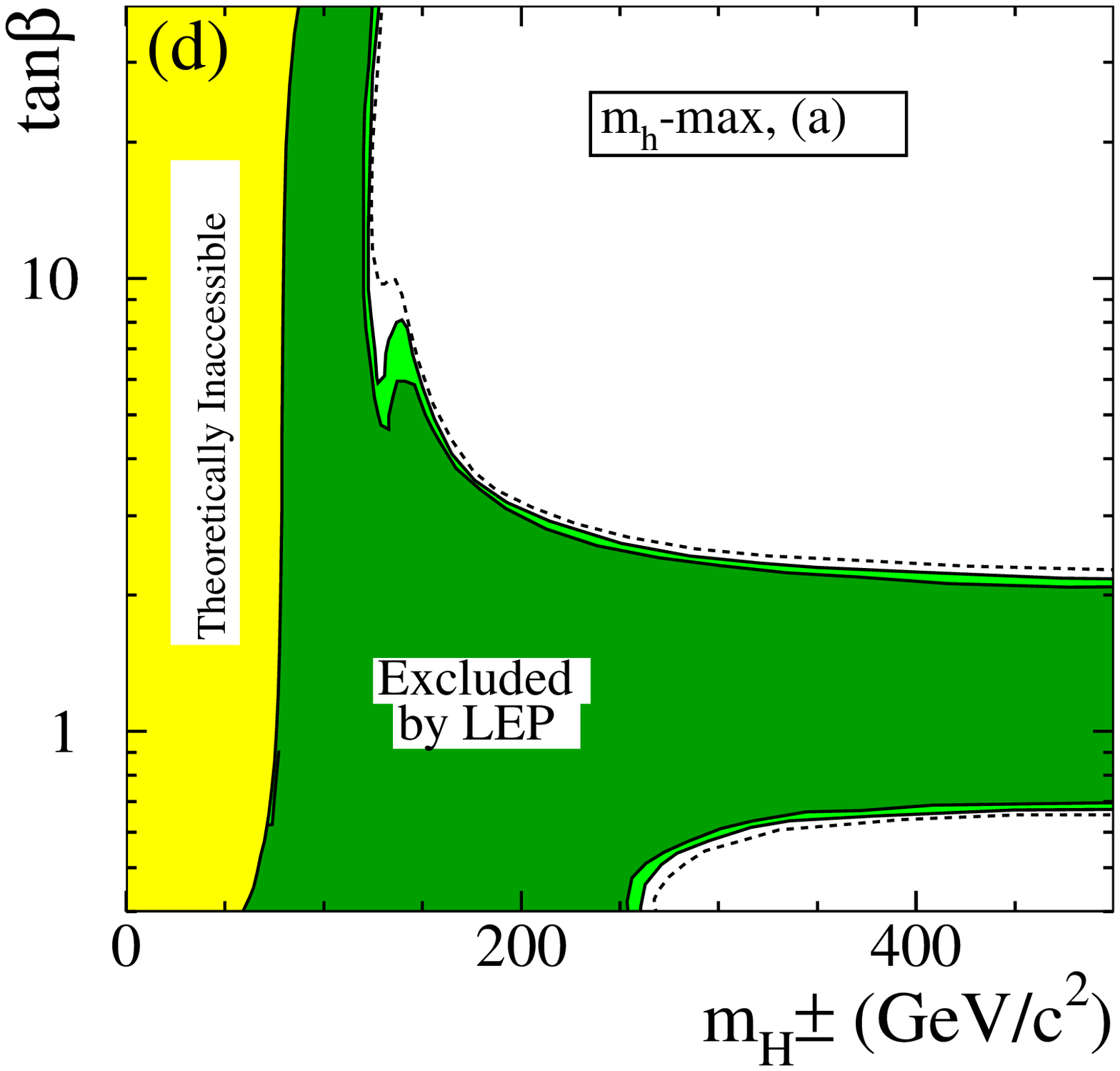,width=0.49\textwidth}
}
\caption[]{\label{fig:mhmax-a}
\sl  Exclusions in the case of the CP-conserving {\it \mh-max} benchmark scenario,
variant (a) (see Section 2.1.1.).
See the caption of Figure~\ref{fig:mhmax} for the legend.
 Note the small domains at \mh\ between 60 and 75~\Gcs, small \mA\ and \tanb~$<$~0.9 
which, although excluded at the 95\% CL, are not excluded at the 99.7\% CL.
}
\end{figure}

\begin{figure}[p]
\centerline{
\epsfig{file=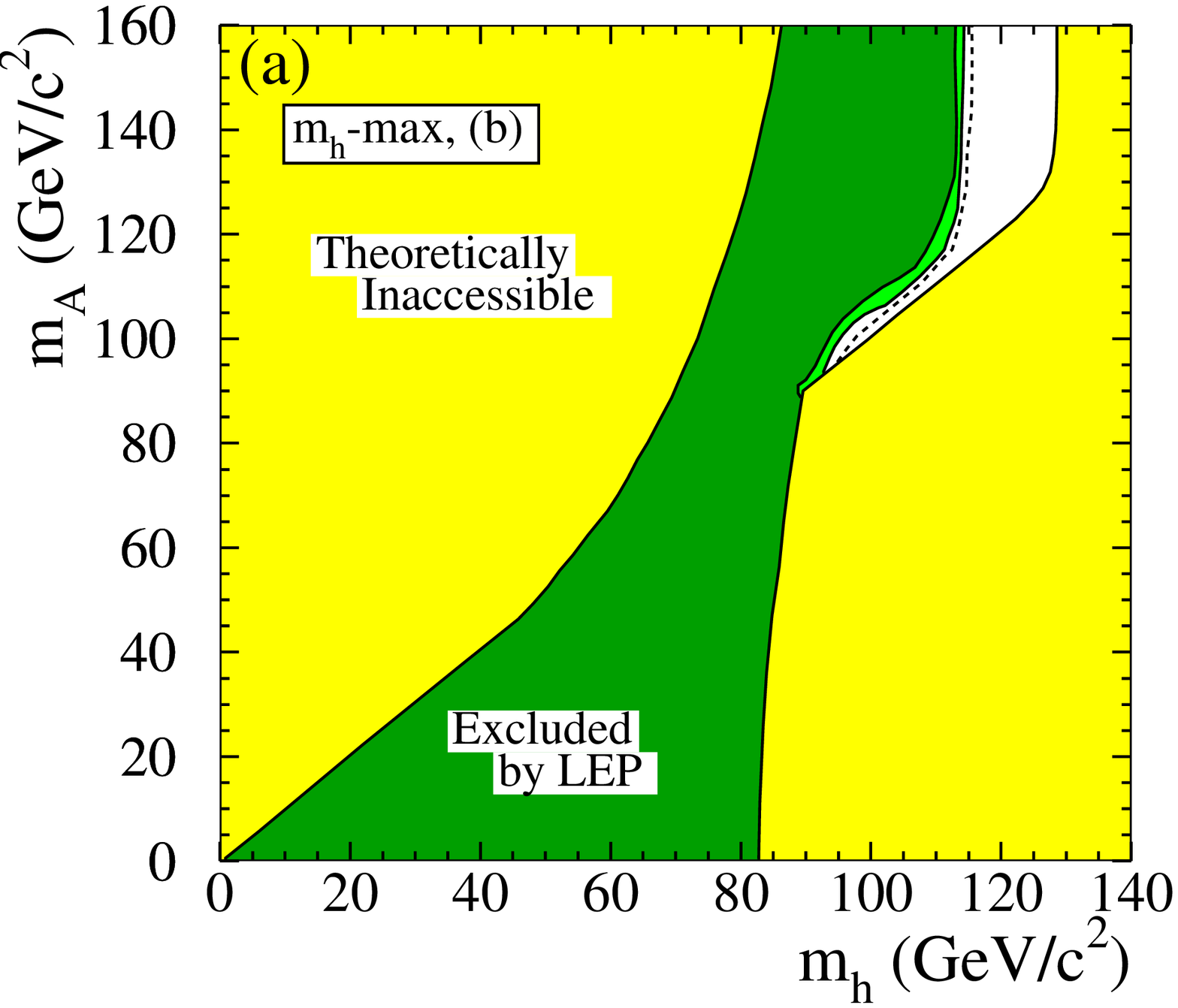,width=0.49\textwidth}
\epsfig{file=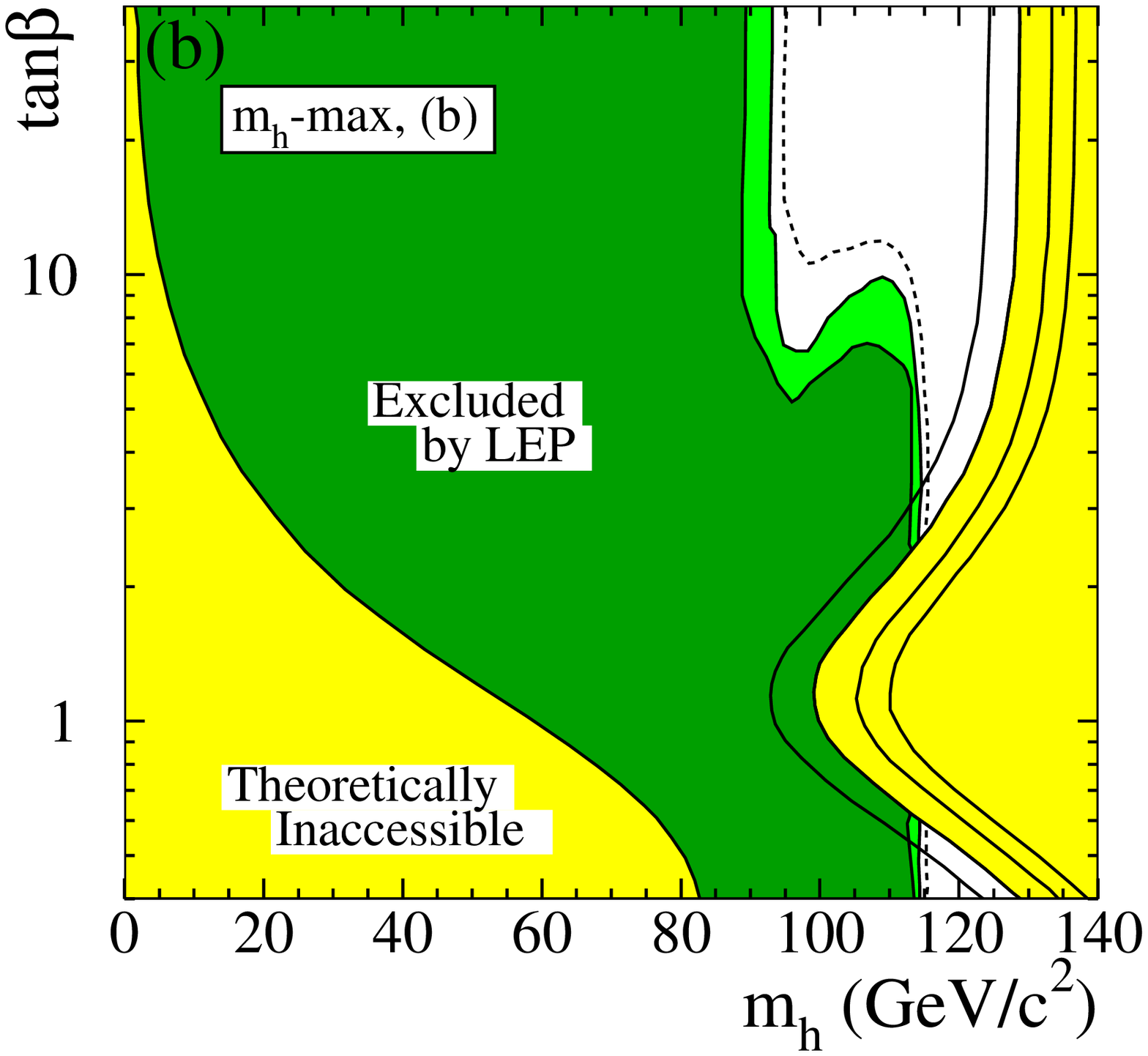,width=0.49\textwidth}
}
\centerline{
\epsfig{file=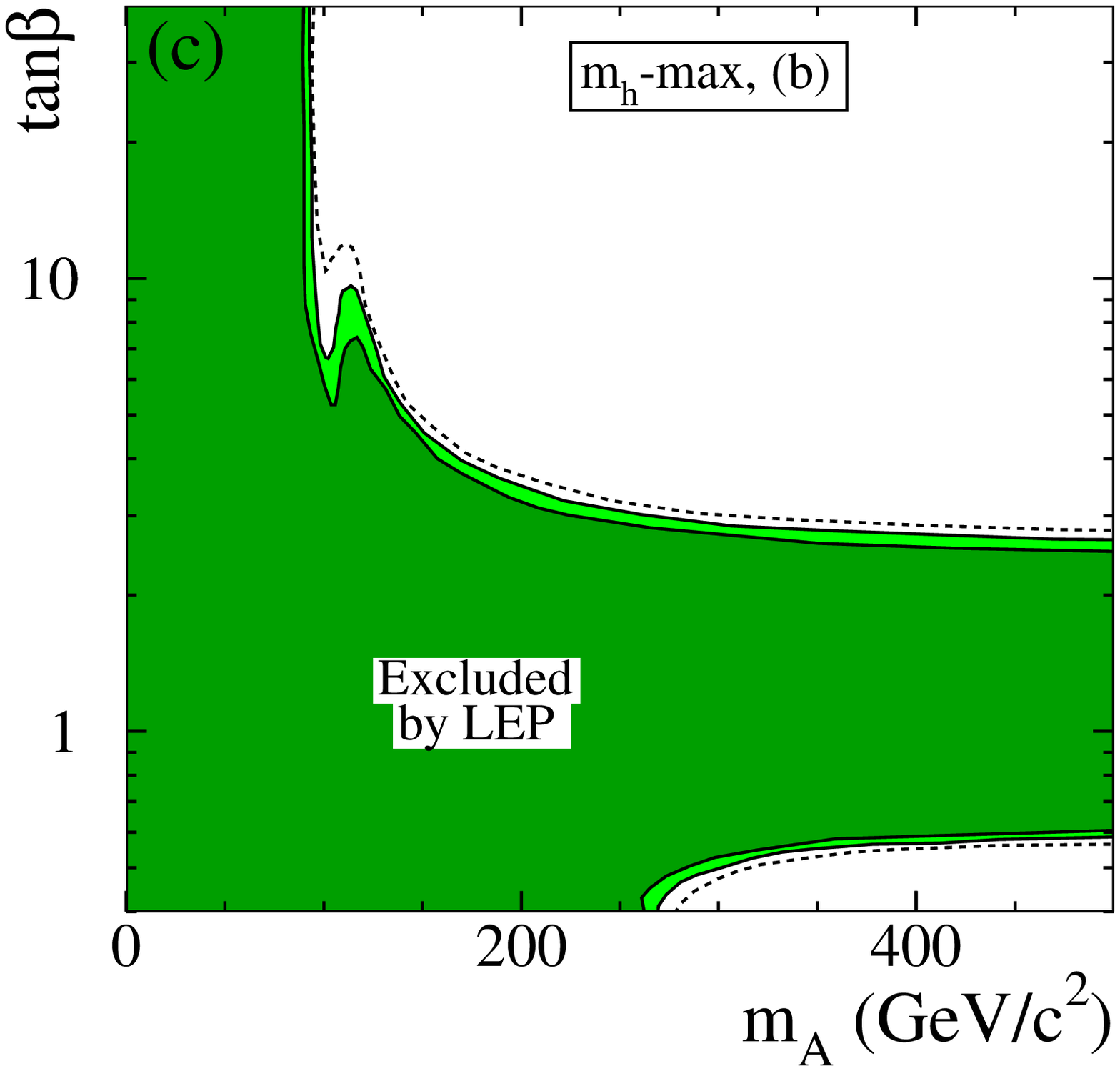,width=0.49\textwidth}
\epsfig{file=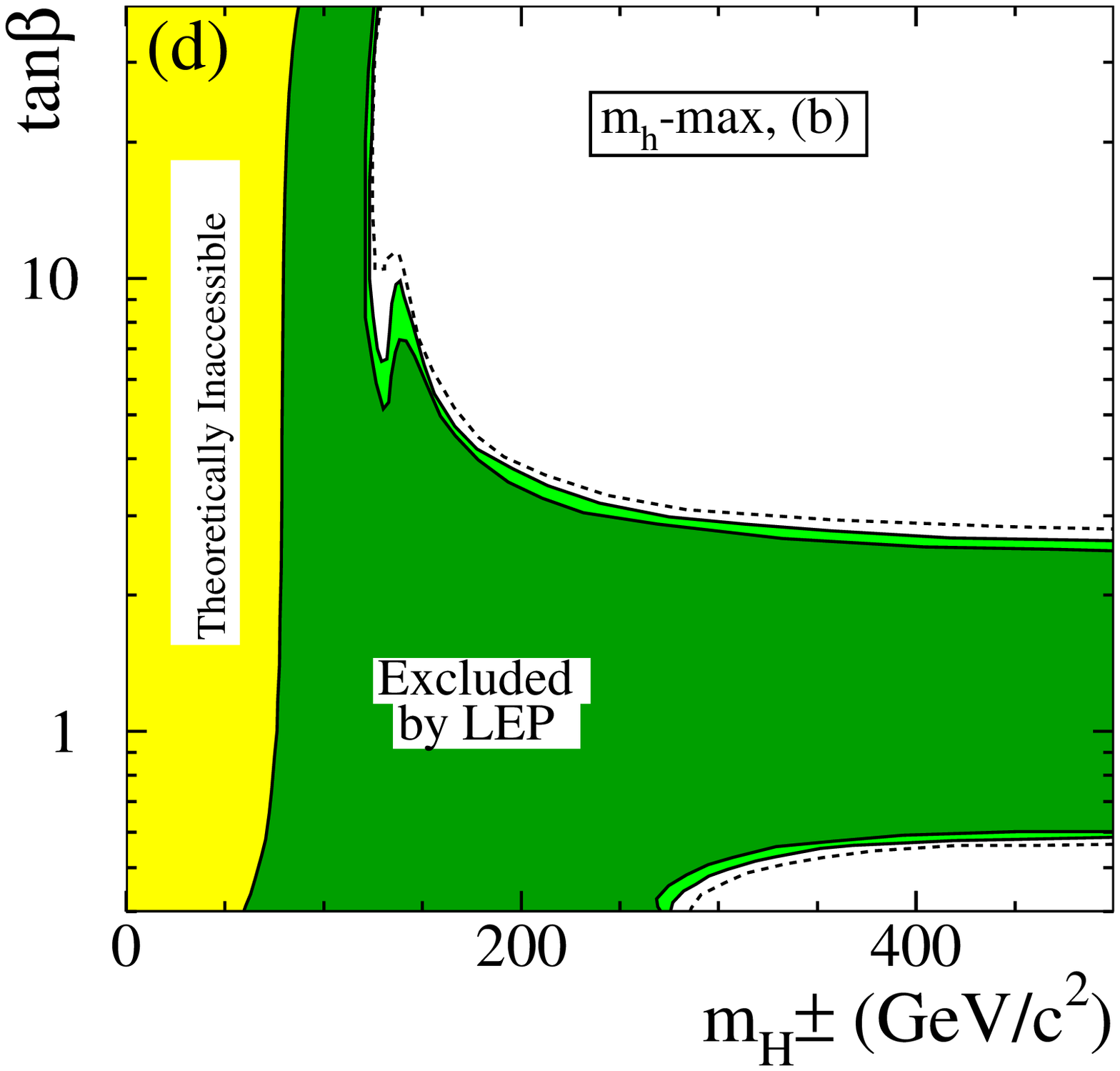,width=0.49\textwidth}
}
\caption[]{\label{fig:mhmax-b}
\sl  Exclusions in the case of the CP-conserving {\it \mh-max} benchmark scenario,
variant (b) (see Section 2.1.1.).
See the caption of Figure~\ref{fig:mhmax} for the legend.
}
\end{figure}

Figures~\ref{fig:mhmax-a} and~\ref{fig:mhmax-b} show the same set of plots for the two variants, (a) and (b), 
of the {\it \mh-max}
scenario introduced in Section 2.1.1. The change of the sign of the Higgs mass parameter $\mu$ or of the
mixing parameter $X_{\rm t}$ barely affect the mass limits; however, sizable differences occur in the exclusions 
of \tanb\ (see Table~\ref{tab:cpc-results}).
For example, in variant (b), a small domain of \tanb\ is excluded even for $m_{\rm t} = 183.0$~\Gcs, which is not the case in the
standard {\it \mh-max} scenario and its variant (a). 
Note, in Figure~\ref{fig:mhmax-a}, the small domains at \mh\ between 60 and 75~\Gcs, small \mA\ and \tanb$<$0.9 
which are excluded at the 95\% CL but not at the 99.7\% CL.

\begin{figure}[p]
\centerline{
\epsfig{file=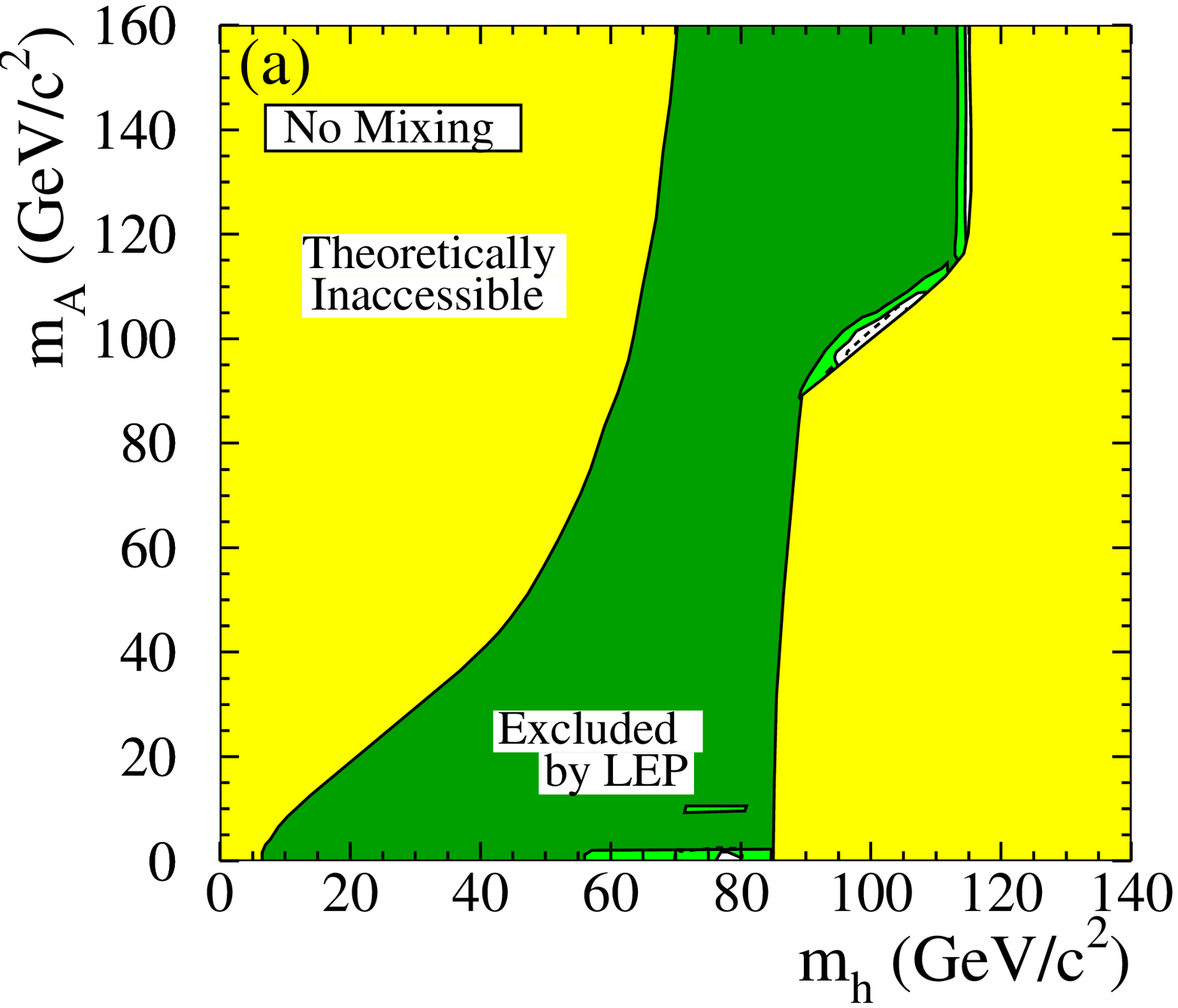,width=0.49\textwidth}
\epsfig{file=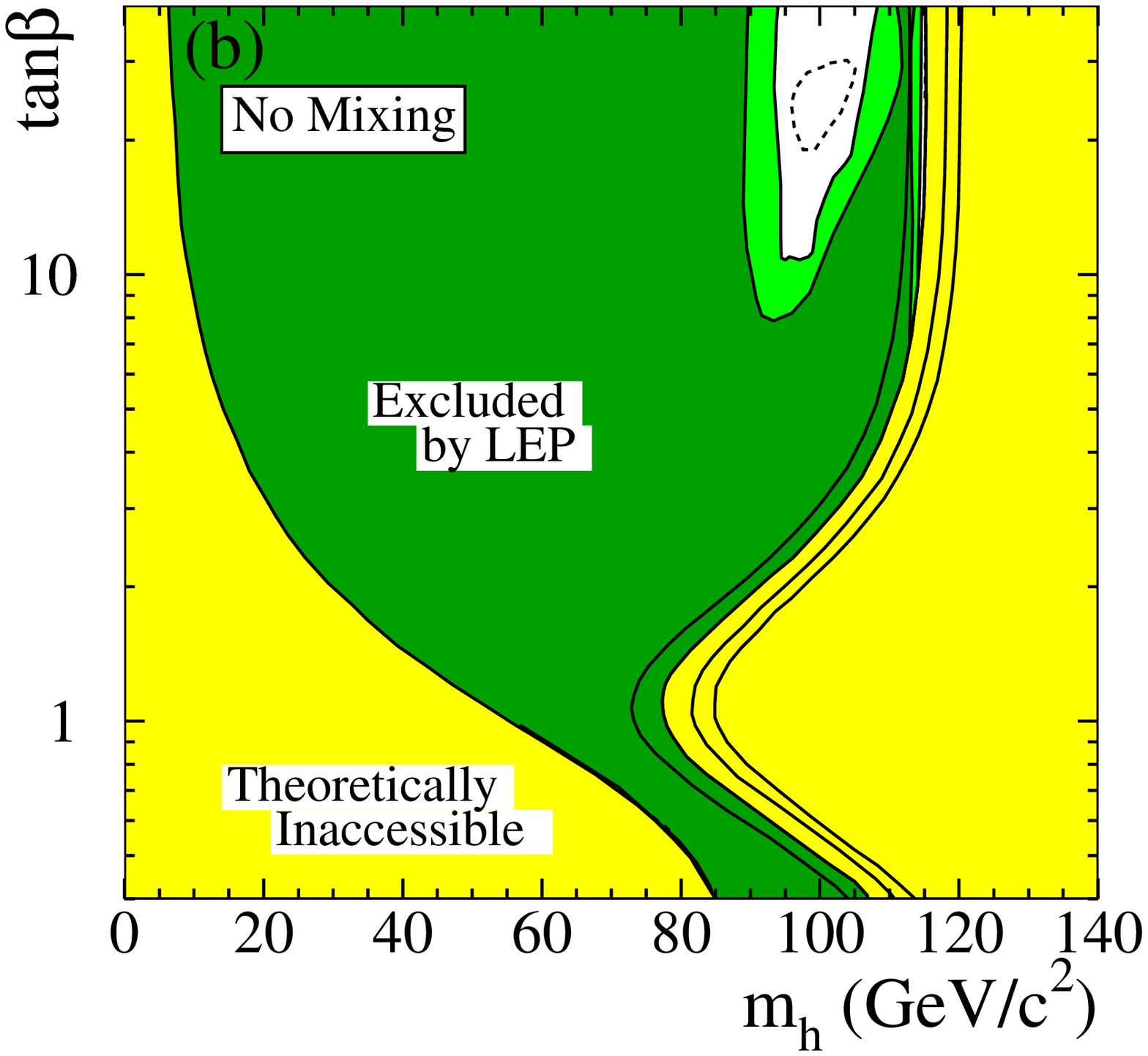,width=0.49\textwidth}
}
\centerline{
\epsfig{file=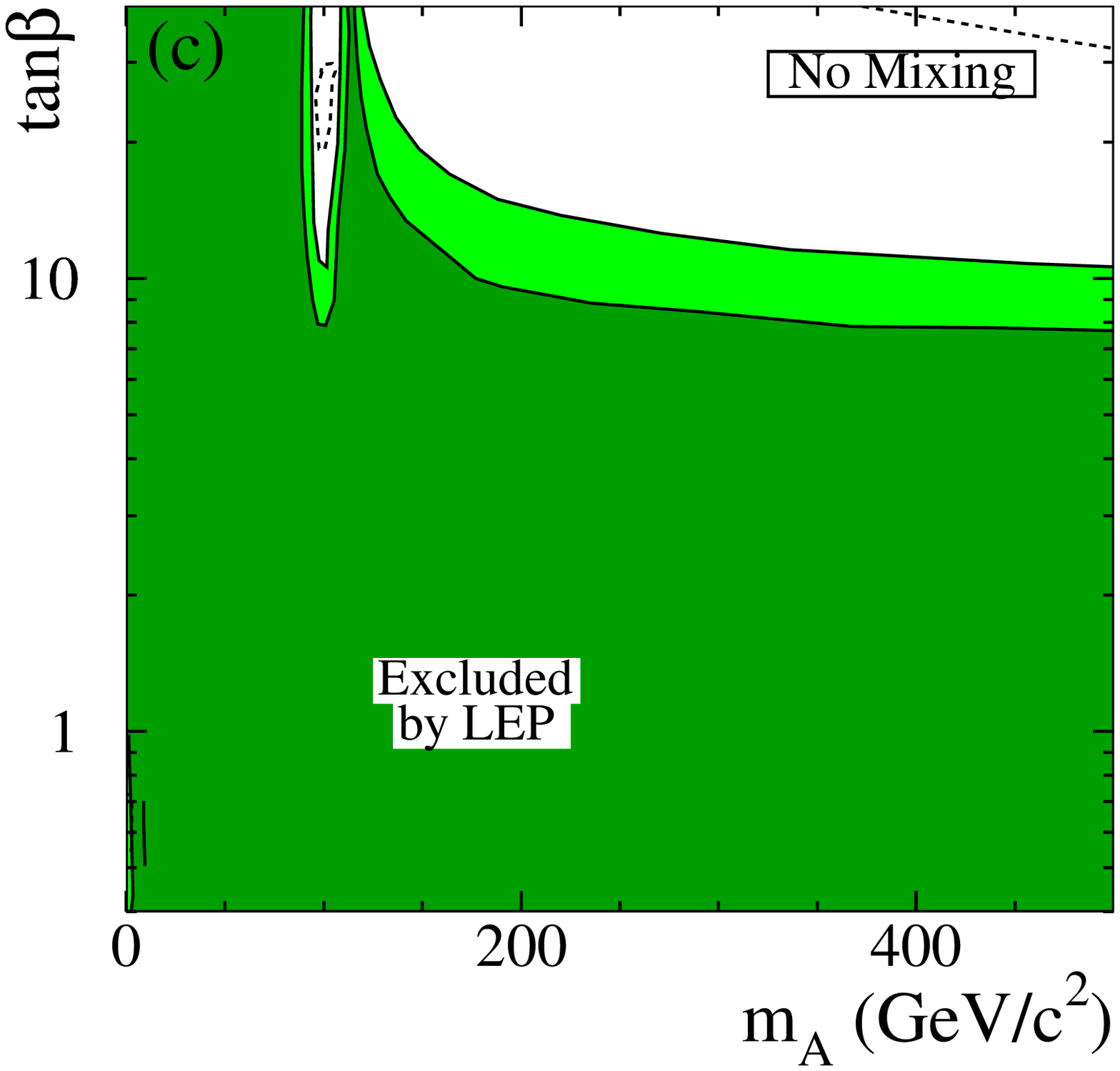,width=0.49\textwidth}
\epsfig{file=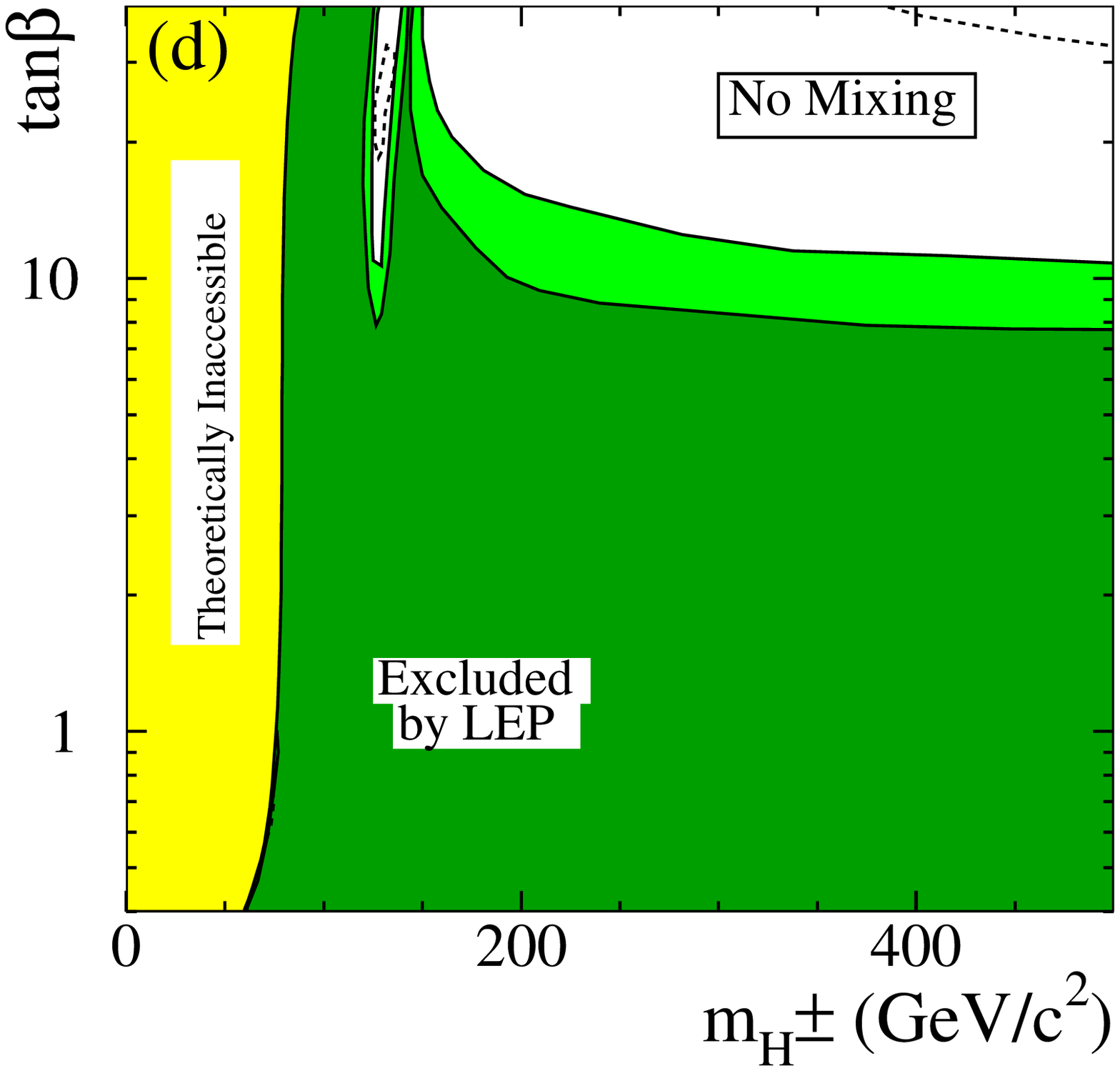,width=0.49\textwidth}
}
\caption[]{\label{fig:nomix}\sl
Exclusions in the case of the CP-conserving {\it no-mixing} benchmark scenario.
 See the caption of Figure~\ref{fig:mhmax} for the legend. 
Note the small domain at \mh\ between 75 and 80~\Gcs, small \mA\ and \tanb~$<$~0.7 
which is not excluded at the 95\% CL.
}
\end{figure}

The exclusions for the CP-conserving {\it no-mixing} benchmark scenario are shown in 
Figure~\ref{fig:nomix}. In this scenario, the theoretical boundaries of the parameter
space are more restricted than in the {\it \mh-max} scenario. As a consequence, large domains of \tanb\ are excluded 
for all the top quark masses considered. Note the relatively strong variation of the exclusion limits with \mt\ 
in this scenario
(see Table~\ref{tab:cpc-results}), which is caused by the proximity of the experimental lower bound of \mh\ from the
Higgsstrahlung searches
and the theoretical upper bound of \mh.

An interesting feature of this scenario is that, for \mh\ larger than about 100~\Gcs\ and large \tanb, the heavy scalar 
boson H is within kinematic reach. Moreover, the cross-section for the process \ee\ra~HZ is increasing with \tanb, 
resulting in an improved search sensitivity; this explains the nearly circular 
shape of the expected limit in Figure~\ref{fig:nomix} (b). 

Note the small domain at \mh\ between 75 and 80~\Gcs, small \mA\ and \tanb~$<$~0.7, 
barely perceptible in the plots, 
which is not excluded in this scenario at 95\% CL (this domain is excluded for \mt~=~169.3~\Gcs). 
The branching ratio for
\h\ra~\bb\ is small and the decay \h\ra~AA is dominant in this region. The A boson, 
with mass below the $\tau^+\tau^-$ threshold, may decay to final states
which are not sufficiently covered by the present searches.
For this reason, the mass limits given in Table~\ref{tab:cpc-results} for this scenario and for \mt\ larger than 169.3~\Gcs\
are valid only for \tanb~$\geq$~0.7. Conversely, for \mt\ larger than 169.3~\Gcs, the quoted exclusion of \tanb\ is valid 
only for \mA\ larger than about 3~\Gcs.

\begin{figure}[p]
\centerline{
\epsfig{file=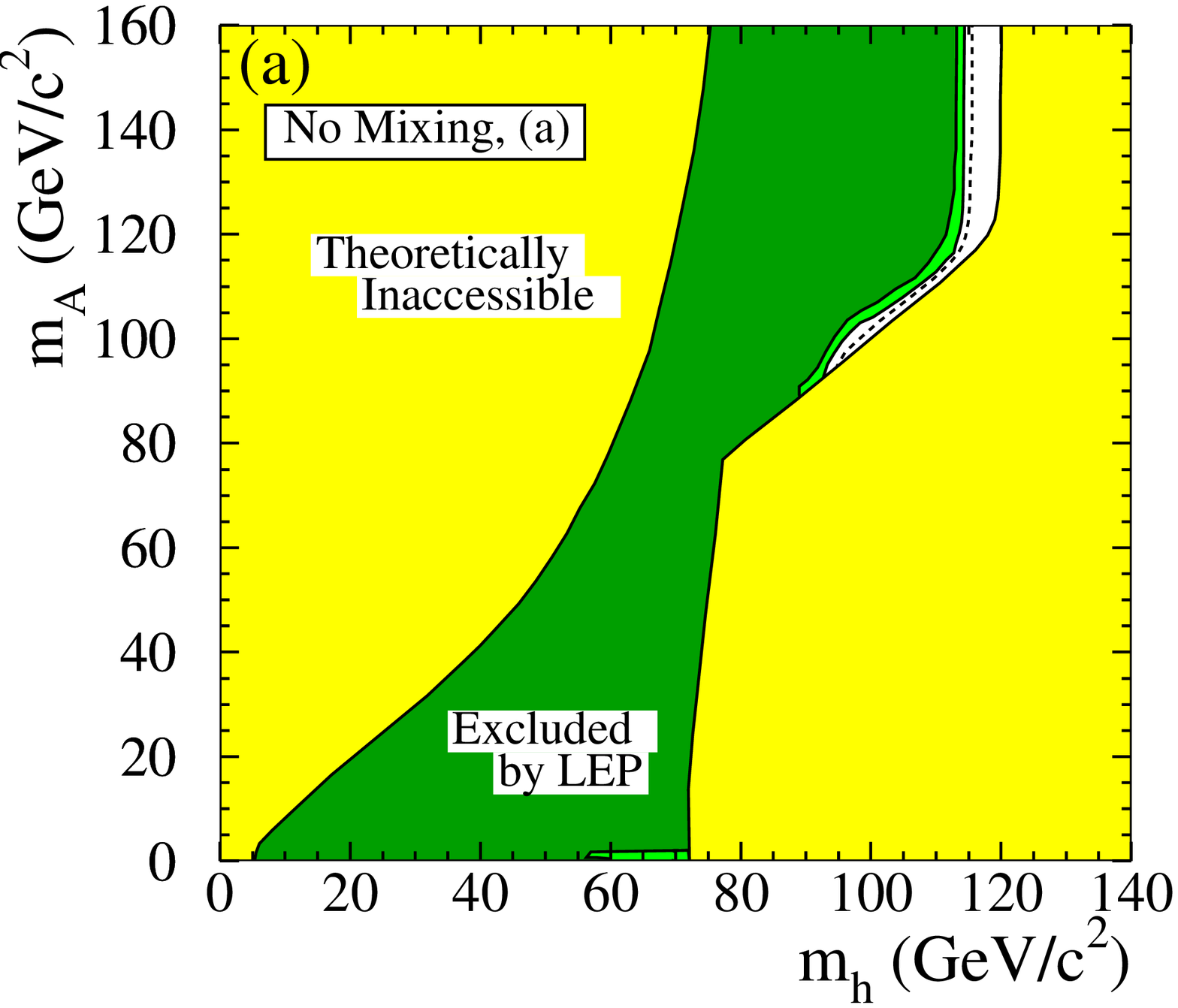,width=0.49\textwidth}
\epsfig{file=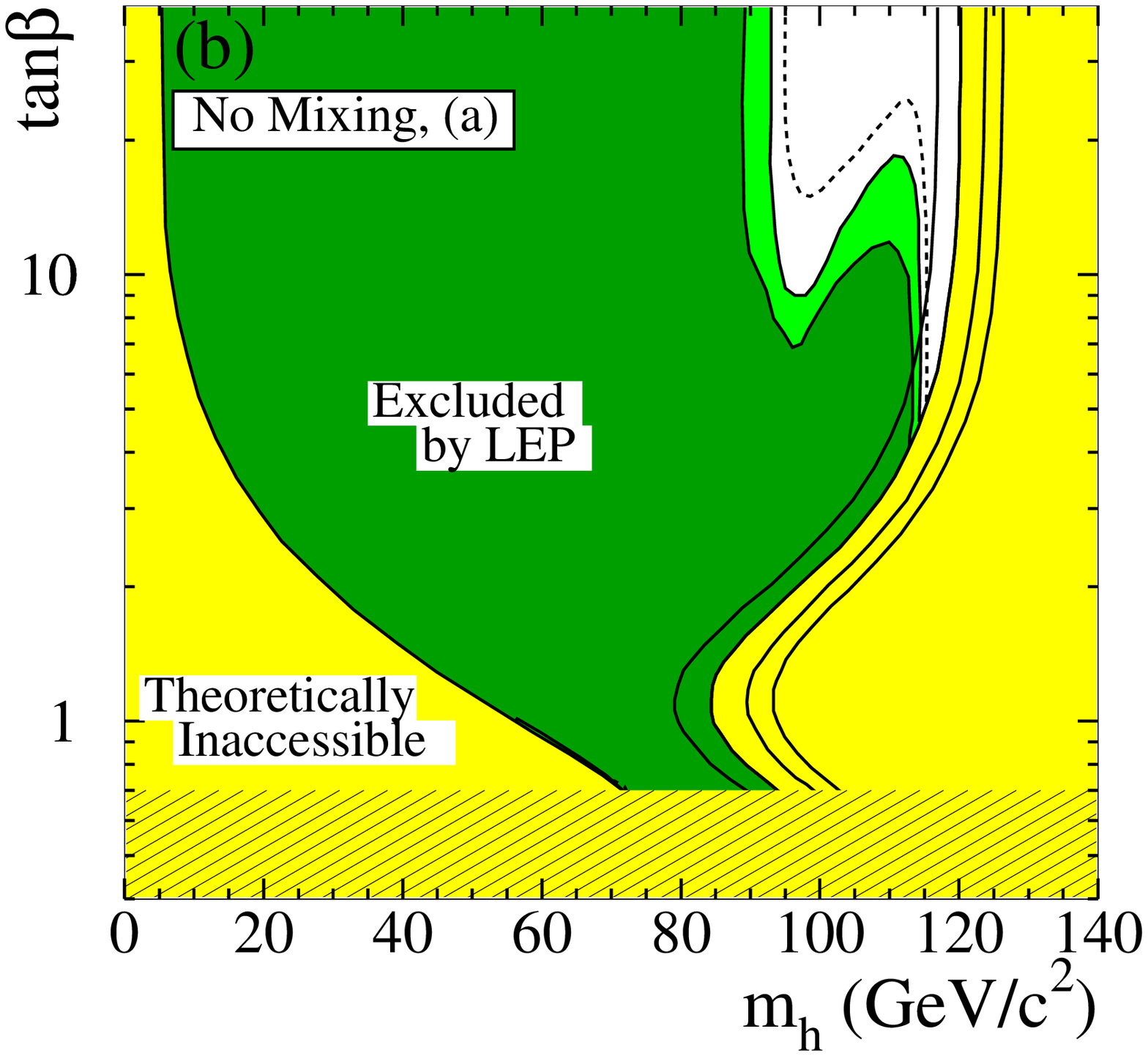,width=0.49\textwidth}
}
\centerline{
\epsfig{file=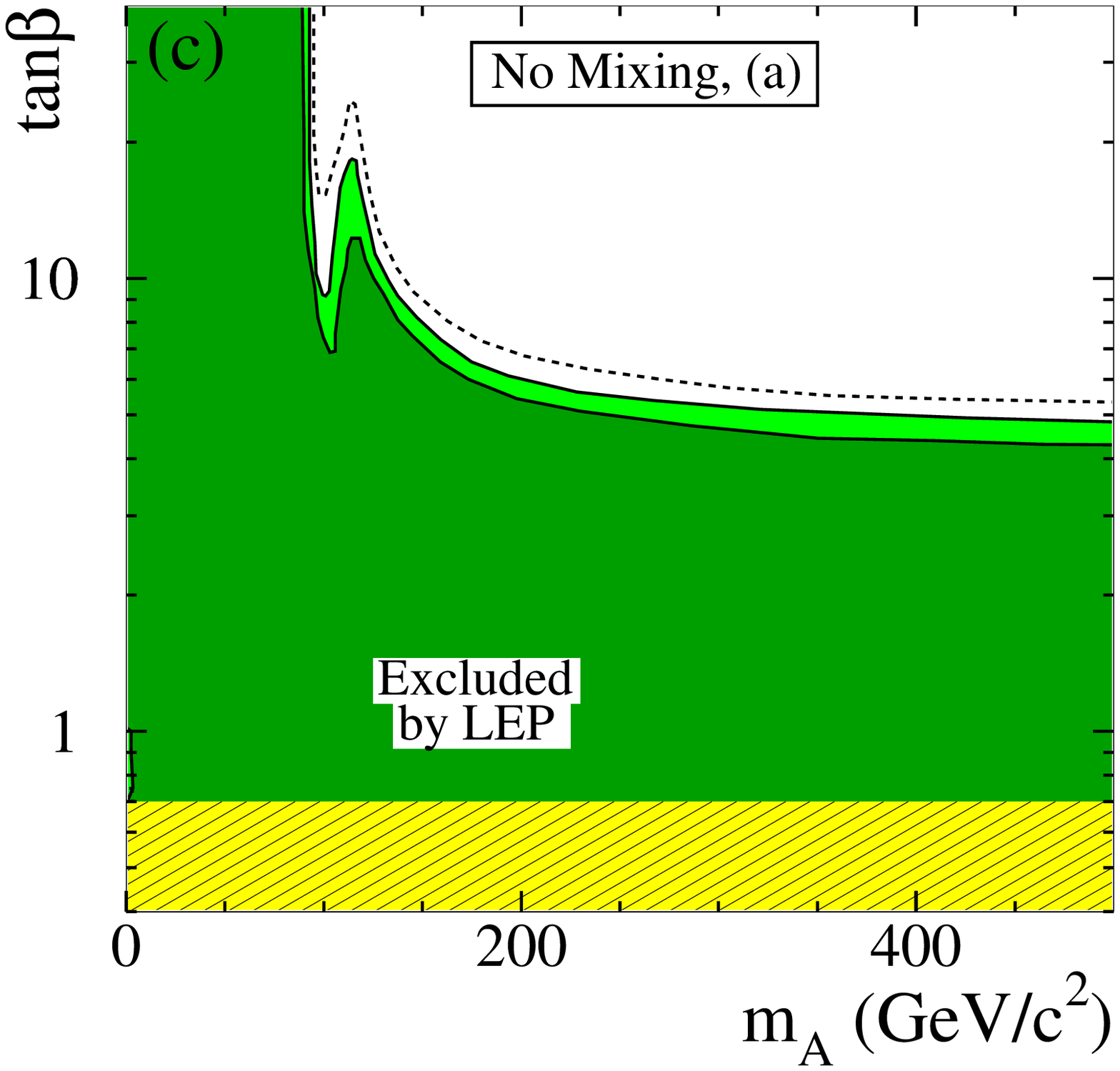,width=0.49\textwidth}
\epsfig{file=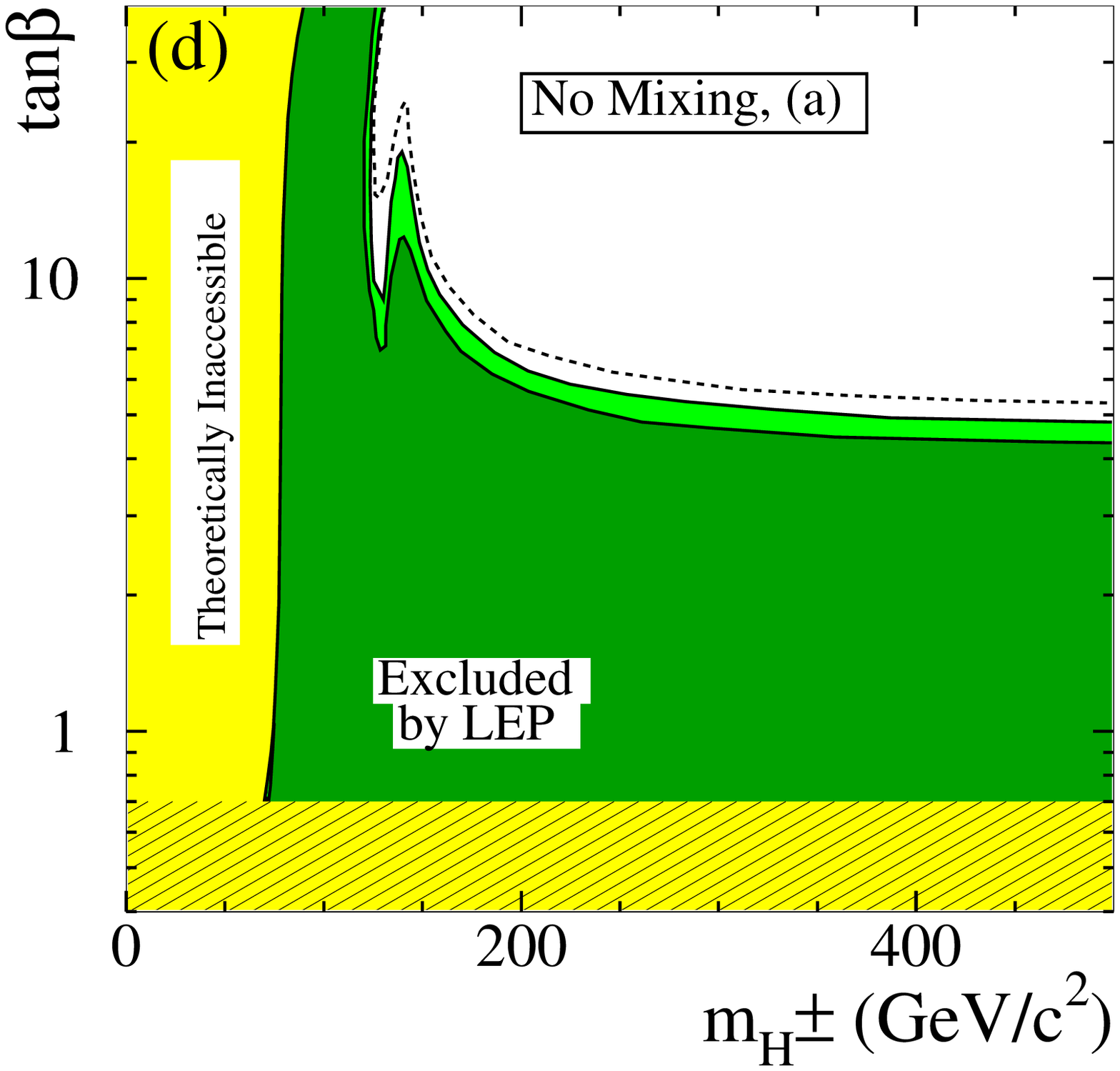,width=0.49\textwidth}
}
\caption[]{\label{fig:nomix-a}
\sl  Exclusions in the case of the CP-conserving {\it no-mixing} benchmark scenario, variant (a) (see Section 2.1.2).
 See the caption of Figure~\ref{fig:mhmax} for the legend. In the hatched domain (\tanb~$<$~0.7), 
 the contributions from top and stop quark loops to the radiative corrections 
 are large and uncertain. 
 Note the small domain at \mh\ between 56 and 72~\Gcs, small \mA\ and \tanb~$<1$
which, although excluded at the 95\% CL, is not excluded at the 99.7\% CL. 
}
\end{figure}

Figure~\ref{fig:nomix-a} shows the exclusion plots for the (a) variant of the {\it no-mixing}
scenario introduced in Section 2.1.2. The change of sign of the Higgs mass parameter $\mu$ 
and the increase of the weak SUSY-breaking scale from 1~TeV to 2~TeV affect only the theoretical bounds
of the parameter space but barely change the mass limits, except for \mt=169.3~\Gcs. There are moderate changes 
though in the exclusions of \tanb. 
In the hatched domain (\tanb$<$0.7), 
the contributions from top and stop quark loops to the radiative corrections 
are large and uncertain; hence, no exclusions can be claimed there.

\begin{figure}[p]
\centerline{
\epsfig{file=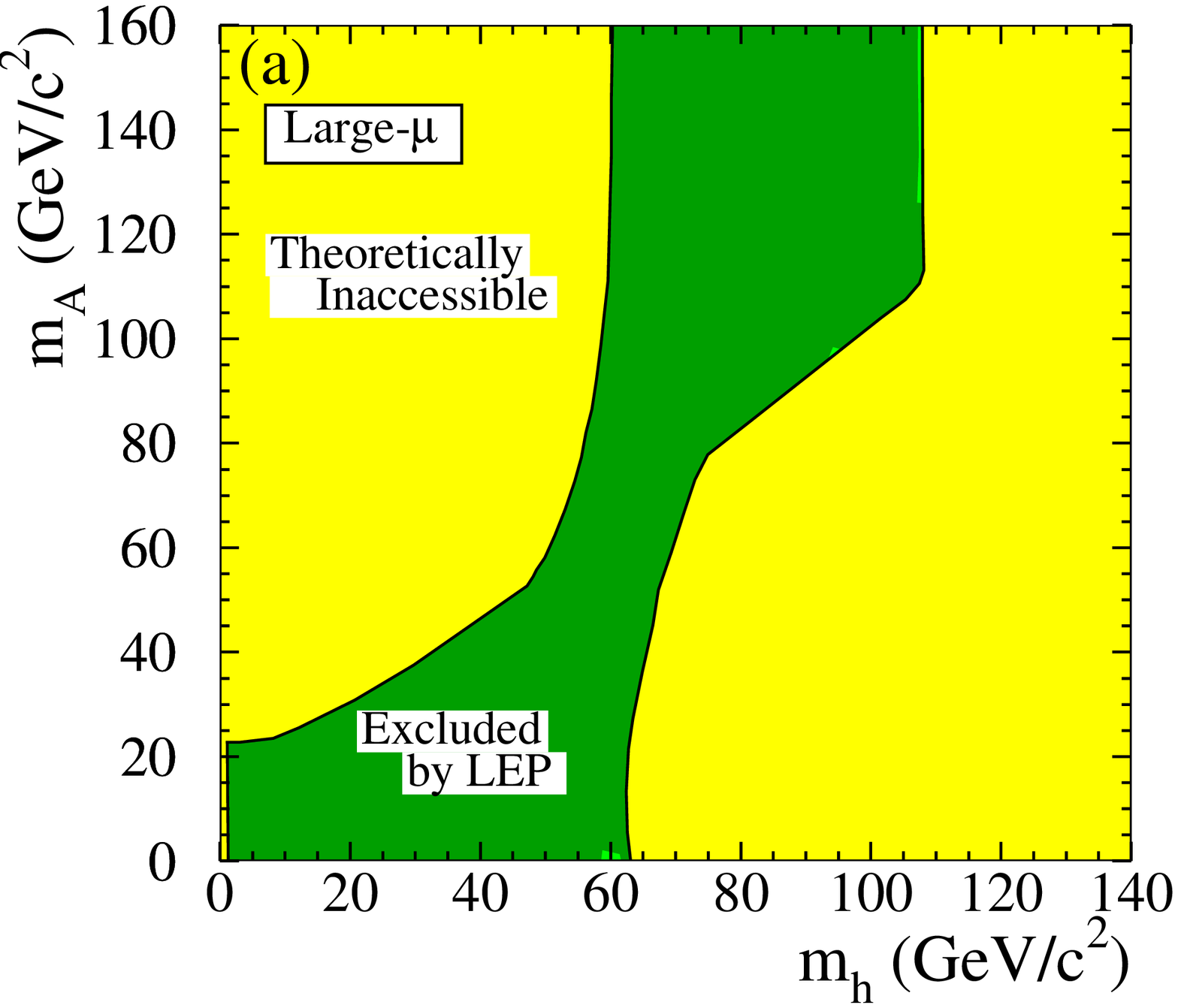,width=0.49\textwidth}
\epsfig{file=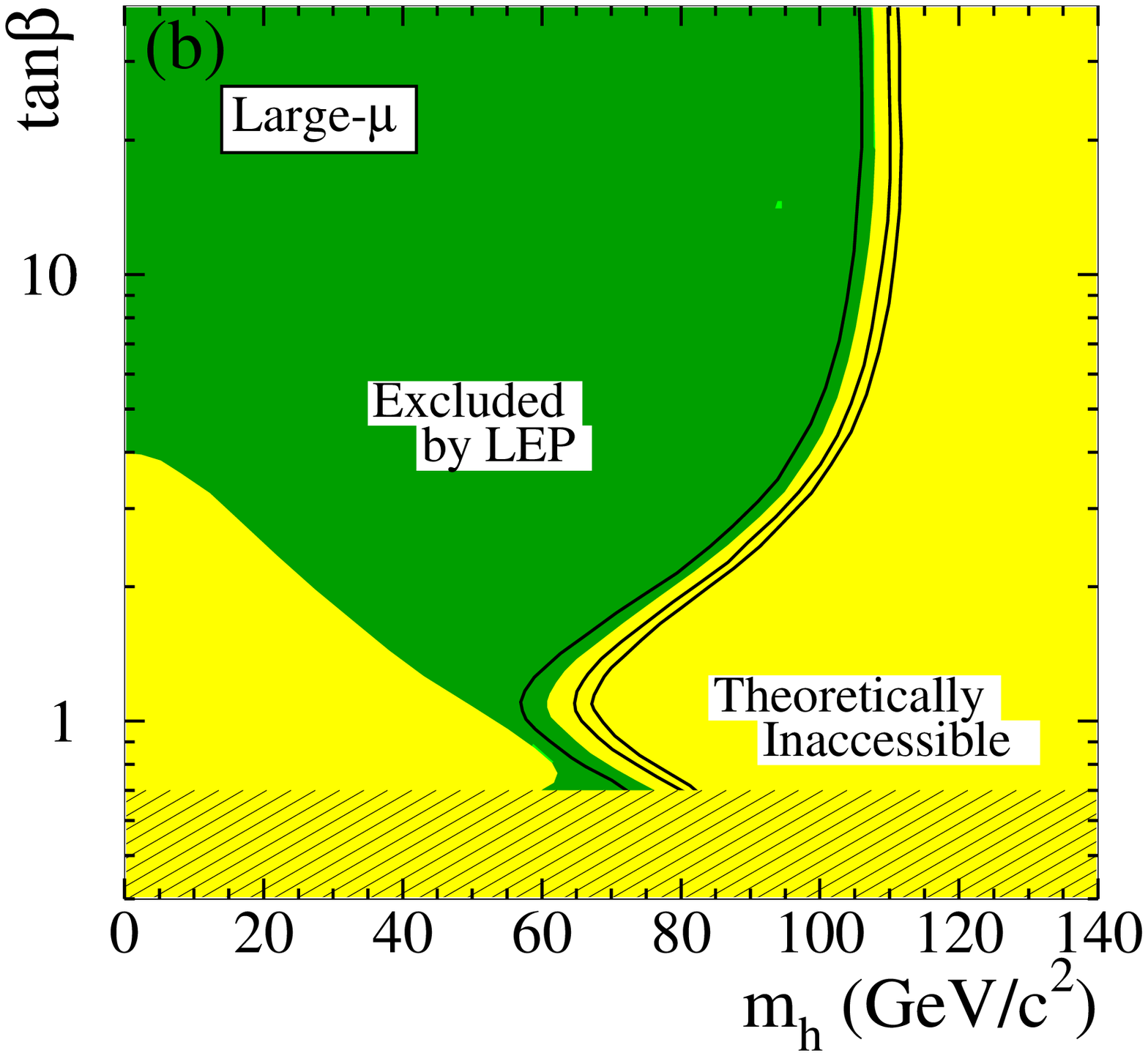,width=0.49\textwidth}
}
\centerline{
\epsfig{file=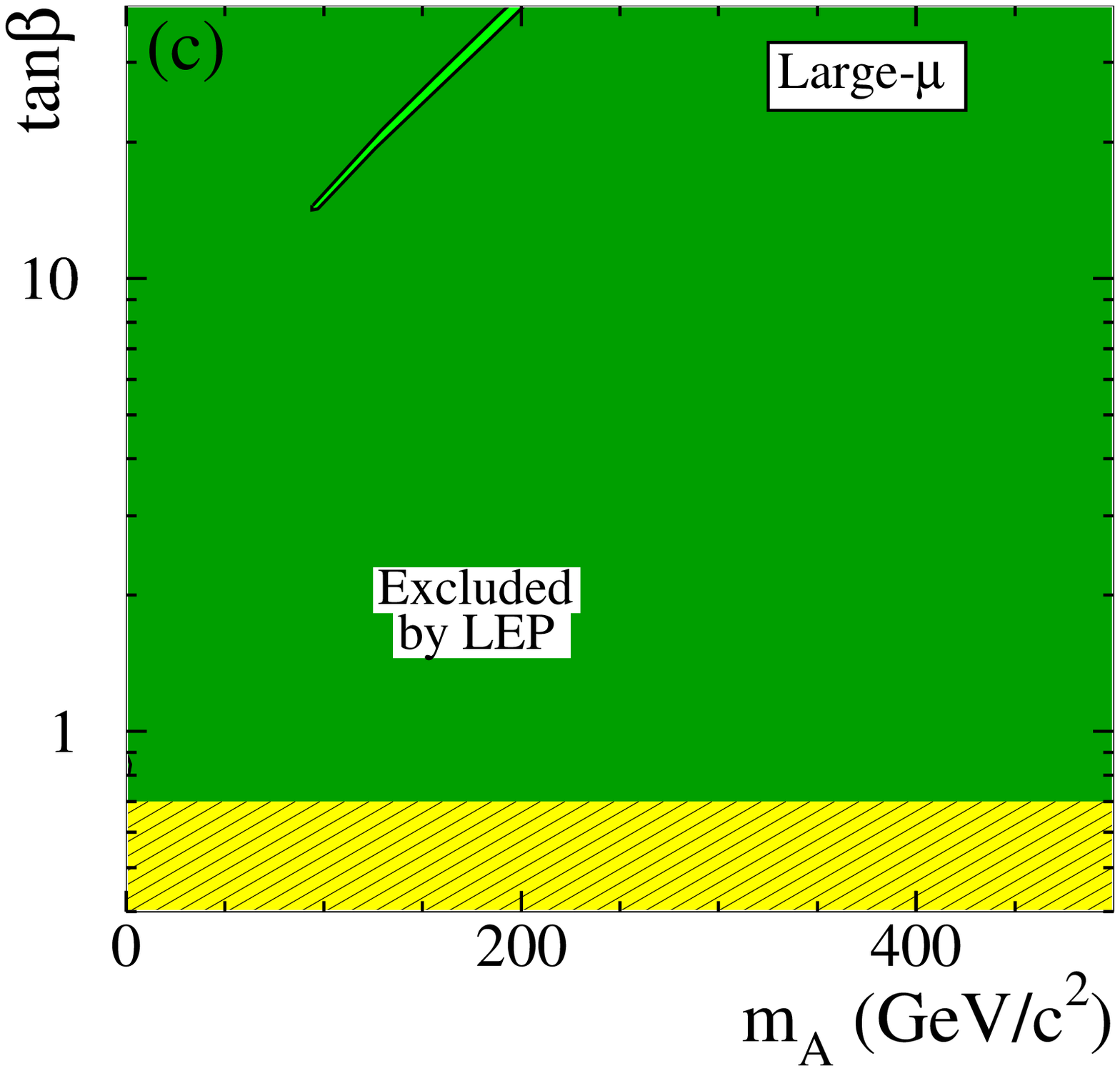,width=0.49\textwidth}
\epsfig{file=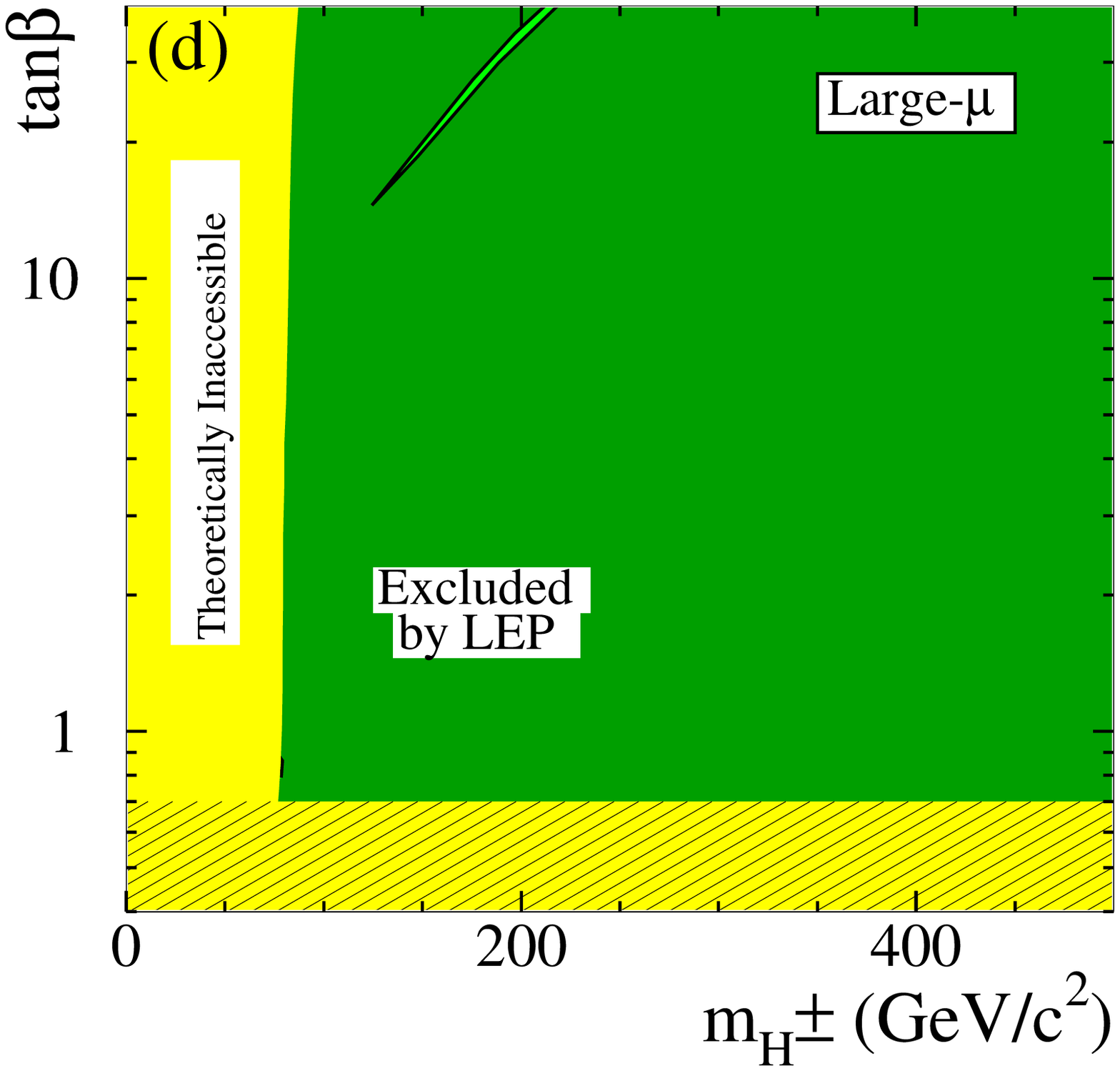,width=0.49\textwidth}
}
\caption[]{\sl
 Exclusions in the case of the CP-conserving large-$\mu$ benchmark scenario (see Section 2.1.3).
 See the caption of Figure~\ref{fig:mhmax} for the legend. In the hatched domain (\tanb~$<$~0.7), 
 the contributions from top and stop quark loops to the radiative corrections 
 become large and uncertain; hence, no exclusions can be claimed there.
\label{fig:largemu}}
\end{figure}

The exclusions for the {\it large-$\mu$} benchmark scenario are shown in Figure~\ref{fig:largemu}.
As mentioned in Section 2.3, this scenario was
constructed to test the sensitivity of LEP to MSSM scenarios which may be {\it a priori}
difficult to handle experimentally since the Higgs boson decays to \bb\ are largely suppressed. 
It turns out that the flavour-independent and decay-mode-independent searches are sufficiently powerful 
to exclude all such situations at 95\% CL, for top quark masses up to 174.3~\Gcs.
There remains a thin strip at \tanb\ larger than about 10 and running from \mA\ of about 100 to about 200~\Gcs, 
which is excluded at 
the 95\% CL but not at 99.7\% CL because the suppression of the \bb\ channel is particularly strong
in that region. This strip is 
found to grow 
with increasing \mt\ and becomes gradually non-excluded at the 95\% CL.
Other small, weakly excluded, regions are located at \mh~$\approx$~60~\Gcs\ and small \mA, and    
along the \mh~$\approx$~\mA\ ``diagonal" of plot (a).

\begin{figure}[p]
\centerline{
\epsfig{file=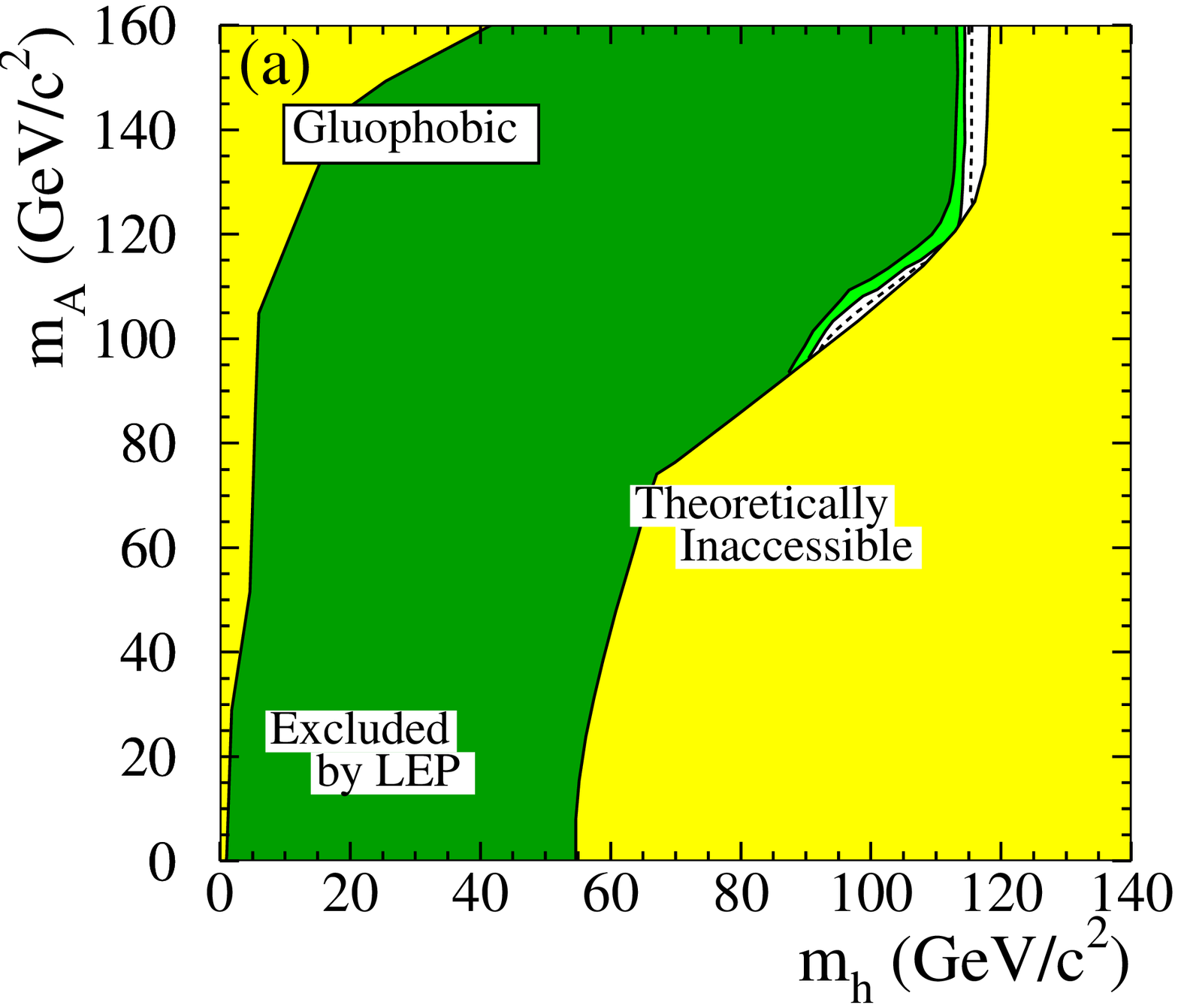,width=0.49\textwidth}
\epsfig{file=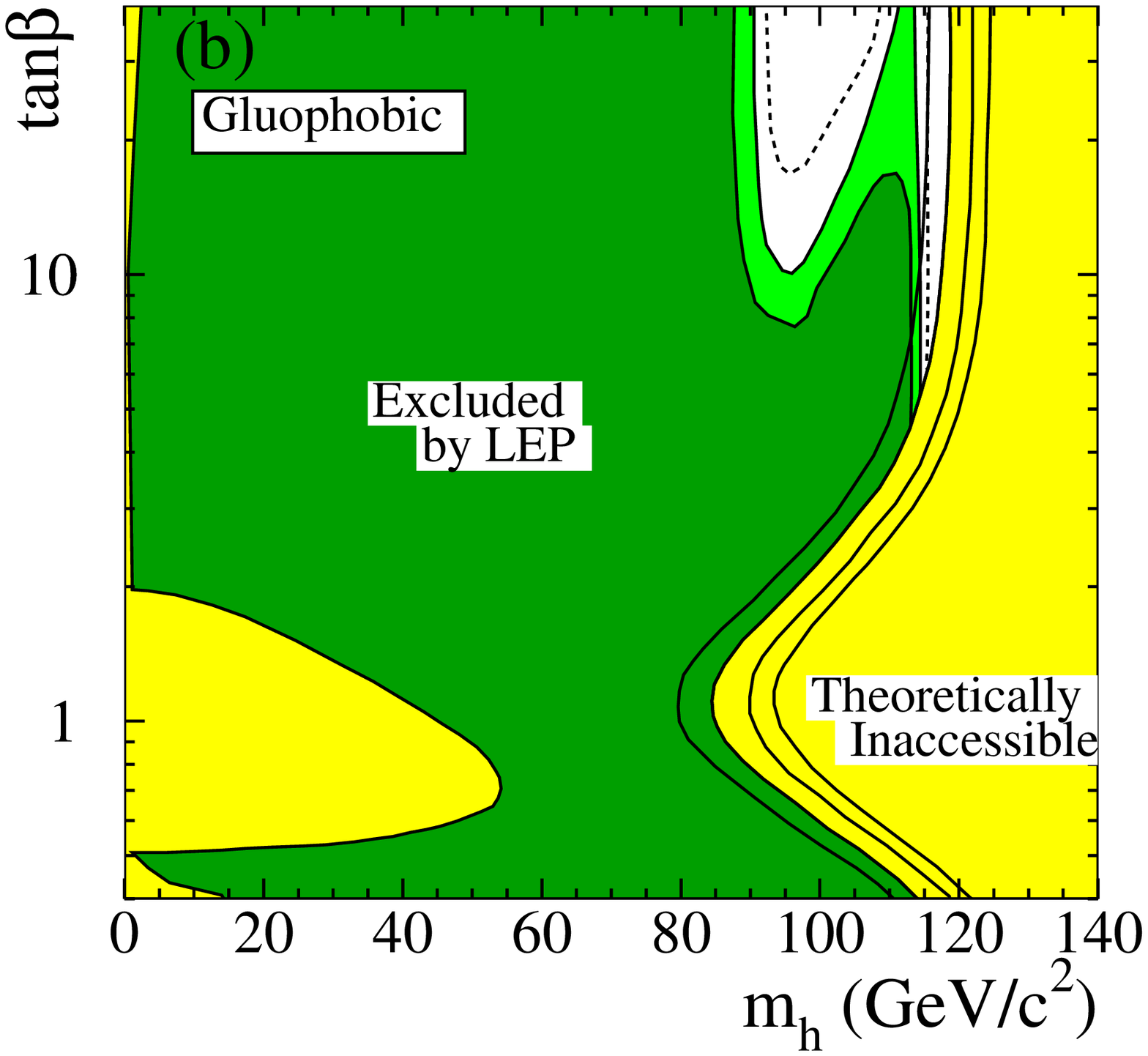,width=0.49\textwidth}
}
\centerline{
\epsfig{file=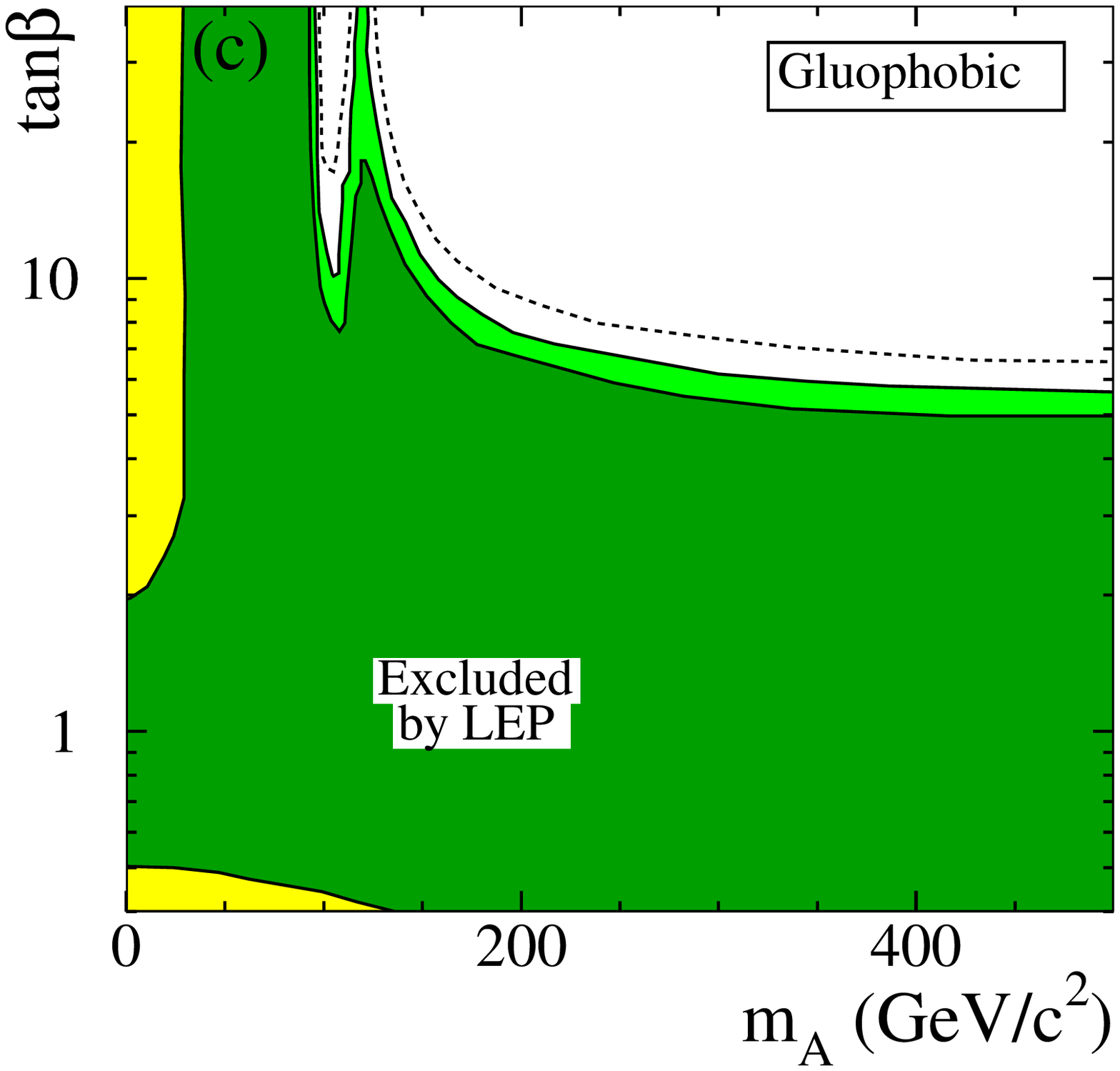,width=0.49\textwidth}
\epsfig{file=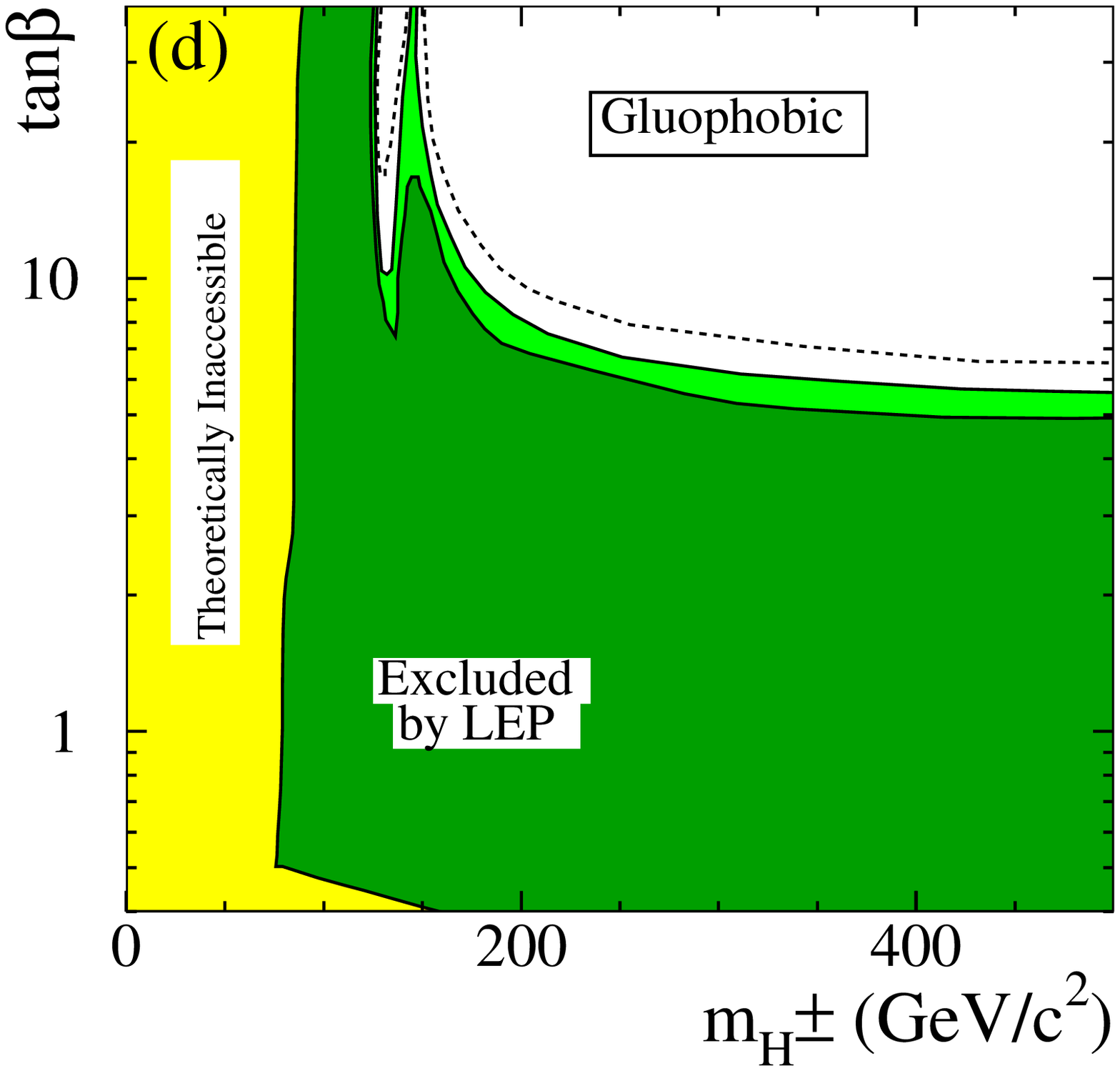,width=0.49\textwidth}
}
\caption[]{\sl
 Exclusions in the case of the {\it gluophobic} benchmark scenario (see Section 2.1.3).
 See the caption of Figure~\ref{fig:mhmax} for the legend.  
\label{fig:gluophob}}
\end{figure}

\begin{figure}[p]
\centerline{
\epsfig{file=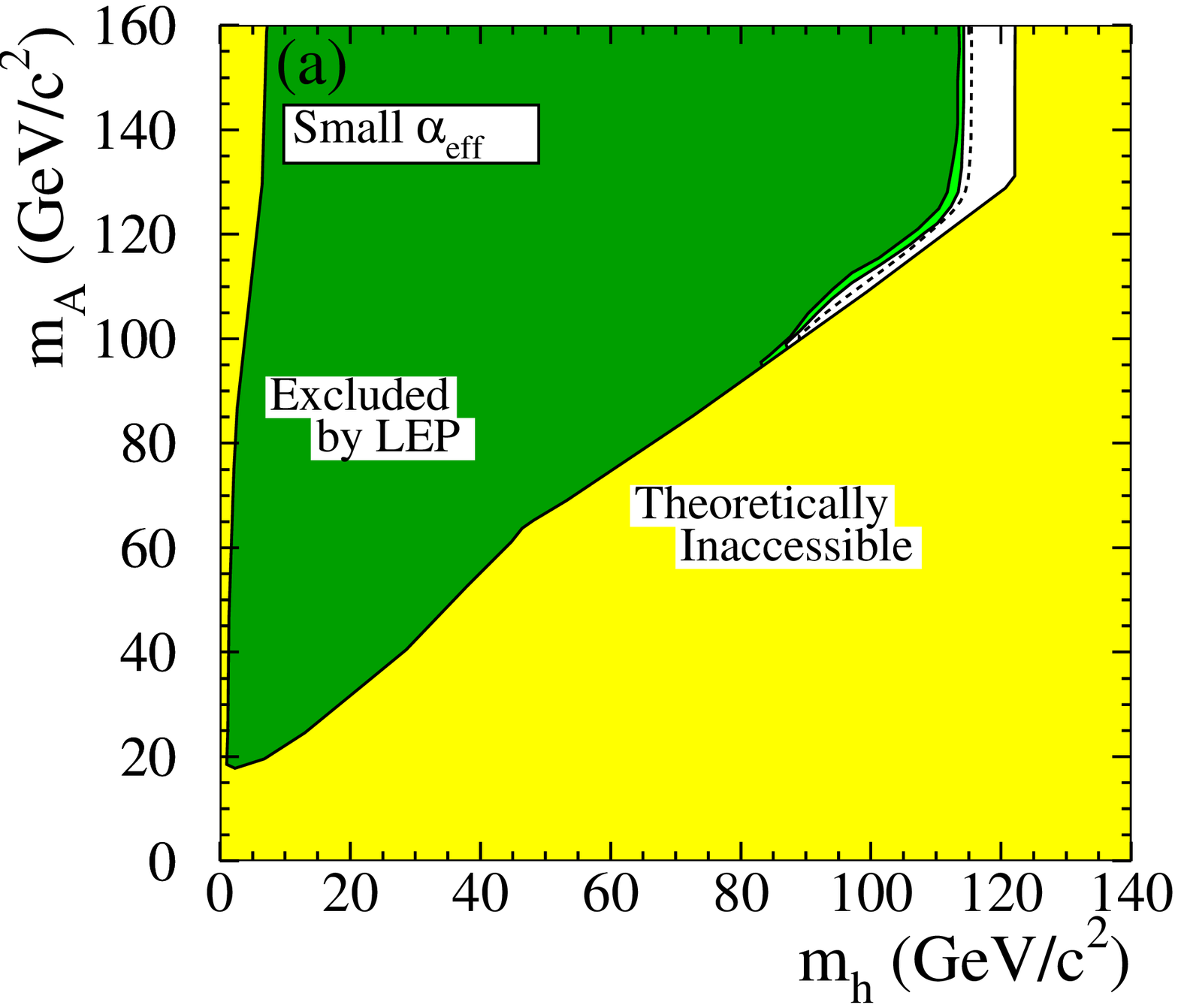,width=0.49\textwidth}
\epsfig{file=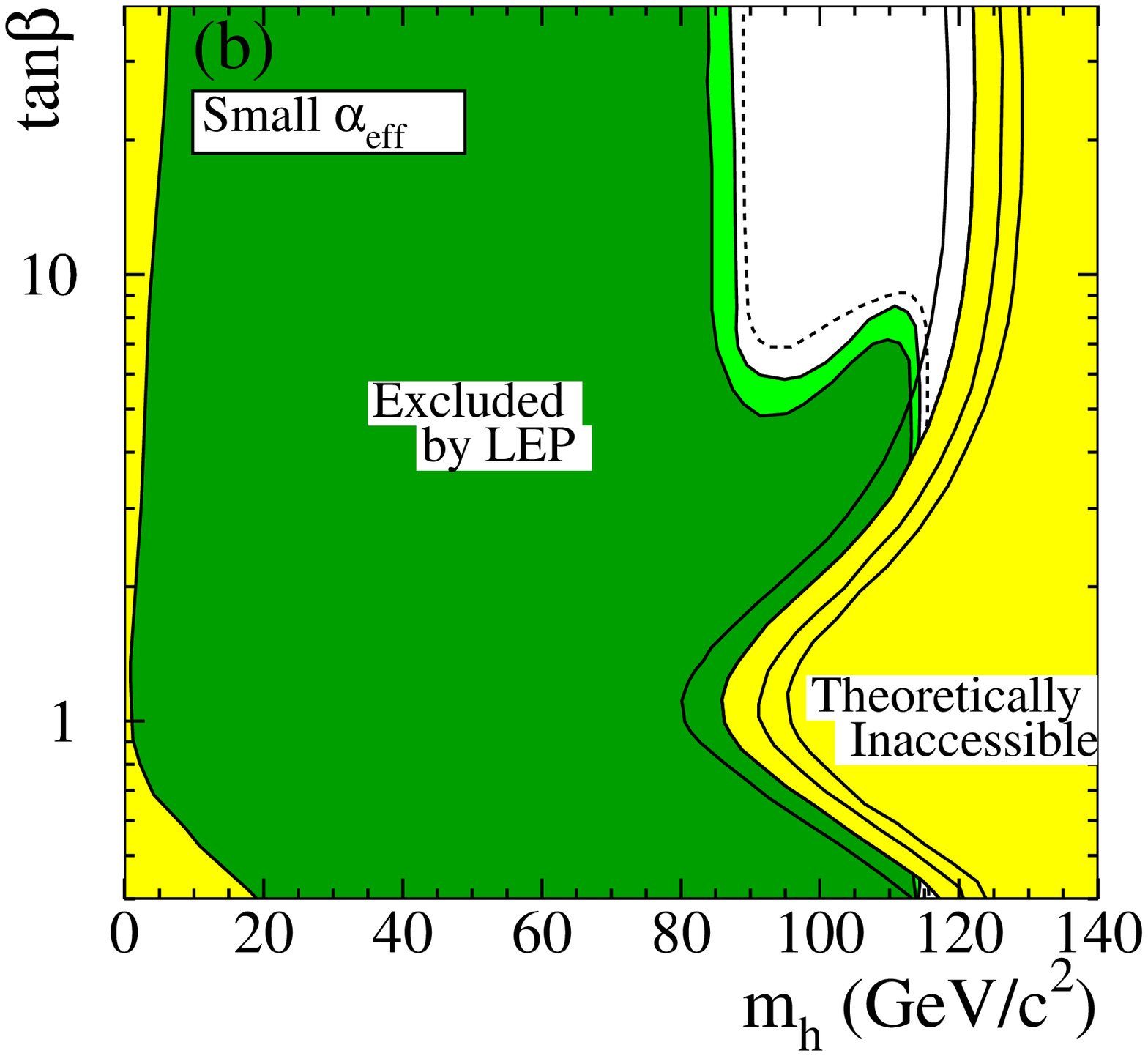,width=0.49\textwidth}
}
\centerline{
\epsfig{file=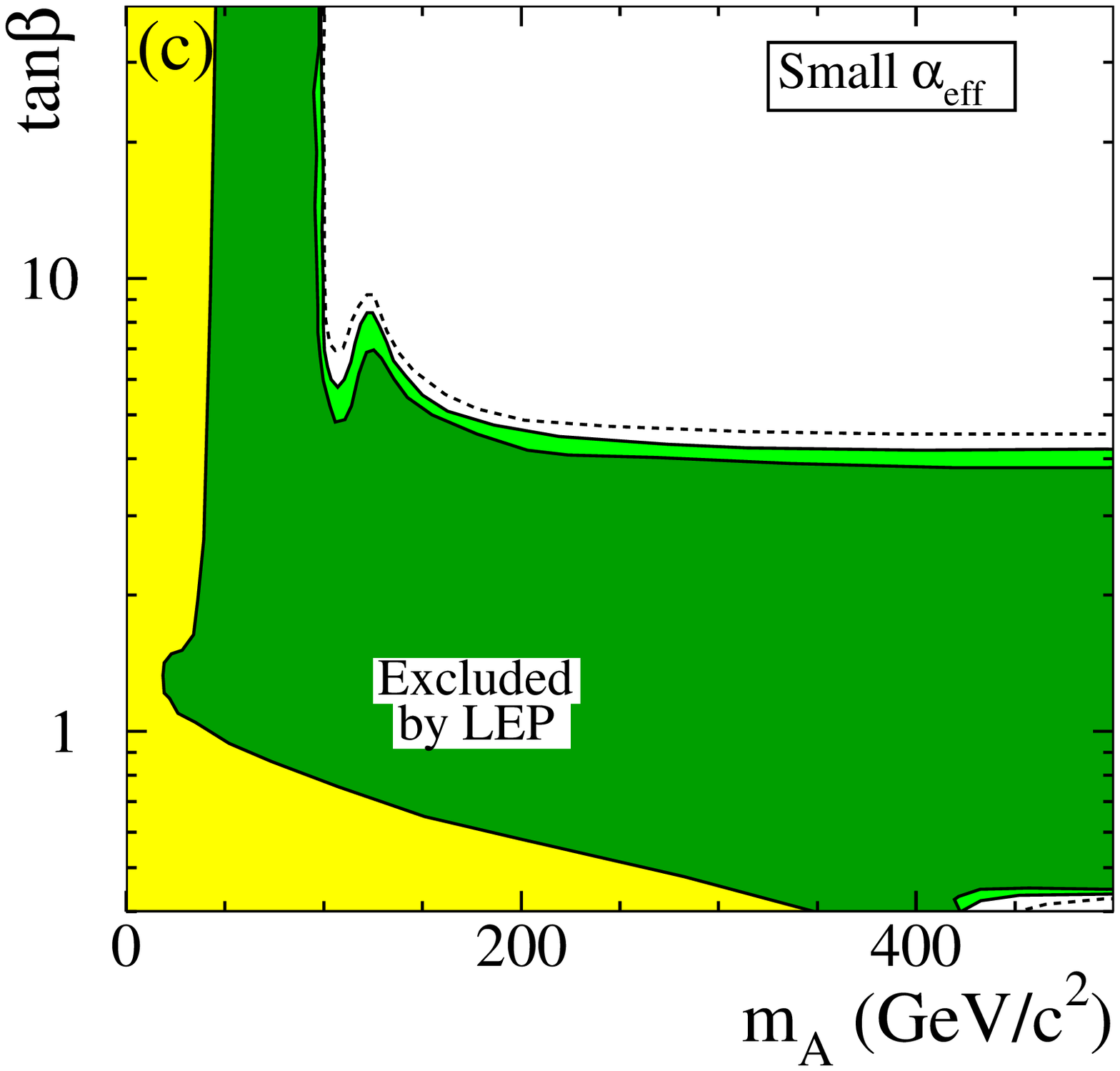,width=0.49\textwidth}
\epsfig{file=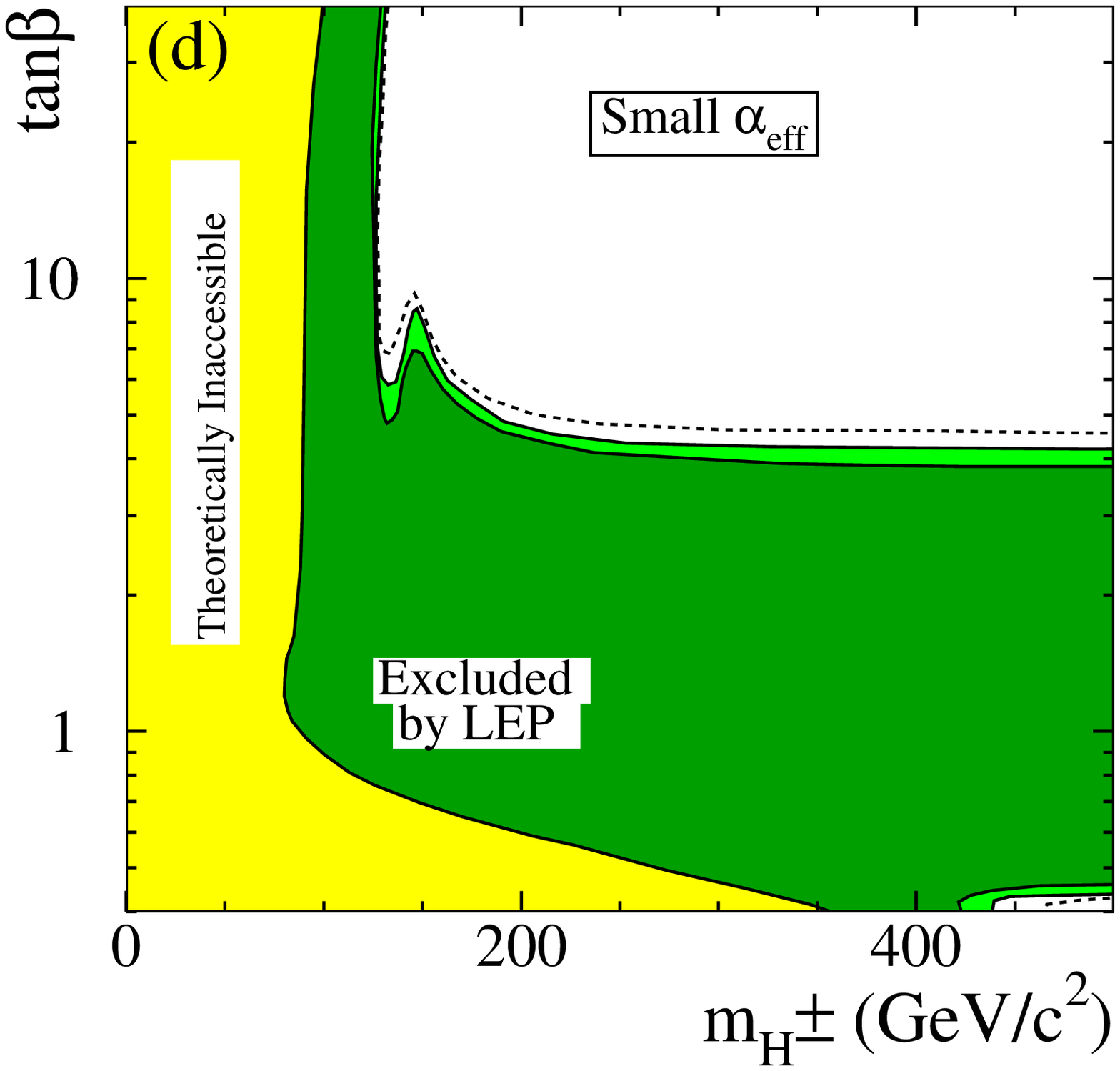,width=0.49\textwidth}
}
\caption[]{\sl
 Exclusions in the case of the CP-conserving {\it small-$\alpha_{eff}$} benchmark scenario (see Section 2.1.3).
 See the caption of Figure~\ref{fig:mhmax} for the legend.  
\label{fig:small-alpha-eff}}
\end{figure}

Similar plots are shown in Figures~\ref{fig:gluophob} and~\ref{fig:small-alpha-eff} for the {\it gluophobic} and {\it
small-$\alpha_{eff}$} scenarios defined in Section 2.1.3. These scenarios were designed to test situations 
which can be problematic at the Tevatron and LHC colliders. In both cases, large domains of the parameter 
space are excluded by the LEP searches.
\begin{table}[ht]
\begin{center}
{\small
\begin{tabular}{|l|c|c|c|} \hline
&  &  &  \\
Benchmark~~~\mt~(\Gcs)                 & \mh\ (\Gcs) & \mA\ (\Gcs) & Exclusions of \\
~~scenario &  &  & \tanb\  \\
\hline
{\it \mh-max}~~~~~~~~~~~~\thinspace169.3                 &  92.9 (94.8) & 93.4 (95.1) &  0.6--2.6 (0.6--2.7)\\
$\phantom{.................}$ ~~~~~~~ 174.3     &  92.8 (94.9) & 93.4 (95.2) &  0.7--2.0 (0.7--2.1) \\
$\phantom{.................}$ ~~~~~~~ 179.3     &  92.9 (94.8) & 93.4 (95.2) &  0.9--1.5 (0.9--1.6)\\
$\phantom{.................}$ ~~~~~~~ 183.0     &  92.8 (94.8) & 93.5 (95.2) &  no excl. (no excl.)  \\
\hline
{\it \mh-max}~~~~~~~~~~~~\thinspace169.3                 & 92.7 (94.9) & 93.1 (95.1)  & 0.7--2.1 (0.7--2.2) \\
$\phantom{.....}$(a)$\phantom{....}$ ~~~~~~~~~~ 174.3     & 92.7 (94.8)  & 93.1 (95.1) & 0.7--2.1 (0.7--2.2) \\
$\phantom{.................}$ ~~~~~~~ 179.3     & 92.6 (94.8)  & 93.1 (95.1) & 0.9--1.6 (0.8--1.7) \\
$\phantom{.................}$ ~~~~~~~ 183.0     & 92.7 (94.8)  & 93.1 (95.1) & no excl. (no excl.) \\
\hline
{\it \mh-max}~~~~~~~~~~~~~169.3                  & 92.8 (94.8) & 93.2 (95.2) & 0.5--3.3 (0.5-3.5) \\
$\phantom{.....}$(b)$\phantom{....}$ ~~~~~~~~~~ 174.3     & 92.6 (94.9)  & 93.4 (95.1) & 0.6--2.5 (0.6--2.7) \\
$\phantom{.................}$ ~~~~~~~ 179.3     & 92.6 (94.8)  & 93.4 (95.1) & 0.7--2.0 (0.7--2.1) \\
$\phantom{.................}$ ~~~~~~~ 183.0     & 92.7 (94.7)  & 93.4 (95.1) & 0.8--1.7 (0.8--1.8) \\
\hline
{\it no-mixing}~~~~~~~~~~169.3                  & excl. (excl.) & excl. (excl.) & excl. (excl.)\\
$\phantom{.................}$  ~~~~~~~~174.3    & 93.6 (96.0)   & 93.6  (96.4)  & 0.4--10.2 (0.4--19.4) \\
$\phantom{.................}$ ~~~~~~~ 179.3     & 93.3 (95.0)   & 93.4  (95.0)  & 0.4--5.5  (0.4--6.5) \\
$\phantom{.................}$ ~~~~~~~ 183.0     & 92.9 (95.0)   & 93.1  (95.0)  & 0.4--4.4  (0.4--4.9) \\
\hline
{\it no-mixing}~~~~~~~~~~169.3                  & 93.2 (95.2) & 93.4 (95.4) & 0.7--7.1 (0.7--9.3) \\
$\phantom{.....}$(a)$\phantom{....}$ ~~~~~~~~~~~174.3  & 92.8 (94.9) & 93.1 (95.1) & 0.7--4.6 (0.7--5.1) \\
$\phantom{.................}$ ~~~~~~~ 179.3     & 92.8 (94.9) & 93.1 (95.0) & 0.7--3.5 (0.7--3.8) \\
$\phantom{.................}$ ~~~~~~~ 183.0     & 92.9 (94.8)  & 93.1 (95.0) & 0.7--3.0 (0.8--3.2) \\
\hline
{\it large-$\mu$}~~~~~~~~~~~~~~169.3            & excl. (excl.) &  excl. (excl.) & excl. (excl.) \\
$\phantom{.................}$ ~~~~~~~~174.3     & excl. (excl.) &  excl. (excl.) & excl. (excl.) \\
$\phantom{.................}$ ~~~~~~~ 179.3     & 109.2 (109.2) & 225.0 (225.0)  & 0.7--43 (0.7--43)    \\
$\phantom{.................}$ ~~~~~~~ 183.0     &  95.6 (95.6)  &  98.9 (98.9)   & 0.7--11.5 (0.7--11.5)  \\
\hline
{\it gluophobic}~~~~~~~~~~169.3                 & 90.6 (93.2) & 95.7 (98.2) & 0.4--10.3 (0.4--21.5) \\
$\phantom{.................}$  ~~~~~~~~174.3    & 90.5 (92.3) & 96.3 (98.0) & 0.4--5.4 (0.4--6.4)\\
$\phantom{.................}$ ~~~~~~~ 179.3     & 90.0 (91.8) & 96.5 (98.2) & 0.4--3.9 (0.4--4.2)\\
$\phantom{.................}$ ~~~~~~~ 183.0     & 89.8 (91.5) & 96.8 (98.7) & 0.5--3.3 (0.5-3.6)\\
\hline
{\it small-$\alpha_{eff}$}~~~~~~~~~169.3       & 88.2 (90.0) & 98.2 (99.6) & 0.4--6.1 (0.4--7.4) \\
$\phantom{.................}$  ~~~~~~~~174.3    & 87.3 (89.0) & 98.8 (100.0)& 0.4--4.2 (0.4--4.5) \\
$\phantom{.................}$ ~~~~~~~ 179.3     & 86.6 (88.0) & 99.8 (100.7)& 0.5--3.2 (0.5--3.4) \\
$\phantom{.................}$ ~~~~~~~ 183.0     & 85.6 (87.5) &101.0 (101.3)& 0.6--2.7 (0.5--2.9) \\
\hline
\end{tabular}
}
\vspace{0.5cm}
\caption{\label{tab:cpc-results} \sl
Lower mass bounds and exclusions in \tanb, at 95\% CL, obtained in the case of the CP-conserving 
MSSM benchmark scenarios, for various values of the top quark mass. In each case, the observed limit is followed, 
between parentheses, by the
value expected on the basis of Monte Carlo simulations with no signal.
In the {\it \mh-max} scenario and its variant (a), there is no exclusion in \tanb\ for \mt\ = 183.0~\Gcs\ or larger.
The {\it no-mixing} scenario is entirely excluded for \mt~=~169.3~\Gcs\ or smaller. In the {\it no-mixing} scenario and
for \mt\ larger than 169.3~\Gcs, the quoted mass limits are only valid for \tanb~$\geq$~0.7 and the exclusion in \tanb\
is only valid for \mA\ larger than about 3~\Gcs.
The {\it large-}$\mu$ scenario is entirely excluded for \mt~=~174.3~\Gcs\ or smaller.
}
\end{center}
\end{table}
\section{Results interpreted in CP-violating MSSM scenarios}
In this section, the search results are interpreted in the CP-violating benchmark scenario {\it CPX} 
presented in Section 2.2, and in some variants of {\it CPX} where the basic model parameters are varied one-by-one.
Note that in these scenarios \mcalHc\ is always larger than 120~\Gcs, except where the CP-violating phases 
$\arg (A) = \arg (m_{\tilde {\rm g}})$  are put to $0^\circ$ or $180^\circ$.

\begin{figure}[htb]
\begin{center}
\centerline{
\epsfig{figure=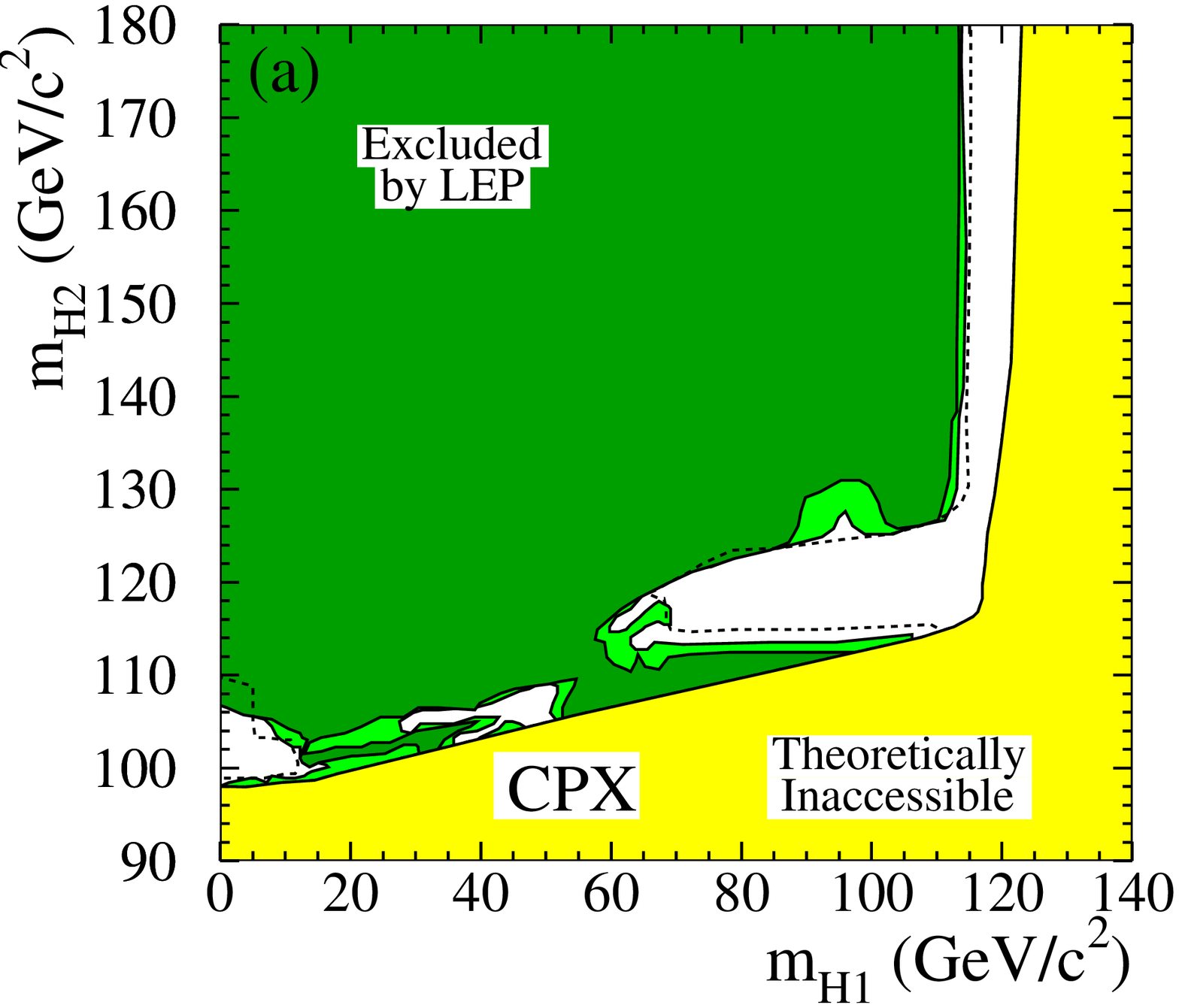,width=0.49\textwidth}
\epsfig{figure=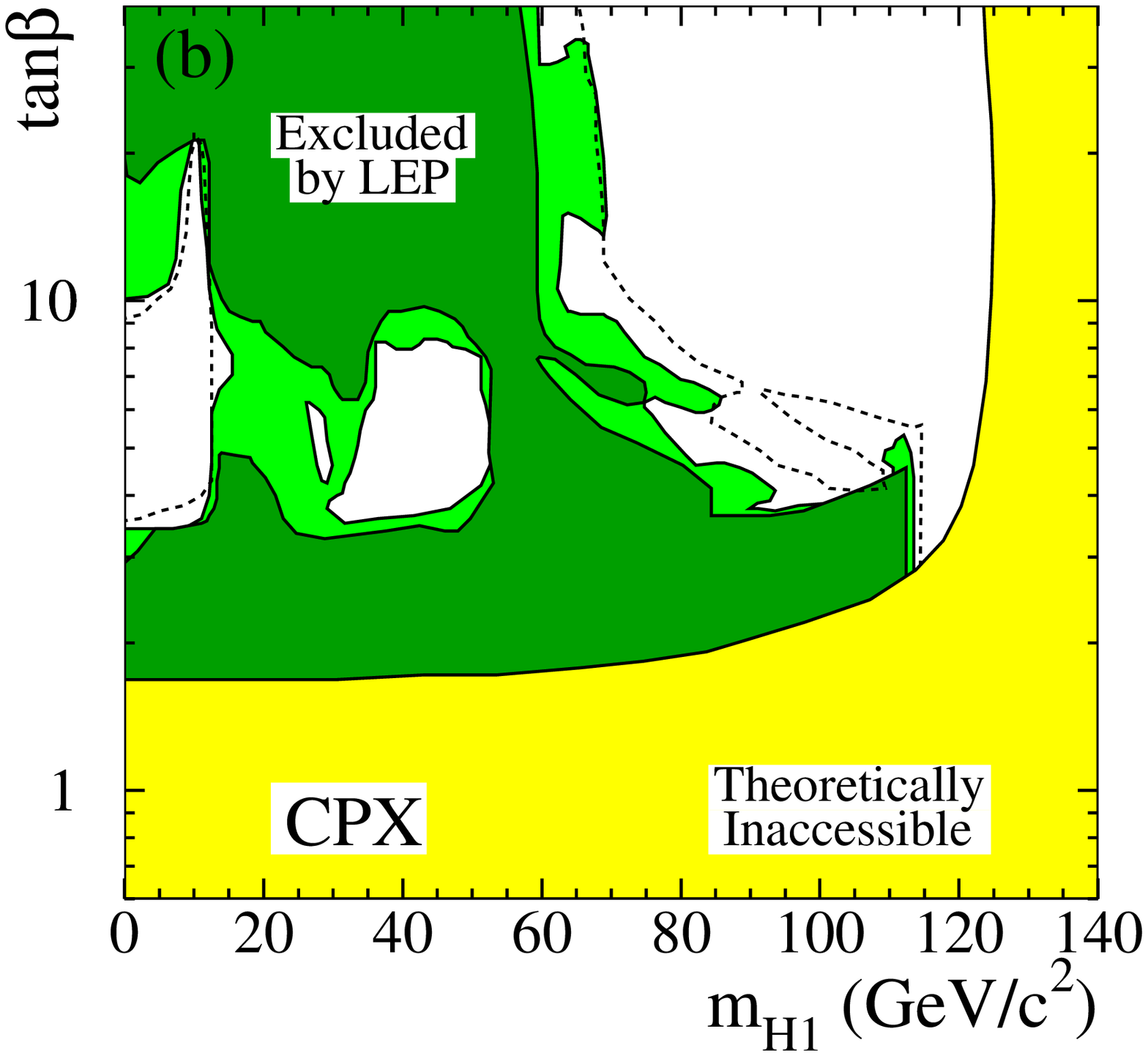,width=0.49\textwidth}
}
\centerline{
\epsfig{figure=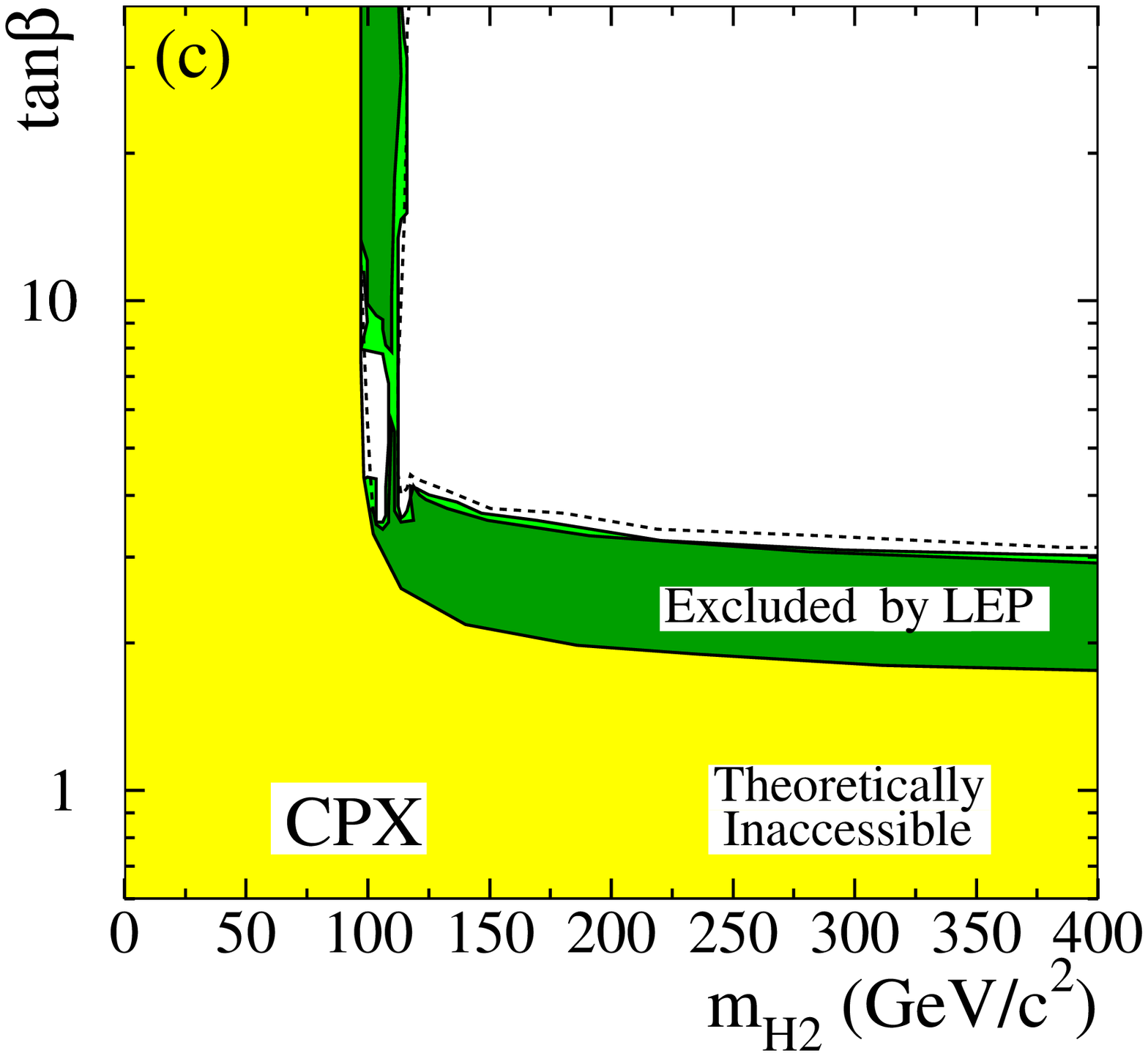,width=0.49\textwidth}
\epsfig{figure=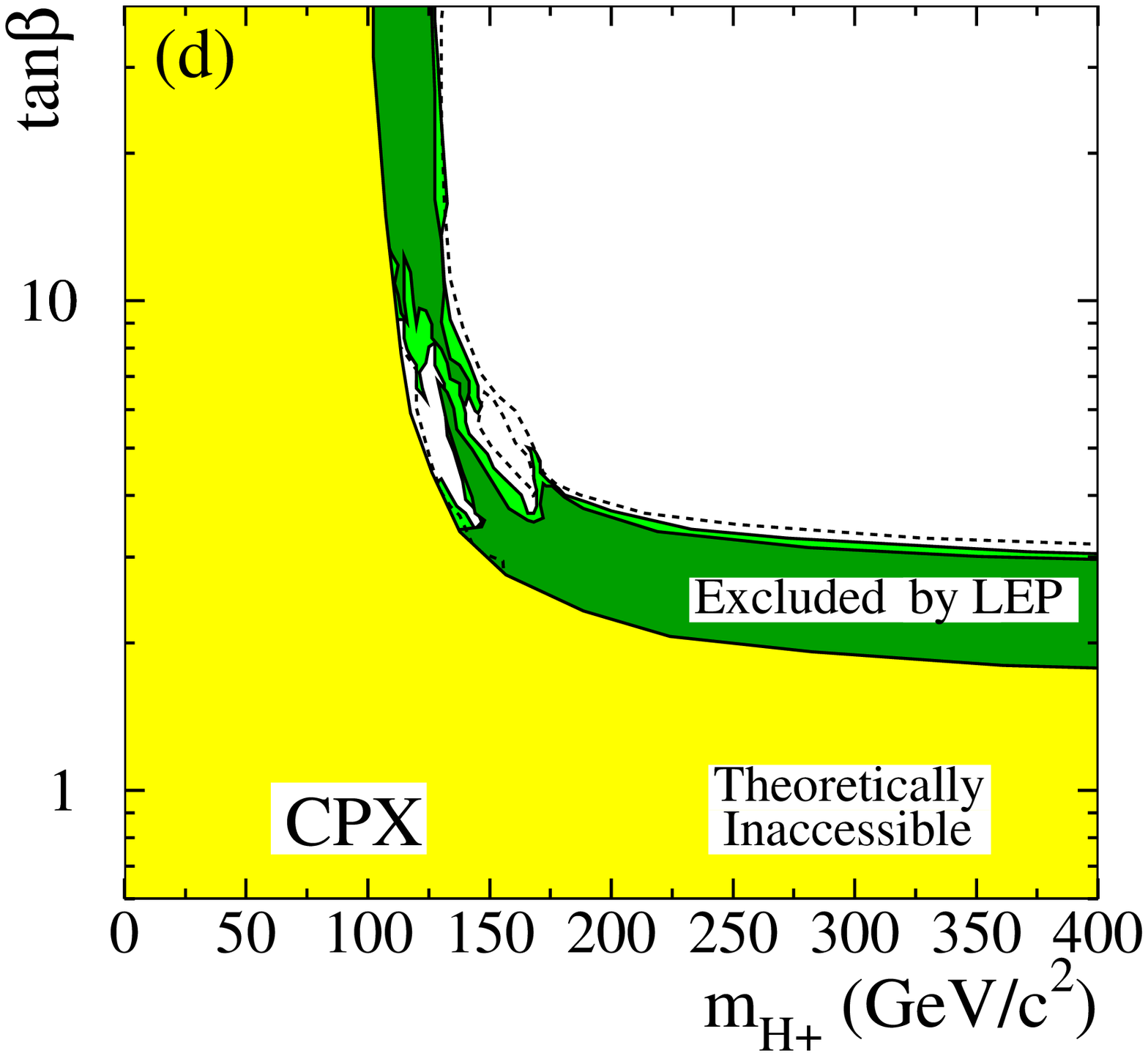,width=0.49\textwidth}
}
\caption[]{\sl Exclusions, at 95\% CL (medium-grey or light-green) and the 99.7\% CL 
(dark-grey or dark-green), for the CP-violating {\it CPX} scenario with $m_{\rm t}=174.3$~\Gcs.
  The figure shows the theoretically inaccessible domains (light-grey or yellow) and the
  regions excluded by the present search, 
         in four projections of the MSSM parameter space:
         (\mcalHa,~\mcalHb), (\mcalHa,~\tanb), (\mcalHb,~\tanb) and (\mHp,~\tanb).
         The dashed lines indicate the boundaries of the
         regions expected to be excluded, at the 95\% CL, 
	 on the basis of Monte Carlo simulations 
	 with no signal. In each scan point, the more conservative
of the two theoretical calculations, {\tt FeynHiggs} or {\tt CPH}, is used.

\label{fig:cpx-179}}
\end{center}
\end{figure}

The experimental exclusions for the {\it CPX} benchmark scenario 
are shown in Figure~\ref{fig:cpx-179}, in four projections. 
For large \mcalHb, the \calHa\ is almost completely CP-even; in this case the limit on \mcalHa\ is close to 114~\Gcs, 
the limit obtained for the Standard Model Higgs boson~\cite{lep-sm}. For example, for \mcalHb\ larger than 133~\Gcs, 
one can quote a lower bound of 113~\Gcs\ for \mcalHa. Large CP-odd admixtures to \calHa\ occur, however,
for smaller \mcalHb, giving rise to domains at lower \mcalHa\ which are not excluded. 

\begin{figure}[htb]
\begin{center}
\centerline{
\epsfig{figure=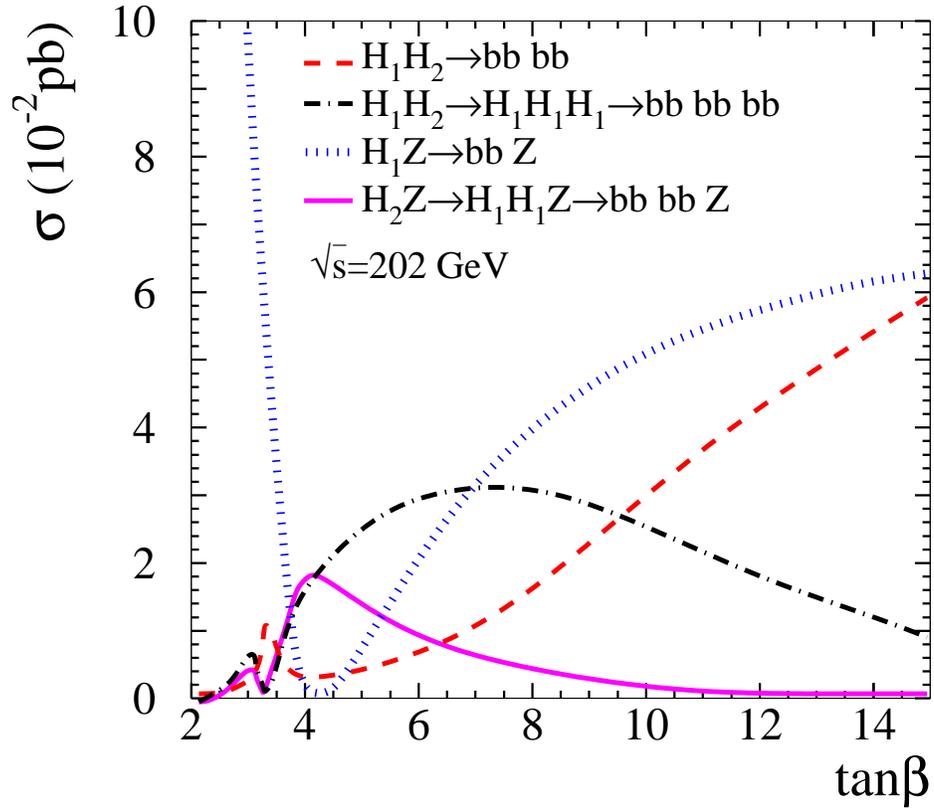,width=0.8\textwidth}
}
\caption[]{\sl Cross-sections, as a function of \tanb, for some of the dominant signal processes,
in the CP-violating scenario {\it CPX},
using the {\tt FeynHiggs} calculation, with a centre-of-mass energy of 202 {\rm GeV}, \mt~=~175~\Gcs, 
and \mcalHa\ between 35 and 45 \Gcs. 
\label{fig:cpv-xsec}}
\end{center}
\end{figure}

The exclusion is particularly weak
for \tanb\ between about 3.5 and 10. Here, 
the signal is spread over several channels arising
from the Higgsstrahlung and pair-production processes, including the \calHb\ra~\calHa\calHa\ cascade decays, 
which give rise to complex final states with six jets. The parameter set of
Table~\ref{tab:cpv-examples} is a typical example of this situation.
%
This is illustrated in
Figure~\ref{fig:cpv-xsec} where the main final-state cross-sections
are plotted as a function of \tanb\ (the {\tt FeynHiggs} calculation is used). 
In general, these signal contributions cannot be added up statistically 
because of a large overlap in the 
selected events; hence, a relatively low overall detection efficiency is expected.
Moreover, one of the experiments presents a local excess of about two standard deviations in this domain
of \tanb\ and for \mcalHa\ of about 45~\Gcs~\cite{mssm-o},
which lowers the exclusion power below the expectation.   
Nonetheless, the region defined by \mcalHa~$<$~114~\Gcs\ and \tanb~$<$~3.0 is 
excluded by the data (see Figure~\ref{fig:cpx-179} (b)) and a 95\% CL lower bound
of 2.9 can be set on \tanb\ in this scenario.

\begin{figure}[htb]
\begin{center}
{\large~~~~~~~~Excluded in {\tt CPH}~~~~~~~~~~~~~~~~~~~~~~~~~~Excluded in {\tt FeynHiggs}}\\
\vspace{-0.5cm}
\epsfig{figure=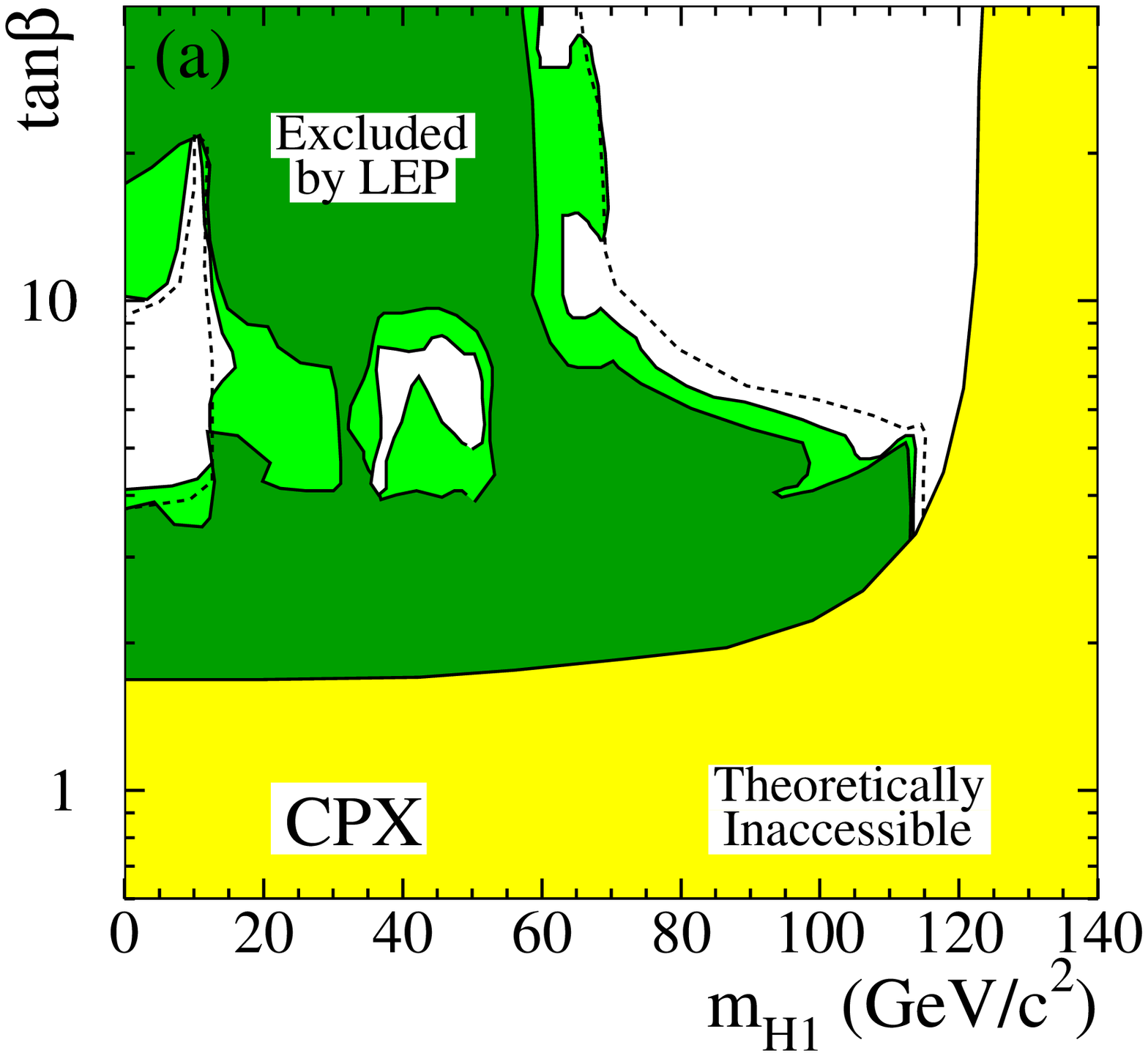,width=0.49\textwidth}
\epsfig{figure=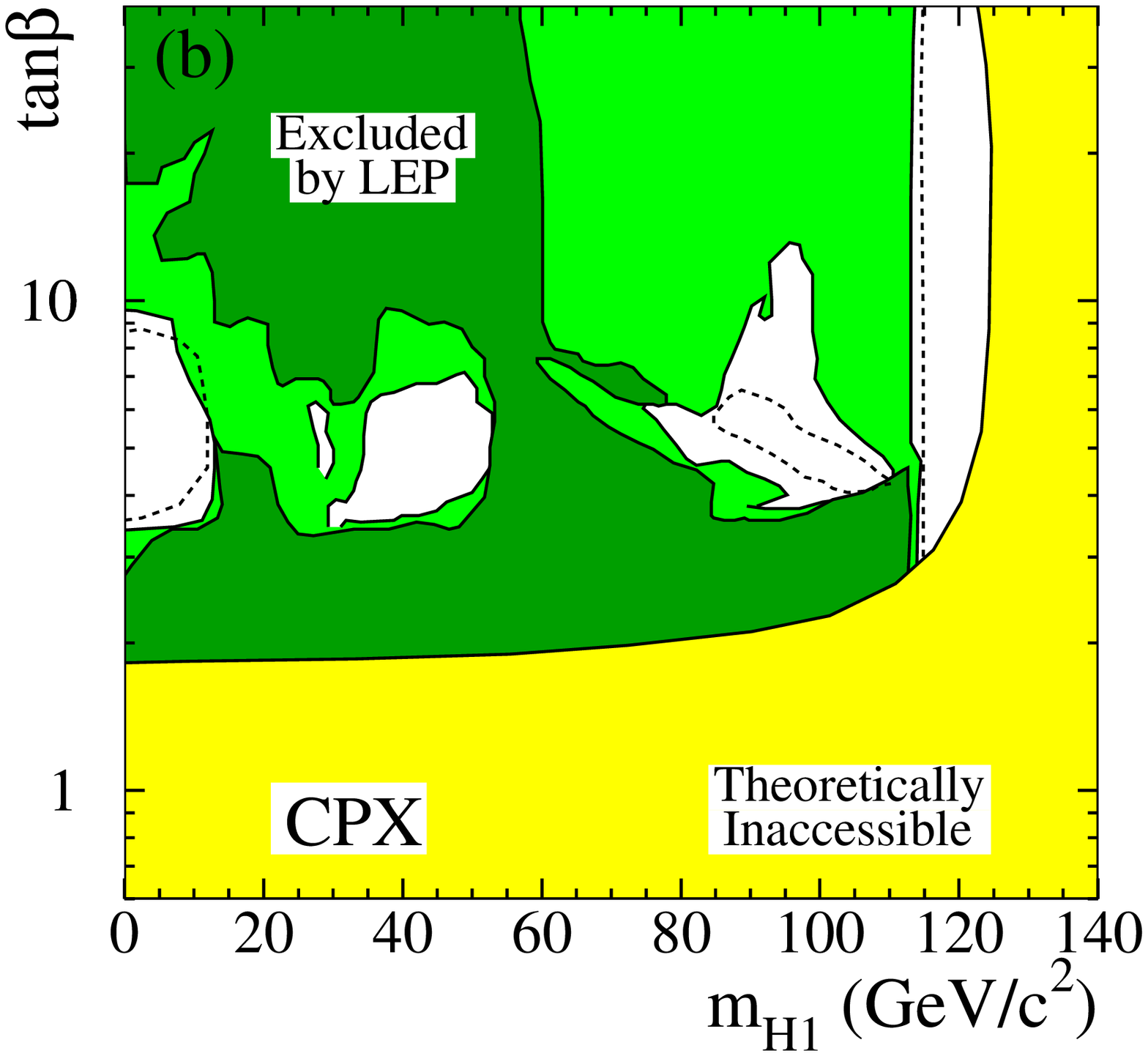,width=0.49\textwidth}\\
$\phantom{...}$\\
$\phantom{...}$\\
{\large Excluded in {\tt CPH} and {\tt FeynHiggs}}\\
\vspace{-0.5cm}
\epsfig{figure=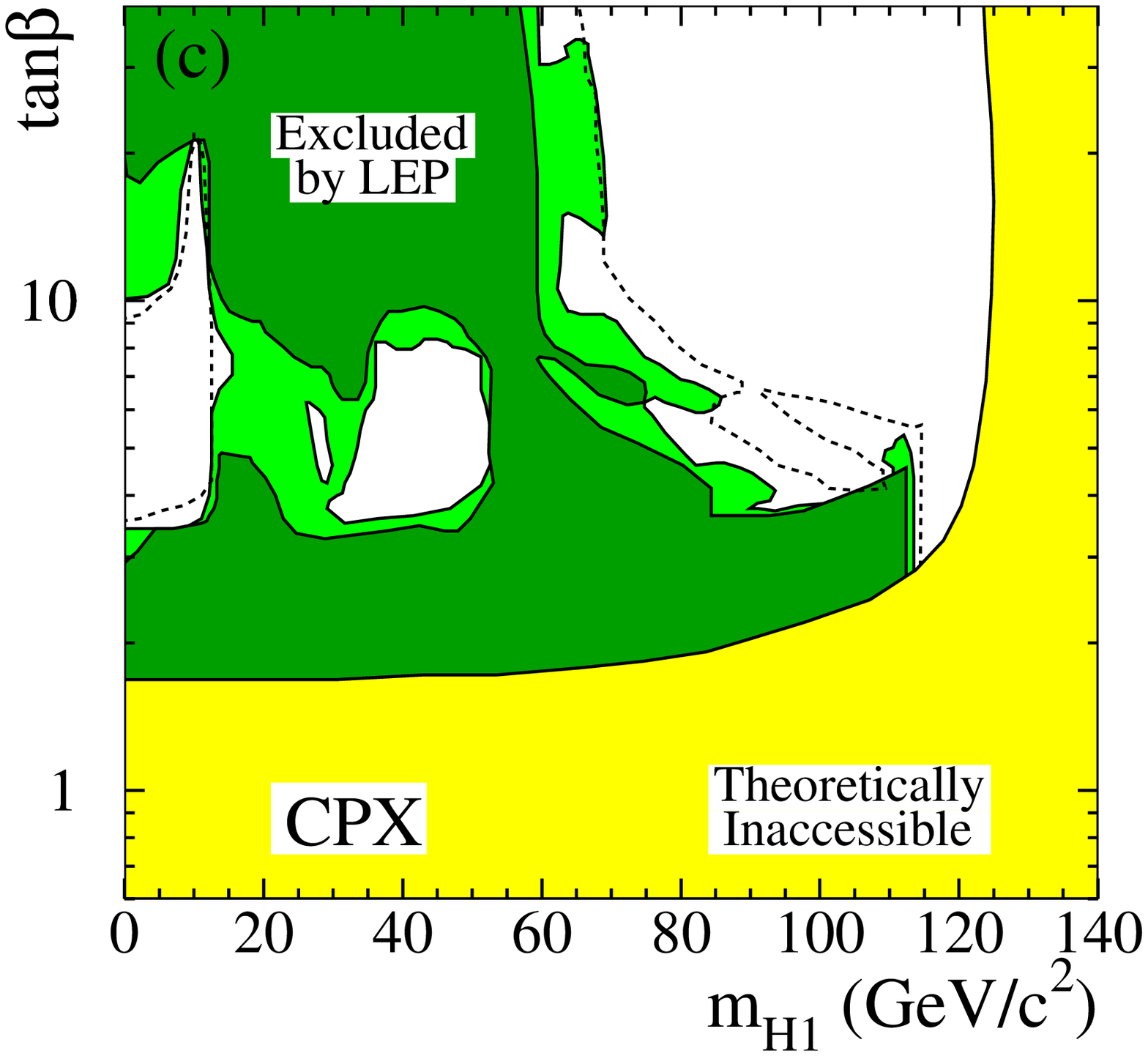,width=0.49\textwidth}
\end{center}
\caption[]{\sl Exclusions, in the case of the CP-violating {\it CPX} scenario, 
for the two theoretical approaches, {\tt CPH} and {\tt FeynHiggs}. See the caption of Figure~\ref{fig:cpx-179} 
for the legend.
In part (a) the {\tt CPH} calculation is used and in part (b) the {\tt FeynHiggs} calculation. 
In part (c) the procedure
is adopted where, in each scan point of the parameter space, the more conservative of the two calculations is used.
\label{fig:cpx-models}}
\end{figure}

Figure~\ref{fig:cpx-models} illustrates the exclusions in the $(\mcalHa,\tanb)$ projection,
using the {\tt CPH} calculation (part (a)) and the {\tt FeynHiggs} calculation (part (b)).
Differences occur mainly at large \tanb\ where the {\tt FeynHiggs} calculation
predicts a larger Higgsstrahlung cross-section and hence a better search sensitivity than the {\tt CPH} calculation.
In parts (a) and (b) of the figure, one observes two distinct domains at moderate \tanb, with \mcalHa$<$15~\Gcs\ and 
30~\Gcs$<$\mcalHa$<$55~\Gcs, which are not excluded at the 95\% CL. The values of $1-CL_b$ indicate that
these domains are excluded, respectively, at the 55\% CL and 77\% CL using the {\tt CPH} calculation, and at the 50\% CL 
and 66\% CL, respectively, 
using the {\tt FeynHiggs} calculation. A third domain appears in part (b) at higher \mcalHa\  
(where the {\tt CPH} calculation indicates no exclusion power at all); this domain is excluded at the 42\% CL using {\tt FeynHiggs}.

As explained in Section 2, neither of the two approaches, {\tt CPH} or {\tt FeynHiggs}, are preferred
on theoretical grounds. 
For this reason, part (c) of this figure was obtained by choosing in each scan point of the parameter space
the more conservative of the two approaches, {\it i.e.}, the one for which the less significant exclusion is observed. 
The same procedure was adopted in Figure~\ref{fig:cpx-179} and in all the figures which follow.

\begin{figure}[htb]
\begin{center}
{\large ~~~~~$m_{\rm t}= 169.3$~\Gcs~~~~~~~~~~~~~~~~~~~~~~~~~~~~~~$m_{\rm t}= 174.3$~\Gcs}\\
\vspace{-0.5cm}
\epsfig{figure=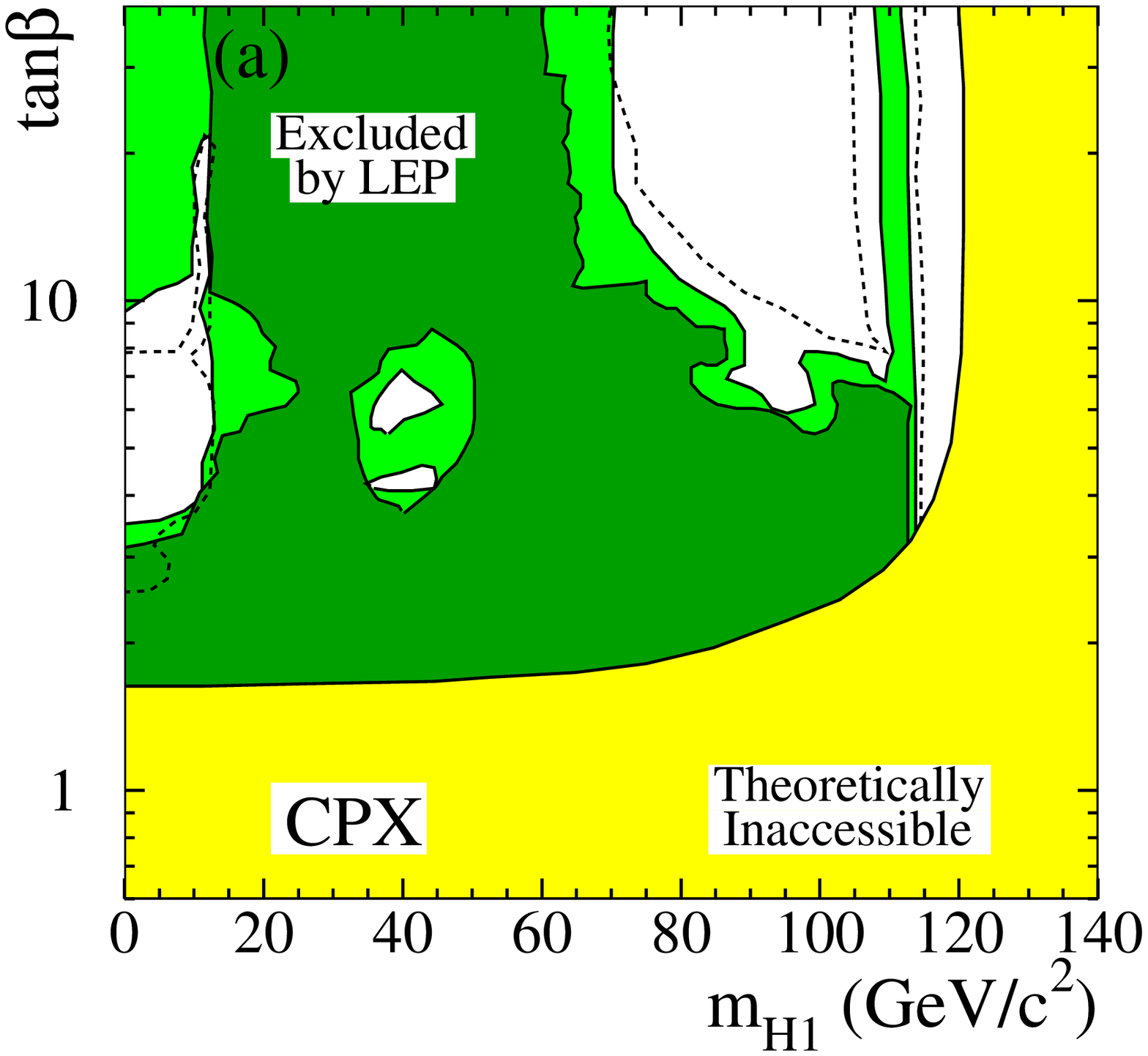,width=0.49\textwidth}
\epsfig{figure=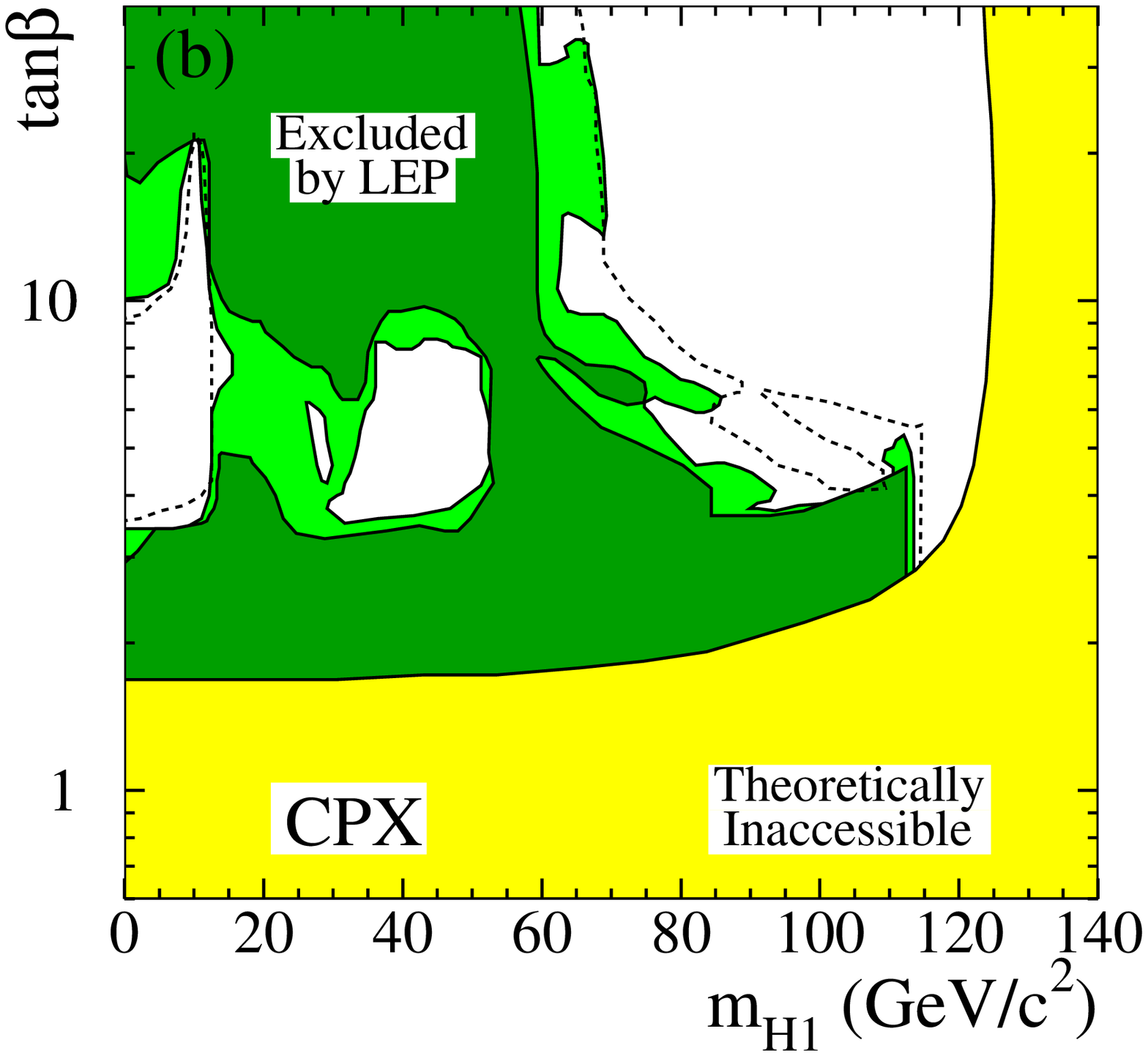,width=0.49\textwidth}\\
$\phantom{...}$\\
$\phantom{...}$\\
{\large ~~~~~$m_{\rm t}= 179.3$~\Gcs~~~~~~~~~~~~~~~~~~~~~~~~~~~~~~$m_{\rm t}= 183.0$~\Gcs}\\
\vspace{-0.5cm}
\epsfig{figure=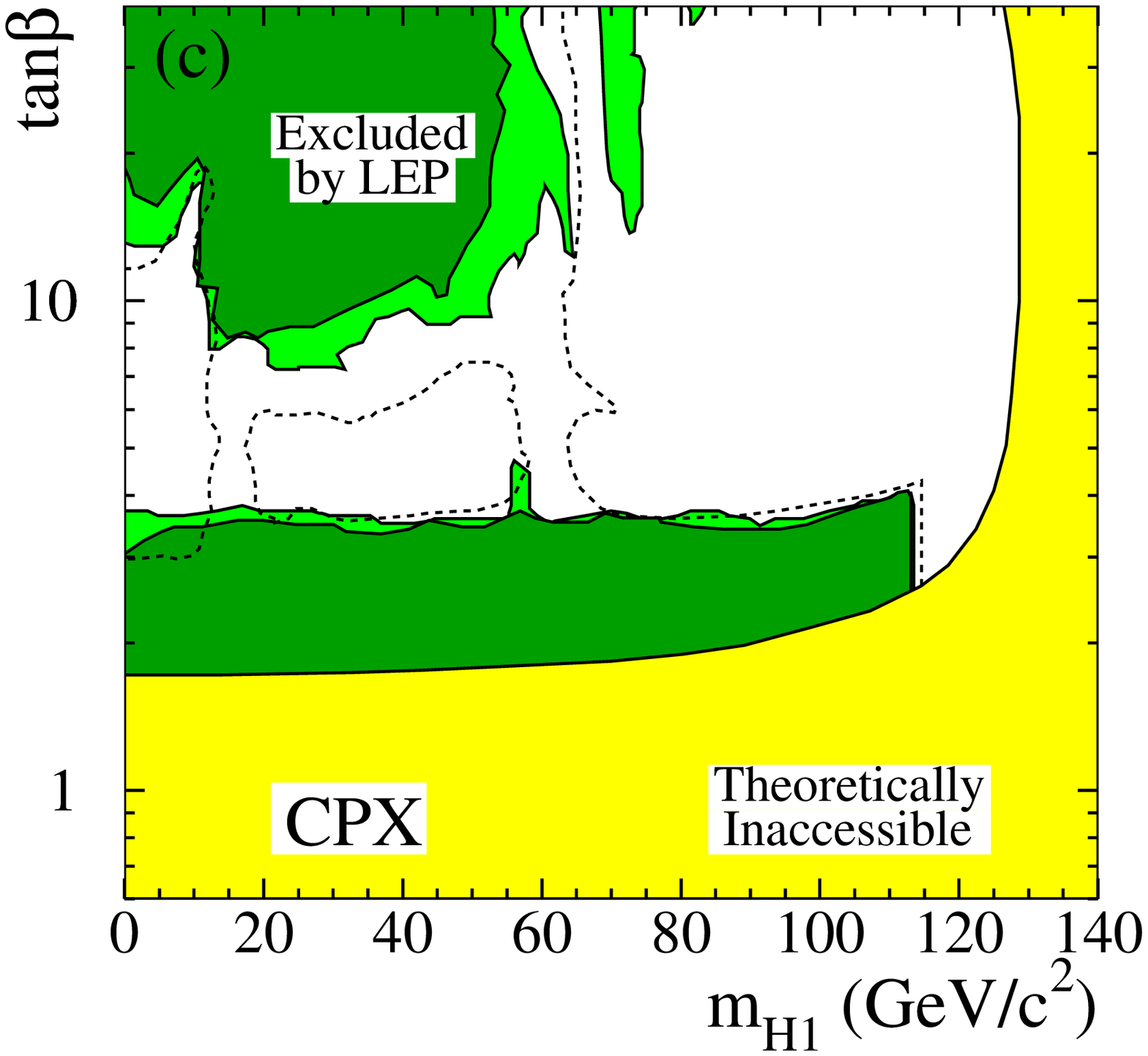,width=0.49\textwidth}
\epsfig{figure=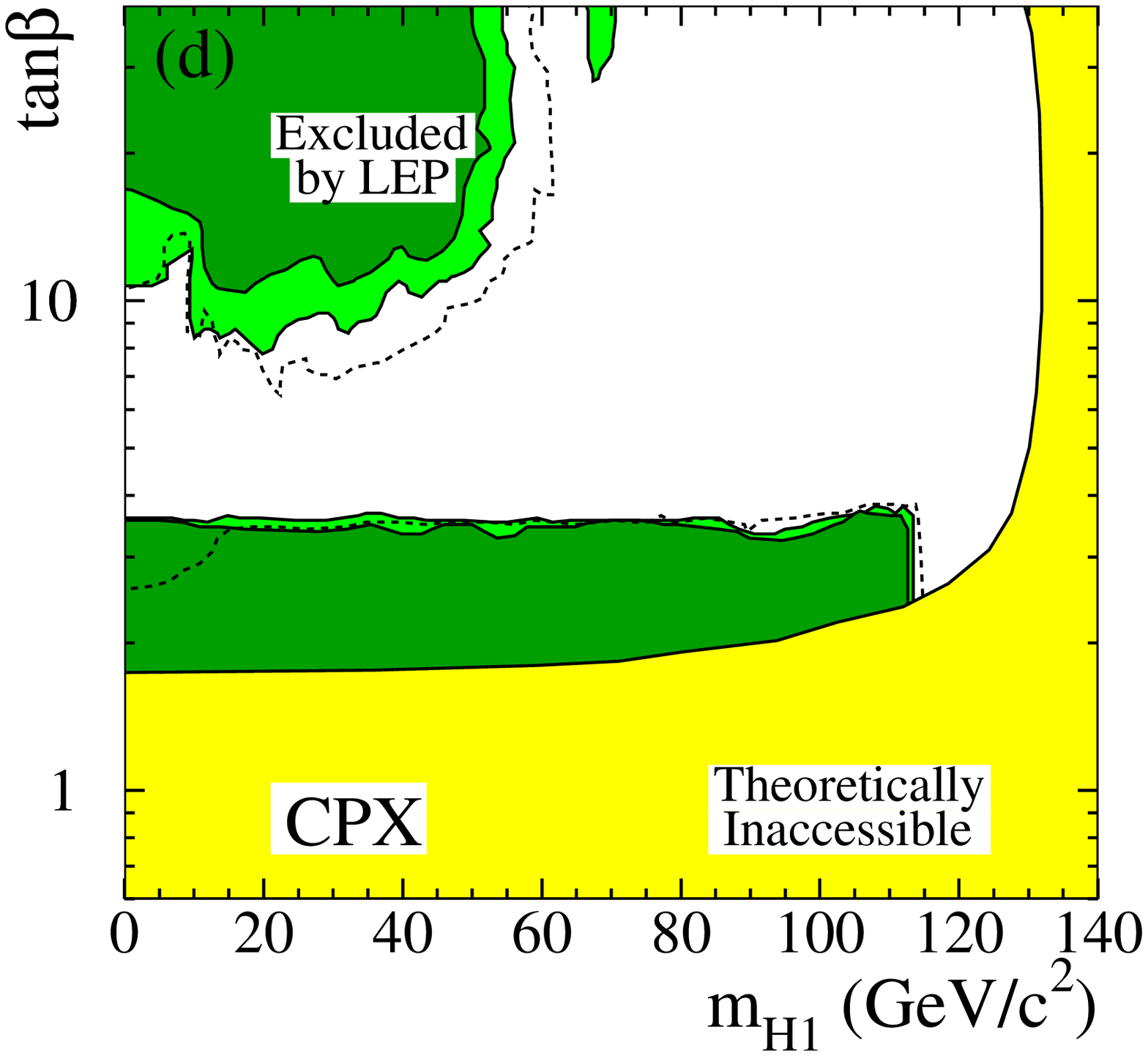,width=0.49\textwidth}
\end{center}
\caption[]{\sl Exclusions, in the case of
the CP-violating {\it CPX} scenario, for four top quark masses:
$m_{\rm t}= 169.3$~\Gcs, $174.3$~\Gcs, $179.3$~\Gcs\ and $183.0$~\Gcs. 
See the caption of Figure~\ref{fig:cpx-179} for the legend.
\label{fig:cpx-topmass}}
\end{figure}

The significant impact of the top mass on the CP-violating effects, indicated by Eq.~11, is illustrated in 
Figure~\ref{fig:cpx-topmass} where the (\mcalHa, \tanb) projection is shown for four values of \mt.
With increasing \mt, one observes a reduction of the
exclusion power, especially in the region of \tanb\ 
between 3.5 and 10. No lower bound on \mcalHa\ can be quoted in this domain.
In plot (a) (for \mt~=~169.3~\Gcs), the two domains with \mcalHa~$<$~15~\Gcs\ and 
30~\Gcs~$<$~\mcalHa~$<$~55~\Gcs\ are excluded at the 60\% CL and 88\% CL, respectively.

\begin{figure}[htb]
\begin{center}
{\large ~~~~~$\arg (A) =\arg (m_{\tilde {\rm g}}) = 0^\circ$~~~~~~~~~~~$\arg (A) =\arg (m_{\tilde {\rm g}}) = 30^\circ$}\\
\vspace{-0.5cm}
\epsfig{figure=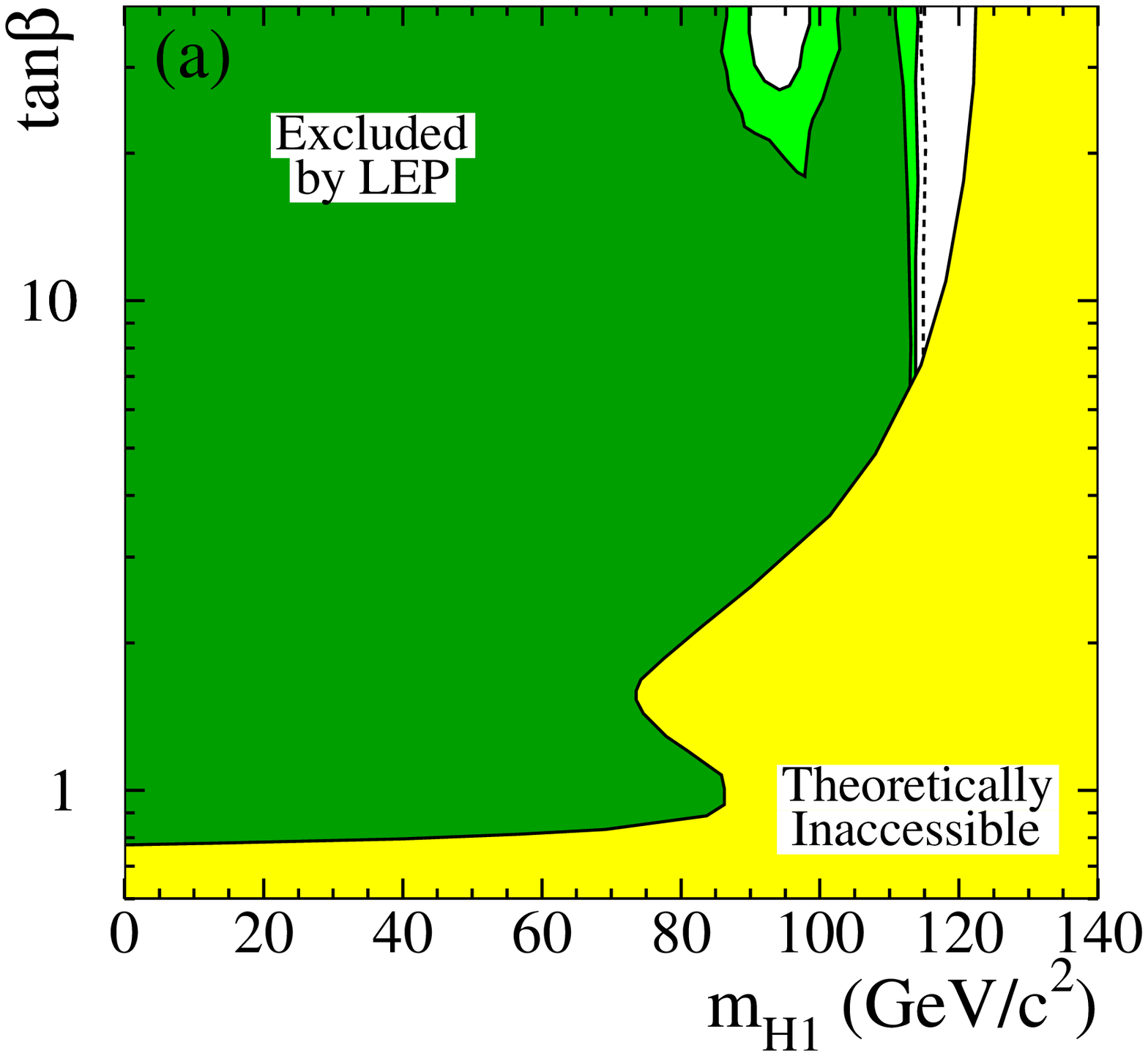,width=0.4\textwidth}
\epsfig{figure=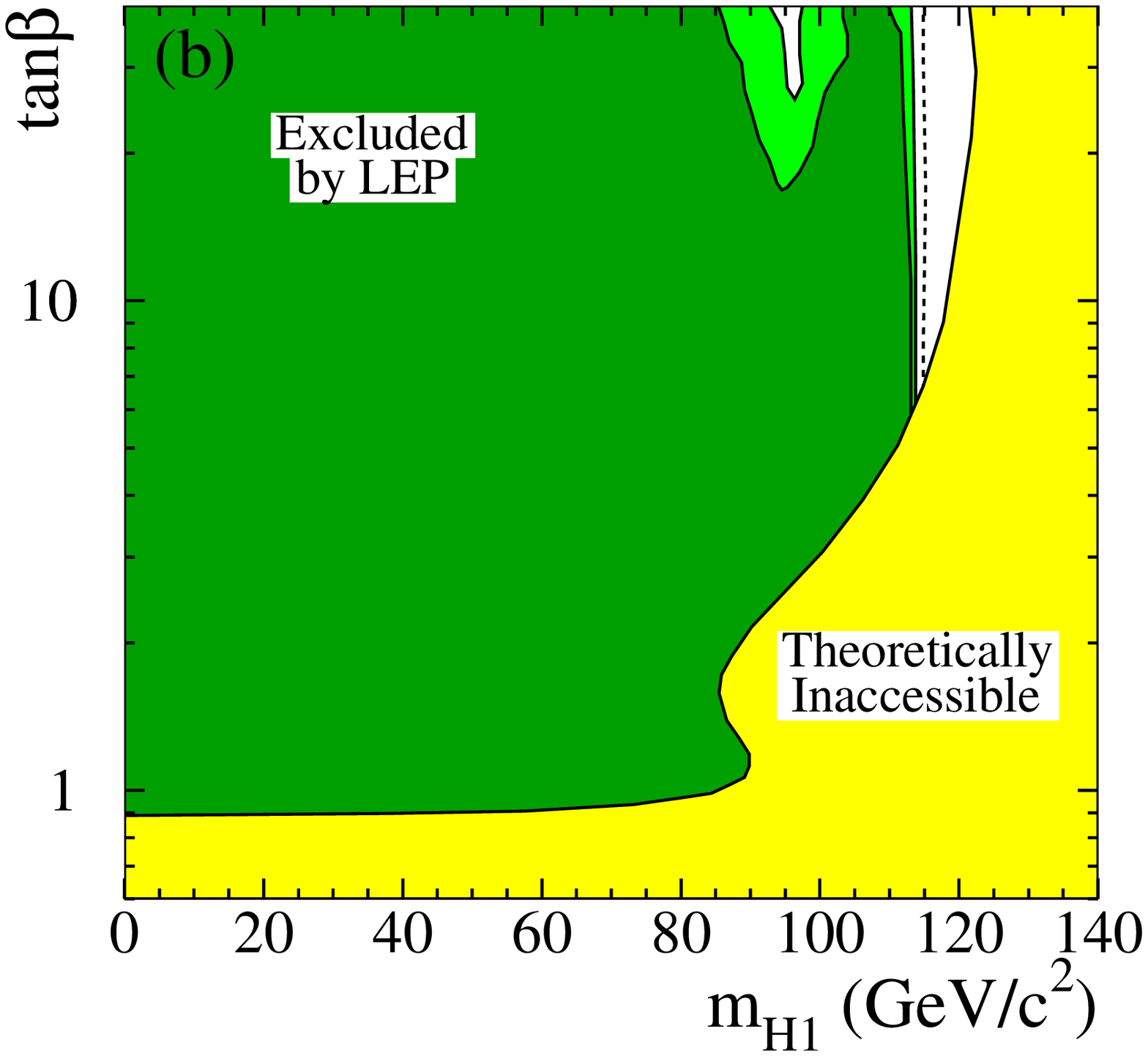,width=0.4\textwidth}\\
$\phantom{...}$\\
{\large ~~~~~$\arg (A) =\arg (m_{\tilde {\rm g}}) = 60^\circ$~~~~~~~~~~~$\arg (A) =\arg (m_{\tilde {\rm g}}) = 90^\circ$}\\
\vspace{-0.5cm}
\epsfig{figure=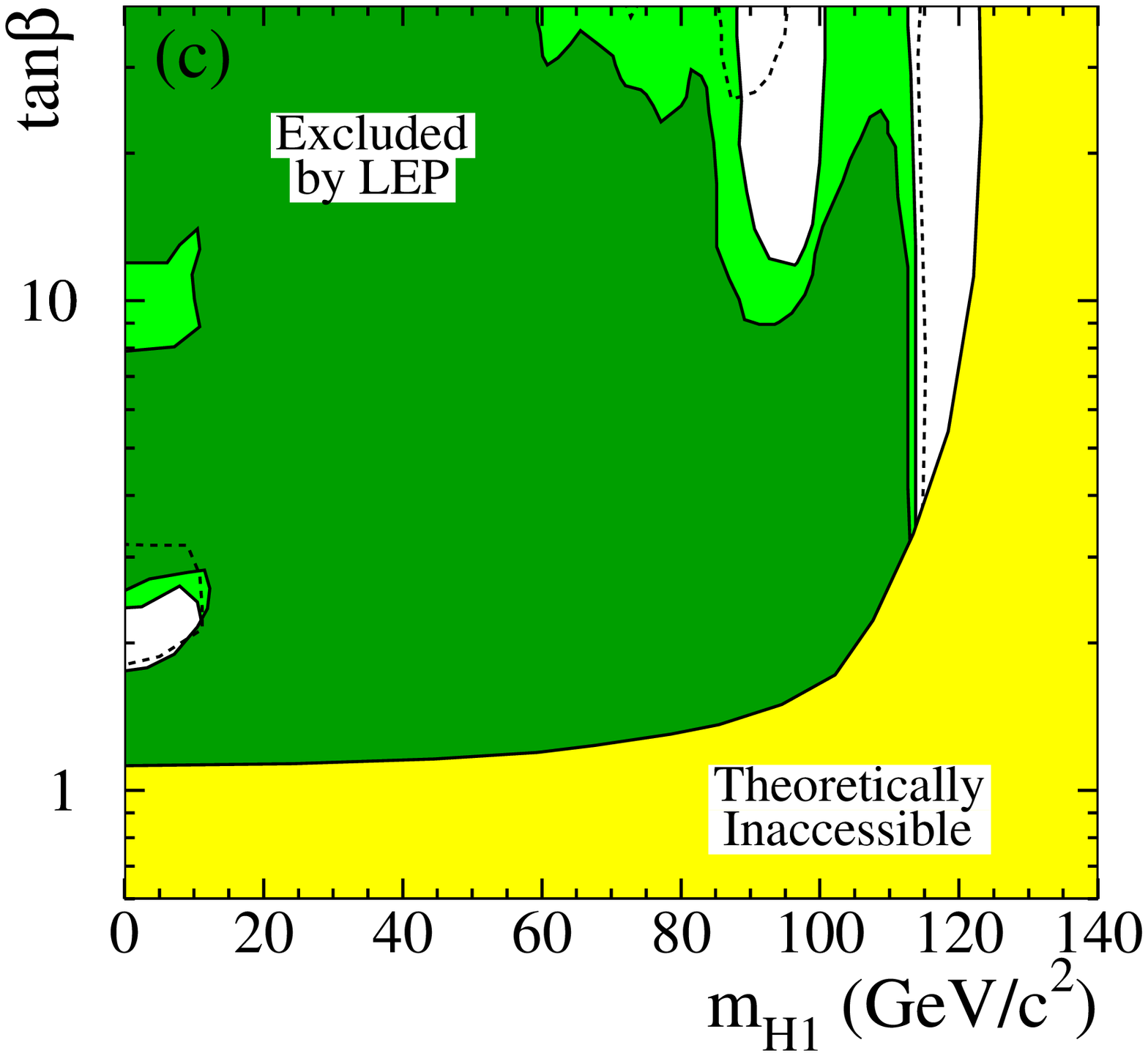,width=0.4\textwidth}
\epsfig{figure=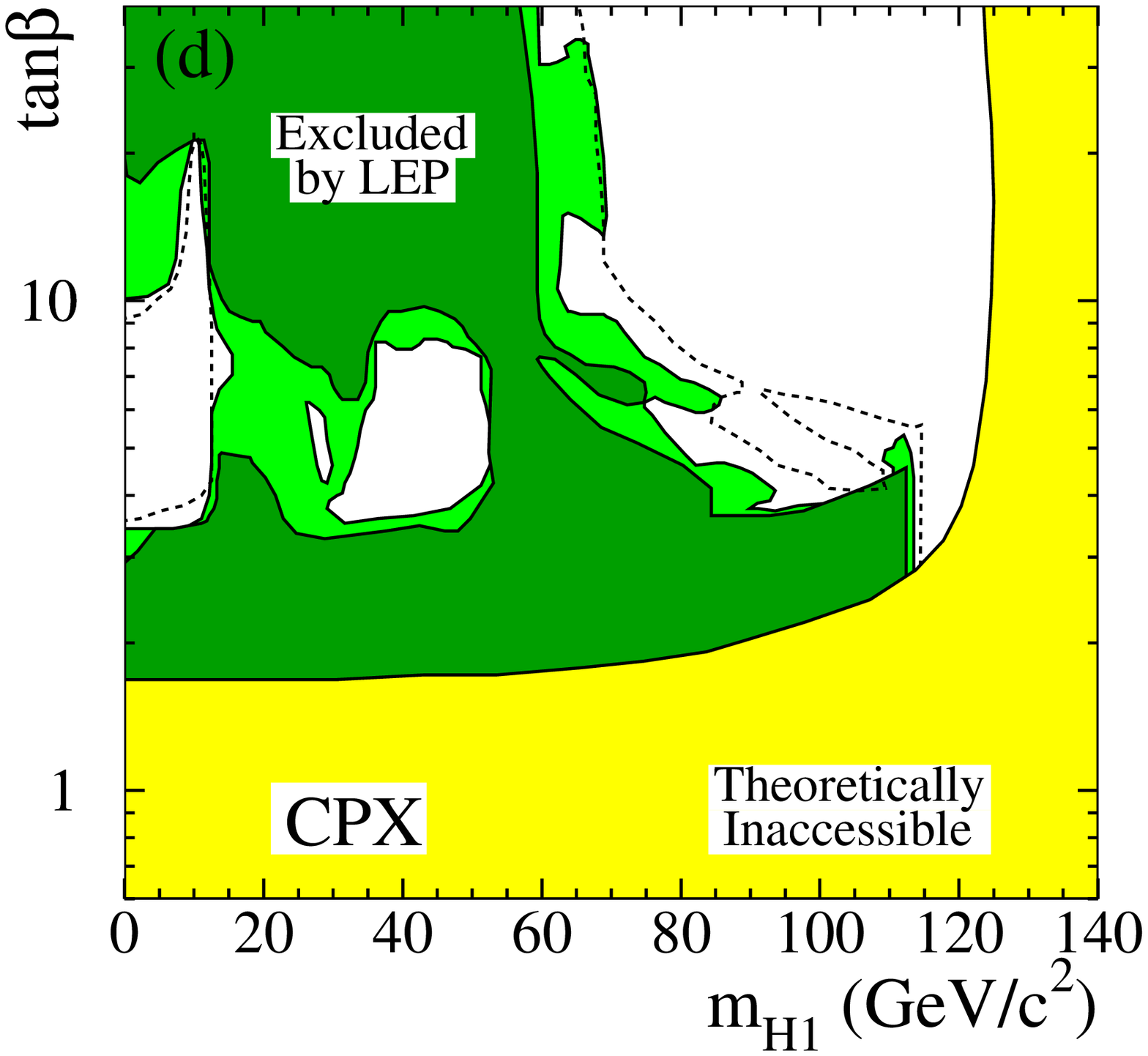,width=0.4\textwidth}\\
$\phantom{...}$\\
{\large ~~~~~$\arg (A) =\arg (m_{\tilde {\rm g}}) = 135^\circ$~~~~~~~~~~~$\arg (A) = \arg (m_{\tilde {\rm g}}) = 180^\circ$}\\
\vspace{-0.5cm}
\epsfig{figure=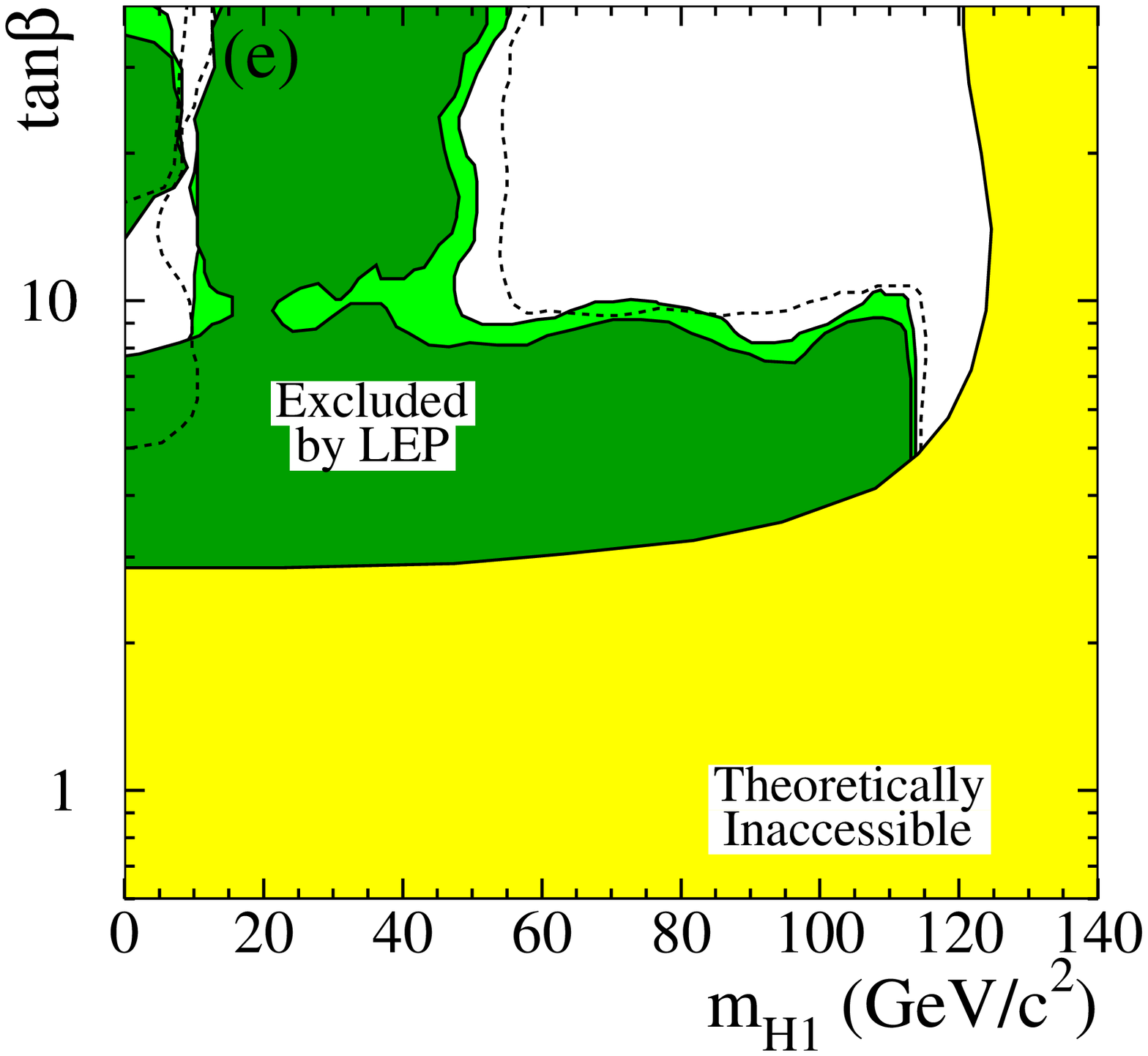,width=0.4\textwidth}
\epsfig{figure=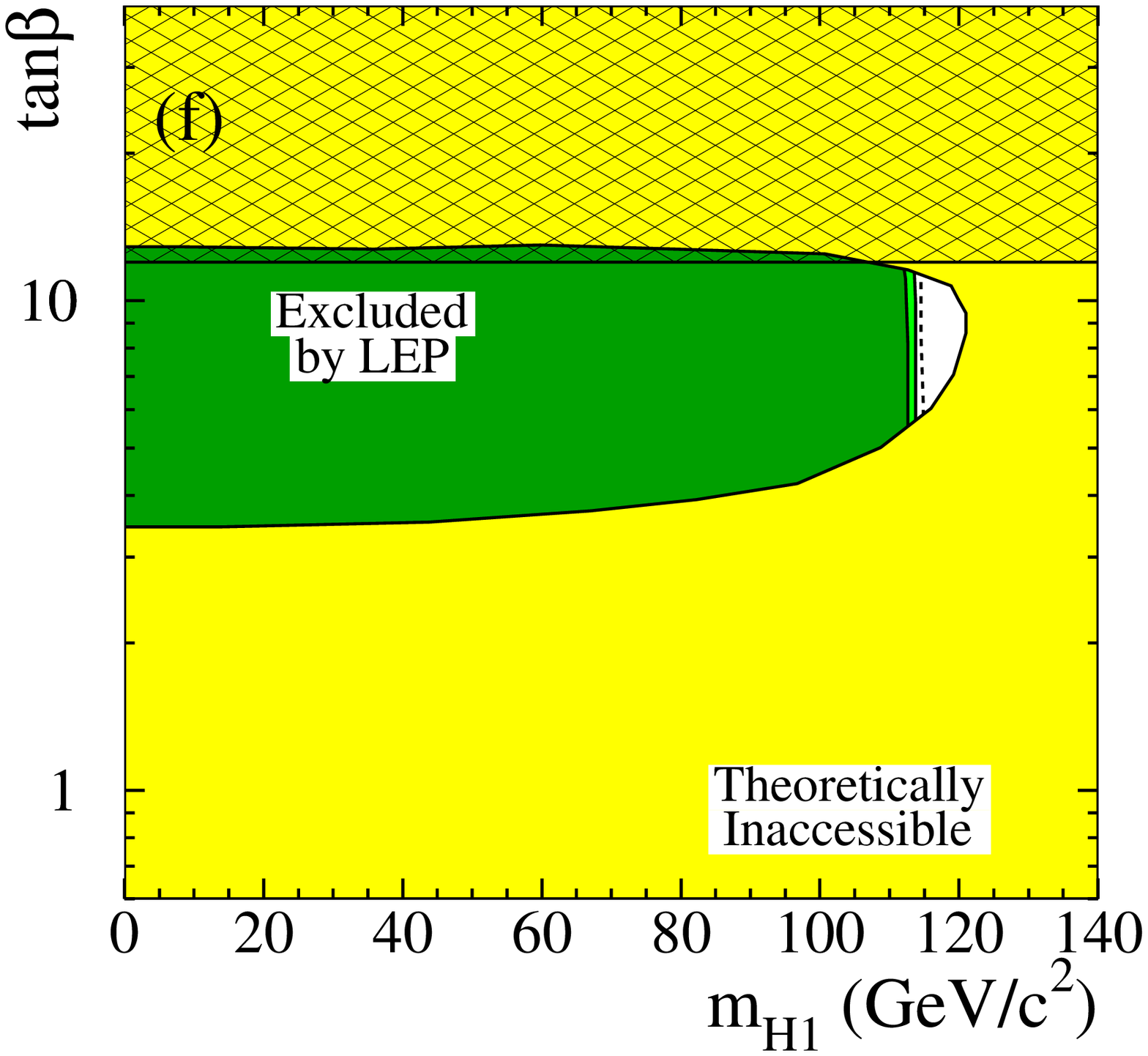,width=0.4\textwidth}
\end{center}
\caption[]{\sl Exclusions, in the case of the {\it CPX} scenario with various CP-violating phases, 
$\arg (A) = \arg (m_{\tilde {\rm g}})$:
$0^\circ,~30^\circ,~60^\circ,~90^\circ$ (the {\it CPX} value),~$135^\circ$ and $180^\circ$. 
See the caption of Figure~\ref{fig:cpx-179} for the legend. In the hatched region in part (f) the 
calculations are uncertain (see text).
\label{fig:cpx-phase}}
\end{figure}

Figure~\ref{fig:cpx-phase} illustrates the exclusion in the (\mcalHa, \tanb) plane as a function of the
CP-violating phases, $\arg (A) = \arg (m_{\tilde {\rm g}})$, which are varied together. 
For phase angles close to $0^\circ$, the experimental exclusions
are similar to those in the CP-conserving scenarios (see, for example, Figure~\ref{fig:mhmax} but note the differences
in the allowed parameter space).
Sizable differences are observed for larger phase angles, especially for $\arg (A) = \arg (m_{\tilde {\rm g}}) = 90^\circ$ 
(the {\it CPX} value).
At $\arg (A) = \arg (m_{\tilde {\rm g}}) = 180^\circ$ (another CP-conserving scenario), the allowed parameter space is excluded 
almost completely. Note however that in the hatched region, with \tanb\ greater than about 12, 
the calculation of the bottom-Yukawa coupling has large theoretical
uncertainties; hence no exclusion can be claimed in this domain.
                                                                                                  
\begin{figure}[htb]
\begin{center}
{\large ~~~~~$\mu = 500$~GeV~~~~~~~~~~~~~~~~~~~~~~~~~~~~~~~$\mu = 1000$~GeV}\\
\vspace{-0.5cm}
\epsfig{figure=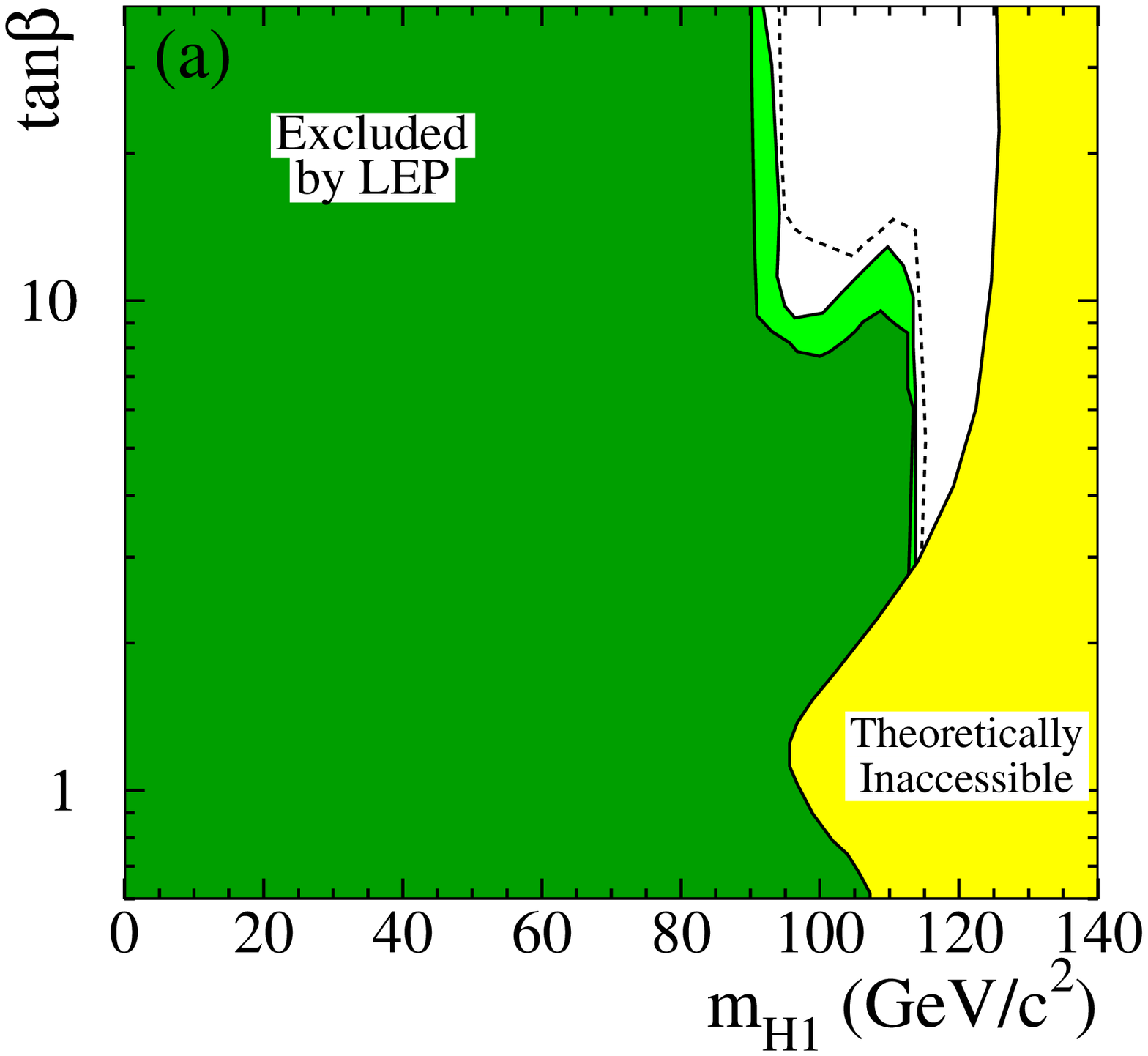,width=0.49\textwidth}
\epsfig{figure=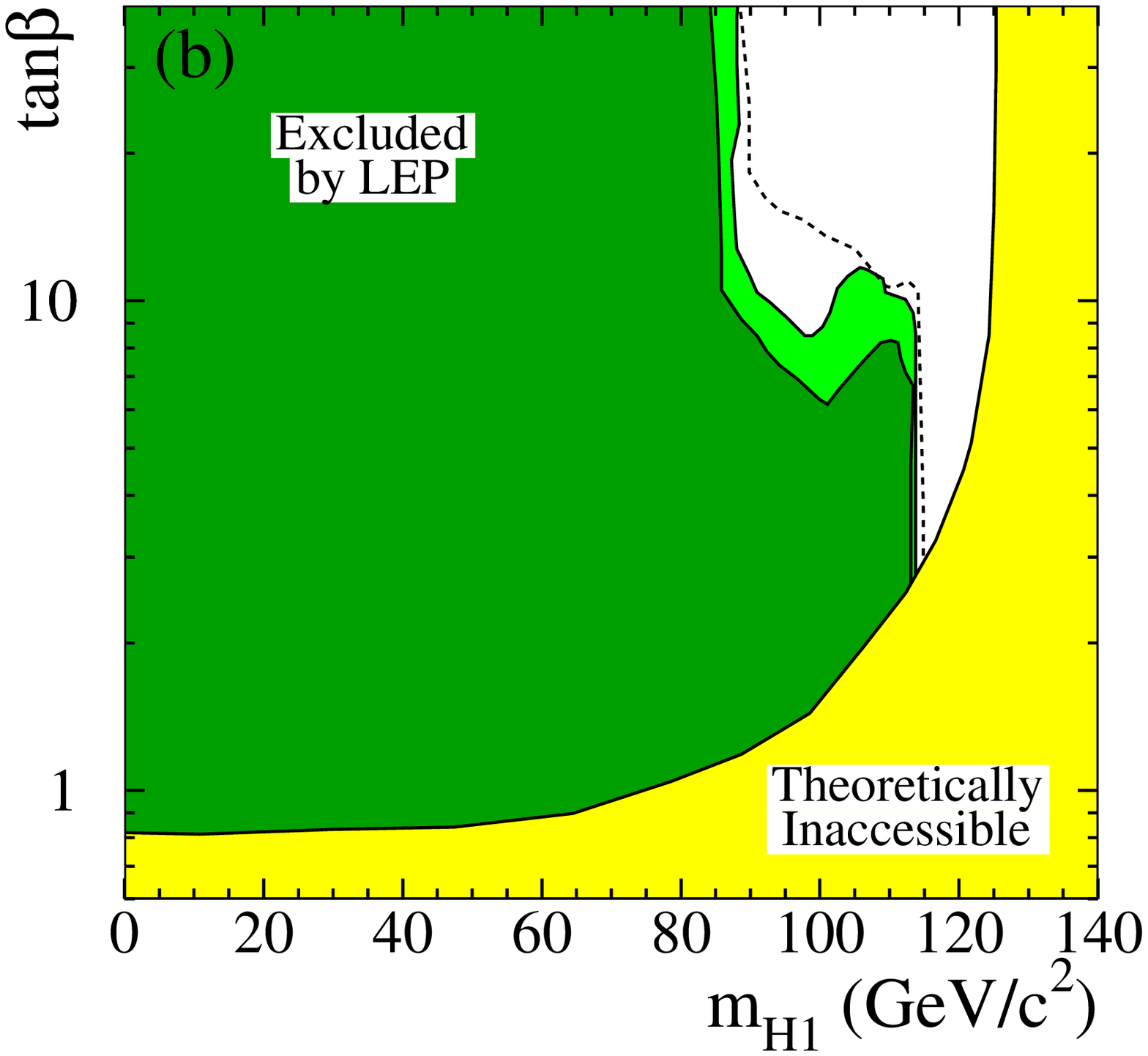,width=0.49\textwidth}\\
$\phantom{...}$\\
$\phantom{...}$\\
{\large ~~~~~$\mu = 2000$~GeV~~~~~~~~~~~~~~~~~~~~~~~~~~~~~~~$\mu = 4000$~GeV}\\
\vspace{-0.5cm}
\epsfig{figure=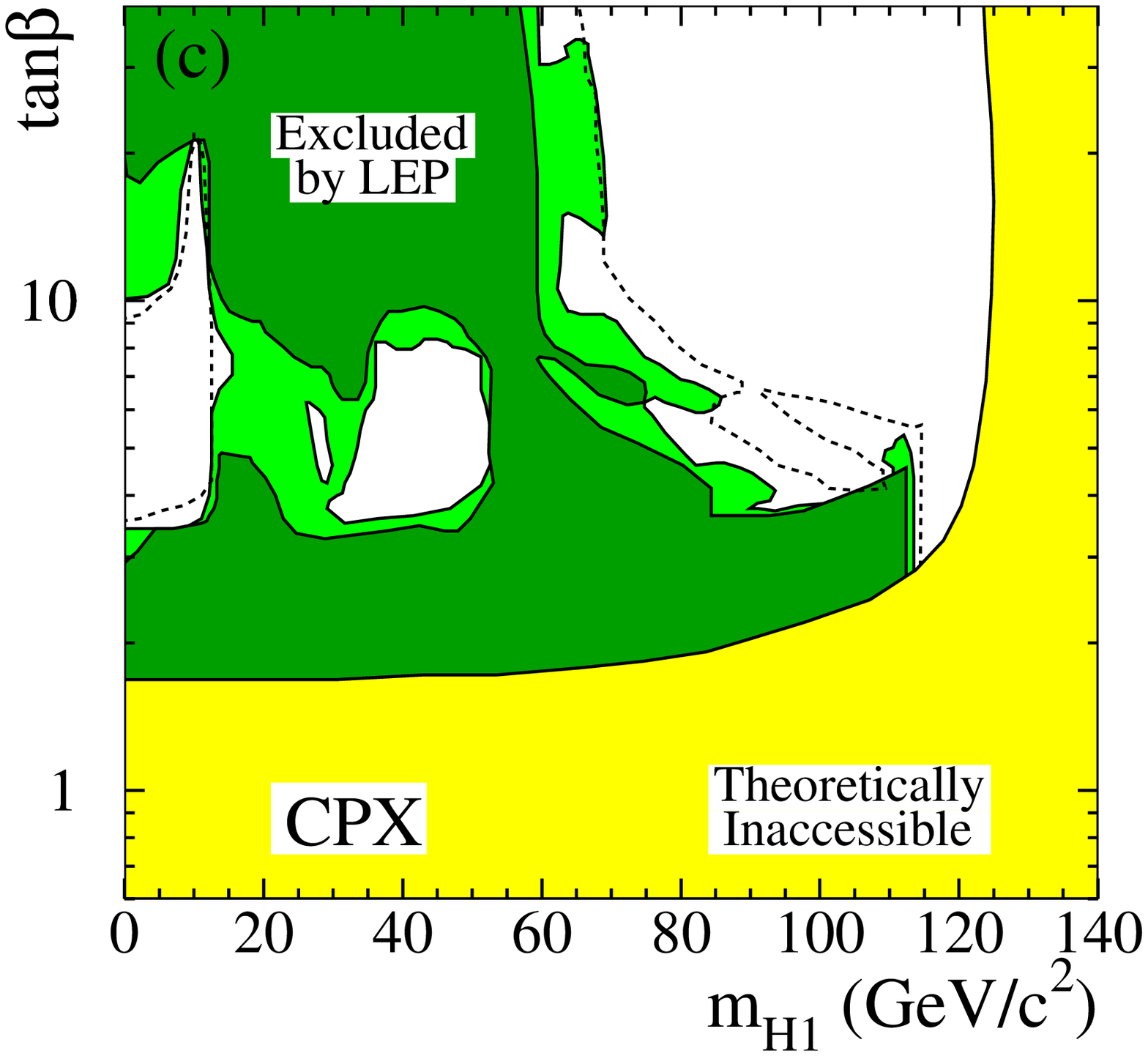,width=0.49\textwidth}
\epsfig{figure=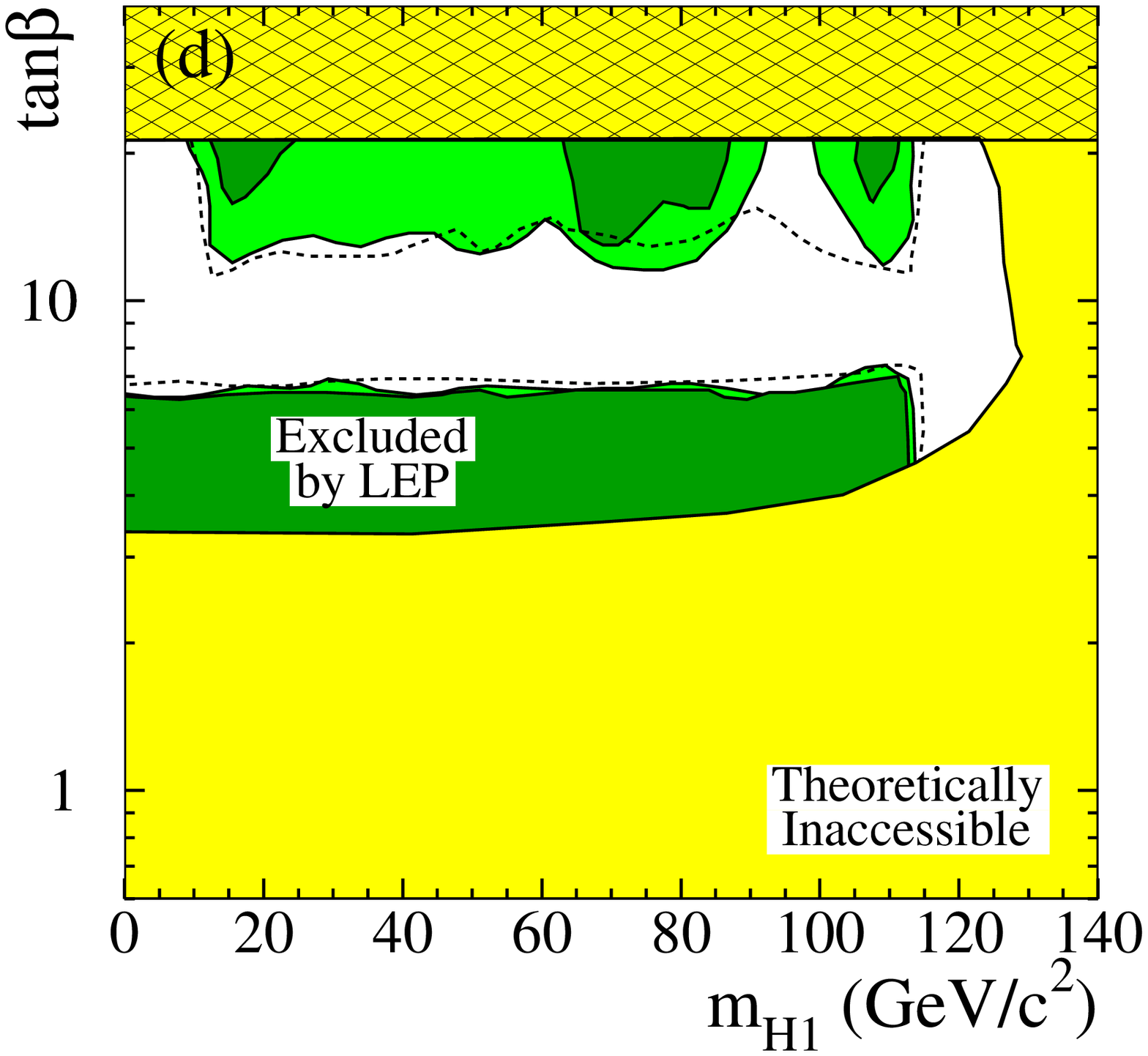,width=0.49\textwidth}
\end{center}
\caption[]{\sl Exclusions, 
for the CP-violating {\it CPX} scenario with various values of the Higgs mass parameter $\mu$:
500~GeV,~1000~GeV,~2000~GeV (the standard {\it CPX} value) and 4000~GeV. 
See the caption of Figure~\ref{fig:cpx-179} for the legend. In the hatched region in part (d) the 
calculations are uncertain (see text).
\label{fig:cpx-mu}}
\end{figure}

In Figure~\ref{fig:cpx-mu}, the value of the Higgs mass parameter $\mu$ is varied from 500~GeV 
through 1000~GeV and 2000~GeV (the {\it CPX} value) to 4000~GeV.
At small values, the CP-violating effects are small (see Eq.~11) and the exclusion power is strong (as in the
CP-conserving case).  
For $\mu$ larger than 2000~GeV and large \tanb,
the {\tt FeynHiggs} and {\tt CPH} calculations both provide bottom-Yukawa coupling 
in the non-perturbative regime, giving rise to negative values for the square of \mcalHa\ and to other unphysical results. 
For $\mu\leq$~2000~GeV this regime sets in only at \tanb\ larger than 40
whereas for $\mu = 4000$~GeV this situation already occurs at \tanb\ abowe 20.
Hence, in Figure~\ref{fig:cpx-mu} (d), the hatched domain should not be
considered as being integrally part of the allowed parameter space. 

\begin{figure}[htb]
\begin{center}
{\large ~~~~~\msusy = 500~GeV }\\
\vspace{-0.5cm}
\epsfig{figure=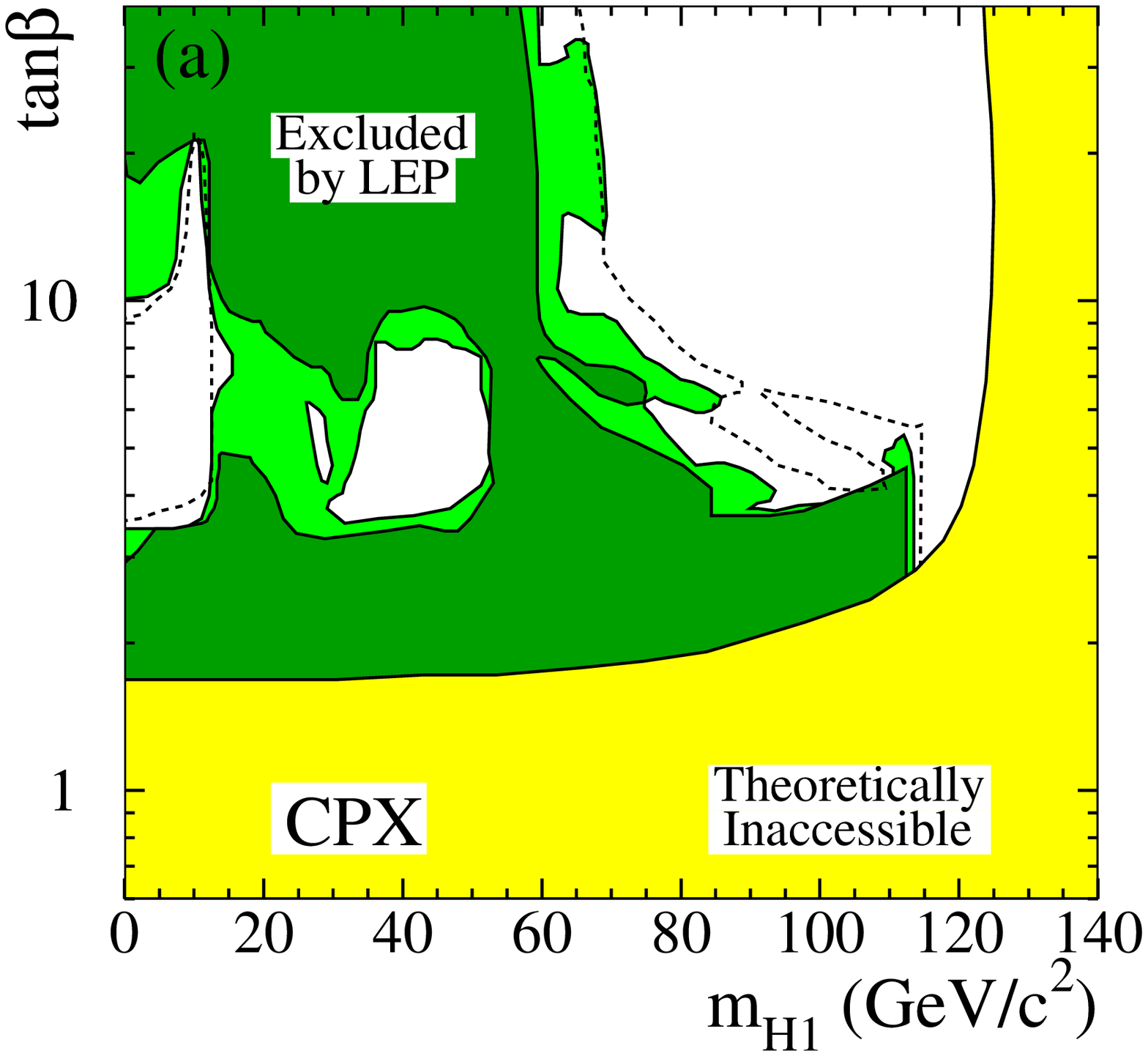,width=0.49\textwidth}\\
$\phantom{...}$\\
$\phantom{...}$\\
{\large ~~~~~\msusy=1000~GeV~~~~~~~~~~~~~~~~~~~~~~~~~~~~~\msusy=1000~GeV, scaled}\\
\vspace{-0.5cm}
\epsfig{figure=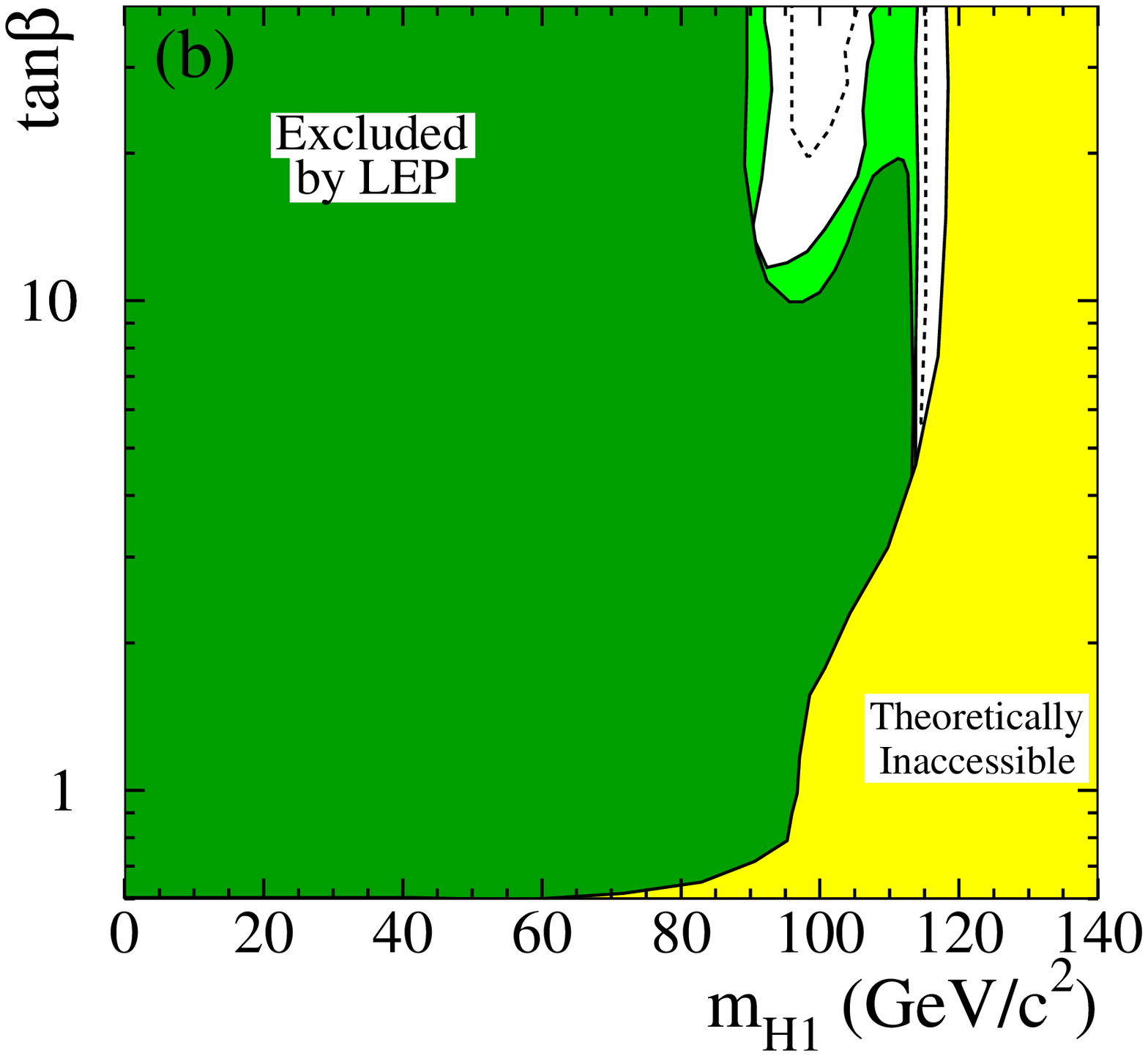,width=0.49\textwidth}
\epsfig{figure=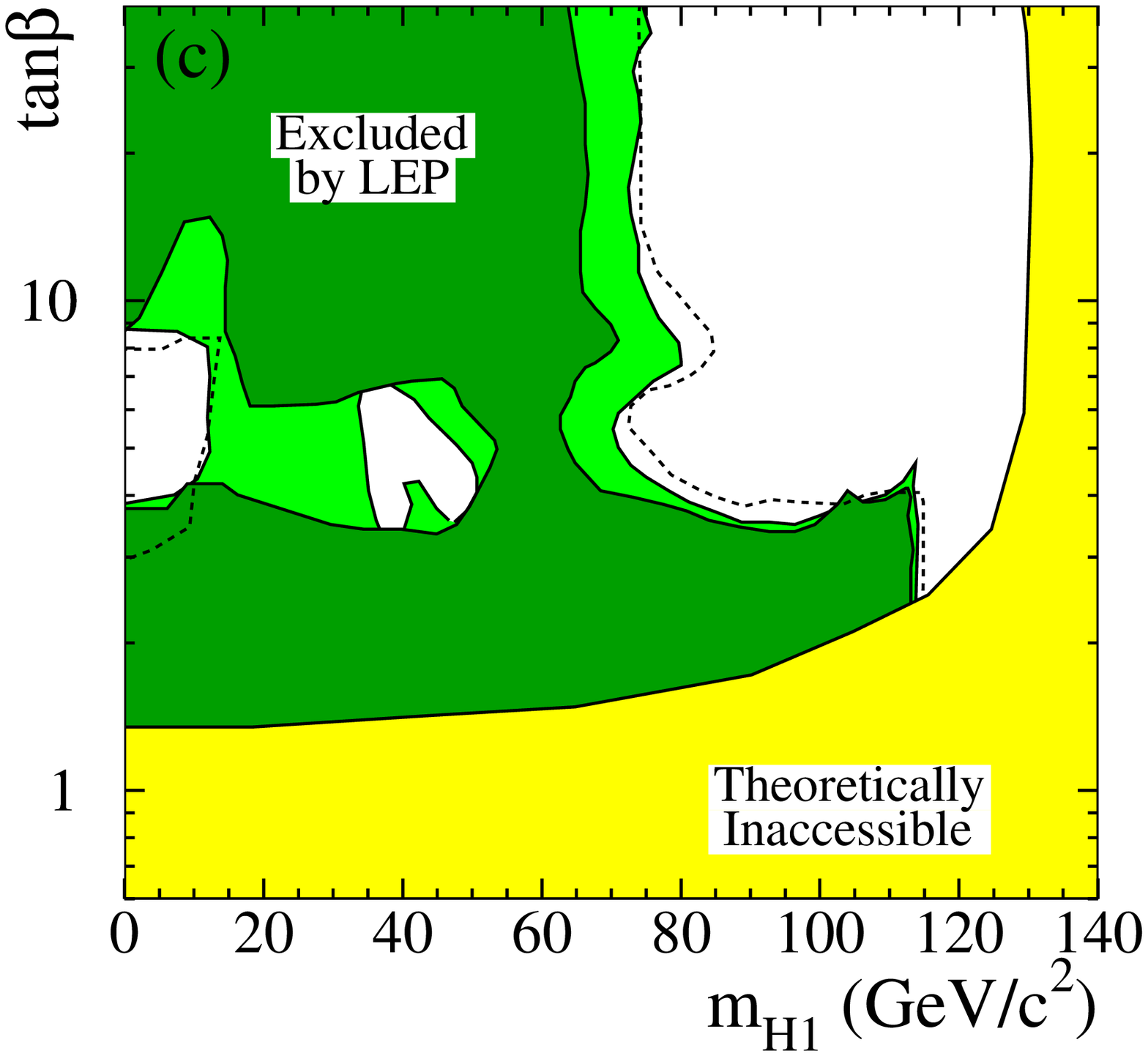,width=0.49\textwidth}
\end{center}
\caption[]{\sl Exclusions, for 
the CP-violating {\it CPX} scenario with various values of the soft SUSY-breaking scale \msusy.
(a): \msusy=500~GeV (the standard {\it CPX} value); (b): \msusy=1000~GeV while all other parameters are kept 
at their standard CPX values; (c): \msusy=1000~GeV while
$A,~m_{\tilde {\rm g}}$ and $\mu$ are ``scaled" to 2000~GeV, 
2000~GeV and 4000~GeV, respectively.
See the caption of Figure~\ref{fig:cpx-179} for the legend.
\label{fig:cpx-msusy}}
\end{figure}

Figure~\ref{fig:cpx-msusy} illustrates the dependence on the soft SUSY-breaking scale parameter, 
$M_{\rm SUSY}$, which is increased from the {\it CPX} value of 500~GeV in part (a) to 1000~GeV in part (b). This 
decreases the CP-violating effects (see Eq.~11) and leads to a larger exclusion.  
The ``scaling" behaviour mentioned in Section 2.3, namely the relative insensitivity of the exclusions 
to changes in \msusy\ as long as the relations $|A_{\rm t,b}| = |m_{\tilde {\rm g}}| = \mu/2 = 2\msusy$
are preserved, is qualitatively confirmed by comparing parts (a) and (c) of the figure.

\section{Summary}
The searches for neutral Higgs bosons described in this paper are based on the data collected by the four 
LEP collaborations, ALEPH, DELPHI, L3 and OPAL, which were statistically combined by the LEP Working Group for Higgs Boson
Searches. The data samples include those collected during 
the LEP 2 phase at \ee\ centre-of-mass energies up to 209 GeV; two experiments also provided LEP 1 data, 
at energies in the vicinity of the \Z\ boson resonance. 
The searches address a large number of
final-state topologies arising from the Higgsstrahlung process \ee\ra~\calHa\Z\ and from the pair production 
process \ee\ra~\calHb\calHa.
The combined LEP data do not reveal any excess of events which would indicate the production of Higgs bosons. The
differences with respect to the background predictions are compatible with statistical fluctuations of the background.

From these results, upper bounds are derived for the cross-sections of a number of Higgs-like event topologies. 
These upper bounds cover a wide range of Higgs boson masses and are typically well below the 
cross-sections predicted within the MSSM framework; these limits can be used to constrain a large number of 
theoretical models.

The combined search results are used to test several MSSM scenarios which include CP-conserving 
and CP-violating benchmark models. These models are motivated mainly by physics arguments but
some of them are constructed to test specific situations where the detection of Higgs bosons at
the Tevatron and LHC colliders might present experimental difficulties. 
It is found that in all these scenarios the searches conducted at LEP 
exclude sizable domains of the theoretically allowed parameter space. 

In the CP-conserving case, lower bounds 
can be set on the masses of neutral Higgs bosons and the value of \tanb\ can be restricted.
Taking, for example, the CP-conserving scenario {\it \mh-max} and a top quark mass of 
174.3~\Gcs, values of \mh\ and \mA\ less than 92.8~\Gcs\ and 
93.4~\Gcs, respectively, are excluded at the 95\% CL. In the same scenario, values of \tanb\ between 
0.7 and 2.0 are excluded, but this range depends considerably on the assumed top quark mass and 
may also depend on \msusy.

In the CP-violating benchmark scenario {\it CPX} and the variants which have been studied, the combined LEP data
show large domains which are not excluded, down to the lowest mass values; hence, no absolute limits can be set
for the Higgs boson masses. The excluded domains
vary considerably with the precise value of the top quark mass and the MSSM model parameters.
For example, in the {\it CPX} scenario with standard parameters and $m_{\rm t} = 174.3$~\Gcs, \tanb\ can be restricted 
to values larger than 2.9 at the 95\% CL.

%
%
$\phantom{.}$\\
$\phantom{.}$\\
\noindent{\large\bf Acknowledgements}\\

We congratulate the LEP Accelerator Division for the successful running of LEP over twelve years, up to the
highest energies. We also would like to express our thanks to the engineers and technicians in all our institutions
for their contributions to the excellent performance of the four LEP experiments. 

\clearpage
\newpage
%

%
\clearpage
\newpage
\noindent {\bf \Large  Appendix A: Catalog of searches}\\

\noindent The searches of the four LEP collaborations which contribute to this combined analysis are listed
in Tables~\ref{tab:aleph-hz} to \ref{tab:opal-hh}. The list is structured into two tables per experiment,
one for the Higgsstrahlung process \ee\ra~\calHa\Z\ and one for the pair production process \ee\ra~\calHb\calHa.
In each of these tables, the upper part contains the final states of the direct process and the lower
part contains, where it applies, those of the cascade process \calHb\ra~\calHa\calHa. 

The final-state topologies are listed in the first column.
In the notation adopted, \calHa\ represents the lightest and \calHb\ the second-lightest neutral Higgs boson.
In the CP-conserving case, \calHa\ is identified with the CP-even eigenstate \h. The \calHb\ is
identified in most cases with the CP-odd eigenstate \A~~(the cascade process 
\calHb\ra~\calHa\calHa\ is identified with \h\ra~\A\A).

The symbol q indicates an arbitrary quark flavour, u, d, s, c or b. ``Hadrons" include quarks and gluons.
In the missing energy channel,
in addition to the \calHa\Z\ra~\calHa\nn\ process, the \W\ fusion process
\calHa$\nu_e\bar{\nu}_e$ (including interference) is also considered; 
similarly, in the leptonic channel, in addition to the \calHa\Z\ra~\calHa\lpm\ process, the
\Z\ fusion process \calHa\ee\ (including interference) is also considered.   

The contributions based on LEP1 data (from two experiments only) can be identified by their value ``91" in the second 
column which indicates the \ee\ collision
energy, \sqrts~(GeV); the LEP1 data used in this combination represent an integrated luminosity $\cal L$ of about 125~\pb. 
The LEP2 data 
span an energy range between 133~GeV and 209~GeV; they represent an integrated luminosity of about 2400~\pb. 
The integrated luminosities for the individual searches are listed in the third column.

Responding to the increasing data samples and 
\ee\ energies, the searches were gradually upgraded or replaced so as to become more efficient 
in detecting Higgs bosons of higher masses. 
The mass ranges where the searches are relevant are listed in the next column(s). In the last column, references 
are given to the publications where the details of the searches can be found. 
\begin{table}
\begin{center}
{\small
\begin{tabular}{|l|cc|c|c|}
\hline
         & $\sqrt{s}$ (GeV)& $\cal{L}$ ($\pb$)& Mass range ($\Gcs$)& Ref.\\
\hline
\calHa\Z\ra~(...)\ (...)  & & & \mcalH$_1$ & \\
\hline
$(\bb)(\qq)$, $(\bb,\cc,\tau\tau,\mathrm{gg})(\nn)$ & 189 & 176.2 & 75 -- 110 & \cite{a189} \\
(any)$(\ee,\,\mumu)$&  189  & 176.2 & 75 -- 110 &  \cite{a189}\\
$(\bb)(\tautau),(\tautau)(\qq)$ & 189 &  176.2 &  65 -- 110 & \cite{a189}\\
$(\bb)(\qq, \nn)$& 192 -- 202 & 236.7 & 60 -- 120 & \cite{a192} \\
$(\bb,\tautau,\cc,\mathrm{gg})(\ee,\mumu)$&192 -- 202  &236.7  &  60 -- 120 & \cite{a192}\\
$(\bb,\tautau,\cc,\mathrm{gg})(\tautau),(\tautau)(\qq)$& 192 -- 202 & 236.7 &  60 -- 120 & \cite{a192}\\
$(\bb)(\qq)$& 199 -- 209 & 217.2 & 75 -- 120 & \cite{a2000a,a2000b} \\
$(\bb,\tautau,\cc,\mathrm{gg},\W\W)(\tautau, \nn)$& 199 -- 209& 217.2& 75 -- 120 & \cite{a2000a,a2000b}\\
$(\bb,\tautau,\cc,\mathrm{gg})(\ee,\mumu)$& 199 -- 209& 217.2&  70 -- 120 & \cite{a2000a,a2000b}\\
\hline
$(\bb,\cc,\mathrm{s\bar{s},gg})(\qq)$ & 189 & 176.2 & 40 -- 100 & \cite{flavind} \\
$(\bb,\cc,\mathrm{s\bar{s},gg})(\nn)$ &  189 & 176.2&  60 -- 100 & \cite{flavind}\\
$(\bb,\cc,\mathrm{s\bar{s},gg})(\ee,\mm)$ & 189  & 176.2&  60 -- 115 & \cite{a189,flavind}\\
$(\tautau)(\qq)$ & 189  & 176.2& 65 -- 110 & \cite{a189}\\
$(\bb,\cc,\mathrm{s\bar{s},gg})(\qq)$ & 192 -- 202 & 236.7 &  40 -- 110& \cite{flavind} \\
$(\bb,\cc,\mathrm{s\bar{s},gg})(\nn)$ & 192 -- 202 & 236.7 &  60 -- 116 & \cite{flavind} \\
$(\bb,\cc,\mathrm{s\bar{s},gg})(\ee,\mm)$ & 192 -- 202 & 236.7 &  60 -- 115 & \cite{a192,flavind}\\
$(\tautau)(\qq)$ & 192 -- 202 &236.7  &  60 -- 120 & \cite{a192}\\
$(\bb,\cc,\mathrm{s\bar{s},gg})(\qq)$ & 199 -- 209 & 217.2 &  40 -- 115 & \cite{flavind} \\
$(\bb,\cc,\mathrm{s\bar{s},gg})(\nn)$ & 199 -- 209 & 217.2 &  75 -- 120 & \cite{flavind}\\
$(\bb,\cc,\mathrm{s\bar{s},gg})(\ee,\mm)$ & 199 -- 209 & 217.2 &  70 -- 120 & \cite{a2000a,a2000b,flavind}\\
$(\tautau)(\qq)$ & 199 -- 209 & 217.2 &  60 -- 120 & \cite{a2000a,a2000b}\\
\hline
\end{tabular}
}
\caption{\sl
Summary of the ALEPH searches for the Higgsstrahlung process \ee\ra~\calHa\Z. The top part of the table lists the searches
originally developed for the Standard Model Higgs boson. The bottom part lists flavour-independent searches where
the decays of the Higgs boson into a quark pair of any flavour, a gluon pair or a tau pair were considered;
the signal efficiencies were evaluated for all indicated hadronic decays of the Higgs boson. In the cases of 
the $~(\tautau)(\qq)$ and leptonic
channels listed in the flavour-independent part, the event selections of the Standard Model Higgs boson searches were used.
}
\label{tab:aleph-hz}
\end{center}
\end{table}
\clearpage 
\begin{table}
\begin{center}
{\small
\begin{tabular}{|l|cc|c|c|}
\hline
  & $\sqrt{s}$\,(GeV)& $\cal{L}$ ($\pb$)& Mass range ($\Gcs$)& Ref.\\
\hline
\calHb\calHa\ra~(...)(...)  &  &  &  $(\mcalHb+\mcalHa)/2$ &  \\
\hline
$(\bb)(\bb)$, $(\tautau)(\bb)$, $ (\bb)(\tautau)$  & 189 & 176.2 &  65 --  95 & \cite{a189} \\
$(\bb)(\bb)$, $(\bb,\tautau,\cc,\mathrm{gg})(\tautau)$,& & & & \\
~~~~~~~~~~~~~~$(\tautau)(\bb,\tautau,\cc,\mathrm{gg})$  & 192 -- 202& 236.7 &  60 -- $\sqrt{s}/2$& \cite{a192}\\
$(\bb)(\bb)$, $(\bb,\tautau,\cc,\mathrm{gg})(\tautau)$,   &  &  & & \\
~~~~~~~~~~~~~~$(\tautau)(\bb,\tautau,\cc,\mathrm{gg})$&  199 -- 209 & 217.2 & 75 -- $\sqrt{s}/2$& \cite{a2000a,a2000b}\\
\hline
\end{tabular}
}
\caption{\sl
Summary of the ALEPH searches for the pair production process \ee\ra~\calHb\calHa.
The searches are restricted to $|$\mcalHb~$-$~\mcalHa$|$ less than about 20~\Gcs.}
\label{tab:aleph-hh}
\end{center}
\end{table}
\clearpage 
\begin{table}
\begin{center}
{\small
\begin{tabular}{|l|cc|cc|c|}     
\hline 
  & $\sqrt{s}$ (GeV)& $\cal{L}$ ($\pb$) & \multicolumn{2}{|c|}{Mass ranges (\Gcs)} & Ref.\\
\hline
\ee\ra~\calHa\Z\ra~(...)(...) & & & \multicolumn{2}{|c|}{\mcalHa}   & \\ 
\hline
 (any)(\ee, \mm), (V$^0$)(any)
          & 91 &   2.5& \multicolumn{2}{|c|}{$< 0.21$}  & \cite{ref:vlow} \\
  (2 prongs)(\qq) 
          & 91 & 0.5&  \multicolumn{2}{|c|}{$ 0.21 - 2$}    & \cite{ref:low} \\
  (jet)(\ee, \mm)
          & 91 &  0.5&  \multicolumn{2}{|c|}{$1 - 20$}  &  \cite{ref:low} \\
  (jet jet)($\ell^+\ell^-$, \nunubar)
          & 91 &  3.6&  \multicolumn{2}{|c|}{$12 - 50$}  & \cite{ref:ha92} \\
  (jet jet)(\ee, \mm, \nunubar)
          & 91 &  33.4&  \multicolumn{2}{|c|}{$35 - 70$}  &  \cite{ref:4prongs} \\
  (\bb)(any), ($\tau^+\tau^-$)(\qq) 
            & 161,172 &  19.9&  \multicolumn{2}{|c|}{$40 - 80$}  & \cite{ref:pap96} \\
  (\bb)(any), (\tautau)(\qq)  
            & 183 &  52.0&  \multicolumn{2}{|c|}{$45- 95$}  & \cite{ref:pap97} \\
  (\bb)(any), ($\tau^+\tau^-$)(qq)   
            & 189 &  158.0&  \multicolumn{2}{|c|}{$65 - 100$}   &\cite{ref:pap98} \\
  (\bb)(any)   
            & 192-209  &  452.4&  \multicolumn{2}{|c|}{$12 - 120$} &\cite{ref:pap99,ref:pap00} \\
  ($\tau^+\tau^-$)(qq)  
            & 192-209 &  452.4 &  \multicolumn{2}{|c|}{$45 - 120$}  &\cite{ref:pap99,ref:pap00} \\
\hline
  (\qq, {\rm gg})(\qq, \nunubar, \ee, \mm)
            & 189-209 &  610.4 & \multicolumn{2}{|c|}{$4 - 116$}  &\cite{ref:fl-blind} \\	    
	    
\hline
\ee\ra~\calHb\Z\ra~(\calHa\calHa)\Z\ra~(...)(...)& & & \mcalHb\ & \mcalHa\  & \\
 \hline
 (any)(\qq)                 
       & 91  &16.2                      & $12 - 70$   & $<0.21$         & \cite{ref:dallas} \\
(V$^0$V$^0$)(any but $\tau^+\tau^-$)                  
       & 91  &9.7                       & $0.5 - 55$  & $<0.21$         & \cite{ref:dallas} \\
($\gamma\gamma$)(any)                  
       &  91  &12.5                     & $0.5 - 60$  & $<0.21$         & \cite{ref:dallas} \\
(4 prongs)(any)                  
       & 91  &12.9                      & $0.5 - 60$  & $0.21 - 10$     & \cite{ref:dallas} \\
(hadrons)(\nn)                  
       &  91  &15.1                     & $1 - 60$    & $0.21 - 30$     & \cite{ref:dallas} \\
($\tau^+\tau^-\tau^+\tau^-$)(\nn)                  
       &  91   &15.1                    & $9 - 73$    & $3.5 - 12$      & \cite{ref:dallas} \\
(any)(\qq, \nn)    
      & 161,172  &20.0                  & $40 - 70$   & $20 - 35$       & \cite{ref:pap96} \\
(\bb\bb)(\qq)
      & 183    &54.0                    & $45 - 85$   & $12 - 40$       & \cite{ref:pap97} \\
(\bb\bb, \bb\cc, \cc\cc)(\qq)   
      & 192-208  &452.4                 & $30 - 105$  & $12 - 50$       & \cite{ref:pap99,ref:pap00} \\
(\cc\cc)(\qq)    
      &  192-208  &452.4                & $10 - 105$  & $4 - 12$        & \cite{ref:pap03} \\
\hline
\end{tabular}
}
\caption{\sl List of the DELPHI searches for the Higgsstrahlung processes \ee\ra~\calHa\Z\ and \calHb\Z.}
\label{tab:delphi-hz}
\end{center}
\end{table}
\clearpage 
\begin{table}
\begin{center}
{\small
\begin{tabular}{|l|cc|cc|c|}     
\hline 
  & $\sqrt{s}$ (GeV)& $\cal{L}$ ($\pb$)& \multicolumn{2}{|c|}{Mass ranges ($\Gcs$)}  & Ref.\\
\hline
 \ee\ra~\calHb\calHa\ra~(...)(...) & & & \mcalHb\ & \mcalHa\  &\\
 \hline
   4 prongs       
 
        & 91  & 5.3  & $0.2 - 10$ & $0.2 - 10$   & \cite{ref:4prongs} \\
   (\toto)(hadrons)       
        & 91 & 0.5   & $4 - 35$ & $4 - 35$  & \cite{ref:ha89} \\
   (\toto)(jet jet)       
        & 91   &3.6  & $25 - 42$ & $25 - 42$ & \cite{ref:ha90} \\
   (\bbbar)(\bbbar), (\bbbar)(\ccbar) 
       & 91    &33.4   & $15 - 46$ & $15 - 46$  & \cite{ref:ha92} \\
   \toto \bbbar 
      & 91    &79.4    & $4 - 70$ &$4 - 70$  & \cite{ref:pap03} \\
   \bbbar \bbbar 
      & 91    &79.4   & $12 - 40$ &$20 - 70$  & \cite{ref:2hdm} \\
   \bbbar \bbbar
      &  133  &6.0     & $40 - 68$ & $35 - 73$   & \cite{ref:pap95} \\
   \bbbar \bbbar, \toto \bbbar   
      & 161,172    &20.0      & $40 - 70$ & $35 - 75$  & \cite{ref:pap96} \\
   \bbbar \bbbar, \toto \bbbar   
     & 183   & 54.0     & $50 - 80$ & $25 - 105$ & \cite{ref:pap97} \\
   \bbbar \bbbar, \toto \bbbar  
      & 189    & 158.0   & $65 - 90$ & $40 - 115$  &\cite{ref:pap98} \\
   \toto  \bbbar  
      & 192-208  & 452.4    & $50 - 100$ & $60 - 150$  &\cite{ref:pap99,ref:pap00} \\
   \bbbar \bbbar  
      & 192-208  &452.4   & $12 - 100$ &$40 - 190$   &\cite{ref:pap99,ref:pap00} \\
   \toto \toto     
      & 189-208  &570.9    & $4 - 90$ &$4 - 170$  & \cite{ref:2hdm} \\
   \bbbar \bbbar    
      & 189-208  &610.2         & $12 - 70$&$30 - 170$  & \cite{ref:2hdm} \\
    quarks or gluons
      & 189-208  &610.4   & $4 - 170 $ &    $4 - 170 $ & \cite{ref:fl-blind}\\
      
\hline
\ee\ra~\calHb\calHa\ra~(\calHa\calHa)\calHa\ra~(...)(...)  & & & \mcalHb\ & \mcalHa\  & \\
\hline
($\gamma\gamma$)($\gamma\gamma$)                  
       &  91  &12.5                  & $0.5 - 60$  & $<0.21$       & \cite{ref:dallas} \\
(4 prongs)(2 prongs)                  
       & 91  &12.9                   & $0.5 - 60$  & $0.21 - 10$   & \cite{ref:dallas} \\
(hadrons)(hadrons)                  
       &  91  &15.1                  & $1 - 60  $  & $0.21 - 30$   & \cite{ref:dallas} \\
(\tautau\tautau)(\tautau)                  
       &  91   &15.1                 & $9 - 60$    & $3.5 - 12$    & \cite{ref:dallas} \\
(any)(any)    
      & 161,172  &20.0               & $40 - 70$   & $20 - 35$     & \cite{ref:pap96} \\
\hline
\end{tabular}
}
\caption{\sl List of the DELPHI searches for the pair production process \ee\ra~\calHb\calHa.}
\label{tab:delphi-hh}
\end{center}
\end{table}
\clearpage 
\begin{table}
\begin{center}
{\small
\begin{tabular}{|l|cc|cc|c|}     
\hline 
  & $\sqrt{s}$ (GeV)& $\cal{L}$ ($\pb$)& \multicolumn{2}{|c|}{Mass ranges ($\Gcs$)}  & Ref.\\
\hline
\ee\ra~\calHa\Z\ra~(...)(...)                &    &    &   \multicolumn{2}{|c|}{\mcalHa }    &                          \\
\hline
(\bb)(any),($\tau^+\tau^-$)(\qq) &  189   & 176.4          & \multicolumn{2}{|c|}{60 -- 100}  & \cite{sm_1998} \\
(\bb)(any),($\tau^+\tau^-$)(\qq) & 192 -- 202  & 233.2      &  \multicolumn{2}{|c|}{60 -- 110} & \cite{sm_1999} \\
(\bb)(any),($\tau^+\tau^-$)(\qq) & 203 -- 209  & 217.3     &  \multicolumn{2}{|c|}{60 -- 120}  & \cite{sm_2000} \\         
\hline
(\bb, \cc, gg)(any)  & 189      & 176.4       &  \multicolumn{2}{|c|}{60 -- 100}  & \cite{flindep} \\
(\bb, \cc, gg)(any)  & 192 -- 202  & 233.2      &  \multicolumn{2}{|c|}{60 -- 110} & \cite{flindep} \\
(\bb, \cc, gg)(any)  & 204 -- 209    & 214.5      &  \multicolumn{2}{|c|}{60 -- 120}& \cite{flindep} \\
\hline
\ee\ra~\calHb\Z\ra~(\calHa\calHa)\Z\ra~(...)(...)                &   &    &   \mcalHb\ &   \mcalHa\          & \\ 
\hline
(\calHa\ra~\bb,cc,gg)(\qq)   & 189 -- 209  & 626.9     &  30 -- 85 & 10 -- 42  & \cite{mssm_2000} \\ 
\hline
\end{tabular}
}
\caption{\sl List of the L3 searches for the Higgsstrahlung processes \ee\ra~\calHa\Z\ and \calHb\Z.}
\label{tab:L3-hz}
\end{center}
\end{table}
\begin{table}
\begin{center}
{\small
\begin{tabular}{|l|cc|cc|c|}     
\hline 
  & $\sqrt{s}$ (GeV)& $\cal{L}$ ($\pb$)& \multicolumn{2}{|c|}{Mass ranges ($\Gcs$)}  & Ref.\\
\hline
\ee\ra~\calHb\calHa\ra~(...)(...)                &    &    &   \mcalHb\  & \mcalHa\    &                          \\
\hline
(\bb)(\bb), (\bb)($\tau^+\tau^-$), ($\tau^+\tau^-$)(\bb) & 189   & 176.4           & 50 -- 95  & 50 -- 95  & \cite{mssm_1998} \\
(\bb)(\bb), (\bb)($\tau^+\tau^-$), ($\tau^+\tau^-$)(\bb)  & 192 -- 202   & 233.2    & 50 -- 105 & 50 -- 105  & \cite{mssm_1999} \\     
(\bb)(\bb), (\bb)($\tau^+\tau^-$), ($\tau^+\tau^-$)(\bb)  & 204 -- 209   & 216.6    & 50 -- 110 & 50 -- 110  & \cite{mssm_2000} \\
\hline
\end{tabular}
}
\caption{\sl List of the L3 searches for the pair production process \ee\ra~\calHb\calHa.}
\label{tab:L3-hh}
\end{center}
\end{table}
\clearpage 
\begin{table}
{\small
\begin{center}
\begin{tabular}{|l|cc|cc|c|}     
\hline 
  & $\sqrt{s}$\,(GeV)& $\cal{L}$\,($\pb$)& \multicolumn{2}{|c|}{Mass ranges ($\Gcs$)} & Ref.\\
\hline 
\calHa\Z\ra~(...) (...)   & & &  \multicolumn{2}{|c|}{\mcalHa} &  \\
\hline
(\bb)(\qq)                                 &  161--172  & 20.4  &  \multicolumn{2}{|c|}{ $40 - 80  $}  & \cite{bib:opalhiggsold2,bib:opalhiggsold3} \\ 
(\bb)(\qq)                                 &    183     & 54.1  &  \multicolumn{2}{|c|}{ $40 - 95  $}  & \cite{bib:opalhiggsold1} \\
(\bb)(\qq)                                 &    189     & 172.1 &   \multicolumn{2}{|c|}{$40 - 100 $}  & \cite{pr285} \\ 
(\bb)(\qq)                                 &  192--209  & 421.2 &   \multicolumn{2}{|c|}{$80-120   $}  & \cite{OPALSMPAPER}           \\

(\bb)(\nn)                                 &  161--172  & 20.4  &  \multicolumn{2}{|c|}{ $50 - 70  $}  & \cite{bib:opalhiggsold2,bib:opalhiggsold3} \\ 
(\bb)(\nn)                                 &    183     & 53.9  &  \multicolumn{2}{|c|}{ $50 - 95  $}  & \cite{bib:opalhiggsold1}\\
(\bb)(\nn)                                 &    189     & 171.4 &  \multicolumn{2}{|c|}{ $50 - 100 $}  & \cite{pr285} \\ 
(\bb)(\nn)                                 &  192--209  & 419.9 &  \multicolumn{2}{|c|}{ $30-120   $}  & \cite{OPALSMPAPER}            \\     

(\bb)(\tautau), (\tautau)(\qq)             &  161--172  & 20.4  &  \multicolumn{2}{|c|}{ $30 - 95  $}  & \cite{bib:opalhiggsold2,bib:opalhiggsold3} \\ 
(\bb)(\tautau), (\tautau)(\qq)             &    183     & 53.7  &  \multicolumn{2}{|c|}{ $30 - 100 $}  & \cite{bib:opalhiggsold1}\\
(\bb)(\tautau), (\tautau)(\qq)             &    189     & 168.7 &  \multicolumn{2}{|c|}{ $30 - 100 $}  & \cite{pr285} \\ 
(\bb)(\tautau), (\tautau)(\qq)             &  192--209  & 417.4 &  \multicolumn{2}{|c|}{ $80-120   $}  & \cite{OPALSMPAPER}           \\     

(\bb)(\ee), (\bb)($\mu^+\mu^-$)            &    183     & 55.9  &  \multicolumn{2}{|c|}{ $60 - 100 $}  & \cite{bib:opalhiggsold1}\\
(\bb)(\ee), (\bb)($\mu^+\mu^-$)            &    189     & 170.0 &   \multicolumn{2}{|c|}{$70 - 100 $}  & \cite{pr285} \\ 
(\bb)(\ee), (\bb)($\mu^+\mu^-$)            &  192--209  & 418.3 &   \multicolumn{2}{|c|}{$40-120   $}  & \cite{OPALSMPAPER}           \\     

        \hline

(\qq, gg)(\tautau, \nn), (\tautau)(\qq) &    91    & 46.3  &  \multicolumn{2}{|c|}{ $0 - 70$}  & \cite{bib:opalhiggsold4,bib:opalhiggsold5}\\ 
(\qq, gg)(\ee, $\mu^+\mu^-$)             &    91    & 46.3  &  \multicolumn{2}{|c|}{ $20 - 70 $}   & \cite{bib:opalhiggsold4,bib:opalhiggsold5}\\ 
(any)(\ee, $\mu^+\mu^-$)            &  161--172  & 20.4  &  \multicolumn{2}{|c|}{ $35 - 80  $}  & \cite{bib:opalhiggsold2,bib:opalhiggsold3} \\

(\qq, gg)(\qq)                                  &    189      & 174.1 &  \multicolumn{2}{|c|}{$60 - 100 $}   & \cite{Abbiendi:2000ug}    \\ 
(\qq, gg)(\qq)                                  & 192--209    & 424.2 &  \multicolumn{2}{|c|}{$60 - 120$}    & \cite{2HDMFINAL}                  \\ 

(\qq, gg)(\nn)                                  &    189      & 171.8 &  \multicolumn{2}{|c|}{$30 - 100 $}   & \cite{Abbiendi:2000ug}     \\ 
(\qq, gg)(\nn)                                  & 192--209    & 414.5 &  \multicolumn{2}{|c|}{$30 - 110$}    & \cite{2HDMFINAL}                   \\ 

(\qq, gg)(\tautau), (\tautau)(\qq)              &    189      & 168.7 &  \multicolumn{2}{|c|}{$30 - 100 $}   & \cite{Abbiendi:2000ug}     \\ 
(\qq, gg)(\tautau), (\tautau)(\qq)              & 192--209    & 418.9 &  \multicolumn{2}{|c|}{$60 - 115$}    & \cite{2HDMFINAL}                  \\ 

(\qq, gg)(\ee, $\mu^+\mu^-$)            &    189      & 170.0 &  \multicolumn{2}{|c|}{$70 - 100 $}   & \cite{Abbiendi:2000ug}     \\ 
(\qq, gg)(\ee, $\mu^+\mu^-$)             & 192--209    & 422.0 &  \multicolumn{2}{|c|}{$60 - 120$}    & \cite{2HDMFINAL}                  \\ 
        \hline
        \ee\ra~\calHb\Z\ra~(\calHa\calHa)\Z\ra~(...)(...)    &   &    &   \mcalHb\ &   \mcalHa\  & \\ 
	\hline
        (\qq\qq)(\nn)                   & 91     & 46.3  &  $10 - 75$  & $0 - 35  $    & \cite{bib:opalhiggsold4,bib:opalhiggsold5}\\ 
        (\bb\bb)(\qq)                   & 183      & 54.1  &  $40 - 80$  & $10.5-38 $    & \cite{bib:opalhiggsold1}\\
        (\bb\bb)(\qq)                   & 189      & 172.1 &  $40 - 100$ & $10.5-48$     & \cite{pr285} \\ 
        (\bb\bb)(\qq)                   & 192--209 & 421.2 &  $80 - 120$ & $12-\mcalHb/2$ & \cite{mssm-o}   \\     

        (\bb\bb)(\nn)                   & 183      & 53.9  &  $50 -  95$ &  $10.5-\mcalHb/2$              & \cite{bib:opalhiggsold1}\\
        (\qq\qq)(\nn)                   & 189      & 171.4 &  $50 - 100$ &  $10.5-\mcalHb/2$              & \cite{pr285} \\ 
        (\bb\bb)(\nn)                   & 199--209 & 207.2 &  $100 -110$ & $12-\mcalHb/2$ &  \cite{mssm-o}     \\     

        (\bb\bb)(\tautau)               & 183      & 53.7  &  $30 - 100$ &  $10.5-\mcalHb/2$              & \cite{bib:opalhiggsold1}                   \\
        (\bb\bb)(\tautau)               & 189      & 168.7 &  $30 - 100$ &   $10.5-\mcalHb/2$             & \cite{pr285}                     \\

        (\bb\bb,~\bb\tautau,~\tautau\tautau) &         &       &          &                  &              \\
	~~~~~~~~~~~~~~~~~~~~(\nn,~\ee,~$\mu^+\mu^-$)  & 189--209 & 598.5 &  $45 - 90$  & $2-10.5 $     & \cite{lowma} \\ 
        
        \hline        
\end{tabular}
\caption{\sl List of the OPAL searches for the Higgsstrahlung processes \ee\ra~\calHa\Z\ and \calHb\Z.
} 
\label{tab:oppal-hz}
\end{center}
}
\end{table}
\clearpage 
\begin{table}
{\small
\begin{center}
\begin{tabular}{|l|cc|cc|c|}     
\hline 
  & $\sqrt{s}$\,(GeV)& $\cal{L}$\,($\pb$)&  \multicolumn{2}{|c|}{ Mass ranges ($\Gcs$)}  & Ref.\\
\hline
\calHb\calHa\ra~(...) (...)   & & & \mcalHb\  &  \mcalHa\  & Ref.\\
\hline
(\bb)(\bb)                      & 130--136 & 5.2   & $\Sigma=80-130$  & $\Delta=0-50$  & \cite{bib:opalhiggsold3} \\
(\bb)(\bb)                      & 161      & 10.0  & $\Sigma=80-130$  & $\Delta=0-60$  & \cite{bib:opalhiggsold2,bib:opalhiggsold3} \\ 
(\bb)(\bb)                      & 172      & 10.4  & $\Sigma=80-130$  & $\Delta=0-60$  & \cite{bib:opalhiggsold2,bib:opalhiggsold3} \\ 
(\bb)(\bb)                      & 183      & 54.1  & $\Sigma=80-150$  & $\Delta=0-60$  & \cite{bib:opalhiggsold1}\\
(\bb)(\bb)                      & 189      & 172.1 & $\Sigma=80-180$  & $\Delta=0-70$  & \cite{pr285} \\ 
(\bb)(\bb)                      & 192      & 28.9  & $\Sigma=83-183$  & $\Delta=0-70 $  & \cite{mssm-o} \\     
(\bb)(\bb)                      & 196      & 74.8  & $\Sigma=80-187 $ & $\Delta=0-70 $  & \cite{mssm-o} \\     
(\bb)(\bb)                      & 200      & 77.2  & $\Sigma=80-191 $ & $\Delta=0-70 $  & \cite{mssm-o} \\     
(\bb)(\bb)                      & 202      & 36.1  & $\Sigma=80-193 $ & $\Delta=0-70 $  & \cite{mssm-o} \\     
(\bb)(\bb)                      & 199--209 & 207.3 & $\Sigma=120-190$ & $\Delta=0-70 $  & \cite{mssm-o} \\     
(\bb)(\bb)                      & 199--209 & 207.3 & $\Sigma=100-140$ & $\Delta=60-100$ & \cite{mssm-o} \\     
(\bb)(\tautau), (\tautau)(\bb)  & 161      & 10.0  & $40-160 $        & $52-160$       & \cite{bib:opalhiggsold2,bib:opalhiggsold3} \\ 
(\bb)(\tautau), (\tautau)(\bb)  & 172      & 10.4  & $37-160$         & $28-160$       & \cite{bib:opalhiggsold2,bib:opalhiggsold3} \\ 
(\bb)(\tautau), (\tautau)(\bb)  & 183      & 53.7  & $\Sigma=70-170$  & $\Delta=0-70$  & \cite{bib:opalhiggsold1}\\
(\bb)(\tautau), (\tautau)(\bb)  & 189      & 168.7 & $\Sigma=70-190$  & $\Delta=0-90$  & \cite{pr285} \\ 
(\bb)(\tautau), (\tautau)(\bb)  & 192      & 28.7  & $\Sigma=10-174 $ & $\Delta=0-182 $ & \cite{mssm-o} \\     
(\bb)(\tautau), (\tautau)(\bb)  & 196      & 74.7  & $\Sigma=10-182 $ & $\Delta=0-191 $ & \cite{mssm-o} \\     
(\bb)(\tautau), (\tautau)(\bb)  & 200      & 74.8  & $\Sigma=10-182 $ & $\Delta=0-191 $ & \cite{mssm-o} \\     
(\bb)(\tautau), (\tautau)(\bb)  & 202      & 35.4  & $\Sigma=10-174 $ & $\Delta=0-182 $ & \cite{mssm-o} \\     
(\bb)(\tautau), (\tautau)(\bb)  & 199--209 & 203.6 & $\Sigma=70-190 $ & $\Delta=0-90  $ & \cite{mssm-o} \\     
\hline
(\qq)(\tautau), (\tautau)(\qq)  & 91     & 46.3  & $ 12-75 $        & $10-78$        & \cite{bib:opalhiggsold4,bib:opalhiggsold5}\\

        \hline
        \ee\ra~\calHb\calHa\ra\ &   &    &   &  &  \\
~~~~~~~~~~~~~(\calHa\calHa)\calHa\ra~(...)(...)    &   &    &   \mcalHb\ &   \mcalHa\  & \\ 
	\hline
(\bb\bb)(\bb)                 & 91    & 27.6 &  $40-70 $ & $ 5-35 $ & \cite{bib:opalhiggsold4,bib:opalhiggsold5} \\ 
(\bb\bb)(\bb)                 & 130--136& 5.2  &  $55-65 $ & $>27.5 $ & \cite{bib:opalhiggsold3} \\
(\bb\bb)(\bb)                 & 161     & 10.0 &  $55-65 $ & $>20.0 $ & \cite{bib:opalhiggsold2,bib:opalhiggsold3} \\ 
(\bb\bb)(\bb)                 & 172     & 10.4 &  $55-65 $ & $25-35 $ & \cite{bib:opalhiggsold2,bib:opalhiggsold3} \\ 
(\bb\bb)(\bb)                 & 183     & 54.1 &  $30-80 $ & $12-40 $ & \cite{bib:opalhiggsold1}\\
(\bb\bb)(\bb)                 & 189     &172.1 &  $24-80 $ & $12-40 $ & \cite{pr285} \\ 
(\bb\bb)(\bb)                 & 199--209& 207.3 & $\Sigma = 90-200$ & $\Delta=40-160$   & \cite{mssm-o} \\     
\hline
6$\tau$, 4$\tau$2q, 2$\tau$4q & 91    & 46.3 &  $30-75 $ & $ 4-30 $ & \cite{bib:opalhiggsold4,bib:opalhiggsold5}\\ 
\hline     
\end{tabular}
\caption{\sl List of the OPAL searches for the pair production process \ee\ra~\calHb\calHa.
The symbols $\Sigma$ and $\Delta$ stand for  
the mass sum \mcalHb~$+$~\mcalHa\ and mass difference $|$\mcalHb~$-$~\mcalHa$|$.} 
\label{tab:opal-hh}
\end{center}
}
\end{table}
\clearpage
\newpage
{\bf \Large Appendix B: Limits on topological cross-sections}\\

The tables presented below summarise the 95\% CL upper bounds, as a function of the Higgs boson masses,
of the scaling factor $S_{\rm 95}$ defined in the text (see Eq.~16). Tables~\ref{tab:forFig3}, 
\ref{tab:forFig2a} and~\ref{tab:forFig2b} refer to  final-state topologies arising from the 
Higgsstrahlung processes \ee\ra~\calHa\Z\ and \mbox{\ee\ra~(\calHb\ra~\calHa\calHa)\Z}; 
Tables~\ref{tab:forFig4a} to \ref{tab:forFig5b} refer to those arising from the
pair production processes \ee\ra~\calHb\calHa\ and \ee\ra~(\calHb\ra~\calHa\calHa)\calHa.
The corresponding figures, showing the same results, are mentioned in the table captions.
\begin{table}[ht] 
\begin{center}
{\small
\begin{tabular}{|c|ccc||c|ccc|}
\hline
&&&&&&& \\
\mcalHa\  &    (a)  &  (b)    &  (c) &  \mcalHa\  &    (a)  &  (b)    &  (c)\\
(\Gcs)    &         &         &      &    (\Gcs)  &         &         &     \\   
&&&&&&& \\
\hline
     12  &   0.0204   &  0.0154  &   0.0925 &  66  &   0.0236   &    0.0218  &  0.0287 \\ 
     14  &   0.0176   &  0.0143  &   0.0899 &  68  &   0.0236   &    0.0218  &  0.0287 \\
     16  &   0.0158   &  0.0134  &   0.0923 &  70  &   0.0271   &    0.0246  &  0.0287 \\
     18  &   0.0150   &  0.0131  &   0.0933 &  72  &   0.0291   &    0.0274  &  0.0271 \\
     20  &   0.0156   &  0.0139  &   0.1060 &  74  &   0.0320   &    0.0301  &  0.0297 \\
     22  &   0.0177   &  0.0156  &   0.1080 &  76  &   0.0421   &    0.0380  &  0.0351 \\
     24  &   0.0194   &  0.0174  &   0.1110 &  78  &   0.0469   &    0.0424  &  0.0350 \\
     26  &   0.0207   &  0.0186  &   0.1140 &  80  &   0.0435   &    0.0410  &  0.0316 \\
     28  &   0.0223   &  0.0195  &   0.1110 &  82  &   0.0467   &    0.0475  &  0.0281 \\
     30  &   0.0203   &  0.0181  &   0.0893 &  84  &   0.0539   &    0.0585  &  0.0222 \\ 
     32  &   0.0193   &  0.0173  &   0.0796 &  86  &   0.0762   &    0.0816  &  0.0257 \\
     34  &   0.0191   &  0.0172  &   0.0682 &  88  &   0.112    &    0.118   &  0.0296 \\  
     36  &   0.0241   &  0.0187  &   0.0653 &  90  &   0.153    &    0.152   &  0.0331 \\   
     38  &   0.0299   &  0.0235  &   0.0634 &  92  &   0.179    &    0.175   &  0.0354 \\   
     40  &   0.0333   &  0.0267  &   0.0615 &  94  &   0.229    &    0.214   &  0.0491 \\   
     42  &   0.0367   &  0.0297  &   0.0599 &  96  &   0.239    &    0.220   &  0.0570 \\   
     44  &   0.0378   &  0.0310  &   0.0594 &  98  &   0.256    &    0.233   &  0.0565 \\   
     46  &   0.0387   &  0.0328  &   0.0572 &  100 &   0.244    &    0.216   &  0.0582 \\   
     48  &   0.0391   &  0.0337  &   0.0575 &  102 &   0.237    &    0.216   &  0.0588 \\   
     50  &   0.0363   &  0.0316  &   0.0445 &  104 &   0.255    &    0.227   &  0.0704 \\
     52  &   0.0386   &  0.0344  &   0.0454 &  106 &   0.263    &    0.223   &  0.0896 \\   
     54  &   0.0387   &  0.0349  &   0.0464 &  108 &   0.266    &    0.227   &  0.110  \\ 
     56  &   0.0384   &  0.0360  &   0.0403 &  110 &   0.297    &    0.244   &  0.144  \\
     58  &   0.0390   &  0.0367  &   0.0427 &  112 &   0.435    &    0.343   &  0.212  \\
     60  &   0.0398   &  0.0365  &   0.0456 &  114 &   0.824    &    0.640   &  0.410  \\  
     62  &   0.0293   &  0.0264  &   0.0444 &  116 &   1.41     &    1.79    &  1.79   \\    
     64  &   0.0278   &  0.0258  &   0.0394 &      &            &            &         \\    
\hline
\end{tabular}
}
\end{center}
\caption{\sl The 95\% CL upper bound, $S_{95}$, obtained for the normalised 
cross-section (see text) of the 
Higgsstrahlung process \ee\ra~${\cal H}_1$Z, as a function of the Higgs boson mass. 
The numbers listed in this table correspond to the observed limit (full line) in Figure~\ref{sm-xi2},
which is reproduced from Ref.~\cite{lep-sm}.
In the columns labelled (a) the Higgs boson is assumed to decay 
as in the Standard Model; in columns (b) it is
assumed to decay exclusively to \bb\ and in columns (c) exclusively to \tautau.}
\label{tab:forFig3}
\end{table} 
\newpage
\begin{table}[ht] 
\begin{center}
{\small
\begin{tabular}{|c|cccccccccc|}
\hline
& & & & & & & & & &   \\
 {\small \mcalHb (\Gcs)} & \multicolumn{10}{|c|}{{\small \mcalHa (\Gcs)}} \\
& & & & & & & & & &   \\
\hline
& & & & & & & & & &   \\
        & 10       & 15       & 20       & 25       & 30       & 35       & 40       & 45       & 50       & 55 \\
& & & & & & & & & &   \\
\hline
20 & 0.020    &       &       &       &       &       &       &       &       &  \\
25 & 0.026    &       &       &       &       &       &       &       &       &  \\
30 & 0.037    & 0.046 &       &       &       &       &       &       &       &  \\
35 & 0.048    & 0.042 &       &       &       &       &       &       &       &  \\
40 & 0.053    & 0.056 & 0.051 &       &       &       &       &       &       &  \\
45 & 0.066    & 0.059 & 0.046 &       &       &       &       &       &       &  \\
50 & 0.087    & 0.058 & 0.048 & 0.049 &       &       &       &       &       &  \\
55 & 0.11     & 0.055 & 0.050 & 0.050 &       &       &       &       &       &  \\
60 & 0.29     & 0.103 & 0.094 & 0.094 & 0.053 &       &       &       &       &  \\
65 & 0.30     & 0.099 & 0.091 & 0.088 & 0.084 &       &       &       &       &  \\
70 & 0.25     & 0.098 & 0.097 & 0.095 & 0.083 & 0.059 &       &       &       &  \\
75 & 0.34     & 0.11  & 0.10  & 0.11  & 0.10  & 0.096 &       &       &       &  \\
80 & 0.39     & 0.13  & 0.14  & 0.14  & 0.13  & 0.12  & 0.13     &      &       &  \\
85 & 0.52     & 0.20  & 0.20  & 0.20  & 0.21  & 0.19  & 0.18     &      &       &  \\
90 & $\geq 1$ & 0.23  & 0.23  & 0.23  & 0.27  & 0.26  & 0.24     & 0.28      &       &  \\
95 & $\geq 1$ & 0.29  & 0.27  & 0.29  & 0.31  & 0.29  & 0.28     & 0.30      &       &  \\
100& $\geq 1$ & 0.30  & 0.29  & 0.31  & 0.30  & 0.27  & 0.28     & 0.29 & 0.29      &  \\
105& $\geq 1$ & 0.27  & 0.32  & 0.36  & 0.40  & 0.36  & 0.31     & 0.35 & 0.35      &  \\
110& $\geq 1$ & 0.44  & 0.54  & 0.55  & 0.96  & 0.97  & $\geq 1$ & $\geq 1$ & 0.89 & $\geq 1$ \\
\hline
\end{tabular}
}
\end{center}
\caption{\sl The 95\% CL upper bound, $S_{95}$, obtained for the normalised 
cross-section (see text) of the 
Higgsstrahlung cascade process \mbox{\ee\ra~(\calHb\ra~\calHa\calHa)\Z\ra~(\bb\bb)\Z}, as
a function of the Higgs boson masses \mcalHa\ and \mcalHb. 
The numbers correspond to the contours shown in Figure~\ref{s2-h2h1} (a). 
}
\label{tab:forFig2a}
\end{table} 
\newpage
\begin{table}[ht] 
\begin{center}
{\small
\begin{tabular}{|c|ccccccccc|}
\hline
& & & & & & & & & \\
       & 5       & 10       & 15       & 20       & 25       & 30       & 35       & 40       & 45 \\
& & & & & & & & & \\
\hline
10 & 0.26 &      &       &       &       &       &       &       &  \\
15 & 0.033&      &       &       &       &       &       &       &  \\
20 & 0.048& 0.32 &       &       &       &       &       &       &  \\
25 & 0.070& 0.076&       &       &       &       &       &       &  \\
30 & 0.10 & 0.11 & 0.38     &          &          &       &       &       &  \\
35 & 0.18 & 0.19 & 0.51     &          &          &       &       &       &  \\
40 & 0.22 & 0.22 & 0.40     & 0.39     &          &       &       &       &  \\
45 & 0.30 & 0.31 & 0.49     & 0.49     &          &       &       &       &  \\
50 & 0.18 & 0.38 & 0.66     & 0.66     & 0.63     &       &       &       &  \\
55 & 0.18 & 0.37 & 0.68     & 0.69     & 0.68     &       &       &       &  \\
60 & 0.20 & 0.38 & 0.95     & 0.96     & 0.96     & 0.94      &       &       &  \\
65 & 0.20 & 0.38 & $\geq 1$ & $\geq 1$ & $\geq 1$ & $\geq 1$      &       &       &  \\
70 & 0.21 & 0.43 & $\geq 1$ & $\geq 1$ & $\geq 1$ & $\geq 1$ & $\geq 1$      &       &  \\
75 & 0.19 & 0.46 & $\geq 1$ & $\geq 1$ & $\geq 1$ & $\geq 1$ & $\geq 1$      &       &  \\
80 & 0.20 & 0.44 & 0.83     & 0.83     & 0.83     & 0.83 & 0.84 & 0.84      &  \\
85 & 0.25 & 0.56 & $\geq 1$ & $\geq 1$ & $\geq 1$ & $\geq 1$ & $\geq 1$ & $\geq 1$      &  \\
\hline
\end{tabular}
}
\end{center}
\caption{\sl The 95\% CL upper bound, $S_{95}$, obtained for the normalised
cross-section (see text) of the 
Higgsstrahlung cascade process \mbox{\ee\ra~(\calHb\ra~\calHa\calHa)\Z\ra~(\tautau\tautau)\Z}, as
a function of the Higgs boson masses \mcalHa\ and \mcalHb. 
The numbers correspond to the contours shown in Figure~\ref{s2-h2h1} (b). 
}
\label{tab:forFig2b}
\end{table} 
\begin{table}[ht] 
\begin{center}
{\small
\begin{tabular}{|c|cccc||c|cccc|}
\hline
&&&&&&&&&\\
\mcalHa+\mcalHb\ &    (a)  &  (b)    &  (c) &   (d)  & \mcalHa+\mcalHb\ &   (a)  &  (b)  &  (c)   &  (d)  \\
(\Gcs)           &         &         &      &        &  (\Gcs)          &        &       &        &       \\    
&&&&&&&&&\\
\hline
     0&  0.0237     & 0.0237     & 0.0237    &  0.0237 & 105&  0.0243     & 0.0213     & 0.0354 &  0.0300 \\ 
     5&  0.0238     & 0.0238     & 0.0238    &  0.0238 & 110&  0.0297     & 0.0250     & 0.0418 &  0.0313 \\     
    10&  0.0242     & 0.0242     & 0.0242    &  0.0242 & 115&  0.0472     & 0.0387     & 0.0484 &  0.0332 \\    
    15&  0.0248     & 0.0248     & 0.0248    &  0.0248 & 120&  0.0682     & 0.0599     & 0.0409 &  0.0348 \\    
    20&  0.0255     & 0.0255     & 0.0255    &  0.0255 & 125&  0.0676     & 0.0542     & 0.0493 &  0.0387 \\     
    25&  0.0266     & 0.0266     & 0.0266    &  0.0042 & 130&  0.0688     & 0.0541     & 0.0524 &  0.0429 \\     
    30&  0.0054     & 0.0054     & 0.0018    &  0.0043 & 135&  0.0618     & 0.0478     & 0.0571 &  0.0604 \\  
    35&  0.0044     & 0.0041     & 0.0018    &  0.0043 & 140&  0.0669     & 0.0524     & 0.0660 &  0.0665 \\ 
    40&  0.0029     & 0.0026     & 0.0021    &  0.0048 & 145&  0.0600     & 0.0540     & 0.0506 &  0.0739 \\  
    45&  0.0033     & 0.0030     & 0.0021    &  0.0051 & 150&  0.0798     & 0.0726     & 0.0591 &  0.0847 \\ 
    50&  0.0036     & 0.0034     & 0.0017    &  0.0055 & 155&  0.0967     & 0.0895     & 0.0696 &  0.0995 \\  
    55&  0.0043     & 0.0042     & 0.0016    &  0.0067 & 160&  0.136      & 0.125      & 0.0847 &  0.118  \\  
    60&  0.0055     & 0.0057     & 0.0016    &  0.0083 & 165&  0.179      & 0.122      & 0.175  &  0.144  \\   
    65&  0.0073     & 0.0070     & 0.0010    &  0.0097 & 170&  0.323      & 0.237      & 0.234  &  0.188  \\   
    70&  0.0097     & 0.0106     & 0.0021    &  0.0117 & 175&  0.352      & 0.294      & 0.245  &  0.269  \\   
    75&  0.0142     & 0.0163     & 0.0029    &  0.0134 & 180&  0.765      & 0.596      & 0.408  &  0.391  \\  
    80&  0.0203     & 0.0227     & 0.0043    &  0.0165 & 185&  0.838      & 0.702      & 0.582  &  0.700  \\  
    85&  0.0357     & 0.0383     & 0.0101    &  0.0198 & 190&  1.04       & 0.855      & 0.764  &  1.07   \\
    90&  0.0527     & 0.0522     & 0.0292    &  0.0247 & 195&  1.93       & 1.81       & 1.10   &  2.88   \\ 
    95&  0.0520     & 0.0493     & 0.0400    &  0.0266 & 200&  6.97       & 6.47       & 3.49   &  5.29   \\ 
   100&  0.0298     & 0.0257     & 0.0370    &  0.0283 &    &             &            &        &         \\
\hline
\end{tabular}
}
\end{center}
\caption{\sl The 95\% CL upper bound, $S_{95}$, obtained for the normalised
cross-section (see text) of the 
pair production  process \ee\ra~\calHb\calHa, as a function of the Higgs boson 
mass sum \mcalHa~+~\mcalHb. The bounds are given 
for the particular case where $\mcalHb$ and $\mcalHa$ are approximately equal. This occurs, for example, 
in the CP-conserving MSSM scenario {\it \mh-max} for \tanb\ greater than 10 and small \mcalHb ($\equiv$~\mA). 
The numbers listed in this table correspond 
to the four plots in Figure~\ref{eta-h2h1} (see the corresponding labels).
For \mcalHa~+~\mcalHb\ less than 30~\Gcs, the bounds are derived from the measured decay width of the Z boson, see Section
3.2.  
Columns labelled (a): the Higgs boson decay branching ratios correspond to the {\it \mh-max} benchmark scenario with \tanb=10, 
giving
94\% for \calHa\ra~\bb, 6\% for \calHa\ra~\tautau, 92\% for \calHb\ra~\bb\
and 8\% for \calHb\ra~\tautau; 
columns (b): both Higgs bosons are assumed to decay exclusively to \bb; 
columns (c): one Higgs boson is assumed to decay exclusively to \bb\ only and the other exclusively to \tautau;
columns (d): both Higgs bosons are assumed to decay exclusively to \tautau.
}
\label{tab:forFig5}
\end{table} 
\newpage
\begin{table}[ht] 
\begin{center}
{\tiny
\begin{tabular}{|c|ccccccccc|}
\hline
& & & & & & & & &     \\
{\small \mcalHb (\Gcs)} & \multicolumn{9}{|c|}{{\small \mcalHa (\Gcs)}} \\
& & & & & & & & &    \\
\hline
& & & & & & & & &     \\
   &    10    &    15  &   20   &   25     &   30   &   35     &    40   &    45   &    50   \\
& & & & & & & & &      \\
\hline
15 & $\geq 1$ & 0.012 &       &         &       &         &         &         &       \\
20 & $\geq 1$ & 0.013 & 0.010 &         &       &         &         &         &       \\
25 & $\geq 1$ & 0.017 & 0.013 & 0.011   &       &         &         &         &       \\
30 & $\geq 1$ & 0.015 & 0.013 & 0.012   & 0.020 & 0.023   &         &         &       \\
40 & $\geq 1$ & 0.016 & 0.018 & 0.022   & 0.028 & 0.039   & 0.043   &         &       \\
45 & $\geq 1$ & 0.029 & 0.029 & 0.026   & 0.037 & 0.048   & 0.067   & 0.041   &       \\
50 & $\geq 1$ & 0.035 & 0.026 & 0.042   & 0.044 & 0.069   & 0.043   & 0.035   & 0.028 \\
55 & $\geq 1$ & 0.063 & 0.056 & 0.076   & 0.071 & 0.058   & 0.050   & 0.038   & 0.030 \\
60 & $\geq 1$ & 0.075 & 0.084 & 0.098   & 0.051 & 0.051   & 0.050   & 0.044   & 0.039 \\
65 & $\geq 1$ & 0.14  & 0.13  & 0.10    & 0.065 & 0.064   & 0.070   & 0.068   & 0.069 \\
70 & $\geq 1$ & 0.20  & 0.16  & 0.11    & 0.072 & 0.074   & 0.066   & 0.072   & 0.071 \\
75 & $\geq 1$ & 0.23  & 0.13  & 0.14    & 0.076 & 0.075   & 0.083   & 0.066   & 0.093 \\
80 & $\geq 1$ & 0.26  & 0.19  & 0.12    & 0.078 & 0.089   & 0.072   & 0.064   & 0.093 \\
85 & $\geq 1$ & 0.26  & 0.17  & 0.13    & 0.095 & 0.080   & 0.070   & 0.071   & 0.10  \\
90 & $\geq 1$ & 0.18  & 0.13  & 0.11    & 0.073 & 0.070   & 0.076   & 0.081   & 0.13  \\
95 & $\geq 1$ & 0.20  & 0.13  & 0.095   & 0.073 & 0.078   & 0.081   & 0.11    & 0.15  \\
100& $\geq 1$ & 0.21  & 0.12  & 0.092   & 0.085 & 0.091   & 0.12    & 0.16    & 0.18  \\
105& $\geq 1$ & 0.16  & 0.13  & 0.12    & 0.11  & 0.13    & 0.18    & 0.20    & 0.20  \\
110& 0.297    & 0.16  & 0.14  & 0.15    & 0.14  & 0.17    & 0.20    & 0.20    & 0.19  \\
115& 0.338    & 0.22  & 0.20  & 0.20    & 0.18  & 0.20    & 0.21    & 0.21    & 0.23  \\
120& 0.355    & 0.28  & 0.25  & 0.23    & 0.22  & 0.22    & 0.23    & 0.27    & 0.36  \\
125& 0.409    & 0.29  & 0.26  & 0.25    & 0.22  & 0.25    & 0.29    & 0.40    & 0.51  \\
130& 0.494    & 0.35  & 0.32  & 0.32    & 0.24  & 0.32    & 0.46    & 0.57    & 0.72  \\
135& 0.617    & 0.44  & 0.42  & 0.42    & 0.36  & 0.51    & 0.67    & 0.84    & 0.98  \\
140& 0.696    & 0.57  & 0.53  & 0.66    & 0.62  & 0.83    & 0.97    & $\geq 1$ & $\geq 1$\\
145& 0.811    & 0.73  & 0.80  & $\geq 1$& 0.94  & $\geq 1$& $\geq 1$ & $\geq 1$ & $\geq 1$\\
\hline
& & & & & & & & &   \\
   &        65  &     70  &    75    &    80    &    85    &     90   &    95    &    100   &  105      \\
& & & & & & & & &   \\
\hline
65 & 0.067     &          &          &          &          &          &          &          &            \\
70 & 0.082     & 0.078    &          &          &          &          &          &          &            \\
75 & 0.10      & 0.10     & 0.098    &          &          &          &          &          &            \\
80 & 0.11      & 0.11     & 0.14     & 0.14     &          &          &          &          &            \\
85 & 0.12      & 0.14     & 0.16     & 0.15     & 0.21     &          &          &          &            \\
90 & 0.17      & 0.16     & 0.17     & 0.25     & 0.24     & 0.41     &          &          &            \\
95 & 0.19      & 0.21     & 0.30     & 0.35     & 0.44     & 0.51     & 0.64     &          &            \\
100& 0.21      & 0.31     & 0.39     & 0.42     & 0.43     & 0.74     & $\geq 1$ & $\geq 1$ &            \\
105& 0.32      & 0.42     & 0.55     & 0.53     & 0.90     & $\geq 1$ & $\geq 1$ & $\geq 1$ & $\geq 1$   \\
110& 0.47      & 0.55     & 0.63     & $\geq 1$ & $\geq 1$ & $\geq 1$ & $\geq 1$ & $\geq 1$ & $\geq 1$   \\
115& 0.56      & 0.65     & $\geq 1$ & $\geq 1$ & $\geq 1$ & $\geq 1$ & $\geq 1$ & $\geq 1$ & $\geq 1$   \\
120& 0.64      & $\geq 1$ & $\geq 1$ & $\geq 1$ & $\geq 1$ & $\geq 1$ & $\geq 1$ & $\geq 1$ & $\geq 1$   \\
125& $\geq 1$  & $\geq 1$ & $\geq 1$ & $\geq 1$ & $\geq 1$ & $\geq 1$ & $\geq 1$ & $\geq 1$ & $\geq 1$   \\
\hline
\end{tabular}
}
\end{center}
\caption{\sl The 95\% CL upper bound, $S_{95}$, obtained for the normalised 
cross-section (see text) of the 
pair production process \mbox{\ee\ra~\calHb\calHa\ra~\bb\bb}, as
a function of the Higgs boson masses \mcalHa\ and \mcalHb. 
The numbers correspond to the contours shown in Figure~\ref{c2-h2h1} (a). 
}
\label{tab:forFig4a}
\end{table} 

\newpage
\begin{table}[ht] 
\begin{center}
{\tiny
\begin{tabular}{|c|cccccccccc|}
\hline
& & & & & & & & & &    \\
{\small \mcalHb (\Gcs)} & \multicolumn{10}{|c|}{{\small \mcalHa (\Gcs)}} \\
& & & & & & & & & &    \\
\hline
& & & & & & & & & &   \\
   &     5  &    10  &    15  &   20   &    25  &   30   &   35  &   40    & 45       & 50     \\
& & & & & & & & & &    \\
\hline
5  & 0.00041&        &        &        &        &        &       &         &          &         \\
10 & 0.00047& 0.00035&        &        &        &        &       &         &          &         \\
15 & 0.0036 & 0.0032 & 0.0032 &        &        &        &       &         &          &         \\
20 & 0.0033 & 0.0035 & 0.0037 & 0.0040 &        &        &       &         &          &         \\
25 & 0.0037 & 0.0039 & 0.0043 & 0.0043 & 0.0046 &        &       &         &          &         \\
30 & 0.0052 & 0.0058 & 0.0045 & 0.0047 & 0.0055 & 0.0060 &       &         &          &         \\
35 & 0.0060 & 0.0058 & 0.0056 & 0.0065 & 0.0070 & 0.0081 & 0.0084&         &          &         \\
40 & 0.0063 & 0.0064 & 0.0071 & 0.0070 & 0.0078 & 0.0092 & 0.011 & 0.0099  &          &         \\
45 & 0.0079 & 0.0068 & 0.0066 & 0.0083 & 0.0088 & 0.011  & 0.011 & 0.012   & 0.016    &         \\
50 & 0.0096 & 0.011  & 0.0086 & 0.0089 & 0.011  & 0.011  & 0.015 & 0.018   & 0.017  & 0.018    \\
60 & 0.013  & 0.012  & 0.011  & 0.014  & 0.017  & 0.019  & 0.022 & 0.022   & 0.024  & 0.024    \\
65 & 0.015  & 0.015  & 0.015  & 0.016  & 0.019  & 0.022  & 0.023 & 0.023   & 0.024  & 0.026     \\
70 & 0.019  & 0.017  & 0.017  & 0.021  & 0.021  & 0.022  & 0.023 & 0.024   & 0.025  & 0.033    \\
75 & 0.023  & 0.023  & 0.021  & 0.023  & 0.024  & 0.024  & 0.025 & 0.028   & 0.031  & 0.035    \\
80 & 0.028  & 0.026  & 0.024  & 0.025  & 0.025  & 0.026  & 0.030 & 0.032   & 0.036  & 0.041    \\
85 & 0.032  & 0.029  & 0.028  & 0.027  & 0.030  & 0.031  & 0.032 & 0.035   & 0.040  & 0.043    \\
90 & 0.033  & 0.031  & 0.028  & 0.030  & 0.031  & 0.032  & 0.035 & 0.041   & 0.045  & 0.049    \\
95 & 0.037  & 0.034  & 0.031  & 0.034  & 0.036  & 0.037  & 0.041 & 0.047   & 0.050  & 0.054    \\
100& 0.040  & 0.036  & 0.036  & 0.037  & 0.038  & 0.044  & 0.048 & 0.053   & 0.059  & 0.062    \\
105& 0.045  & 0.040  & 0.042  & 0.043  & 0.047  & 0.052  & 0.055 & 0.065   & 0.068  & 0.072    \\
110& 0.051  & 0.044  & 0.050  & 0.053  & 0.054  & 0.057  & 0.062 & 0.076   & 0.081  & 0.085    \\
115& 0.055  & 0.050  & 0.060  & 0.065  & 0.064  & 0.069  & 0.074 & 0.083   & 0.089  & 0.105     \\
120& 0.067  & 0.060  & 0.071  & 0.075  & 0.077  & 0.083  & 0.085 & 0.093   & 0.12   & 0.145     \\
125& 0.075  & 0.071  & 0.086  & 0.084  & 0.089  & 0.097  & 0.109 & 0.12    & 0.17   & 0.198     \\
130& 0.085  & 0.088  & 0.10   & 0.10   & 0.11   & 0.13   & 0.14  & 0.16    & 0.20   & 0.317     \\
135& 0.11   & 0.11   & 0.13   & 0.15   & 0.16   & 0.16   & 0.19  & 0.23    & 0.31   & 0.436     \\
140& 0.14   & 0.13   & 0.18   & 0.19   & 0.21   & 0.23   & 0.26  & 0.30    & 0.50   & $\geq 1$  \\
145& 0.18   & 0.18   & 0.25   & 0.26   & 0.28   & 0.33   & 0.42  & 0.59    & $\geq 1$ & $\geq 1$\\
150& 0.25   & 0.26   & 0.35   & 0.37   & 0.42   & 0.51   & 0.69  & $\geq 1$& $\geq 1$ & $\geq 1$\\
\hline
& & & & & & & & & &   \\
   &   55   &    60    &    65    &    70    &    75    &    80    &    85    &    90    &    95    &    100      \\
& & & & & & & & & &    \\
\hline
55 & 0.021 &          &          &          &          &          &          &          &          &             \\
60 & 0.025 & 0.028    &          &          &          &          &          &          &          &             \\
65 & 0.030 & 0.033    & 0.036    &          &          &          &          &          &          &             \\
70 & 0.033 & 0.039    & 0.039    & 0.042    &          &          &          &          &          &             \\
75 & 0.038 & 0.040    & 0.044    & 0.048    & 0.049    &          &          &          &          &             \\
80 & 0.043 & 0.049    & 0.047    & 0.051    & 0.057    & 0.064    &          &          &          &             \\
85 & 0.048 & 0.050    & 0.055    & 0.061    & 0.071    & 0.075    & 0.097    &          &          &             \\
90 & 0.053 & 0.059    & 0.065    & 0.077    & 0.080    & 0.10     & 0.14     & 0.21     &          &             \\
95 & 0.059 & 0.067    & 0.076    & 0.080    & 0.10     & 0.14     & 0.21     & 0.38     & 0.70     &             \\
100& 0.069 & 0.077    & 0.086    & 0.11     & 0.15     & 0.21     & 0.39     & 0.71     & $\geq 1$ & $\geq 1$    \\
105& 0.083 & 0.096    & 0.11     & 0.15     & 0.22     & 0.39     & 0.73     & $\geq 1$ & $\geq 1$ & $\geq 1$    \\
110& 0.11  & 0.13     & 0.16     & 0.21     & 0.39     & 0.76     & $\geq 1$ & $\geq 1$ & $\geq 1$ & $\geq 1$    \\
115& 0.14  & 0.20     & 0.27     & 0.36     & 0.79     & $\geq 1$ & $\geq 1$ & $\geq 1$ & $\geq 1$ & $\geq 1$    \\
120& 0.19  & 0.28     & 0.49     & 0.83     & $\geq 1$ & $\geq 1$ & $\geq 1$ & $\geq 1$ & $\geq 1$ & $\geq 1$    \\
125& 0.26  & 0.53     & 0.65     & $\geq 1$ & $\geq 1$ & $\geq 1$ & $\geq 1$ & $\geq 1$ & $\geq 1$ & $\geq 1$    \\
130& 0.46  & 0.85     & $\geq 1$ & $\geq 1$ & $\geq 1$ & $\geq 1$ & $\geq 1$ & $\geq 1$ & $\geq 1$ & $\geq 1$    \\
135& 0.89  & $\geq 1$ & $\geq 1$ & $\geq 1$ & $\geq 1$ & $\geq 1$ & $\geq 1$ & $\geq 1$ & $\geq 1$ & $\geq 1$    \\
\hline
\end{tabular}
}
\end{center}
\caption{\sl The 95\% CL upper bound, $S_{95}$, obtained for the normalised 
cross-section (see text) of the 
pair production process \mbox{\ee\ra~\calHb\calHa\ra~\tautau\tautau}, as
a function of the Higgs boson masses \mcalHa\ and \mcalHb. 
The numbers correspond to the contours shown in Figure~\ref{c2-h2h1} (b). 
}
\label{tab:forFig4b}
\end{table} 
\newpage
\begin{table}[ht] 
\begin{center}
{\small
\begin{tabular}{|c|ccccccccccc|}
\hline
& & & & & & & & & & &     \\
{\small \mcalHb (\Gcs)} & \multicolumn{11}{|c|}{{\small \mcalHa (\Gcs)}} \\
& & & & & & & & & & &     \\
\hline
& & & & & & & & & & &     \\
& 10 & 15 & 20 & 25 & 30 & 35 & 40 & 45 & 50 & 55 & 60    \\
& & & & & & & & & & &   \\
\hline
20 & $\geq 1$ &         &       &        &       &          &          &          &          &          &                   \\
25 & 0.096    &         &       &        &       &          &          &          &          &          &                   \\
30 & 0.11     & 0.17    &       &        &       &          &          &          &          &          &                   \\
35 & 0.13     & 0.075   &       &        &       &          &          &          &          &          &                   \\
40 & 0.028    & 0.034   & 0.19  &        &       &          &          &          &          &          &                   \\
45 & 0.15     & 0.047   & 0.034 &        &       &          &          &          &          &          &                   \\
50 & 0.063    & 0.063   & 0.029 & 0.039  &       &          &          &          &          &          &                   \\
55 & 0.074    & 0.087   & 0.042 & 0.055  &       &          &          &          &          &          &                   \\
60 & 0.11     & 0.12    & 0.099 & 0.086  & 0.12  &          &          &          &          &          &                   \\
65 & 0.25     & 0.17    & 0.13  & 0.14   & 0.13  &          &          &          &          &          &                   \\
70 & $\geq 1$ & 0.15    & 0.14  & 0.13   & 0.12  & 0.13     &          &          &          &          &                   \\
75 & 0.72     & 0.17    & 0.16  & 0.14   & 0.13  & 0.13     &          &          &          &          &                   \\
80 & 0.99     & 0.18    & 0.14  & 0.11   & 0.11  & 0.12     & 0.13     &          &          &          &                   \\
85 & $\geq 1$ & 0.19    & 0.15  & 0.13   & 0.14  & 0.14     & 0.16     &          &          &          &                   \\
90 & $\geq 1$ & 0.19    & 0.15  & 0.15   & 0.15  & 0.16     & 0.15     & 0.17     &          &          &                   \\  
95 & $\geq 1$ & 0.20    & 0.17  & 0.17   & 0.16  & 0.15     & 0.17     & 0.20     &          &          &                   \\
100& $\geq 1$ & 0.22    & 0.20  & 0.19   & 0.18  & 0.18     & 0.20     & 0.23     & 0.30     &          &                   \\
105& $\geq 1$ & 0.26    & 0.22  & 0.21   & 0.23  & 0.24     & 0.27     & 0.32     & 0.38     &          &                   \\
110& $\geq 1$ & 0.30    & 0.26  & 0.27   & 0.28  & 0.31     & 0.36     & 0.40     & 0.43     & 0.55     &                   \\
115& $\geq 1$ & 0.35    & 0.32  & 0.33   & 0.33  & 0.38     & 0.43     & 0.47     & 0.54     & 0.70     &                   \\
120& $\geq 1$ & 0.43    & 0.39  & 0.39   & 0.42  & 0.46     & 0.50     & 0.58     & 0.71     & 0.93     & $\geq 1$          \\
125& $\geq 1$ & 0.53    & 0.49  & 0.48   & 0.51  & 0.56     & 0.63     & 0.77     & 0.99     & $\geq 1$ & $\geq 1$          \\
130& $\geq 1$ & 0.66    & 0.59  & 0.62   & 0.64  & 0.72     & 0.86     & $\geq 1$ & $\geq 1$ & $\geq 1$ & $\geq 1$  \\
135& $\geq 1$ & 0.82    & 0.66  & 0.75   & 0.84  & 0.98     & $\geq 1$ & $\geq 1$ & $\geq 1$ & $\geq 1$ & $\geq 1$  \\
140& $\geq 1$ & $\geq 1$ & 0.90 & 0.96   & 0.98  & $\geq 1$ & $\geq 1$ & $\geq 1$ & $\geq 1$ & $\geq 1$ & $\geq 1$  \\

\hline
\end{tabular}
}
\end{center}
\caption{\sl The 95\% CL upper bound, $S_{95}$, obtained for the normalised 
cross-section (see text) of the 
pair production cascade process \mbox{\ee\ra~(\calHb\ra~\calHa\calHa)\calHa\ra~(\bb\bb)\bb}, as
a function of the Higgs boson masses \mcalHa\ and \mcalHb. 
The numbers correspond to the contours shown in Figure~\ref{c2-h1h1h1} (a). 
}
\label{tab:forFig5a}
\end{table} 
\newpage
\begin{table}[ht] 
\begin{center}
{\small
\begin{tabular}{|c|cccc|}
\hline
& & & &     \\
{\small \mcalHb (\Gcs)} & \multicolumn{4}{|c|}{{\small \mcalHa (\Gcs)}} \\
& & & &     \\
\hline
& & & &   \\
   &    5     &   10     &   15     &   20      \\
& & & &   \\
\hline
10 & 0.0006  &         &          &            \\
15 & 0.0016  &         &          &             \\
20 & 0.0017  & 0.011   &          &              \\
25 & 0.0018  & 0.0019  &          &          \\
30 & 0.0021  & 0.0021  & 0.013   &           \\
35 & 0.0024  & 0.0025  & 0.017   &          \\
40 & 0.0009  & 0.0016  & $\geq 1$ & $\geq 1$    \\
45 & 0.0010  & 0.0019  & $\geq 1$ & $\geq 1$ \\
50 & 0.0013  & 0.0023  & $\geq 1$ & $\geq 1$    \\
55 & 0.0017  & 0.0029  & $\geq 1$ & $\geq 1$  \\
60 & 0.0024  & 0.0043  & $\geq 1$ & $\geq 1$  \\
65 & 0.0058  & 0.014   & $\geq 1$ & $\geq 1$  \\
\hline
\end{tabular}
}
\end{center}
\caption{\sl The 95\% CL upper bound, $S_{95}$, obtained for the normalised 
cross-section (see text) of the 
pair production cascade process \mbox{\ee\ra~(\calHb\ra~\calHa\calHa)\calHa\ra~(\tautau\tautau)\tautau}, as
a function of the Higgs boson masses \mcalHa\ and \mcalHb. 
The numbers correspond to the contours shown in Figure~\ref{c2-h1h1h1} (b). 
}
\label{tab:forFig5b}
\end{table} 
%
%
\clearpage
\newpage
\noindent
{\bf \Large Appendix C: List of authors}\\
$\phantom{.}$\\
$\phantom{.}$\\
\noindent
The ALEPH, DELPHI, L3 and OPAL Collaborations have provided the inputs for the combined results presented in this
paper. The LEP Working Group for Higgs Boson Searches has performed the combinations. The Working Group consists of
members of the four collaborations and of theorists among whom S.~Heinemeyer\footnote{Departamento de Fisica Teorica,
Facultad de Ciencias, Universidad de Zaragoza, 50009 Zaragoza, Spain;\\
$\phantom{.......}$CERN TH division, Dept. of Physics,
1211 Geneva 23, Switzerland},
A.~Pilaftsis\footnote{Department of Physics and Astronomy, University of Manchester, Manchester M13 9PL, UK} 
and G.~Weiglein\footnote{Institute for Particle Physics Phenomenology, University of Durham, Durham DH1 3LE, UK} 
are authors of this paper. The lists of authors from the collaborations follow.
\clearpage
\newpage
\noindent{\bf\large Appendix C1: The ALEPH Collaboration}\\
$\phantom{.}$\\
$\phantom{.}$\\
{\small\raggedright
\noindent
S.\thinspace Schael$^1$,
R.\thinspace Barate$^2$,
R.\thinspace Bruneli\`ere$^2$,
I.\thinspace De~Bonis$^2$,
D.\thinspace Decamp$^2$,
C.\thinspace Goy$^2$,
S.\thinspace J\'ez\'equel$^2$,
J.\thinspace -P.\thinspace ~Lees$^2$,
F.\thinspace Martin$^2$,
E.\thinspace Merle$^2$
% ...
\\
}
$\phantom{.}$\\
$\phantom{.}$\\
{\small
~1~~Physikalisches Institut das RWTH-Aachen, D-52056 Aachen, Germany\\
~2~~Laboratoire de Physique des Particules (LAPP), IN$^{2}$P$^{3}$-CNRS,\\
$\phantom{....}$F-74019 Annecy-le-Vieux Cedex, France\\
~3~~Institut de F\'{i}sica d'Altes Energies, Universitat Aut\`{o}noma de Barcelona,\\
$\phantom{....}$E-08193 Bellaterra (Barcelona), Spain$^\alpha$\\
~4~~Dipartimento di Fisica, INFN Sezione di Bari, I-70126 Bari, Italy\\
~5~~Institute of High Energy Physics, Academia Sinica, Beijing, The People's Republic of China$^\beta$\\
~6~~European Laboratory for Particle Physics (CERN), CH-1211 Geneva 23, Switzerland\\
~7~~Laboratoire de Physique Corpusculaire, Universit\'e Blaise Pascal, IN$^{2}$P$^{3}$-CNRS,\\
$\phantom{....}$Clermont-Ferrand, F-63177 Aubi\`{e}re, France\\
~8~~Niels Bohr Institute, 2100 Copenhagen, DK-Denmark$^\gamma$\\
~9~~Nuclear Research Center Demokritos (NRCD), GR-15310 Attiki, Greece\\
10~~Laoratoire Leprince-Ringuet, Ecole Polytechnique, IN$^{2}$P$^{3}$-CNRS,\\
$\phantom{......}$F-91128 Palaiseau Cedex, France\\
11~~Dipartimento di Fisica, Universit\`a di Firenze, INFN Sezione di Firenze,\\
$\phantom{......}$I-50125 Firenze, Italy\\
12~~Laboratori Nazionali dell'INFN (LNF-INFN), I-00044 Frascati, Italy\\
13~~Department of Physics and Astronomy, University of Glasgow,\\
$\phantom{......}$Glasgow G128QQ,United Kingdom$^\delta$\\
14~~Utah Valley State College, Orem, UT 84058, U.S.A.\\
15~~Kirchhoff-Institut f\"ur Physik, Universit\"at Heidelberg, D-69120 Heidelberg, Germany$^\epsilon$\\
16~~Department of Physics, Imperial College, London SW7 2BZ, United Kingdom$^\delta$\\
17~~Institut f\"ur Experimentalphysik, Universit\"at Innsbruck, A-6020 Innsbruck, Austria$^\nu$\\
18~~Department of Physics, University of Lancaster, Lancaster LA1 4YB, United Kingdom$^\delta$\\
19~~Institut de Physique Nucl\'eaire, D\'epartement de Physique,\\
$\phantom{......}$Universit\'e Catholique de Louvain, 1348 Louvain-la-Neuve, Belgium\\
20~~Institut f\"ur Physik, Universit\"at Mainz, D-55099 Mainz, Germany$^\lambda$\\
21~~Centre de Physique des Particules de Marseille, Univ M\'editerran\'ee, IN$^{2}$P$^{3}$-CNRS,\\
$\phantom{......}$F-13288 Marseille, France\\
22~~Dipartimento di Fisica, Universit\`a di Milano e INFN Sezione di Milano,\\
$\phantom{......}$I-20133 Milano, Italy\\
23~~Max-Planck-Institut f\"ur Physik, Werner-Heisenberg-Institut, D-80805 M\"unchen, Germany$^\lambda$\\
24~~Laboratoire de l'Acc\'el\'erateur Lin\'eaire, Universit\'e de Paris-Sud, IN$^{2}$P$^{3}$-CNRS,\\
$\phantom{......}$F-91898 Orsay Cedex, France\\
25~~Dipartimento di Fisica dell'Universit\`a, INFN Sezione di Pisa, e Scuola Normale Superiore,\\
$\phantom{......}$I-56010 Pisa, Italy\\
26~~Department of Physics, Royal Holloway \& Bedford New College, University of London,\\
$\phantom{......}$Egham, Surrey TW20 OEX, United Kingdom$^\delta$\\
27~~Particle Physics Dept., Rutherford Appleton Laboratory, Chilton, Didcot, \\
$\phantom{......}$Oxon OX11 OQX, United Kingdom$^\delta$\\
28~~CEA, DAPNIA/Service de Physique des Particules, CE-Saclay,\\
$\phantom{......}$F-91191 Gif-sur-Yvette Cedex, France$^\mu$\\
29~~Institute for Particle Physics, University of California at Santa Cruz, Santa Cruz,\\
$\phantom{......}$CA 95064, USA$^\pi$\\
30~~Department of Physics, University of Sheffield, Sheffield S3 7RH, United Kingdom$^\delta$\\
31~~Fachbereich Physik, Universit\"at Siegen, D-57068 Siegen, Germany$^\lambda$\\
32~~Dipartimento di Fisica, Universit\`a di Trieste e INFN Sezione di Trieste,\\
$\phantom{......}$I-34127 Trieste, Italy\\
33~~Experimental Elementary Particle Physics, University of Washington,\\
$\phantom{......}$Seattle, WA 98195 U.S.A.\\
34~~Department of Physics, University of Wisconsin, Madison, WI 53706, USA$^\epsilon$\\
35~~Institute for Particle Physics, ETH H\"onggerberg, 8093 Z\"urich, Switzerland.\\
$\phantom{.}$\\
$\phantom{.}$\\
a~~{Also at CERN, 1211 Geneva 23, Switzerland}\\
b~~{Now at Fermilab, PO Box 500, MS 352, Batavia, IL 60510, USA}\\
c~~{Also at Dipartimento di Fisica di Catania and INFN Sezione di Catania,\\ 
$\phantom{....}$95129 Catania, Italy.}\\
d~~{Now at University of Florida, Department of Physics, Gainesville, Florida 32611-8440, USA}\\
e~~{Also IFSI sezione di Torino, INAF, Italy.}\\
f~~{Also at Groupe d'Astroparticules de Montpellier, Universit\'{e} de Montpellier II,\\
$\phantom{....}$34095 Montpellier, France.}\\
g~~{Now at Departement de Physique Corpusculaire, Universit\'e de Gen\`eve,\\
$\phantom{....}$1211 Gen\`eve 4, Switzerland.}\\
h~~{Also at Department of Physics, Tsinghua University, Beijing, \\
$\phantom{....}$The People's Republic of China.}\\
i~~{Supported by the Leverhulme Trust.}\\
j~~{Permanent address: Universitat de Barcelona, 08208 Barcelona, Spain.}\\
k~~{Now at SAP AG, 69185 Walldorf, Germany}\\
l~~{Now at Groupe d' Astroparticules de Montpellier, Universit\'e de Montpellier II,\\
$\phantom{....}$34095 Montpellier, France.}\\
m~~{Now at BNP Paribas, 60325 Frankfurt am Mainz, Germany}\\
n~~{Now at Institut Inter-universitaire des hautes Energies (IIHE), CP 230, \\
$\phantom{....}$Universit\'{e} Libre de Bruxelles, 1050 Bruxelles, Belgique}\\
o~~{Now at Dipartimento di Fisica e Tecnologie Relative, Universit\`a di Palermo,\\
$\phantom{....}$Palermo, Italy.}\\
q~~{Now at SLAC, Stanford, CA 94309, U.S.A}\\
r~~{Now at CERN, 1211 Geneva 23, Switzerland}\\
s~~{Research Fellow of the Belgium FNRS}\\
t~~{Research Associate of the Belgium FNRS} \\
u~~{Now at Liverpool University, Liverpool L69 7ZE, United Kingdom}\\ 
v~~{Supported by the Federal Office for Scientific, Technical and Cultural Affairs\\
$\phantom{....}$through the Interuniversity Attraction Pole P5/27}\\ 
w~~{Now at Henryk Niewodnicznski Institute of Nuclear Physics, Polish Academy of Sciences,\\ 
$\phantom{....}$Cracow, Poland} \\  
$\dagger$~~{Deceased}\\
$\phantom{.}$\\
$\phantom{.}$\\
$\alpha$~~{Supported by CICYT, Spain.}\\
$\beta$~~{Supported by the National Science Foundation of China.}\\
$\gamma$~~{Supported by the Danish Natural Science Research Council.}\\
$\delta$~~{Supported by the UK Particle Physics and Astronomy Research Council.}\\
$\epsilon$~~{Supported by the US Department of Energy, grant DE-FG0295-ER40896.}\\
$\lambda$~~{Supported by Bundesministerium f\"ur Bildung und Forschung, Germany.}\\
$\mu$~~{Supported by the Direction des Sciences de la Mati\`ere, C.E.A.}\\
$\nu$~~{Supported by the Austrian Ministry for Science and Transport.}\\
$\pi$~~{Supported by the US Department of Energy, grant DE-FG03-92ER40689.}
}
\clearpage
\newpage
\noindent{\bf\large Appendix C2: The DELPHI Collaboration}\\
$\phantom{.}$\\
$\phantom{.}$\\
{\small\raggedright
\noindent
{J.\thinspace Abdallah}$^{25}$,
{P.\thinspace Abreu}$^{22}$,
{W.\thinspace Adam}$^{51}$,
{P.\thinspace Adzic}$^{11}$,
{T.\thinspace Albrecht}$^{17}$,
{T.\thinspace Alderweireld}$^{2}$,
{R.\thinspace Alemany-Fernandez}$^{8}$,
{T.\thinspace Allmendinger}$^{17}$,
{P.\thinspace P.\thinspace Allport}$^{23}$,
{U.\thinspace Amaldi}$^{29}$
% ...
\\
}
$\phantom{.}$\\
$\phantom{.}$\\
{\small
1~~~Department of Physics and Astronomy, Iowa State
     University, Ames IA 50011-3160, USA\\
2~~~Physics Department, Universiteit Antwerpen,
     Universiteitsplein 1, B-2610 Antwerpen, Belgium \\
     \indent and IIHE, ULB-VUB,
     Pleinlaan 2, B-1050 Brussels, Belgium \\
     \indent and Facult\'e des Sciences,
     Univ. de l'Etat Mons, Av. Maistriau 19, B-7000 Mons, Belgium\\
3~~~Physics Laboratory, University of Athens, Solonos Str.
     104, GR-10680 Athens, Greece\\
4~~~Department of Physics, University of Bergen,
     All\'egaten 55, NO-5007 Bergen, Norway\\
5~~~Dipartimento di Fisica, Universit\`a di Bologna and INFN,
     Via Irnerio 46, IT-40126 Bologna, Italy\\
6~~~Centro Brasileiro de Pesquisas F\'{\i}sicas, rua Xavier Sigaud 150,
     BR-22290 Rio de Janeiro, Brazil \\
     \indent and Depto. de F\'{\i}sica, Pont. Univ. Cat\'olica,
     C.P. 38071, BR-22453 Rio~de~Janeiro, Brazil \\
     \indent and Inst. de F\'{\i}sica, Univ. Estadual do Rio de Janeiro,
     rua S\~{a}o Francisco Xavier 524,\\
     \indent Rio de Janeiro, Brazil\\
7~~~Coll\`ege de France, Lab. de Physique Corpusculaire, IN2P3-CNRS,\\
     \indent FR-75231 Paris Cedex 05, France\\
8~~~CERN, CH-1211 Geneva 23, Switzerland\\
9~~~Institut de Recherches Subatomiques, IN2P3 - CNRS/ULP - BP20,
     FR-67037 Strasbourg Cedex,\\
     \indent  France\\
10~~Now at DESY-Zeuthen, Platanenallee 6, D-15735 Zeuthen, Germany\\
11~~Institute of Nuclear Physics, N.C.S.R. Demokritos,
     P.O. Box 60228, GR-15310 Athens, Greece\\
12~~FZU, Inst. of Phys. of the C.A.S. High Energy Physics Division,
     Na Slovance 2,\\
     \indent CZ-180 40 Praha 8, Czech Republic\\
13~~Dipartimento di Fisica, Universit\`a di Genova and INFN,
     Via Dodecaneso 33, IT-16146 Genova,\\
     \indent Italy\\
14~~Institut des Sciences Nucl\'eaires, IN2P3-CNRS, Universit\'e
     de Grenoble 1,\\
     \indent FR-38026 Grenoble Cedex, France\\
15~~Helsinki Institute of Physics and Department of Physical Sciences,
     P.O. Box 64,\\
     \indent FIN-00014 University of Helsinki, Finland\\
16~~Joint Institute for Nuclear Research, Dubna, Head Post
     Office, P.O. Box 79,\\
     \indent RU-101 000 Moscow, Russian Federation\\
17~~Institut f\"ur Experimentelle Kernphysik,
     Universit\"at Karlsruhe, Postfach 6980,\\
     \indent DE-76128 Karlsruhe, Germany\\
18~~Institute of Nuclear Physics PAN,Ul. Radzikowskiego 152,
     PL-31142 Krakow, Poland\\
19~~Faculty of Physics and Nuclear Techniques, University of Mining
     and Metallurgy,\\
     \indent PL-30055 Krakow, Poland\\
20~~Universit\'e de Paris-Sud, Lab. de l'Acc\'el\'erateur
     Lin\'eaire, IN2P3-CNRS, B\^{a}t. 200,\\
     \indent FR-91405 Orsay Cedex, France\\
21~~School of Physics and Chemistry, University of Lancaster,
     Lancaster LA1 4YB, UK\\
22~~LIP, IST, FCUL - Av. Elias Garcia, 14-$1^{o}$,
     PT-1000 Lisboa Codex, Portugal\\
23~~Department of Physics, University of Liverpool, P.O.
     Box 147, Liverpool L69 3BX, UK\\
24~~Dept. of Physics and Astronomy, Kelvin Building,
     University of Glasgow, Glasgow G12 8QQ\\
25~~LPNHE, IN2P3-CNRS, Univ.~Paris VI et VII, Tour 33 (RdC),
     4 place Jussieu,\\
     \indent FR-75252 Paris Cedex 05, France\\
26~~Department of Physics, University of Lund,
     S\"olvegatan 14, SE-223 63 Lund, Sweden\\
27~~Universit\'e Claude Bernard de Lyon, IPNL, IN2P3-CNRS,
     FR-69622 Villeurbanne Cedex, France\\
28~~Dipartimento di Fisica, Universit\`a di Milano and INFN-MILANO,
     Via Celoria 16,\\
     \indent IT-20133 Milan, Italy\\
29~~Dipartimento di Fisica, Univ. di Milano-Bicocca and
     INFN-MILANO, Piazza della Scienza 2,\\
     \indent  IT-20126 Milan, Italy\\
30~~IPNP of MFF, Charles Univ., Areal MFF,
     V Holesovickach 2, CZ-180 00, Praha 8, Czech Republic\\
31~~NIKHEF, Postbus 41882, NL-1009 DB
     Amsterdam, The Netherlands\\
32~~National Technical University, Physics Department,
     Zografou Campus, GR-15773 Athens, Greece\\
33~~Physics Department, University of Oslo, Blindern,
     NO-0316 Oslo, Norway\\
34~~Dpto. Fisica, Univ. Oviedo, Avda. Calvo Sotelo
     s/n, ES-33007 Oviedo, Spain\\
35~~Department of Physics, University of Oxford,
     Keble Road, Oxford OX1 3RH, UK\\
36~~Dipartimento di Fisica, Universit\`a di Padova and
     INFN, Via Marzolo 8, IT-35131 Padua, Italy\\
37~~Rutherford Appleton Laboratory, Chilton, Didcot
     OX11 OQX, UK\\
38~~Dipartimento di Fisica, Universit\`a di Roma II and
     INFN, Tor Vergata, IT-00173 Rome, Italy\\
39~~Dipartimento di Fisica, Universit\`a di Roma III and
     INFN, Via della Vasca Navale 84,\\
     \indent IT-00146 Rome, Italy\\
40~~DAPNIA/Service de Physique des Particules,
     CEA-Saclay, FR-91191 Gif-sur-Yvette Cedex, France\\
41~~Instituto de Fisica de Cantabria (CSIC-UC), Avda.
     los Castros s/n, ES-39006 Santander, Spain\\
42~~Inst. for High Energy Physics, Serpukov
     P.O. Box 35, Protvino, (Moscow Region),\\
     \indent Russian Federation\\
43~~J. Stefan Institute, Jamova 39, SI-1000 Ljubljana, Slovenia
     and Laboratory for Astroparticle\\
     \indent Physics, Nova Gorica Polytechnic, Kostanjeviska 16a, 
     SI-5000 Nova Gorica, Slovenia, \\
     \indent and Department of Physics, University of Ljubljana,
     SI-1000 Ljubljana, Slovenia\\
44~~Fysikum, Stockholm University,
     Box 6730, SE-113 85 Stockholm, Sweden\\
45~~Dipartimento di Fisica Sperimentale, Universit\`a di
     Torino and INFN, Via P. Giuria 1,\\
     \indent IT-10125 Turin, Italy\\
46~~INFN,Sezione di Torino and Dipartimento di Fisica Teorica,
     Universit\`a di Torino, Via Giuria 1,\\
     \indent IT-10125 Turin, Italy\\
47~~Dipartimento di Fisica, Universit\`a di Trieste and
     INFN, Via A. Valerio 2, IT-34127 Trieste, Italy \\
     \indent and Istituto di Fisica, Universit\`a di Udine,
     IT-33100 Udine, Italy\\
48~~Univ. Federal do Rio de Janeiro, C.P. 68528
     Cidade Univ., Ilha do Fund\~ao\\
     \indent BR-21945-970 Rio de Janeiro, Brazil\\
49~~Department of Radiation Sciences, University of
     Uppsala, P.O. Box 535,\\
     \indent SE-751 21 Uppsala, Sweden\\
50~~IFIC, Valencia-CSIC, and D.F.A.M.N., U. de Valencia,
     Avda. Dr. Moliner 50,\\
     \indent ES-46100 Burjassot (Valencia), Spain\\
51~~Institut f\"ur Hochenergiephysik, \"Osterr. Akad.
     d. Wissensch., Nikolsdorfergasse 18,\\
     \indent AT-1050 Vienna, Austria\\
52~~Inst. Nuclear Studies and University of Warsaw, Ul.
     Hoza 69, PL-00681 Warsaw, Poland\\
53~~Fachbereich Physik, University of Wuppertal, Postfach
     100 127, DE-42097 Wuppertal, Germany \\
\noindent {$~\dagger$~~~Deceased}
}
\clearpage
\newpage
\noindent{\bf\large Appendix C3: The L3 Collaboration}\\
$\phantom{.}$\\
$\phantom{.}$\\
{\small\raggedright
\noindent
P.\thinspace Achard$^{20}$, 
O.\thinspace Adriani$^{17}$, 
M.\thinspace Aguilar-Benitez$^{25}$, 
J.\thinspace Alcaraz$^{25}$,
G.\thinspace Alemanni$^{23}$,
J.\thinspace Allaby$^{18}$,
A.\thinspace Aloisio$^{29}$, 
M.\thinspace G.\thinspace Alviggi$^{29}$,
H.\thinspace Anderhub$^{49}$, 
V.\thinspace P.\thinspace Andreev$^{6,34}$
% ...
\\
}
$\phantom{.}$\\
$\phantom{.}$\\
{\small
1~~~III. Physikalisches Institut, RWTH, D-52056 Aachen, Germany$^{\S}$\\
2~~~National Institute for High Energy Physics, NIKHEF, 
     and University of Amsterdam,\\
     \indent NL-1009 DB Amsterdam, The Netherlands\\
3~~~University of Michigan, Ann Arbor, MI 48109, USA\\
4~~~Laboratoire d'Annecy-le-Vieux de Physique des Particules, 
     LAPP,IN2P3-CNRS, BP 110,\\
     \indent F-74941 Annecy-le-Vieux CEDEX, France\\
5~~~Institute of Physics, University of Basel, CH-4056 Basel,
     Switzerland\\
6~~~Louisiana State University, Baton Rouge, LA 70803, USA\\
7~~~Institute of High Energy Physics, IHEP, 
  100039 Beijing, China$^{\triangle}$ \\
8~~~University of Bologna and INFN-Sezione di Bologna, 
     I-40126 Bologna, Italy\\
9~~~Tata Institute of Fundamental Research, Mumbai (Bombay) 400 005, India\\
10~~Northeastern University, Boston, MA 02115, USA\\
11~~Institute of Atomic Physics and University of Bucharest,
     R-76900 Bucharest, Romania\\
12~~Central Research Institute for Physics of the 
     Hungarian Academy of Sciences,\\
     \indent H-1525 Budapest 114, Hungary$^{\ddag}$\\
13~~Massachusetts Institute of Technology, Cambridge, MA 02139, USA\\
14~~Panjab University, Chandigarh 160 014, India\\
15~~KLTE-ATOMKI, H-4010 Debrecen, Hungary$^\P$\\
16~~UCD School of Physics, University College Dublin, 
 Belfield, Dublin 4, Ireland\\
17~~INFN Sezione di Firenze and University of Florence, 
     I-50125 Florence, Italy\\
18~~European Laboratory for Particle Physics, CERN, 
     CH-1211 Geneva 23, Switzerland\\
19~~World Laboratory, FBLJA  Project, CH-1211 Geneva 23, Switzerland\\
20~~University of Geneva, CH-1211 Geneva 4, Switzerland\\
21~~University of Hamburg, D-22761 Hamburg, Germany\\
22~~Chinese University of Science and Technology, USTC,
      Hefei, Anhui 230 029, China$^{\triangle}$\\
23~~University of Lausanne, CH-1015 Lausanne, Switzerland\\
24~~Institut de Physique Nucl\'eaire de Lyon, 
     IN2P3-CNRS,Universit\'e Claude Bernard, \\
     \indent F-69622 Villeurbanne, France\\
25~~Centro de Investigaciones Energ{\'e}ticas, 
     Medioambientales y Tecnol\'ogicas, CIEMAT,\\
     \indent E-28040 Madrid, Spain${\flat}$ \\
26~~Florida Institute of Technology, Melbourne, FL 32901, USA\\
27~~INFN-Sezione di Milano, I-20133 Milan, Italy\\
28~~Institute of Theoretical and Experimental Physics, ITEP, 
     Moscow, Russia\\
29~~INFN-Sezione di Napoli and University of Naples, 
     I-80125 Naples, Italy\\
30~~Department of Physics, University of Cyprus,
     Nicosia, Cyprus\\
31~~Radboud University and NIKHEF, 
     NL-6525 ED Nijmegen, The Netherlands\\
32~~California Institute of Technology, Pasadena, CA 91125, USA\\
33~~INFN-Sezione di Perugia and Universit\`a Degli 
     Studi di Perugia, I-06100 Perugia, Italy   \\
34~~Nuclear Physics Institute, St. Petersburg, Russia\\
35~~Carnegie Mellon University, Pittsburgh, PA 15213, USA\\
36~~INFN-Sezione di Napoli and University of Potenza, I-85100 Potenza, Italy\\
37~~Princeton University, Princeton, NJ 08544, USA\\
38~~University of Californa, Riverside, CA 92521, USA\\
39~~INFN-Sezione di Roma and University of Rome, ``La Sapienza",
     I-00185 Rome, Italy\\
40~~University and INFN, Salerno, I-84100 Salerno, Italy\\
41~~University of California, San Diego, CA 92093, USA\\
42~~Bulgarian Academy of Sciences, Central Lab.~of 
     Mechatronics and Instrumentation,\\
     \indent  BU-1113 Sofia, Bulgaria\\
43~~The Center for High Energy Physics, 
     Kyungpook National University, 702-701 Taegu,\\
     \indent Republic of Korea\\
44~~National Central University, Chung-Li, Taiwan, China\\
45~~Department of Physics, National Tsing Hua University,
      Taiwan, China\\
46~~Purdue University, West Lafayette, IN 47907, USA\\
47~~Paul Scherrer Institut, PSI, CH-5232 Villigen, Switzerland\\
48~~DESY, D-15738 Zeuthen, Germany\\
49~~Eidgen\"ossische Technische Hochschule, ETH Z\"urich,
     CH-8093 Z\"urich, Switzerland\\
     $\phantom{.}$\\
     $\phantom{.}$\\
$\S$~~~Supported by the German Bundesministerium 
        f\"ur Bildung, Wissenschaft,\\
	\indent Forschung und Technologie.\\
$\ddag$~~~Supported by the Hungarian OTKA fund under contract
          numbers T019181,\\
	  \indent F023259 and T037350.\\
$\P$~~~Also supported by the Hungarian OTKA fund under contract
           number T026178.\\
$\flat$~~~Supported also by the Comisi\'on Interministerial de Ciencia y 
        Tecnolog{\'\i}a.\\
$\sharp$~~~Also supported by CONICET and Universidad Nacional de La Plata,
        CC 67,\\
	\indent 1900 La Plata, Argentina.\\
$\triangle$~~~Supported by the National Natural Science
  Foundation of China.\\
}

\clearpage
\newpage
\noindent{\bf\large Appendix C4: The OPAL Collaboration}\\

{\small\raggedright
\noindent
G.\thinspace Abbiendi$^{  2}$,
C.\thinspace Ainsley$^{  5}$,
P.\thinspace F.\thinspace {\AA}kesson$^{  3,  y}$,
G.\thinspace Alexander$^{ 22}$,
J.\thinspace Allison$^{ 16}$,
P.\thinspace Amaral$^{  9}$, 
G.\thinspace Anagnostou$^{  1}$,
K.\thinspace J.\thinspace Anderson$^{  9}$,
S.\thinspace Asai$^{ 23}$,
D.\thinspace Axen$^{ 27}$
% ...
\\
}
$\phantom{.}$\\
$\phantom{.}$\\
{\small
1~~~School of Physics and Astronomy, University of Birmingham,
Birmingham B15 2TT, UK\\
2~~~Dipartimento di Fisica dell' Universit\`a di Bologna and INFN,
I-40126 Bologna, Italy\\
3~~~Physikalisches Institut, Universit\"at Bonn,
D-53115 Bonn, Germany\\
4~~~Department of Physics, University of California,
Riverside CA 92521, USA\\
5~~~Cavendish Laboratory, Cambridge CB3 0HE, UK\\
6~~~Ottawa-Carleton Institute for Physics,
Department of Physics, Carleton University, Ottawa,\\
\indent Ontario K1S 5B6, Canada\\
8~~~CERN, European Organisation for Nuclear Research,
CH-1211 Geneva 23, Switzerland\\
9~~~Enrico Fermi Institute and Department of Physics,
University of Chicago, Chicago IL 60637, USA\\
10~~Fakult\"at f\"ur Physik, Albert-Ludwigs-Universit\"at 
Freiburg, D-79104 Freiburg, Germany\\
11~~Physikalisches Institut, Universit\"at
Heidelberg, D-69120 Heidelberg, Germany\\
12~~Indiana University, Department of Physics,
Bloomington IN 47405, USA\\
13~~Queen Mary and Westfield College, University of London,
London E1 4NS, UK\\
14~~Technische Hochschule Aachen, III Physikalisches Institut,
Sommerfeldstrasse 26-28,\\
\indent D-52056 Aachen, Germany\\
15~~University College London, London WC1E 6BT, UK\\
16~~Department of Physics, Schuster Laboratory, The University,
Manchester M13 9PL, UK\\
17~~Department of Physics, University of Maryland,
College Park, MD 20742, USA\\
18~~Laboratoire de Physique Nucl\'eaire, Universit\'e de Montr\'eal,
Montr\'eal, Qu\'ebec H3C 3J7, Canada\\
19~~University of Oregon, Department of Physics, Eugene
OR 97403, USA\\
20~~CCLRC Rutherford Appleton Laboratory, Chilton,
Didcot, Oxfordshire OX11 0QX, UK\\
21~~Department of Physics, Technion-Israel Institute of
Technology, Haifa 32000, Israel\\
22~~Department of Physics and Astronomy, Tel Aviv University,
Tel Aviv 69978, Israel\\
23~~International Centre for Elementary Particle Physics and
Department of Physics,\\ 
\indent University of Tokyo, Tokyo 113-0033, and
Kobe University, Kobe 657-8501, Japan\\
24~~Particle Physics Department, Weizmann Institute of Science,
Rehovot 76100, Israel\\
25~~Universit\"at Hamburg/DESY, Institut f\"ur Experimentalphysik, 
Notkestrasse 85,\\
\indent D-22607 Hamburg, Germany\\
26~~University of Victoria, Department of Physics, P O Box 3055,
Victoria BC V8W 3P6, Canada\\
27~~University of British Columbia, Department of Physics,
Vancouver BC V6T 1Z1, Canada\\
28~~University of Alberta,  Department of Physics,
Edmonton AB T6G 2J1, Canada\\
29~~Research Institute for Particle and Nuclear Physics,
H-1525 Budapest, P O  Box 49, Hungary
30~~Institute of Nuclear Research,
H-4001 Debrecen, P O  Box 51, Hungary\\
31~~Ludwig-Maximilians-Universit\"at M\"unchen,
Sektion Physik, Am Coulombwall 1,\\
\indent D-85748 Garching, Germany\\
32~~Max-Planck-Institute f\"ur Physik, F\"ohringer Ring 6,
D-80805 M\"unchen, Germany\\
33~~Yale University, Department of Physics, New Haven, 
CT 06520, USA\\
$\phantom{.}$\\
$\phantom{.}$\\
${  a}$~~~and at TRIUMF, Vancouver, Canada V6T 2A3\\
${  b}$~~~now at SLAC\\
${  c}$~~~and Institute of Nuclear Research, Debrecen, Hungary\\
${  e}$~~~and Department of Experimental Physics, University of Debrecen, Hungary\\
${  f}$~~~and MPI M\"unchen\\
${  g}$~~~and Research Institute for Particle and Nuclear Physics,
Budapest, Hungary\\
${  h}$~~~now at University of Liverpool, Dept of Physics,
Liverpool L69 3BX, U.K.\\
${  i}$~~~now at Dept. Physics, University of Illinois at Urbana-Champaign, 
U.S.A.\\
${  j}$~~~and Manchester University\\
${  k}$~~~now at University of Kansas, Dept of Physics and Astronomy,
Lawrence, KS 66045, U.S.A.\\
${  l}$~~~now at University of Toronto, Dept of Physics, Toronto, Canada \\
${  m}$~~current address Bergische Universit\"at, Wuppertal, Germany\\
${  n}$~~~now at University of Mining and Metallurgy, Cracow, Poland\\
${  o}$~~~now at University of California, San Diego, U.S.A.\\
${  p}$~~~now at The University of Melbourne, Victoria, Australia\\
${  q}$~~~now at IPHE Universit\'e de Lausanne, CH-1015 Lausanne, Switzerland\\
${  r}$~~~now at IEKP Universit\"at Karlsruhe, Germany\\
${  s}$~~~now at University of Antwerpen, Physics Department, B-2610 Antwerpen, Belgium;\\
\indent supported by Interuniversity Attraction Poles Programme -- Belgian
Science Policy\\
${  t}$~~~now at University of Nijmegen, Nijmegen, The Netherlands\\ 
${  u}$~~~and High Energy Accelerator Research Organisation (KEK), Tsukuba,
Ibaraki, Japan\\
${  v}$~~~now at University of Pennsylvania, Philadelphia, Pennsylvania, USA\\
${  w}$~~~now at TRIUMF, Vancouver, Canada\\
${  x}$~~~now at DESY Zeuthen\\
${  y}$~~~now at CERN\\
${  z}$~~~now at DESY\\
${  \dagger}$~~~~Deceased
}

\clearpage

%
%
%
\newpage
%
%
%
\clearpage 
\newpage
\begin{figure}[htb]
\begin{center}
\epsfig{figure=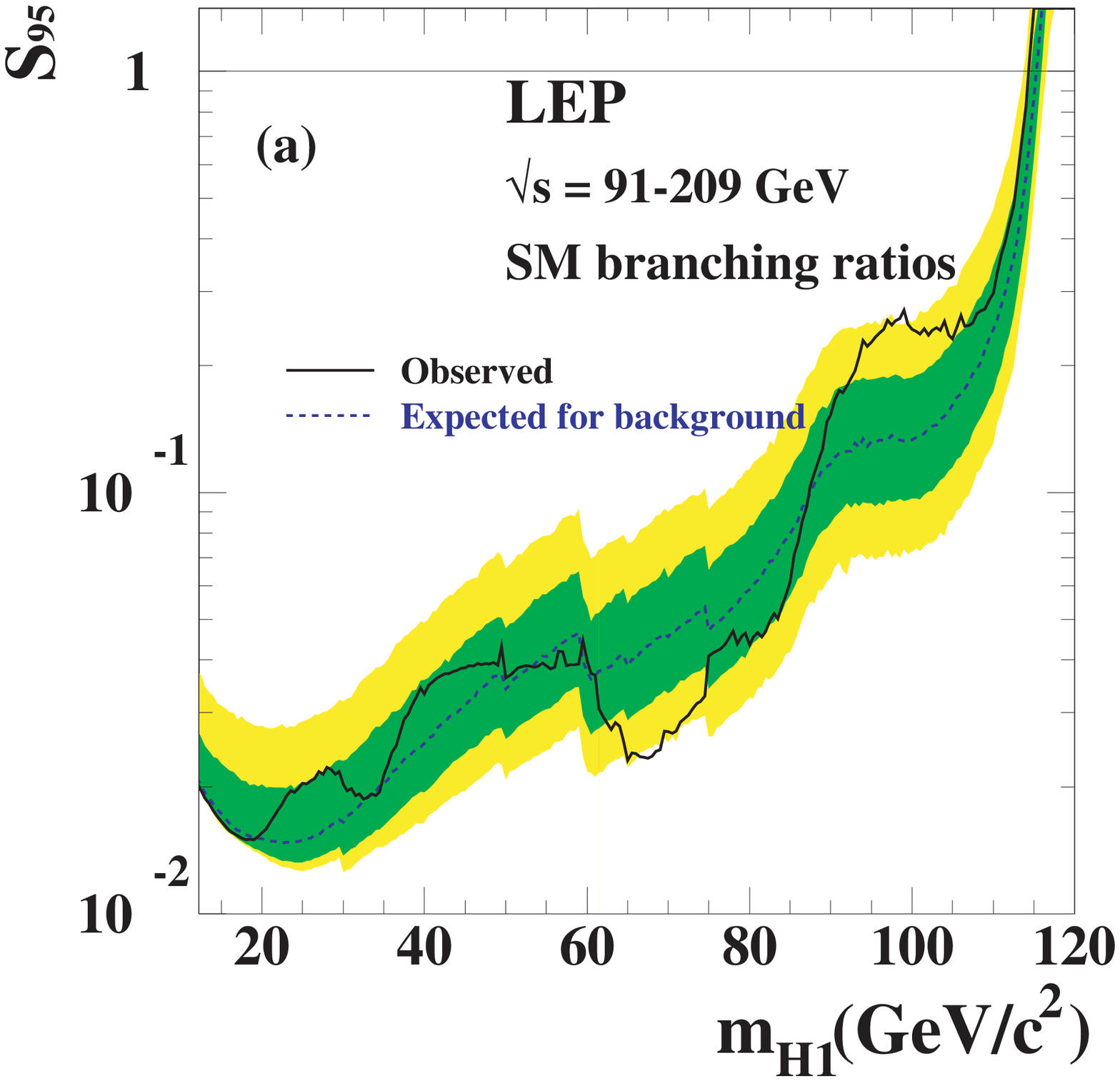,width=0.49\textwidth} \\
\epsfig{figure=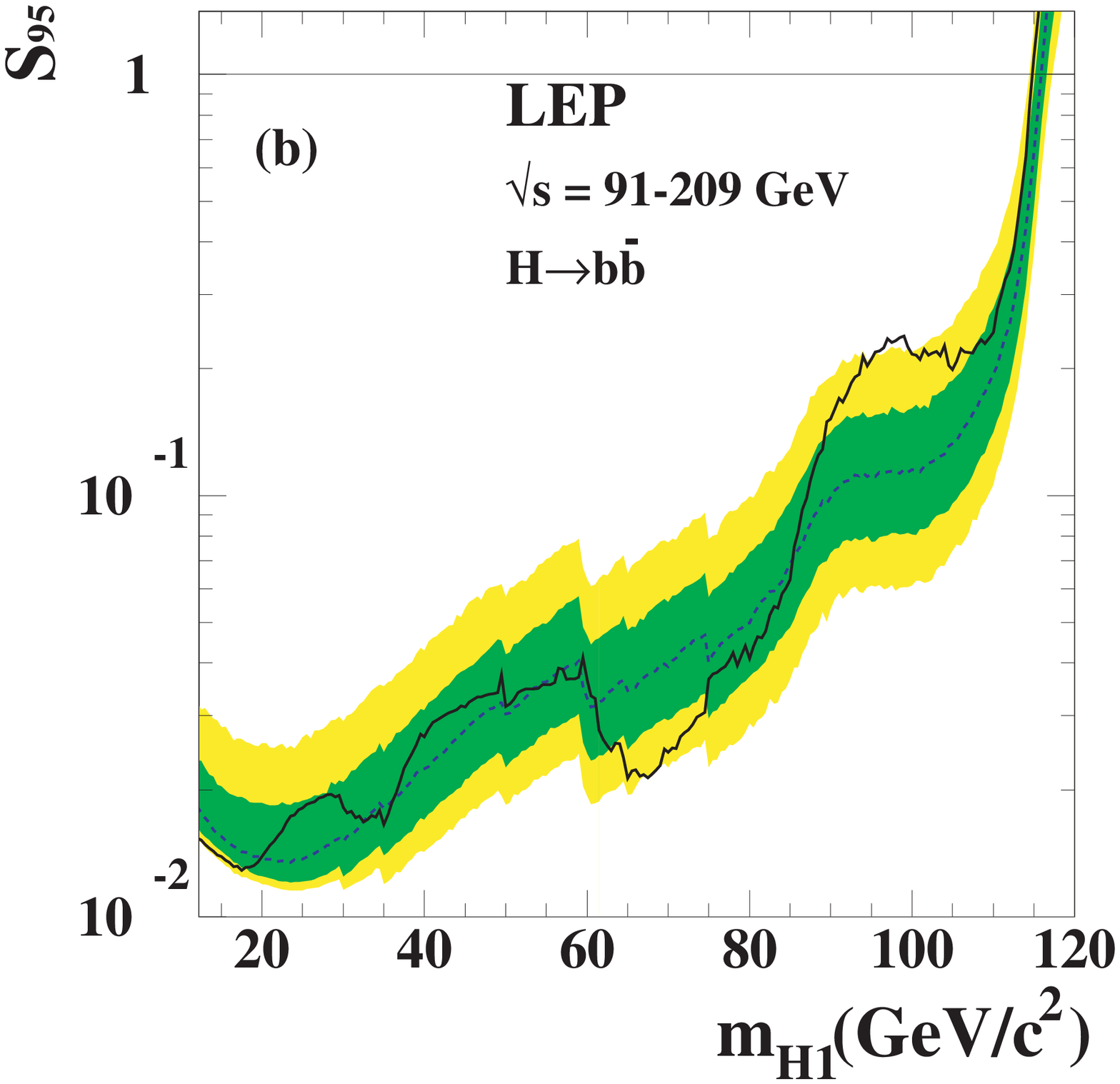,width=0.49\textwidth} 
\epsfig{figure=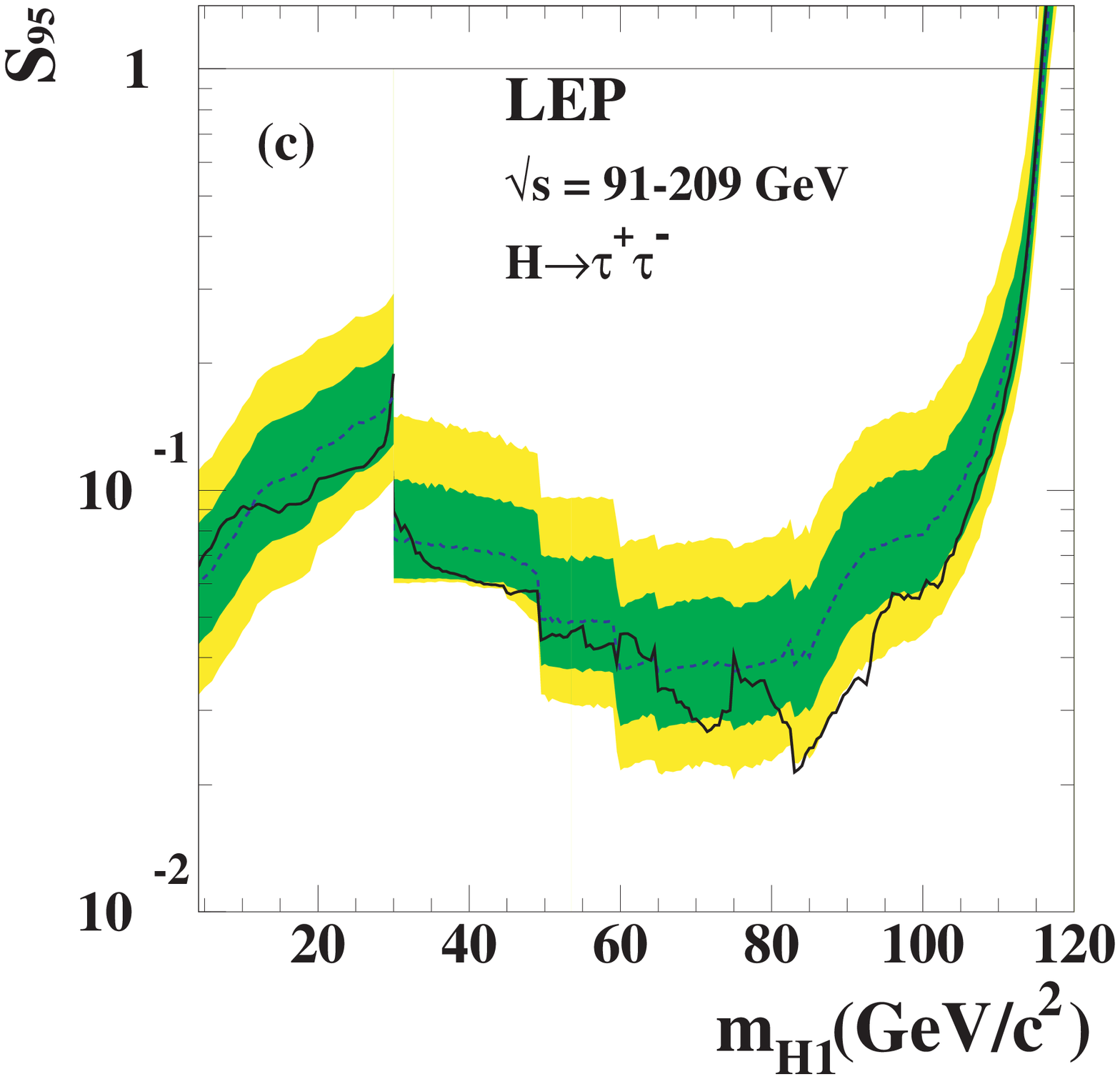,width=0.49\textwidth} 
\end{center}
\caption[]{\sl The 95\% CL upper bounds, $S_{95}$ (see text), 
for various topological cross-sections motivated by the Higgsstrahlung process \ee\ra~\calHa\Z, 
as a function of the Higgs boson mass (the figure is reproduced from Ref.~\cite{lep-sm}).
The full lines represent the observed limits.
The dark (green) and light (yellow) shaded bands around the median
expectations (dashed lines) correspond to the 68\% and 95\% probability bands.
The horizontal lines correspond to the Standard Model cross-sections. 
In part (a) the Higgs boson decay branching ratios are assumed to be those predicted by the Standard Model; 
in part (b) the Higgs boson is assumed to decay exclusively to \bb\ and in part 
(c) exclusively to \tautau. 
\label{sm-xi2}}
\end{figure}
%
\clearpage 
\newpage
%
\clearpage 
\newpage
%
\clearpage 
\newpage
%
\clearpage
\newpage
%
\clearpage
\newpage
%
\clearpage
\newpage
%
\clearpage
\newpage
%
\clearpage
\newpage
%
%
\clearpage
\newpage
%
\clearpage
\newpage
%
\clearpage
\newpage
%
\newpage
%
\newpage
%
%
\clearpage
\newpage
%
\clearpage
\newpage
%
\clearpage
\newpage
%
\clearpage
\newpage
%
\clearpage
\newpage
%
\clearpage
\newpage
%
\clearpage
\newpage
\end{document}